%% file: main.tex
\documentclass[%
 reprint,
superscriptaddress,
 amsmath,amssymb, 
 aps,
prb,
floatfix,
]{revtex4-2}

\usepackage[normalem]{ulem}
\usepackage{xcolor}
\usepackage{graphicx}% Include figure files
\usepackage{dcolumn}% Align table columns on decimal point
\usepackage{bm}% bold math
\usepackage{hyperref}% add hypertext capabilities
\usepackage{braket}
\usepackage[utf8]{inputenc}
\usepackage[T1]{fontenc}
\usepackage{mathrsfs}
\usepackage{longtable}
\usepackage{mathtools}

\newcommand{\fe}[1]{\varphi^{(#1)}}
\newcommand{\fef}[1]{\Tilde{\varphi}^{(#1)}}
\newcommand{\fmp}{a^{(\phi)}}
\newcommand{\fefm}[1]{\Tilde{a}^{(#1)}}
\newcommand{\fem}[1]{a^{(#1)}}
\newcommand{\CFT}[1]{\text{CFT}_{#1}}
\newcommand{\Hp}{\mathcal{H}_{\text{phys}}}

\begin{document}

\preprint{APS/123-QED}

\title{Conformal field theory approach to parton fractional quantum Hall trial wave functions}

\author{Greg J. Henderson}
\affiliation{Rudolf Peierls Centre for Theoretical Physics, Parks Road, Oxford OX1 3PU, United Kingdom}
\author{G. J. Sreejith}
\affiliation{%
 Indian Institute of Science Education and Research, Pune 411008, India
}
\author{Steven H. Simon}
\affiliation{Rudolf Peierls Centre for Theoretical Physics, Parks Road, Oxford OX1 3PU, United Kingdom}

\date{\today}% It is always \today, today,

\begin{abstract}
    We show that all lowest Landau level projected and unprojected chiral parton type fractional quantum Hall ground and edge state trial wave functions, which take the form of products of integer quantum Hall wave functions, can be expressed as conformal field theory (CFT) correlation functions, where we can associate a chiral algebra to each parton state such that the CFT defined by the algebra is the ``smallest'' such CFT that can generate the corresponding ground and edge state trial wave functions (assuming that the corresponding chiral algebra does indeed define a physically ``sensable'' CFT). A field-theoretic generalisation of Laughlin's plasma analogy, known as generalised screening, is formulated for these states. If this holds, along with an additional assumption, we argue that the inner products of edge state trial wave functions, for parton states where the ``densest'' trial wave function is unique, can be expressed as matrix elements of an exponentiated local action operator of the CFT, generalising the result of Dubail et al. [Phys. Rev. B 85, 11531 (2012)], which implies the equality between edge state and entanglement level counting to state counting in the corresponding CFT. We numerically test this result in the case of the unprojected $\nu = 2/5$ composite fermion state and the bosonic $\nu = 1$ $\phi_2^2$ parton state. We discuss how Read's arguments [Phys. Rev. B 79, 045308 (2009)] still apply, implying that conformal blocks of the CFT defined by the corresponding chiral algebra are valid quasi-hole trial wave functions, with the adiabatic braiding statistics given by the monodromy of these functions, assuming the existence of a quasi-particle trapping Hamiltonian. Generalisations of these constructions are discussed, with particular attention given to simple current constructions. It is shown that all chiral composite fermion wave functions can be expressed as CFT correlation functions without explicit symmetrisation or anti-symmetrisation and that the ground, edge, and certain quasi-hole trial wave functions of the $\phi_n^m$ parton states can be expressed as the conformal blocks of the $U(1) \otimes SU(n)_m$ WZW models. Finally, we discuss the relation of the $\phi_2^k$ series with the Read-Rezayi series, where several examples of quasi-hole braiding statistics calculations are given.
\end{abstract}

\maketitle

\tableofcontents

\section{Introduction} \label{Intro}

\subsection{Motivation}
Several decades on, the fractional quantum Hall effect (FQHE) still stands as one of the prime examples of topologically ordered matter \cite{Tsui, prange_quantum_2012, Duncan2017, wen_colloquium_2017, Nayak}. Each plateau in the Hall resistivity corresponds to a different phase of matter, with any pair not being distinguishable by their symmetry properties. As well as their filling fractions $\nu$, each phase of matter can instead be characterized by fractional charge and braiding statistics of their quasi-particles \cite{Laughlin1983, arovas_fractional_1984, Halperin1984}, a ground state degeneracy that only depends on the topology of space they exist within \cite{haldane_many-particle_1985, haldane_periodic_1985, Wen1990a}, and patterns of long-range entanglement \cite{Li2008, Lauchli2010, Papic2011, Sterdyniak2012, Dubail2012a, Rodriguez2012, Levin2006, Kitaev2006, Chen}.

One of the main themes in understanding the FQHE has been the use of zero-parameter trial wave functions. The main idea of this is that, if a trial wave function is \textit{adiabatically connected} to the physical system of interest, then one can predict the universal properties of the physical system by calculating those of the trial wave function. Each trial wave function is then seen as being representative of a particular phase of matter. Given an explicit trial wave function, extracting the universal properties is, however, highly non-trivial.

In the case of Laughlin's trial wave functions at filling factors $\nu = 1/m$, these properties can be calculated by mapping to a one-component plasma that is in a \textit{screening} phase. This is the, now famous, Laughlin plasma analogy. By rather straightforward calculations, this can be used to determine the fractional charge \cite{Laughlin1983} and adiabatic braiding statistics \cite{arovas_fractional_1984} of the quasi-particles, as well as the properties of the edge \cite{wen_theory_1992}.

A broader class of wave functions, to which Laughin's belong, are those that can be written \textit{directly} as conformal blocks of a particular (unitary) \textit{Conformal Field Theory} (CFT). This includes, for example, the Moore-Read \cite{Moore1991} and Read-Rezayi \cite{Read1999} series wave functions. Within the construction, quasi-particle states are generated by the insertion of CFT primary fields within the conformal block. As shown by Read, inner-products between these wave functions can be mapped to correlation functions of a \textit{perturbed} CFT \cite{Read2009}. 
Assuming this perturbed CFT is short-range correlated and there exists a trapping Hamiltonian for which these quasi-particle states are the lowest energy states, Read was then able to show that the adiabatic quasi-particle braiding is given by the \textit{monodromy} of the conformal blocks\footnote{This is up to an area dependant phase factor.} (i.e. the analytic continuation of the conformal block along the quasi-particle paths) (see also Bonderson et al. \cite{bonderson_plasma_2011}). The assumption of short-range correlations in the perturbed CFT has been termed the \textit{generalised screening hypothesis} and can be thought of as a field-theoretic generalization of Laughlin's plasma analogy.

The CFT construction does not only give ground and quasi-particle states. By putting a CFT state that isn't the vacuum on one boundary of the conformal block one can generate \textit{edge} state wave functions, in such a way that this gives a linear map from the CFT Hilbert space to edge state wave functions \cite{Xiao-GangWen1994}. By applying the generalised screening hypothesis Dubail, Read and Rezayi (DRR) were able to show that this linear map preserves the inner product of the two Hilbert spaces (an isometric isomorphism in more technical language) in the thermodynamic limit, with finite size correction being given by the exponentiation of a local action within the CFT \cite{Dubail2012}. Thus, generalised screening then gives rise to a precise bulk-edge correspondence. Using this result for edge-state inner-products DRR were then able to understand the structure of the real-space entanglement spectrum (RSES) as the spectrum of some integral of local operators within the CFT, which is also termed the $\textit{entanglement action}$.

The implications of generalised screening have been directly verified numerically for certain Laughlin, Moore-Read and Read-Rezayi wave functions in various works \cite{Dubail2012,bernevig_screening_2012, Fern2018, wu_braiding_2014, tserkovnyak_monte_2003, baraban_numerical_2009}. Interestingly, in the work of Wu et al. \cite{wu_braiding_2014} the adiabatic braiding of the quasi-particle excitations was computed using matrix product state (MPS) representations of the corresponding wave function, where the MPS representations were directly derived from the CFT construction of these wave functions \cite{Zaletel2012, estienne_matrix_2013, estienne_fractional_2013}. 

In short, if the generalised screening hypothesis holds for a given wave function expressed as a conformal block then the universal or \textit{topological} properties are \textit{manifest}. There are, of course, many wave functions that cannot directly be written as a conformal block.

The phenomenologically successful \textit{composite fermion} (\cite{Jain2007, Jain1989}) trial wave functions are a prominent example. It has previously been shown that they can be expressed as \textit{anti-symmetrized} conformal blocks \cite{Hansson, Suorsa2011a, Kvorning2013, Hansson2017}, however, this anti-symmetrization procedure meant that it was not obvious if some form of a generalised screening hypothesis could be formulated. Furthermore, to the best of the authors' knowledge, the structure of composite fermion edge state trial wave function inner products has not yet been studied from this CFT perspective and it is not clear if results similar to those found for other states \cite{Dubail2012, Fern2018} should hold.

A more general class of trial wave functions, that cannot be \textit{directly} written as conformal blocks and that includes the composite fermion wave functions, are the \textit{parton} trial wave functions \cite{jain_incompressible_1989} which are formed as products of integer quantum Hall wave functions. The parton wave functions have seen renewed interest in recent years as possible candidates in the second Landau level in GaAs heterostructures \cite{balram_parton_2018, balram_fractional_2018, balram_parton_2019, balram_fractional_2020, coimbatore_balram_non-abelian_2021}, in graphene systems \cite{faugno_non-abelian_2020, faugno_unconventional_2021, wu_non-abelian_2017, kim_even_2019, timmel_non-abelian_2023} and even for some states in the lowest Landau level (LLL) \cite{balram_abelian_2021, balram_parton_2021, balram_very-high-energy_2022, dora_nature_2022}. Using a slave particle approach combined with a mean-field approximation, Wen derived an effective low-energy field theory in terms of a \textit{non-abelien} Chern-Simons gauge theory for the symmetric parton states (an integer quantum Hall wave function to an integer power), which strongly indicated these states can host non-Abelian anyons \cite{Wen1991a}. More recently, the non-Abelian statistics of low-energy quasi-particles of some parton states have been directly calculated using an approach based on the Fock-space description of the corresponding wave functions\cite{bandyopadhyay_entangled_2018, ahari_partons_2022}. The non-Abelian nature of these states has been further demonstrated in a recent work by one of the authors, where the real-space entanglement spectra state counting was empirically found to match the state counting of $U(1)\otimes SU(n)_m$ CFT for various cases \cite{anand_real-space_2022}. Such CFTs have conformal blocks that, in principle, can be used to represent non-Abelian anyons \cite{fuchs_braiding_1994, tsuchiya_vertex_1987} and are related to non-Abelian Chern-Simons theory \cite{Witten1989}. To the best of the authors' knowledge, the parton wave functions have not yet been directly related to CFT correlation functions and hence it is not at all obvious if these CFT methods are a viable approach to extract their topological properties.

\subsection{Summary of the paper}
In the following work, we show that all lowest Landau level projected and unprojected chiral parton ground and edge state trial wave functions can be expressed using CFT correlation functions, where to each parton state one can associate a chiral algebra, $\mathcal{A}$, such that the CFT defined by it $\CFT{\mathcal{A}}$ is, in some sense, the ``smallest'' CFT that can be used to generate the ground and edge state trial wave functions of the given parton state, where we are assuming that there indeed exists a CFT with the chiral algebra $\mathcal{A}$. A generalised screening hypothesis is formulated for these CFT constructions. We argue that if generalised screening holds for a given parton state then it's topological properties can be directly related to certain properties of the corresponding chiral algebra.

We first give a review of how these CFT methods have previously been used to understand the topological properties of certain trial wave functions in Sec. \ref{Sec:CFTMthods}. The parton and specifically the composite fermion trial wave functions are reviewed in Sec. \ref{Sec:CFandPartonWF}. 

In Sec. \ref{Sec:integerQHCFT} we show how all integer quantum Hall ground and edge state wave functions can be generated using CFT. This then allows us to show how the parton wave functions can be expressed as CFT correlation functions in Sec. \ref{Sec:CFTPartonBig}. We give two detailed examples of this construction in the case of the $\nu = 2/5$ composite fermion state and the bosonic $\nu = 1$ $\phi_2^2$ parton state. It is shown that all chiral composite fermion states (i.e. those without reverse flux attachment or negative effective magnetic field) can be expressed as CFT correlation functions without explicit symmetrisation or anti-symmetrisation in Sec. \ref{Sec:CFGenCaseCon}. Then in Sec. \ref{Sec:symmetricPartons} it is shown that the symmetric parton states, which take the form of a $\nu = n$ integer quantum Hall ground state raised to the $m^{\text{th}}$ power $\phi_n^m$, can be expressed in terms of the conformal blocks of the $\hat{\mathfrak{u}}(1)\oplus \widehat{\mathfrak{su}}(n)_m$ WZW models. This allows us to give rigorous upper bounds of the state counting of edge state trial wave functions in terms of the state counting of the $\hat{\mathfrak{u}}(1)\oplus \widehat{\mathfrak{su}}(n)_m$ WZW models. 

In Sec. \ref{Sec:PartonGenWFCon} it is shown that if it is known how to generate two wave functions $\Psi_1$ and $\Psi_2$ using CFT with corresponding chiral algebras $\mathcal{A}_1$ and $\mathcal{A}_2$ then the product wave function $\Psi_1\Psi_2$ can also be generated using CFT with the corresponding chiral algebra being a chiral subalgebra of $\mathcal{A}_1 \otimes \mathcal{A}_2$. This is the inductive step that implies that all chiral parton wave functions can be expressed as CFT correlation functions. For each parton state the generation of the trial edge state wave functions can be understood as a linear map from the vacuum representation of the corresponding chiral algebra. This, in general, allows for rigorous upper bounds to be given for the edge state counting in terms of the state counting in the vacuum representation of the corresponding chiral algebra. The general construction for generating trial wave functions with CFT, that these parton states point towards, is discussed in Sec. \ref{Sec:CFTGenStructure}, which allows one to generate trial wave functions using chiral algebras that are generated by multiple fields and their conjugates. This includes a large variety of chiral algebras that can be understood as simple current constructions \cite{schoutens_simple-current_2016}.

A generalised screening hypothesis for these CFT constructions is then given in Sec. \ref{Sec:GenScreening}. Then in Sec. \ref{Sec:CFTInnerProducts} it is argued that if generalised screening holds, along with an additional assumption, that the inner products of edge state trial wave functions, for parton states where the ``densest'' (i.e. lowest angular momentum for a given number of particles) trial wave function is unique, can be expressed as matrix elements of the exponentiation of a CFT operator known as the inner product action which takes the form of a sum of integrals of local operators, thus generalising the result of Dubail et al. \cite{Dubail2012}. The coefficients of the terms in the inner product action scale with the system size in such a way that implies the existence of a linear map from CFT states to edge state trial wave functions that preserves the inner product. Thus, given generalised screening and the additional assumption, the aforementioned upper bounds on the edge state counting will be saturated. We further show in Sec. \ref{Sec:RSES} how the result of Ref. \cite{Dubail2012} can easily be extended to these CFT constructions which is that if generalised screening holds for a given wave function then the real-space entanglement spectrum of the wave function can be expressed as the spectrum of a CFT operator known as the entanglement action which takes the same form as the inner product action, again for the parton states where the ``densest'' trial wave function is unique. This also implies that the entanglement level counting also matches the CFT state counting which gives a partial explanation for some of the observations of Ref. \cite{anand_real-space_2022}.

In Sec. \ref{Sec:numericalTests} we present numerical tests of this edge state inner product result in the cases of the $\nu = 2/5$ composite fermion state and the bosonic $\nu = 1$ $\phi_2^2$ parton state.

Finally, in Sec. \ref{Sec:QPBraiding} it is shown how the quasi-particle trial wave functions of a given state, with corresponding chiral algebra $\mathcal{A}$, can be generated by conformal blocks of $\CFT{\mathcal{A}}$. It is discussed how the arguments of Read \cite{Read2009} still apply in these cases, which implies that if generalised screening holds the adiabatic braiding of the quasi-particle wave functions is given by the monodromy of the corresponding conformal block, assuming that there exists a trapping Hamiltonian for which these quasi-particle trial states have the lowest energy. This then allows one to relate the various anyon types that can be generated for a given state using $\CFT{\mathcal{A}}$ to the irreducible representation of $\mathcal{A}$. In Sec. \ref{Sec:simpleCurrentBraid} for the case when $\mathcal{A}$ can be understood as a simple current construction we show how certain representations of $\mathcal{A}$ can be constructed. In these cases, the corresponding quasi-particle conformal block trial wave functions can be decomposed into a product of a chiral boson conformal block and a conformal block of the simple current CFT. It is then discussed how this implies that for the symmetric parton state, $\phi_n^m$, certain quasi-particle trial wave functions can be expressed in terms of products of chiral boson conformal blocks and conformal blocks of the $\widehat{\mathfrak{su}}(n)_m$ WZW models. The $\phi_2^k$ series is then considered in detail in Sec. \ref{Sec:su2Series}, where the relation to the Read-Rezayi series is discussed and several examples of computing quasi-particle adiabatic statistics are given.

\section{Background}
For the relevant background we  first cover, very briefly, rational conformal field theory in Sec. \ref{Sec:rationalCFT}. Then in Sec. \ref{Sec:CFTMthods} we discuss how rational conformal field theory has already been used to generate certain trial wave functions such as the Laughlin and Moore-Read wave functions, and how quasi-particle braiding statistics and edge state inner-products can be understood assuming the generalised screening hypothesis. The composite fermion and parton trial wave functions will then be briefly discussed in Sec. \ref{Sec:CFandPartonWF}. Finally, in Sec. \ref{Sec:integerQHCFT} we show how integer quantum Hall ground state and edge excitation wave functions can be generated using CFT methods, which will be used later to construct parton and composite fermion wave functions using CFT. 

Throughout this work, we use the following notation. A generic fractional quantum Hall wave function of $N$ particles, in a \textit{finite} number of Landau levels, will take the form $\Psi(z_1, \bar{z}_1, z_2, \bar{z}_2, \dots , z_N, \bar{z}_N,)$, where $\Psi$ is a \textit{finite} degree polynomial in its arguments and $z_j = x_j + iy_j$ with $(x_j,y_j)$ being the usual two-dimensional Cartesian coordinates of the position of particle $j$. For any such wave function, the corresponding ket vector will be denoted by $\ket{\Psi}\rangle$. The Gaussian factor $\exp\big ( - \sum_{i=1}^N |z|^2/(4l_B^2) \big )$, with $l_B$ being the magnetic length, is moved into the integration measure of inner products so that the inner product between two wave functions $\Psi_1$ and $\Psi_2$ is given by,
\begin{equation}
    \begin{split}
        \langle \braket{\Psi_2 | \Psi_1} \rangle = \frac{1}{N!} \int \prod_{i=1}^N D^2z_i & \overline{ \Psi_2(z_1, \bar{z}_1, z_2, \bar{z}_2, \dots , z_N, \bar{z}_N,) } \\
        \times &\Psi_1(z_1, \bar{z}_1, z_2, \bar{z}_2, \dots , z_N, \bar{z}_N,)
    \end{split}
\end{equation}
where $D^2z = d^2 z \exp ( - |z|^2/(2l_B^2) )$. One should note the $1/(N!)$ factor which is used here. We also work in units such that $l_B = 1$.

In second quantized notation, the wave function takes the form,
\begin{equation}
    \begin{split}
        \ket{\Psi}\rangle = \frac{1}{N!} \int & \prod_{i=1}^N d^2z_i e^{-|z_i|^2/4} \\
        &\times \Psi(z_1, \bar{z}_1, z_2, \bar{z}_2, \dots , z_N, \bar{z}_N,) \\
        &\times c^\dagger(z_1, \bar{z}_1) c^\dagger(z_2, \bar{z}_2) \dots c^\dagger(z_N, \bar{z}_N) \ket{0}\rangle 
    \end{split}
\end{equation}
where $\ket{0}\rangle$ is the vacuum state and $c^\dagger(z,\bar{z})$ are the usual real space creation operators with $\{ c(z_1, \bar{z}_1), c^\dagger(z_2, \bar{z}_2) \} = \delta^2(z_1 - z_2, \bar{z_1} - \bar{z}_2)$ for fermions and $[ c(z_1, \bar{z_1}), c^\dagger(z_2, \bar{z}_2) ] = \delta^2(z_1 - z_2, \bar{z}_1 - \bar{z}_2)$ for bosons.

One should also note further shorthand notations used in this paper. If two CFT fields, $\phi(z)$ and $\psi(z)$, have an operator product expansion (OPE) with no singular terms then we denote this by $\phi(z)\psi(w) \sim 0$. If a $\sim$ symbol is not used when expressing any OPE then we are referring to the full OPE which includes both the singular and non-singular terms.

\subsection{Rational Conformal Field Theory} \label{Sec:rationalCFT}
We now give a very brief overview of rational CFT (RCFT), which will lay the foundation for this work. For a fuller account of CFT in general we refer the reader to Refs. \cite{DiFrancesco1997, Ginsparg1988, Cardy2008}.

Following Moore and Seiberg \cite{Moore1988, Moore1989}, the \textit{chiral algebra}, $\mathcal{A}$, of a CFT is the set of all holomorphic fields of the theory (i.e. all the fields which have purely holomorphic dependence on position within any correlation function). In the mathematics literature these algebras correspond to \textit{vertex operator algebras} \cite{huang2005vertex, frenkel1989vertex, kac1994vertex}. In any CFT, $\mathcal{A}$ must at least contain both the identity field, $\mathbf{1}(z)$, and the energy-momentum tensor $T(z)$. We define the modes of a holomorphic field, $\phi(z)$, with scaling dimension $h$, by $\phi_n = \oint \frac{dz}{2\pi i} z^{n + h - 1} \phi(z)$, where $n + h \in \mathbb{Z}$. As the operator product expansion (OPE) between any two holomorphic fields must only contain other holomorphic fields, the commutation, or anti-commutation, relations between the modes of these fields must be expressable as a sum of modes of fields belonging to $\mathcal{A}$. The reader should note that if the fields have integral or half-integral conformal dimensions then one should consider commutation or anti-commutation relations respectively so as to be consistent with the spin-statistics theorem.

A CFT is \textit{rational} iff its Hilbert space can be decomposed as $\mathcal{H} = \oplus_{ij} \mathcal{H}_i \otimes \overline{\mathcal{H}}_j$, where the direct sum of vector spaces is \textit{finite} and, $\mathcal{H}_i$ and $\overline{\mathcal{H}}_j$ are irreducible representations of the chiral algebra and its anti-chiral copy respectively. In other words, by state operator correspondence, the fields of the theory can be organized into a finite number of primary fields, with respect to $\mathcal{A}$, and their descendants. For the remainder of this work, we  only consider the case of a \textit{diagonal unitary} RCFT where the decomposition of the Hilbert space is of the form $\mathcal{H} = \oplus_j \mathcal{H}_j \otimes \overline{\mathcal{H}}_j$ and each $\mathcal{H}_j$ is a unitary representation of the chiral algebra. The WZW models are classic examples of rational CFTs, where the currents $J^i(z)$ generate the chiral algebra of the theory.

The information on what primaries appear in the OPE of two primary fields, $\phi_i$ and $\phi_j$, is encoded in the \textit{fusion rules}, $\phi_i \times \phi_j = \sum_{k} N^k_{ij} \phi_k$. The $N_{ij}^k$ are the number of ``distinct ways'' field $\phi_i$ and $\phi_j$ can fuse to field $\phi_k$. If a field $J$ can only have one result when fusing with any other field (i.e. $J\times \phi = \phi'$) then $J$ is said to be a \textit{simple current}. 

As usual, the correlation function of a set of primary fields can be expressed as,
\begin{equation} \label{Eq:CFTCorr}
    \braket{\phi_{j_1}(w_1, \bar{w}_1)  \phi_{j_2}(w_2, \bar{w_2}) \dots} = \sum_a | \mathcal{F}_a(w_1, w_2, \dots ) |^2
\end{equation}
where $\mathcal{F}_a(w_1, w_2, \dots)$ is a conformal block. Every possible way the fields can fuse to the identity are in one-to-one correspondence with the $\mathcal{F}_a$'s. For example, in the Ising model, we have the fields $\sigma$ and $\psi$ with the fusion rules $\sigma\times \sigma = \mathbf{1} + \psi$, $\psi\times \sigma = \sigma$ and $\psi\times \psi = \mathbf{1}$. Thus, a correlation function of four $\sigma$'s can have two conformal blocks corresponding to the fusion channels $(\sigma ( \sigma ( \sigma \sigma))) \rightarrow (\sigma ( \sigma \psi )) \rightarrow (\sigma \sigma ) \rightarrow \mathbf{1}$ and $(\sigma ( \sigma ( \sigma \sigma))) \rightarrow (\sigma (\sigma \mathbf{1})) \rightarrow (\sigma \sigma) \rightarrow \mathbf{1}$.

In general, the conformal blocks are not single-valued. Under a monodromy transformation, where the paths of the $z_i$ form a \textit{braid}, the conformal blocks transform as,
\begin{equation}
    \mathcal{F}_a(w_1, w_2, \dots) \rightarrow \sum_b B_{ba}\mathcal{F}_b(w_1, w_2, \dots)
\end{equation}
where $B_{ba}$ only depends on the primary fields that are in the correlation function and the \textit{topology} of the braid. This follows from the fact that the correlation function of Eq. \ref{Eq:CFTCorr} must be single-valued. Thus, the action of these monodromy transformations on the conformal blocks gives a representation of the \textit{braid group}.

Let $V_i(z)$ be a set of fields that belong to the chiral algebra, $V_i(z) \in \mathcal{A}$. In general, a correlation function involving some primary fields $\phi_j(w, \bar{w})$ and some $V_i(z)$ takes the form,
\begin{equation} \label{Eq:CFTCorrMod}
    \begin{split}
        &\braket{\phi_{j_1}(w_1, \bar{w}_1)  \phi_{j_2}(w_2, \bar{w_2}) \dots \phi_{j_m}(w_m,\bar{w}_m) V_1(z_1) V_2(z_2) \dots } \\
        =& \sum_a  \mathcal{F}_a(w_1, w_2, \dots w_m; \mathbf{z}) \\
        &\times \overline{\mathcal{F}_a(w_1, w_2, \dots, w_m)}
    \end{split}
\end{equation}
where we refer to $\mathcal{F}_a(w_1, w_2, \dots w_m; \mathbf{z})$ as the modified conformal blocks, $\mathbf{z} = z_1,z_2,\dots $ and $\mathcal{F}_a(w_1, w_2, \dots, w_m)$ are the same conformal blocks appearing in Eq. \ref{Eq:CFTCorr}. We can then see that the $\mathcal{F}_a(w_1, w_2, \dots w_m; \mathbf{z})$ are in one-to-one correspondence with $\mathcal{F}_a(w_1, w_2, \dots w_m)$. Furthermore, under monodromy transformations involving the $w$'s the $\mathcal{F}_a(w_1, w_2, \dots w_m; z_1, z_2, \dots )$ transform the same way as the conformal blocks $\mathcal{F}_a(w_1, w_2, \dots, w_m)$, and $\mathcal{F}_a(w_1, w_2, \dots, w_m; \mathbf{z})$ are single valued in the $z$'s. In other words, the presence of the fields of $\mathcal{A}$ does not affect the ``topological'' properties of the conformal blocks.

\subsection{CFT approach to FQH trial wave functions and the generalised screening hypothesis: a review} \label{Sec:CFTMthods}
As given in previous works, \cite{Moore1991, Read1999, Read2009, Dubail2012a, Fern2018, Fern2018a, ino1997chiral}, many FQHE trial wave functions can be expressed as a CFT correlation function of an operator $\Omega(z)$. In each case, the CFT in question is that defined by the chiral algebra $\mathcal{A}$ generated by OPEs of $\Omega(z)$ and it's conjugate $\Omega^\dagger(z)$, $\CFT{\mathcal{A}}$. Furthermore, each $\CFT{\mathcal{A}}$ can be \textit{represented} by fields in $\CFT{U(1)}\otimes \CFT{\chi}$, where $\CFT{U(1)}$ is a chiral boson CFT and $\CFT{\chi}$ is a CFT referred to as the statistics sector. With this representation $\Omega(z)$ can be expressed as,
\begin{equation} \label{Eq:simpleOmegaDefinition}
    \Omega(z) = :e^{i\varphi (z)/\sqrt{\nu}}: \chi(z)
\end{equation}
where $\varphi(z)$ is the chiral boson field, $:*:$ denotes normal ordering, $\chi(z)$ is a primary field of $\CFT{\chi}$, and $\nu$ is the filling fraction of the wave function that will be generated. We would like to emphasize to the reader that $\CFT{\mathcal{A}}$ and $\CFT{U(1)}\otimes\CFT{\chi}$ are not, in general, equivalent. $\CFT{\mathcal{A}}$ can be conformally embedded in $\CFT{U(1)}\otimes\CFT{\chi}$ in that their energy-momentum tensors are equivalent and all primaries of $\CFT{\mathcal{A}}$ correspond to primaries of $\CFT{U(1)}\otimes\CFT{\chi}$. However, not all primaries of $\CFT{U(1)}\otimes\CFT{\chi}$ correspond to primaries of $\CFT{\mathcal{A}}$.

We now summarize how, given such a CFT, the ground, edge and quasi-particle state ansatzes can be generated. The generalised screening hypothesis is then detailed along with how the topological properties of the trial wave function can be extracted from the properties of the CFT if this holds. Constructions involving explicit anti-symmetrization or symmetrization of correlation functions will not be discussed here.

\subsubsection{CFT construction of ground, edge and quasi-particle state ansatzes}
Starting from the $\Omega(z)$ defined in Eq. \ref{Eq:simpleOmegaDefinition}, we can write the ground state wave function of $N$ particles as,
\begin{equation}
    \Psi_{\bra{0}} (\mathbf{z}) = \bra{0} C(N) \prod_{i=1}^N \Omega(z_i) \ket{0}
\end{equation}
where $\mathbf{z}$ denotes $z_1, z_2, \dots, z_N$ and $\bra{0}C(N) = \lim_{z \rightarrow \infty} z^{-N^2/\nu} \bra{0}e^{-iN\varphi(z)/\sqrt{\nu}}$ is a background charge that is used so that the correlator has a net zero $U(1)$ charge (which is required for it to be non-zero). In each case, $\bra{0}C(N)$ can also be expressed as the Hermitian conjugate of a product of modes of $\Omega(z)$ applied on $\bra{0}$. Strictly speaking, this correlation function is computed within the Hilbert space of the vacuum representation of $\mathcal{A}$, $\mathcal{H}_0$.

As detailed in Ref. \cite{Dubail2012}, we can generate edge-state excitations by replacing $\bra{0}$ in the correlation function with another state $\bra{v} \in \mathcal{H}_0^*$, where the resulting state is given by,
\begin{equation}
    \Psi_{\bra{v}}(\mathbf{z}) = \bra{v} C(N) \prod_{i=1}^N \Omega(z_i) \ket{0}
\end{equation}

Clearly, this will be non-zero only if $\bra{v}$ has zero $U(1)$ charge. We can, however, modify this construction so that $\bra{v}$ with non-zero $U(1)$ charge will generate a wave function of a different number of particles. We first write the wave function in second-quantized notation,
\begin{equation} \label{Eq:secondQuantized}
    \begin{split}
        \ket{\Psi_{\bra{v}}}\rangle =& \frac{1}{N!} \int \prod_{i=1}^N d^2z_i e^{-|z_i|^2/4} \bra{v} C(N) \prod_{i=1}^N \Omega(z_i) \ket{0} \\ 
        &\times\prod_{i = 1}^N c^\dagger(z_i, \bar{z}_i)\ket{0}\rangle
    \end{split}
\end{equation}
where $c^\dagger(z,\bar{z})$ are the creation operators and $\ket{0}\rangle$ is vacuum state with no particles. The modified construction is defined by expressing Eq. \ref{Eq:secondQuantized} as,
\begin{equation} \label{Eq:edgeMapDefinition}
    \ket{\Psi_{\bra{v}}}\rangle \equiv \bra{v} C(N) e^{\int d^2z e^{-|z|^2/4} \Omega(z)\otimes c^\dagger(z,\bar{z}) } \ket{0}\otimes\ket{0}\rangle
\end{equation}
As the CFT correlation function must have net zero $U(1)$ charge the only term that contributes from the exponential is one which has the opposite $U(1)$ charge as $\bra{v} C(N)$. This construction allows us to generate edge excitations with a variable number of particles. 

Roughly speaking, states with quasi-particles at positions $w_1, w_2, \dots$ can be generated by inserting primary fields at the quasi-particle positions into the correlation function,
\begin{equation}
    \Psi(w_1, w_2, \dots; \mathbf{z}) \sim \bra{0} C(N) \phi_1(w_1) \phi_2(w_2) \dots \prod_{i=1}^N \Omega(z_i) \ket{0}
\end{equation}
In general, this expression is only well-defined when the way in which the primaries $\phi_i$ fuse to the identity is specified, with the resulting expression being given by a modified conformal block (see Eq. \ref{Eq:CFTCorrMod}),
\begin{equation}
    \Psi_a(w_1, w_2, \dots; \mathbf{z}) = \mathcal{F}_a(w_1, w_2, \dots; \mathbf{z})
\end{equation}
where $a$ labels the ways the primaries fuse to the identity.

These quasi-particle states are only valid wave functions if there are no singularities with respect to the particle positions $z_i$. In each case, we can define the primaries for each representation of $\mathcal{A}$, $\mathcal{H}_j$, such that the primaries have a completely regular OPE with $\Omega(z)$, $\Omega(z)\phi_j(w) \sim 0$. These primary fields will then generate valid quasi-particle wave functions. 

\subsubsection{The generalised screening hypothesis}
In Laughlin's plasma analogy, one maps the inner products of wave functions to expectation values of observables of a one-component plasma. A field-theoretic generalization of this involves mapping wave function inner products to correlation functions of a field theory that is a ``perturbation'' of $\CFT{\mathcal{A}}$. To express inner products in terms of just the CFT we first need to be able to generate the complex conjugate of these wave functions. To this end, we denote the anti-chiral copy of $\Omega(z)$ as $\overline{\Omega}(\bar{z})$. Complex conjugate wave functions can then be expressed with correlation functions within $\overline{\mathcal{H}}_0$, which is the anti-chiral copy of the vacuum representation of $\mathcal{A}$. 

The squared norm of the ground state trial wave function can then be expressed as,
\begin{equation} \label{Eq:partitionFunctionBackgroundSection}
    \begin{split} 
        Z_N \equiv& \langle \braket{\Psi_{\bra{0}} | \Psi_{\bra{0}}} \rangle \\
        =& \frac{1}{N!} \int D^2\mathbf{z} |\Psi_{\bra{0}}(\mathbf{z})|^2 \\
        =& \frac{1}{N!} \int D^2\mathbf{z} \braket{ \overline{C}(N) \prod_{i=1}^N \overline{\Omega}(\bar{z_i}) C(N) \prod_{i = 1}^N \Omega(z_i) } \\
        =& \braket{ \overline{C}(N) C(N) e^{\int D^2z \overline{\Omega}(\bar{z}) \Omega(z) } }
    \end{split}
\end{equation}
One should note that when $\Omega(z)$ has half integral conformal dimension then one can pick up certain minus signs in rearranging the $\Omega$'s as we should enforce that $\Omega$ anti-cummutes with $\overline{\Omega}$ to be consistent with the spin statistics theorem. We suppress such minus signs for eas of exposition. Once again, as the correlator must have no net $U(1)$ charge the background charges ensure that only one term contributes from the exponential. We have thus expressed the ground state norm as the partition function of a perturbed CFT. As $\bra{v}$ can be expressed as modes of local field acting on $\bra{0}$, edge-state inner products can be mapped to correlation functions of this field theory. Furthermore, inner products of quasi-particle states can be mapped to correlation functions involving the primary fields $\phi_i$. We denote correlation functions of this field theory by $\braket{\dots}_* \equiv \braket{ \overline{C}(N) C(N) \mathcal{R} \dots e^{\int D^2z \overline{\Omega}(\bar{z}) \Omega(z) } }/Z_N$, where $\mathcal{R}$ indicates radial ordering.

\begin{figure}
    \centering
    \includegraphics[scale=0.24]{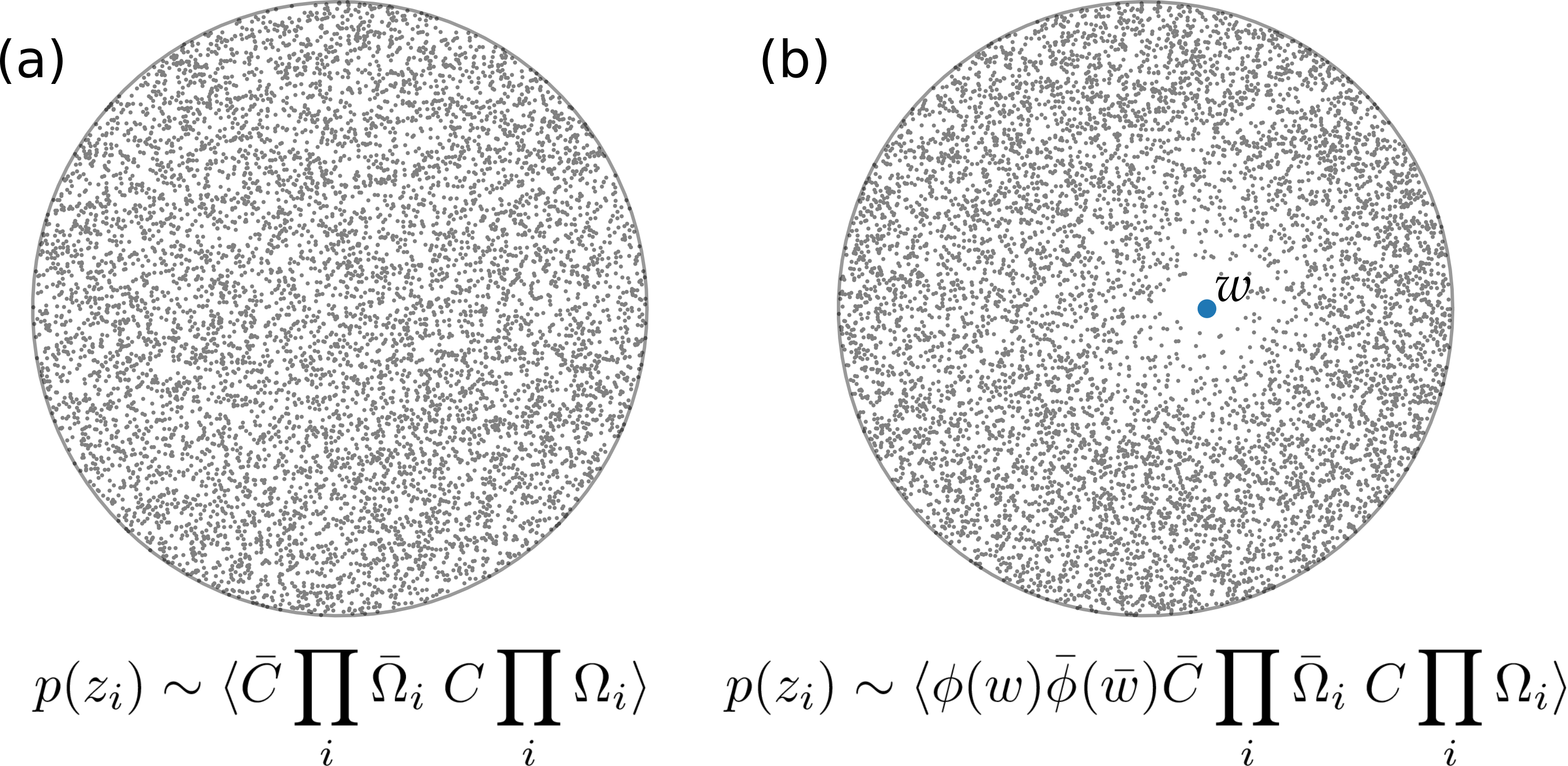}
    \caption{Typical configurations of the $\overline{\Omega}\Omega$ insertions that appear in the partition functions of these ``generalised plasmas'' for (a) the ground state wave function and (b) with the insertion of a quasi-hole operator in the case of a Laughlin wave function. }
    \label{fig:omegaInsertion}
\end{figure}

In the large $N$ limit the partition function of Eq. \ref{Eq:partitionFunctionBackgroundSection} will be dominated by configurations of the $N$ $\overline{\Omega}(\bar{z}) \Omega(z)$ fields insertions where the density of these insertions, on length scales much larger than the magnetic length, is given by one particular $\rho(z,\bar{z})$. As each configuration of the $\overline{\Omega}(\bar{z}) \Omega(z)$ corresponds to a configuration of the actual $N$ particles, with the contribution to the partition function being proportional to the probability density of that configuration, $\rho(z,\bar{z})$ is the density of particles in the ground state trial wave function $\Psi_{\bra{0}}(\mathbf{z})$. As pointed out in Ref. \cite{Dubail2012}, $\rho(z,\bar{z})$ can be computed from a saddle-point approximation, with the density being entirely determined by the $U(1)$ sector. The density profile one finds is that of a droplet of radius $R = \sqrt{\frac{2N}{\nu}}$ with a uniform density of $\frac{\nu}{2\pi l_B^2}$ inside the droplet, precisely what one would expect for a fractional quantum Hall ground state wave function. See Fig. \ref{fig:omegaInsertion} (a) for an example of a typical configuration of the $\overline{\Omega}(\bar{z}) \Omega(z)$ insertions that can appear in the partition function in the case of a Laughlin wave function.

The \textit{generalised screening hypothesis} is that under renormalization group transformations (RG), the field theory flows to a massive infra-red fixed point inside the droplet and flows back to $\CFT{\mathcal{A}}$ outside the droplet. Flow towards a massive infra-red fixed point, inside the droplet, implies that there must only be short-range correlations inside the droplet (i.e. connected correlation functions decay exponentially).

We now wish to point out in what sense \textit{screening} is occurring. Consider now some disk $D$ inside the droplet, with the distance from any point in the disk to the edge of the droplet being much larger than the magnetic length. Within $\CFT{\mathcal{A}}$ we have the OPE, $i\partial\varphi(z)\Omega(w) \sim \sqrt{\nu}^{-1}\Omega(w)/(z-w)$. We must then have,
\begin{equation}
    \begin{split}
        &\frac{\sqrt{\nu}}{2}\times \bigg \langle \oint_{\partial D} \frac{dz}{2\pi i} i\partial\varphi(z) + \oint_{\partial D} \frac{d\bar{z}}{2\pi i} i\overline{\partial}\overline{\varphi}(\bar{z}) \bigg \rangle_* \\
        &= ( \text{Average number of particles in } D) \\
        &= \frac{\nu A_D}{2\pi l_B^2}
    \end{split} 
\end{equation}
where $A_D$ and $\partial D$ is the area and boundary of $D$ respectively. Let $\braket{\phi(w)\bar{\phi}(\bar{w}) \dots }_*$ be a correlation involving the fields $\phi(w)\bar{\phi}(\bar{w})$ and some other fields, where $w \in D$ and all the other field insertions are outside $D$ at distances away from $\partial D$ much larger than the magnetic length. This correlation function could represent, for example, the norm squared of some quasi-particle state. Just as for the partition function, the dominant contributions to this correlation function will be from $\overline{\Omega}(\bar{z}) \Omega(z)$ insertions with a particular density profile, where the number of these insertions in $D$ given by this density profile will be denoted by $N_D$. Let the $U(1)$ charge of $\phi(w)$ be written as $Q_{U(1)}$. We then have,
\begin{equation} \label{Eq:CFTCharge}
    \begin{split}
        &\frac{\sqrt{\nu}}{2} \bigg \langle \bigg ( \oint_{\partial D} \frac{dz}{2\pi i} i\partial\varphi(z) + \oint_{\partial D} \frac{d\bar{z}}{2\pi i} i\overline{\partial}\overline{\varphi}(\bar{z}) \bigg ) \\
        &\times\phi(w)\bar{\phi}(\bar{w}) \dots \bigg \rangle_* \\
        &= ( \sqrt{\nu}Q_{U(1)} + N_D ) \braket{ \phi(w)\bar{\phi}(\bar{w}) \dots }_* \\
    \end{split} 
\end{equation}
Provided the distance from $w$ to $\partial D$ is much larger than the magnetic length then short-range correlation within the droplet implies,
\begin{equation} \label{Eq:shortRangeCharge}
    \begin{split}
        \frac{\sqrt{\nu}}{2}& \bigg \langle \bigg ( \oint_{\partial D} \frac{dz}{2\pi i} i\partial\varphi(z) + \oint_{\partial D} \frac{d\bar{z}}{2\pi i} i\overline{\partial}\overline{\varphi}(\bar{z}) \bigg ) \\
        \times&\phi(w)\bar{\phi}(\bar{w}) \dots  \ \bigg \rangle_* \\
        =& \frac{\sqrt{\nu}}{2} \bigg \langle \oint_{\partial D} \frac{dz}{2\pi i} i\partial\varphi(z) + \oint_{\partial D} \frac{d\bar{z}}{2\pi i} i\overline{\partial}\overline{\varphi}(\bar{z}) \bigg \rangle_* \\
        &\times\braket{ \phi(w)\bar{\phi}(\bar{w}) \dots }_* \\
        =& \frac{\nu A_D}{2\pi l_B^2} \braket{ \phi(w)\bar{\phi}(\bar{w}) \dots }_*
    \end{split} 
\end{equation}
Equating Eq. \ref{Eq:CFTCharge} with Eq. \ref{Eq:shortRangeCharge} we find,
\begin{equation} \label{Ea:quasiparticleCharge}
    N_D = \frac{\nu A_D}{2\pi l_B^2} - \sqrt{\nu}Q_{U(1)}
\end{equation}
In words, the dominant configurations of the $\overline{\Omega}(\bar{z}) \Omega(z)$ insertions that contribute to the correlation function have $\sqrt{\nu}Q_{U(1)}$ fewer insertions in $D$ compared with the number in $D$ in the partition function. See Fig. \ref{fig:omegaInsertion} (b) for an example of these typical $\overline{\Omega}(\bar{z}) \Omega(z)$ insertions that appear in the case of a Laughlin quasi-hole. Thus, the $\overline{\Omega}(\bar{z}) \Omega(z)$ insertions behave analogously to a screened plasma, where field insertions such as $\phi(w)\overline{\phi}(\bar{w})$ can be thought of as fixed charges with an electric charge given by the $U(1)$ charge of the field. 

In the case where the correlation function $\braket{ \phi(w)\bar{\phi}(\bar{w}) \dots }_*$ corresponds to the norm squared of a quasi-particle wave function, the density of the configurations of $\overline{\Omega}(\bar{z}) \Omega(z)$ insertions that are the dominant contributions to the correlation function, will be equal to the particle density in the quasi-particle wave function. Therefore, the physical charge of the quasi-particle generated by $\phi(w)$ is simply given by $-\sqrt{\nu}Q_{U(1)}$ (i.e. it is completely determined by the $U(1)$ sector).

\subsubsection{Edge state inner-products and the RSES}
In Ref. \cite{Dubail2012}, by slightly modifying the edge state mapping of Eq. \ref{Eq:edgeMapDefinition} to,
\begin{equation}
    \ket{\Psi_{\bra{v}}}\rangle \equiv \bra{v} C(N) e^{\lambda_N^{1/2}\int d^2z e^{-|z|^2/4} \Omega(z)\otimes c^\dagger(z,\bar{z}) } \ket{0}\otimes\ket{0}\rangle
\end{equation}
where $\lambda_N$ is a real number adjusted so that $ \lim_{N \rightarrow \infty} \langle \braket{ \Psi_{\bra{0}(\Omega_{-h})^\dagger} | \Psi_{\bra{0}(\Omega_{-h})^\dagger}} \rangle/Z_N \sim R^{2h}$, it was shown that the generalised screening hypothesis implied the edge-state inner products could be expressed as,
\begin{equation} \label{Eq:edgeInnerProdSimple}
    \langle \braket{ \Psi_{\bra{w}} | \Psi_{\bra{v}}} \rangle/Z_N = \bra{v}R^{2L_0} e^S \ket{w}
\end{equation}
where $L_0$ is the zeroth Virasoro mode, and $S$ is an operator referred to as the inner product action that takes the form,
\begin{equation}
    S = \sum_j \frac{\alpha_j}{\sqrt{N}^{h_j - 1}} \oint_{|z| = 1} \frac{dz}{2\pi i} z^{h_j - 1} \phi_j(z)
\end{equation}
where $\alpha_j$ are constants and $\phi_j(z)$ fields belonging to $\mathcal{A}$ with corresponding scaling dimension $h_j$ such that $h_j > 1$. 

From this, it could then be shown that the RSES is given by the spectrum of an operator $S_{ES}$ called the entanglement action that takes the form,
\begin{equation}
    S = \sum_j \frac{\beta_j}{\sqrt{N}^{h_j - 1}} \oint_{|z| = 1} \frac{dz}{2\pi i} z^{h_j - 1} \phi_j(z)
\end{equation}
That is, it takes the same form as an inner-product action. This allows one to understand the structure of the RSES at large but finite system sizes.

In Sec. \ref{Sec:GenScreening} we show how this result can be generalised to composite fermion and parton states using the same methods as DRR.

\subsubsection{Adiabatic quasi-particle braiding statistics}
Consider now a state, $\ket{\Psi(\mathbf{w})}\rangle$, with quasi-particles located at positions $\mathbf{w} = (w_1, w_2, \dots)$ and some path $\mathbf{w}(\tau)$, such that $\mathbf{w}(0) = \mathbf{w}(1)$ and the distance between any $w_i(\tau)$ and $w_j(\tau)$, for $i \neq j$, is much larger than the magnetic length. We assume that there exists a Hamiltonian $\hat{H}(\mathbf{w})$ for which $\ket{\Psi(\mathbf{w})}\rangle$ is a ground state and which is gapped in the bulk of the system with the quasi-particles not coupling the edge of the system (see Ref. \cite{Read2009} for further details). If there is only one state for a given set of quasi-particle positions, then if the quasi-particles are moved adiabatically along the path $\mathbf{w}(\tau)$ then up to a time dependant phase factor the wave functions transforms as, $\ket{\Psi(\mathbf{w})}\rangle \rightarrow e^{i\gamma}\ket{\Psi(\mathbf{w})}\rangle$, where $\gamma$ is the time independent Berry phase \cite{berry_quantal_1997}. If there are multiple states corresponding to quasi-particles located at $\mathbf{w}$ then we use $\ket{\Psi_a(\mathbf{w})}\rangle$ to denote an orthonormal basis for this space of states. In this case, after adiabatically moving the quasi-particles along the path $\mathbf{w}(\tau)$ the wave function will, up to a time dependant phase factor, transform as $\ket{\Psi_a(\mathbf{w})}\rangle \rightarrow \sum_b U_{ba} \ket{\Psi_b(\mathbf{w})}\rangle$, where $U_{ba}$ is a time independent unitary matrix, which we refer to as the Wilczek-Zee matrix \cite{wilczek_appearance_1984}. In Ref. \cite{Read2009} Read argues that if the generalised screening hypothesis holds, then both the Berry phase and the Wilczek-Zee matrix are given by the monodromy of the quasi-particle wave functions along the path $\mathbf{w}(\tau)$, up to an area dependant multiplicative phase factor (i.e. the usual phases for adiabatically moving charges in a magnetic field). We  now summarise Read's argument.

First, let us consider the \textit{Abelian} case where there is only one conformal block for a given set of quasi-particles. We write the normalized wave function as,
\begin{equation}
    \Psi(\mathbf{w}; \mathbf{z}) = \bra{0}C(N) \phi_1(w_1)\phi_2(w_2) \dots \prod_{i=1}^N \Omega(z) \ket{0} / Z_N
\end{equation}
The norm squared of this wave function is then given by,
\begin{equation}
    Z(\mathbf{w}, \bar{\mathbf{w}}) = \braket{ \overline{\phi}_1(\bar{w}_1)\phi(w_1) \overline{\phi}_2(\bar{w}_2)\phi(w_2) \dots }_*
\end{equation}

The wave function is not, in general, single-valued in the quasi-particle coordinates. Furthermore, as we require $\ket{\Psi(\mathbf{w}(\tau))} \rangle$ to be a smooth function of $\tau$, we must have that $\ket{\Psi(\mathbf{w}(1)}\rangle$ is a monodromy transformation of $\ket{\Psi(\mathbf{w}(0))}\rangle$. In the Abelian case, this can be expressed as $\ket{\Psi(\mathbf{w}(1))}\rangle = e^{i\alpha} \ket{\Psi(\mathbf{w}(0))}\rangle$, where $\alpha \in \mathbb{R}$. Thus, the total phase picked up by the wave function after the adiabatic transformation is a combination of this phase and the standard Berry phase,
\begin{equation}
    \gamma = \alpha + \sum_j \oint_{\mathbf{w}(\tau)} (dw_j A_{w_j} + d\bar{w} A_{\bar{w}_j} )
\end{equation}
where,
\begin{equation}
    A_{w_j} = i\sqrt{Z(\mathbf{w}, \bar{\mathbf{w}})}^{-1}\langle\bra{\Psi(\mathbf{w})} \partial_{w_j} [ \sqrt{Z(\mathbf{w}, \bar{\mathbf{w}})}^{-1} \ket{\Psi(\mathbf{w})}\rangle ]
\end{equation}
As the unormalised $\ket{\Psi(\mathbf{w})}\rangle$ is holomorphic in $w_j$ we can simplify this expression to, 
\begin{equation}
    \begin{split}
        A_{w_j} =& i\sqrt{Z(\mathbf{w}, \bar{\mathbf{w}})}^{-1} \partial_{w_j} [ \langle\braket{\Psi(\mathbf{w}) | \Psi(\mathbf{w})}\rangle \sqrt{Z(\mathbf{w}, \bar{\mathbf{w}})}^{-1} ] \\
        =& i\sqrt{Z(\mathbf{w}, \bar{\mathbf{w}})}^{-1} \partial_{w_j} [ \sqrt{Z(\mathbf{w}, \bar{\mathbf{w}})} ] \\
        =& \frac{i}{2}\partial_{w_j} \ln Z(\mathbf{w}, \bar{\mathbf{w}}) 
    \end{split}
\end{equation}
Similarly, for $A_{\bar{w}_j}$ we have,
\begin{equation}
    \begin{split}
        A_{\bar{w}_j} &= i\sqrt{Z(\mathbf{w}, \bar{\mathbf{w}})}^{-1}\langle\bra{\Psi(\mathbf{w})} \partial_{\bar{w}_j} [ \sqrt{Z(\mathbf{w}, \bar{\mathbf{w}})}^{-1} \ket{\Psi(\mathbf{w})}\rangle ] \\
        &= -\frac{i}{2}\partial_{\bar{w}_j} \ln Z(\mathbf{w}, \bar{\mathbf{w}})
    \end{split}
\end{equation}

Short-range correlation, combined with the fact that the distance between any two quasi-particles is much larger than the magnetic length, implies that,
\begin{equation}
    Z(\mathbf{w}, \bar{\mathbf{w}}) = \prod_j Z_j(w_j,\bar{w}_j)
\end{equation}
where each $Z_j(w_j,\bar{w}_j)$ cannot depend in any way on the presence of the other quasi-particles. We thus have that,
\begin{equation} \label{Eq:abelianBerryPhase}
    \begin{split}
        \gamma =& \alpha + \frac{i}{2}\sum_j \oint_{\mathbf{w}(\tau)}dw_j\partial_{w_j}\ln Z_j(w_j,\bar{w}_j) \\
        &- d\bar{w}_j \partial_{\bar{w}_j} \ln Z_j(w_j,\bar{w}_j)   
    \end{split}
\end{equation}
As each $Z_j(w_j,\bar{w}_j)$ does not depend on the presence of the other quasi-particles, each term in this sum must be the usual phase for moving charge adiabatically in a magnetic field. That is, each term in the sum must be $2\pi n_j q_j$ where $n_j$ is the number of flux quanta passing through the loop $w_j$ completes in $\mathbf{w}(\tau)$ and $q_j$ is the charge of quasi-particle $j$:
\begin{equation}
    \gamma = \alpha + \sum_j 2 \pi n_j q_j
\end{equation}
Thus, up to an area dependent $n_jq_j$ part, the phase $\gamma$ is given by the phase $\alpha$ of the monodromy transformation of the quasi-particle wave function.

For the \textit{non-Abelian} case there are multiple conformal block wave functions for a given set of quasi-particle positions. Let $\ket{\Psi_a(\mathbf{w})} \rangle$ denote an unnormalized basis for the space of these conformal block wave functions. We first focus on the case when this basis can be picked such that under monodromy transformations, $\ket{\Psi_a(\mathbf{w})} \rangle \rightarrow \sum_b B_{ba}\ket{\Psi_b(\mathbf{w})} \rangle$, where the matrices $B$ form an \textit{irreduciable unitary} representation of the braid group. When the CFT is itself unitary, then there always exists a basis for the conformal blocks where the matrices $B$ form a unitary representation of the braid group. 

We denote the inner product between these wave functions as,
\begin{equation}
    Z_{ab}(\mathbf{w}, \bar{\mathbf{w}}) = \langle \braket{ \Psi_a(\mathbf{w}) | \Psi_b(\mathbf{w}) } \rangle
\end{equation}
Clearly, under a monodromy transformation we have, $Z_{ab} \rightarrow \sum_{cd} B^{\dagger}_{ac}Z_{cd}B_{db}$. However, as the distance between any two quasi-particles is much larger than the magnetic length, the short-range correlations inside the droplet imply that $Z_{ab}$ must be invariant under these monodromy transformations. Thus, $Z_{ab}$ must commute with all the braid matrices $B$. By Schur's lemma it follows that $Z_{ab}(\mathbf{w}, \bar{\mathbf{w}}) = \delta_{ab}Z(\mathbf{w}, \bar{\mathbf{w}})$. 

As with the Abelian case, the wave function at the end of the path $\mathbf{w}(\tau)$ must be a monodromy transformation of the initial wave function, $\ket{\Psi_a(\mathbf{w}(1))} \rangle = \sum_b B_{ba} \ket{\Psi_b(\mathbf{w}(0))} \rangle$ (i.e. the wave function at the end of the path is an analytic continuation along the path of the wave function at the start). If the quasi-particles are moved along this path adiabatically then the wave function will transform as $\ket{\Psi_a(\mathbf{w})}\rangle \rightarrow \sum_b U_{ba} \ket{\Psi_b(\mathbf{w})}\rangle$. As the basis $\sqrt{Z(\mathbf{w}, \bar{\mathbf{w}})}^{-1} \ket{\Psi_a(\mathbf{w})}$ is orthonormal, $U$ is a unitary matrix given by,
\begin{equation}
    U = B \mathcal{P} \exp \bigg ( \sum_j i \oint_{\mathbf{w}(\tau)} (dw_j A_{w_j} + d\bar{w} A_{\bar{w}_j} )  \bigg )
\end{equation}
where $\mathcal{P}$ denotes path ordering, and $A_{w_j}$ and $A_{\bar{w}_j}$ are now matrices given by,
\begin{equation}
    \begin{split}
        A_{w_j, ab} =& i\sqrt{Z(\mathbf{w}, \bar{\mathbf{w}})}^{-1}\langle\bra{\Psi_a(\mathbf{w})} \partial_{w_j} [ \sqrt{Z(\mathbf{w}, \bar{\mathbf{w}})}^{-1} \ket{\Psi_b(\mathbf{w})}\rangle ] \\
        A_{\bar{w}_j, ab} =& i\sqrt{Z(\mathbf{w}, \bar{\mathbf{w}})}^{-1}\langle\bra{\Psi_a(\mathbf{w})} \partial_{\bar{w}_j} [ \sqrt{Z(\mathbf{w}, \bar{\mathbf{w}})}^{-1} \ket{\Psi_b(\mathbf{w})}\rangle ] \\
    \end{split}
\end{equation}
Recalling that the unnormalised wave functions are holomorphic in the quasi-particle coordinates allows the $A$'s to be simplified to,
\begin{equation}
    \begin{split}
        A_{w_j, ab} =& \delta_{ab}\frac{i}{2}\partial_{w_j} \ln Z(\mathbf{w}, \bar{\mathbf{w}}) \\
        A_{\bar{w}_j, ab} =& - \delta_{ab} \frac{i}{2} \partial_{\bar{w}_j} \ln Z(\mathbf{w}, \bar{\mathbf{w}})
    \end{split}
\end{equation}
Once again, the short-range correlations in the bulk imply that $Z(\mathbf{w}, \bar{\mathbf{w}}) = \prod_j Z_j(w_j,\bar{w}_j)$, with each $Z_j$ not depending on the presence of the other quasi-particles. Repeating the same arguments from the Abelian case, we find that,
\begin{equation}
    U = e^{i\sum_j 2 \pi n_j q_j} B
\end{equation}
Thus, up to an area-dependent phase factor, the transformation of the wave function after adiabatically moving the quasi-particles along $\mathbf{w}(\tau)$ is simply given by the monodromy of the quasi-particle wave function along the path.

In the case where the representation of the braid group formed by the matrices $B$ is not irreducible, one can split the reducible representation into a direct sum of irreducible ones. The above argument can then be repeated for each irreducible representation separately. 

It should be noted that these arguments only rely on the generalised screening hypothesis (i.e. short-range correlations in the droplet) and that the unnormalised quasi-particle wave functions are holomorphic in the quasi-particle coordinates.

\subsubsection{Example: the Laughlin wave functions} \label{Sec:LaghlinExample}
We now give a detailed example of this CFT formalism in the case of the Laughlin wave functions. This example should clarify some points made earlier in this background section. We only focus on the fermionic Laughin states where $\nu = 1/m$ with $m$ being an odd integer with $m > 1$. 

As is now well known, one can generate the $\nu = 1/m$ Laughlin trial wave function with the correlation function of the operator,
\begin{equation}
    \Omega(z) = :e^{i\sqrt{m}\varphi(z)}:
\end{equation}
where $\varphi(z)$ is the usual free chiral boson field. Indeed, one can explicitly show that,
\begin{equation} \label{Eq:LaughlinGenerated}
    \bra{0}C(N) \prod_{j = 1}^N e^{\sqrt{m}i\varphi(z_j)}\ket{0} = \prod_{i<j}^N (z_i - z_j)^m
\end{equation}

The conjugate field of $\Omega(z)$ is $\Omega^\dagger(z) = :e^{-i\sqrt{m}\varphi(z)}:$. As the Hermitian conjugate of the operator $\Omega(z)$ is given by $[\Omega(z)]^\dagger = \bar{z}^{-2h}\Omega^\dagger(\bar{z}^{-1})$, where $h = m/2$ is the scaling dimension of $\Omega(z)$, the modes of these fields are related by $(\Omega_n)^\dagger = \Omega^\dagger_{-n}$ with $n \in \mathbb{Z} + m/2$.

In addition, the OPE between $\Omega(z)$ and its conjugate is given by,
\begin{equation}
    \begin{split}
        \Omega(z)\Omega^\dagger(w) =& \frac{1}{(z-w)^m}: e^{i\sqrt{m}\varphi(z) - i\sqrt{m}\varphi(w)} : \\
        \sim& \frac{1}{(z-w)^m} + \frac{i\sqrt{m}\partial\varphi(w)}{(z-w)^{m-1}} + \dots \\
        &+ \frac{1}{(m-1)!}\frac{: e^{-i\sqrt{m}\varphi(w)} \partial^{m-1}e^{i\sqrt{m}\varphi(w))} :}{(z - w)}
    \end{split}
\end{equation}
As $\Omega(z)$ has half integral conformal spin (recalling that $m$ is odd), this OPE gives the anti-commutation relations of the form,
\begin{equation} \label{Eq:LaughlinAlgebra}
    \begin{split}
        \{ \Omega_l, \Omega^\dagger_k \} =& \bigg [ \prod_{j=1}^{m-1}\frac{l + h - j}{j} \bigg ] \delta_{0,l+k} \\
        &+ (\text{normal ordered polynomial in } a_n)
    \end{split}
\end{equation}
where $a_n$ ($n \in \mathbb{Z}$) are the modes of the field $i\partial\varphi(z)$. Furthermore, as $\Omega(z)\Omega(w) \sim 0$ we have $\{ \Omega_k, \Omega_l \} = 0$.

Consider now the set of fields generated by repeated OPEs of $\Omega(z)$ and $\Omega^\dagger(z)$. The OPE between any two such fields must still belong to this set. This set of these fields along with their OPEs defines a chiral algebra $\mathcal{A}$. Given that the OPEs are closed, anti-commutation or commutation relations (depending on the conformal spin) between the modes of these fields must be expressable in terms of modes of fields belonging to this set. We can also think of $\mathcal{A}$ as formal linear combinations of products of modes of these fields with the additional equivalence relation imposed by the requirement that the commutation, anti-commutations and any other relations between the modes dictated by the OPEs must be satisfied. Representations of $\mathcal{A}$ can be defined as the representation of the algebra of these modes. In the particular case we are considering here, this algebra is the fermionic version of the Moore-Seiberg algebra. Importantly, $\mathcal{A}$ contains the Heisenberg algebra $[a_k, a_l] = k\delta_{k+l,0}$ and, thus, the Virasoro algebra.

We can further define $\CFT{\mathcal{A}}$ purely in terms of unitary irreducible representations of $\mathcal{A}$. In Appendix \ref{Sec:LaughlinReps} we show how these representations can be deduced by elementary methods, under certain assumptions. Here we give a summary. Each possible irreducible representation is labeled by a $U(1)$ charge $q_*$, which can take values $q_* = 0, 1/\sqrt{m}, 2/\sqrt{m}, \dots, (m-1)/\sqrt{m}$. The Hilbert space of each irreducible representation, $\mathcal{H}_{q*}$, can be generated by polynomials in $\Omega_k$ and $\Omega_k^\dagger$ applied to a state $\ket{q_*}$, where $a_0\ket{q_*} = q_*\ket{q_*}$ and $a_n\ket{q_*} = 0$ for $n > 0$. A useful basis of $\mathcal{H}_{q_*}$ takes the form $\prod_{n_i}a_{-n_i}\ket{q_* + p\sqrt{m}}$, where $p \in \mathbb{Z}$ (with $p$ taking all integer values for a given irreducible representation), $a_0\ket{q_* + p\sqrt{m}} = (q_* + p\sqrt{m})\ket{q_* + p\sqrt{m}}$ and $a_n\ket{q_* + p\sqrt{m}} = 0$ for $n > 0$. In each case, we can use the $U(1)$ chiral boson to reproduce $\mathcal{H}_{q_*}$, by using the chiral boson representation of $\Omega_k$ and $\Omega_k^\dagger$ and applying polynomials in the modes of these fields on the state $:e^{iq_*\varphi(0)}:\ket{0}$. Thus, the primary field $\phi_{q_*}(z)$ corresponding to $\ket{q_*}$ can be represented by the chiral boson field $\phi_{q_*} = :e^{iq_*\varphi(z)}:$ and so we can calculate all the conformal blocks of $\CFT{\mathcal{A}}$ using this conformal embedding in $\CFT{U(1)}$.

We can now see that the irreducible representations of $\mathcal{A}$ are in one-to-one correspondence with the Laughlin quasi-particles. The quasi-particle wave functions can be generated by the insertion of the $\phi_{q_*}(z)$ into the correlator in Eq. \ref{Eq:LaughlinGenerated}. Let us now consider the specific case of $m$ insertions of the $\phi_{1/\sqrt{m}}$ field at positions $\mathbf{w} = w_1, w_2, \dots, w_m$. For the correlation function to be overall $U(1)$ neutral, the resulting wave function must be for $N-1$ particles and be be written as,
\begin{equation}
    \begin{split}
        \Psi(\mathbf{w}; \mathbf{z}) \propto& \bra{0}C(N) \prod_{l = 1}^m \phi_{1/\sqrt{m}}(w_l) \prod_{j = 1}^{N - 1} e^{\sqrt{m}i\varphi(z_j)}\ket{0} \\
        =& \bigg [ \prod_{l < k}^m (w_l - w_k)^{\frac{1}{m}} \bigg ] \bigg [ \prod_{l=1}^m \prod_{j=1}^{N-1} (z_j - w_l) \bigg ] \\
        &\times \bigg [ \prod_{i<j}^{N - 1} (z_i - z_j)^m \bigg ]
    \end{split}
\end{equation}

From the generalised screening hypothesis, we know that each quasi-particle must have a physical charge $-q_*/\sqrt{m} = - 1/m$. Suppose we now move $w_2$ around a loop that encloses $w_1$ and no other quasi-particle. Clearly, the non-trivial monodromy of the wave functions along this path comes from $(w_1 - w_2)^{\frac{1}{m}}$, where it transforms as $(w_1-w_2)^{\frac{1}{m}} \rightarrow e^{\frac{2\pi i}{m}} (w_1 - w_2)^{\frac{1}{m}}$. Assuming the generalised screening hypothesis, if we physically move these quasi-particles adiabatically along this path then the wave function transforms as $\Psi(\mathbf{w}; \mathbf{z}) \rightarrow e^{\frac{2\pi i}{m}} \Psi(\mathbf{w}; \mathbf{z})$ up to the usual area dependant phase factor. Thus, we can now see how one can obtain the, now well-established, fractional adiabatic statistics of the Laughlin quasi-particles using this CFT formalism.

To generate edge states we replace the state vacuum with some state $\bra{v}$ on the left end of the correlator, where $\ket{v} \in \mathcal{H}_0$, that $\bra{v}$ is in the vacuum representation of $\mathcal{A}$ (i.e. $q_* = 0$). Under this edge state mapping, the states of $\mathcal{H}_0$ that generate edge excitation that keep the number of particles the same, are the states with zero $U(1)$ charge. Such states take the form $\prod_{n_i}a_{-n_i}\ket{0}$ ($n_i > 0$) and they will generate the wave functions of the form,
\begin{equation}
    \begin{split}
        \Psi_{\bra{0}\prod_{n_i}a_{n_i}}(\mathbf{z}) =& \bra{0}\prod_{n_i}a_{n_i} C(N) \prod_{j = 1}^N \Omega(z_j)\ket{0} \\
        =& \prod_{n_i} \sqrt{m}P_{n_i}(\mathbf{z}) \prod_{j<k}^N(z_j - z_k)^m
    \end{split}
\end{equation}
where $P_n(\mathbf{z}) = \sum_{i = 1}^N z_i^n$.

To figure out which edge states the charged states of $\mathcal{H}_0$ generate we can use the general edge state mapping of Eq. \ref{Eq:edgeMapDefinition}. States of $\mathcal{H}_0$ with $U(1)$ charge $p\sqrt{m}$ ($p \in \mathbb{Z}$) are expressed as linear combinations of states of the form $\prod_{n_i}a_{-n_i}\ket{p\sqrt{m}}$ ($n > 0$). These basis elements generate edge states with $N + p$ total particles with the wave function,
\begin{equation}
    \begin{split}
        \Psi_{\bra{p\sqrt{m}} \prod_{n_i}a_{n_i} }( \mathbf{z} ) =& \bra{p\sqrt{m}}\prod_{n_i}a_{n_i} C(N) \prod_{j = 1}^{N+p} \Omega(z_j)\ket{0} \\
        =& \prod_{n_i} \sqrt{m}P_{n_i}(\mathbf{z}) \prod_{j<k}^{N + p}(z_j - z_k)^m  
    \end{split}
\end{equation}
where now we have $P_n(\mathbf{z}) = \sum_{i = 1}^{N+ p} z_i^n$.

Now by adjusting this edge state by using a parameter $\lambda_N$, where we replace $\Omega(z) \rightarrow \sqrt{\lambda_N}\Omega(z)$, and assuming the generalised screening hypothesis, the arguments of DRR give the edge inner products at large system size to be given by Eq. \ref{Eq:edgeInnerProdSimple}. As shown in Ref. \cite{Fern2018}, in the case of the Laughlin wave functions the inner product action, $S$, takes a very simple form to leading order,
\begin{equation}
    S = \frac{1}{6\sqrt{m}N}\oint \frac{dz}{2\pi i} z^2 :(i\partial\varphi(z))^3:
\end{equation}
In the numerical tests of Ref. \cite{Fern2018}, this leading order form of $S$ was found to reproduce the edge state inner products to a high degree of accuracy even at smaller system sizes.

\subsection{Composite fermion and parton trial wave functions} \label{Sec:CFandPartonWF}
Viewed as just trial wave functions, the set of parton wave functions actually includes the composite fermion ones. However, their motivating microscopic models are seemingly quite different. We now give very brief summaries of these microscopic models and detail the resulting trial wave functions for the composite fermion and parton trial wave functions. In each case, we also define the space of \textit{trial edge states}, which naturally arises from a given trial wave function's respective motivating microscopic model.

\subsubsection{Composite fermions}
A wave function is said to have a vortex at a particular location if moving any one of the underlying particles' coordinates around that location makes the phase of the wave function change by an integer number of $2\pi$. As is well known the wave function must go to zero when any particle approaches that point. Thus, a possible mechanism for electrons to lower their energy is for wave function vortices to move with the electrons, which would reduce the probability for two electrons to be near each other. The combination of an electron and an even number of vortices is a quasi-particle known as a \textit{composite fermion} \cite{jain_theory_1990, Jain2007}. 

The composite fermion theory starts by assuming that in the FQHE setting electrons do indeed form composite fermions. At the mean-field level, the effective number of magnetic flux quanta experienced by the composite fermions, for large $N$, is $N_\phi^* = N_\phi - Np$, where $p$ is the number of vortices each electron is attached to ($p$ is even), $N_\phi$ is the actual number of flux quanta and $N$ is the number of particles. If the effective filling fraction $\nu^* = N/N_\phi^*$ is an integer, $n$, then one may expect the composite fermions to form an \textit{integer} quantum Hall state. We would thus have a gapped state at fractional filling $\nu = \frac{n}{pn + 1}$. 

One can then construct a trial wave function from this picture by first including the Jastrow factor $\prod_{i<j}^N(z_i - z_j)^p$ in the wave function, which attaches the wave function vortices. The remaining factor is then interpreted as the wave function of the composite fermions. This naturally gives us,
\begin{equation}
    \Psi(\mathbf{z},\bar{\mathbf{z}}) = \Phi_n(\mathbf{z},\bar{\mathbf{z}})\prod_{i<j}^N(z_i - z_j)^p
\end{equation}
where $\Phi_n(\mathbf{z},\bar{\mathbf{z}})$ is the usual $\nu = n$ integer quantum Hall (IQH) ground state (i.e. the state where all single orbitals in the $n$ lowest Landau levels with angular momenta below a given Fermi level are occupied). This state is not in the lowest Landau level (LLL) and to produce a LLL trial wave function one typically projects this by hand, which gives the wave function,
\begin{equation}
    \Psi(\mathbf{z}) = \mathcal{P}_{LLL} \Phi_n(\mathbf{z},\bar{\mathbf{z}})\prod_{i<j}^N(z_i - z_j)^p
\end{equation}
where $\mathcal{P}_{LLL}$ is LLL projection operator.

Keeping within this mean-field picture, one would expect that the low-energy edge excitations of this state correspond to integer quantum Hall edge states of the composite fermions. Hence, the (unprojected) edge state trial wave functions take the form,
\begin{equation} \label{Eq:CFEdgeExample}
    \Psi(\mathbf{z},\bar{\mathbf{z}}) = \Phi_{edge, n}(\mathbf{z},\bar{\mathbf{z}})\prod_{i<j}^N(z_i - z_j)^p
\end{equation}
where $\Phi_{edge, n}(\mathbf{z},\bar{\mathbf{z}})$ is a $\nu = n$ integer quantum Hall edge state wave function. We define a $\nu = n$ integer quantum Hall edge state to be a state with all particles in the lowest $n$ Landau levels and such that only the single-particle orbitals with angular momenta $m$ in the range $m_F - p < m < m_f + p$ have occupations that differ from the $\nu = n$ ground state, where $m_f$ is the Fermi level and $p$ is such that $p/m_F \ll 1$. 

To produce a LLL edge state trial wave function from the wave function of Eq. \ref{Eq:CFEdgeExample} one simply applies the $\mathcal{P}_{LLL}$ operator. The vector space of wave functions of the form of Eq. \ref{Eq:CFEdgeExample} then defines the space of unprojected edge state trial wave functions for the composite fermion case and the space of wave functions that are LLL projections of the wave function of the form of Eq. \ref{Eq:CFEdgeExample} defines the space of LLL projected edge state trial wave functions.

\subsubsection{Partons}
In the parton theory, one splits the electron into $m$ fictitious partons by mapping the system to that of $mN$ particles, with each parton species having a charge that is a rational multiple $x_i$ of the electron charge, where the electron operator maps as $c(z) = f_1(z)f_2(z)\dots f_m(z)$ and $f_i(z)$ are the operators of each parton species \cite{jain_incompressible_1989}. Clearly, we must restrict the partons to move in groups of $m$ (i.e. as an electron). This constraint can be imposed by adding a dynamical gauge field \cite{Wen1991a}. 

As the parton species $i$ has fractional charege $x_i$, the effective number of magnetic flux quanta for the parton species $i$ must be $N^*_{\phi, i} = x_i N_\phi$. Forgetting about the dynamical gauge field for now, if the effective filling factor for each species is an integer $\nu^*_i = N/N^*_{\phi, i}$ is an integer $n_i$, then one may expect each individual parton species to form a $\nu^*_i = n_i$ integer quantum Hall state. Typically, it is then assumed that adding this gauge field back does not change this fact. This then gives a gapped state of the electrons at filling factor $\nu^{-1} = N_\phi/N = \sum_{i=1}^m x_iN_\phi / N = \sum_{i=1}^m (\nu^*_i)^{-1}$. 

As the electron operator is composed of a product of parton operators, the resulting trial wave function should be a product of the individual parton species' wave functions. We then have the ground state trial wave function,
\begin{equation}
    \Psi(\mathbf{z},\bar{\mathbf{z}}) = \prod_{i=1}^m \Phi_{n_i}(\mathbf{z},\bar{\mathbf{z}}) 
\end{equation}
where, again, $\Phi_n(\mathbf{z},\bar{\mathbf{z}})$ is the $\nu = n$ integer quantum Hall ground state.

Furthermore, one would also expect the edge states of this system to correspond to integer quantum Hall edge states of the partons (again forgetting about the dynamical gauge field). The edge state trial wave functions are then,
\begin{equation}
    \Psi_{edge}(\mathbf{z},\bar{\mathbf{z}}) = \prod_{i = 1}^m \Phi_{i, n_i}(\mathbf{z},\bar{\mathbf{z}})
\end{equation}
where $\Phi_{i, n_i}(\mathbf{z},\bar{\mathbf{z}})$ is some integer quantum Hall edge state of the $i^{\text{th}}$ parton species. The space of edge state trial wave functions is the space of linear combinations of these edge state trial wave functions.

As for the composite fermion case one can produce LLL trial wave functions by simply projected the above wave functions to the LLL.

A short-hand notation is typically used for parton wave functions. For example, $\phi^2_2\phi_1$ is a product of two $\nu = 2$ wave functions and one $\nu = 1$ wave function and $\phi_2^3$ would be a $\nu = 2$ wave function raised to the third power. Another notation of parton states involves labelling them with a series of integers, where $\phi_2^3$ would be labelled $222$, $\phi_2^2\phi_1$ would be labelled $221$, etc.

\subsection{Integer QH wave functions as CFT correlators and edge excitations} \label{Sec:integerQHCFT}
In both the composite fermion and parton theories one understands a particular state as arising from the formation of \textit{integer} quantum Hall states of certain quasi-particles. Hence, to generalise the CFT formalism, outlined in Sec. \ref{Sec:CFTMthods}, to these wave functions we begin by discussing the, seemingly esoteric, task of constructing integer quantum Hall ground and edge state wave functions using CFT. Quasi-particles will not be discussed here, as they are topologically trivial in the integer quantum Hall case. we  first discuss the $\nu = 1$ wave functions, then show how $\nu = 2$ wave functions can be written as CFT correlation functions \textit{without any explicit anti-symmeterization}, and then the generalization to $\nu = n$, $n \in \mathbb{Z}$.

\subsubsection{$\nu = 1$}
The CFT that will be used here is that of the $U(1)$ chiral boson with a compactification radius of one. Within the CFT the chiral boson field, $\varphi(z)$, has the mode expansion,
\begin{equation}
    \varphi(z) = -ia_0 \ln z + i\sum_{n \in \mathbb{Z}/\{ 0 \}} \frac{1}{n}a_n z^{-n}
\end{equation}
where,
\begin{equation}
    [a_n, a_m] = n\delta_{n + m, 0}
\end{equation}
Note that we do not include the usual $\varphi_0$ operator here. This is because we are dealing with a \textit{compactified} boson and so the eigenvalues of the $a_0$ operator belong to a discrete set (not a continuum). Instead, we  use a \textit{Klein factor} $F$ which is a unitary operator satisfying,
\begin{equation} \label{Eq:Compact}
    [a_n, F] = \delta_{n,0}F
\end{equation}

The Hilbert space of this theory is defined as follows. There exists the vacuum state $\ket{0}$, which is the unique state satisfying $a_n\ket{0} = 0$ for $n \geq 0$. Any state in the Hilbert space can be generated by applying polynomials in $a_n$, $F$ and $F^\dagger$ on $\ket{0}$. Hence, from Eq. \ref{Eq:Compact}, the eigenvalue of $a_0$ are integers, which reflects the fact the boson compactification radius is one. The inner-product of the Hilbert space is defined by the fact that $F$ is unitary, the Hermitian conjugates of the field modes are given by $(a_n)^\dagger = a_{-n}$, and $\braket{0 | 0} = 1$. 

The energy-momentum tensor of this theory is given by,
\begin{equation}
    T(z) = \frac{1}{2}:(i\partial\varphi(z))^2:
\end{equation}

We define the vertex operators as,
\begin{equation} \label{Eq:vertexExpansions}
    \begin{split}
        V(z) \equiv& F :e^{i\varphi(z)}: \\
        =& F e^{\sum_{n>0} \frac{1}{n}a_{-n}z^n} e^{a_0\ln z} e^{-\sum_{n>0}\frac{1}{n}a_nz^{-n}} \\
        V^\dagger(z) \equiv& F^\dagger : e^{-i\varphi(z)} : \\
        =& F^\dagger e^{-\sum_{n>0} \frac{1}{n}a_{-n}z^n} e^{-a_0\ln z} e^{\sum_{n>0}\frac{1}{n}a_nz^{-n}}
    \end{split}
\end{equation}
From the above expansions, one can show that $V(z) V^\dagger(w) = (z-w)^{-1}:e^{i\varphi(z) - i\varphi(w)}:$, for $|z|>|w|$, and $V^\dagger(w) V(z) = (w-z)^{-1}:e^{i\varphi(z) - i\varphi(w)}:$, for $|z|<|w|$. Hence, at $|z|=|w|, w\neq z$, we have $V(z)V^\dagger(w) = - V^\dagger(w)V(z)$ (i.e. they anti-commute). By similar arguments we have, $\{ V(z), V(w) \} = \{ V^\dagger(z), V^\dagger(w) \} = 0$ for $|z|=|w|, w\neq z$. Hence, the fields $V(z)$ and $V^\dagger(w)$ are fermionic, with an OPE given by,
\begin{equation}
    V(z)V^\dagger(w) = \frac{1}{z - w} + i\partial\varphi(w) + \dots
 \end{equation}

The modes of these fields are defined by $V_k = \oint \frac{dz}{2 \pi i } z^{k + 1/2 - 1} V(z)$ and $V^\dagger_k = \oint \frac{dz}{2 \pi i } z^{k + 1/2 - 1} V^\dagger(z)$ for $k \in \mathbb{Z} + 1/2$. From the OPE we have,
\begin{equation}
    \{ V_k, V^\dagger_l \} = \delta_{k+l, 0}
\end{equation}
Furthermore, from Eq. \ref{Eq:vertexExpansions} one can directly show that, $[V(z)]^\dagger = \bar{z}^{-2}V^\dagger(\bar{z}^{-1}) \Rightarrow (V_k)^\dagger = V^\dagger_{-k}$ and $V_k\ket{0} = V^\dagger_k\ket{0} = 0$ for $k > 0$. Hence, these vertex operators represent free complex fermions, where $V_{-k}$ and $V^\dagger_{-k}$ for $k>0$ are creation operators for particles and holes respectively. The corresponding annihilation operators are $V^\dagger_k$ and $V_k$ for $k > 0$ for particles and holes respectively.

The modes of these fields can further be used to express the modes of the boson fields. From the OPE it follows that,
\begin{equation} \label{Eq:densityModeExpan}
    a_n = \sum_k :V_{-k}V^\dagger_{k+n}:
\end{equation}
where $:*:$ refers to fermionic normal ordering \footnote{To see this we first not that, $i\partial\varphi(w) = \oint_{w} \frac{dz}{2\pi i} \frac{\mathcal{R} [ V(z) V^\dagger(w)]}{z-w} = \oint_{|z|>|w|} \frac{dz}{2\pi i} \frac{ V(z) V^\dagger(w)}{z-w}  + \oint_{|z|<|w|} \frac{dz}{2\pi i} \frac{ V^\dagger(w) V(z)}{z-w} $ One then expands the $1/(z-w)$ factors to express this as a sum of $V_k V^\dagger(w)$ products. Finally, one can integrate over $w$ to obtain the $a_n$ modes.}. We also have that $F^\dagger V(z)F = zV(z)$ and, hence, $V_kF = FV_{k+1}$. Finally, $F\ket{0} = V_{-1/2}\ket{0}$ and $F^\dagger\ket{0} = V^\dagger_{-1/2}\ket{0}$. These relations allow us to express a polynomial of $a_n$, $F$ and $F^\dagger$ applied on $\ket{0}$ as a polynomial of $V_k$ and $V^\dagger_k$ applied on $\ket{0}$. Hence, the Hilbert space can be generated by the fermion modes. This is merely one way of expressing the usual bosonization relations\cite{VonDelft1998}.

Using standard methods, we have that,
\begin{equation}
    \begin{split}
        \bra{0}(F^\dagger)^N\prod_{i = 1}^N V(z_i) \ket{0} &= \prod_{i<j}^N (z_i-z_j) \\
        &= \begin{vmatrix}
        z_1^{N-1} & z_2^{N-1} & \cdots & z_N^{N-1} \\
        z_1^{N-2} & z_2^{N-2} & \cdots & z_N^{N-2} \\
        \vdots & \vdots & \ddots & \vdots \\
        z_1 & z_2 & \cdots & z_N \\
        1 & 1 & \dots & 1 \\
        \end{vmatrix}
    \end{split}
\end{equation}
We have generated the $\nu = 1$ (non-interacting) ground-state, with $\Omega(z) = V(z)$ and $C(N) = (F^\dagger)^N$. The chiral algebra that we have then used to generate the integer quantum Hall state is that generated by $V(z)$ and $V^\dagger(z)$. This chiral algebra only has one irreducible representation (given by the free fermion Hilbert space) provided we keep periodic boundary conditions as we encircle the origin with any of the particles in this $\nu = 1$ state, which is consistent with the topological triviality of the integer quantum Hall state.

Let us now consider the edge state mapping. Edge states, $\bra{v}(F^\dagger)^N\prod_{i=1}^N V(z_i)\ket{0}$, with a fixed number of particles are generated with a $\bra{v}$ that are sums of states of the form $\bra{0}\prod_{n_i}a_{n_i}$ with $n_i > 0$. From the fact that $[a_n, V(z)] = z^nV(z)$, $[a_n, F] = 0$ and $a_n\ket{0} = 0$ (for $n>0$) it follows that,
\begin{equation}
    \bra{0}\prod_{n_i}a_{n_i} (F^\dagger)^N \prod_{j=1}^N V(z_j)\ket{0} = \prod_{n_i}P_{n_i}(\mathbf{z})\prod_{j<k}(z_j-z_k)
\end{equation}
Hence, this mapping from the CFT can generate the entire lowest Landau level Hilbert space.

Given this, we must have that for any Slater determinant state there exists a CFT state $\bra{v}$ that will generate it. Let $\bra{v}$ be a state that generates the Slater determinant,
\begin{equation}
    \Psi_{\bra{v}}(\mathbf{z}) = \begin{vmatrix}
        z_1^{m_1} & z_2^{m_1} & \cdots & z_N^{m_1} \\
        z_1^{m_2} & z_2^{m_2} & \cdots & z_N^{m_2} \\
        \vdots & \vdots & \ddots & \vdots \\
        z_1^{m_N} & z_2^{m_N} & \cdots & z_N^{m_N} \\
    \end{vmatrix}
\end{equation}
Now we consider the state generated by $\bra{v}V_k$. First, we note that this state will generate a wave function with $N-1$ particles (as the correlation function must be $U(1)$ charge neutral). We then have that,
\begin{equation}
    \begin{split}
        \Psi_{\bra{v}V_k}(\mathbf{z}) =& \bra{v}V_k (F^\dagger)^N \prod_{i=2}^{N}V(z_i)\ket{0} \\
        =& \bra{v}(F^\dagger)^N V_{k-N} \prod_{i=2}^{N}V(z_i)\ket{0} \\
        =& \oint \frac{dz_1}{2\pi i}z_1^{-(N-k-1/2) - 1}\Psi_{\bra{v}}(\mathbf{z})
    \end{split} 
\end{equation}
where we should recall that $\Psi_{\bra{v}}(\mathbf{z})$ is an $N$ particle wave function. If $(N-k-1/2) \in \{ m_1, m_2, \dots , m_N \}$ with $m_j = N - k - 1/2$, then we have (by expanding the Slater determinant along the first column),
\begin{equation}
    \Psi_{\bra{v}V_k}(\mathbf{z}) = (-1)^{j-1}
    \begin{vmatrix}
        z_2^{m_1} & \dots & z_N^{m_1} \\
        \vdots & \ddots & \vdots \\
        z_2^{m_{j-1}} & \dots & z_N^{m_{j-1}} \\
        z_2^{m_{j+1}} & \dots & z_N^{m_{j+1}} \\
        \vdots & \ddots & \vdots \\
        z_2^{m_N} & \dots & z_N^{m_N} \\
    \end{vmatrix}
\end{equation}
Otherwise, if $(N - k - 1/2) \notin \{ m_1, m_2, \dots , m_N \}$ then we have $\Psi_{\bra{v}V_k}(\mathbf{z}) = 0$. 

Now consider the state $\Psi_{\bra{v}V^\dagger_k}(\mathbf{z})$ with $N + k > 0$. By $U(1)$ charge conservation this is an $N+1$ particle wave function. From the anti-commutation relation $\{V^\dagger_k, V(z) \} = z^{k-1/2}$ and $V^\dagger_k\ket{0} = 0$ for $k > 0$ we have that,
\begin{equation}
    \begin{split}
        &\Psi_{\bra{v}V^\dagger_k}(\mathbf{z}) \\
        =& \bra{v}V^\dagger_k (F^\dagger)^N \prod_{i=1}^{N+1} V(z_i) \ket{0} \\
        &= \bra{v}(F^\dagger)^N V^\dagger_{k+N} \prod_{i=1}^{N+1} V(z_i) \ket{0} \\
        &= \sum_{j=1}^{N+1} (-1)^{j-1} z_j^{N+k-1/2} \bra{v} (F^\dagger)^N \prod_{i \neq j}^N V(z_i) \ket{0} \\
        &= 
        \begin{vmatrix}
            z_1^{N+k-1/2} & \dots & z_{N+1}^{N+k-1/2} \\
            z_1^{m_1} & \dots & z_{N+1}^{m_1} \\
            \vdots & \ddots & \vdots \\
            z_1^{m_N} & \dots & z_{N+1}^{m_N}
        \end{vmatrix} \\
    \end{split}
\end{equation}

As $\Psi_{\bra{0}}(\mathbf{z})$ is a Slater determinant, we can then see, inductively, that all Slater determinant states are generated by CFT states of the form $\bra{0}\prod_{i}V^\dagger_{k_i}\prod_j V_{l_j}$ with $k_i, l_j > 0$. 

This mapping is best summarised in the second quantised notation. Firstly, we note that $\ket{\Psi_{\bra{0}}} \rangle / \sqrt{Z_N} = c^\dagger_{N-1} c^\dagger_{N-2} \dots c^\dagger_1 c^\dagger_0\ket{0}\rangle$. Then, for $N - k - 1/2 > 0$ we have,
\begin{equation} \label{Eq:VMapn1}
    \ket{\Psi_{\bra{v}V_k}}\rangle = \sqrt{\mathcal{N}(N-k-1/2)}^{-1} c_{N-k-1/2} \ket{\Psi_{\bra{v}}}\rangle  
\end{equation}
and for $N + k - 1/2 > 0$ we have,
\begin{equation} \label{Eq:VdMapn1}
    \ket{\Psi_{\bra{v}V^\dagger_k}}\rangle =\sqrt{\mathcal{N}(N+k-1/2)} c^\dagger_{N+k-1/2} \ket{\Psi_{\bra{v}}}\rangle 
\end{equation}
where $\mathcal{N}(m) = \int D^2z |z^m|^2 = 2\pi 2^m m!$. This direct relationship between the $V_k$'s and particle creation and annihilation operators indicates this edge state mapping can generally be interpreted as a finite system size non-unitary bosonization. We call this non-unitary bosonization as the map from the boson Hilbert space to the Hilbert space of the actual system is non-unitary.

From Eq. \ref{Eq:VMapn1} and Eq. \ref{Eq:VdMapn1} we can clearly see that this mapping from CFT states to edge states does not preserve inner products. In Appendix \ref{Sec:nu1InnerProd} we show that the CFT generated edge states have inner products given by,
\begin{equation}
    \langle \braket{ \Psi_{\bra{w}} | \Psi_{\bra{v}} } \rangle / Z_N = \bra{v}R^{2L_0}e^S \ket{w}
\end{equation}
where $R = \sqrt{2N}$ (i.e. the radius of the droplet) and,
\begin{equation}
    \begin{split}
        S =& (N\ln N - N + \ln[2\pi\sqrt{\pi}] - 1/(12N))a_0 \\
        &+ \frac{1}{6N}  \oint \frac{dz}{2\pi i} z^2 :(i\partial\varphi(z))^3: 
    \end{split}
\end{equation}
for large $N$ only, which agrees with the more general result of Fern et al. \cite{Fern2018}. We can remove the $a_0$ term by replacing $V(z) \rightarrow \sqrt{\lambda_N}V(z)$ in the generating correlation function, with $\ln \lambda_N = - (N\ln N - N + \ln[2\pi\sqrt{\pi}] - 1/(12N))$.

For any mapping from CFT states to some space of edge state wave functions it is useful to be able know what will be the angular momentum of the wave function a particular CFT state maps to. In general, this is particularly useful for obtaining upper bounds on the number of edge-state wave functions at a given angular momentum. For the IQH cases being considered here these upper bounds are always saturated.

We show in Appendix \ref{Sec:nu1Angular} that the angular momentum operator can be mapped to the CFT, in this particular case, as,
\begin{equation}
    \sum_{i=1}^N z_i\partial_i \Psi_{\bra{v}}(\mathbf{z}) = \Psi_{\bra{v}(L_0 + (2N-1)a_0/2 + N(N-1)/2)}(\mathbf{z})
\end{equation}
With $L_0$ being the zero mode of $T(z)$ (which one should recall has scaling dimension 2). We can then see that for a fixed number of particles (corresponding to eigenstates of $a_0$ in the CFT), all the eigenstates of $L_0$ with a particular eigenvalue will map to wave functions all with the same angular momentum. 

\subsubsection{$\nu = 2$} \label{Sec:nu2CFT}
For $\nu = 2$ the ground state wave function is no longer purely holomorphic (up to Gaussian factors) and, hence, we cannot write it purely as a chiral CFT correlation function. However, for $\nu = 2$ the highest power of any $\bar{z}$ that appears is one. We can then easily separate the anti-holomorphic dependence from the holomorphic dependence, with the latter given by some chiral CFT correlation functions. 

We know that the edge theory must contain two branches of excitations, and so the CFT that we  use is that of two independent chiral bosons, $\fe{i}(z)$ for $i=1,2$, both with compactification radius one and the following mode expansions,
\begin{equation}
    \fe{j}(z) = -i \fem{j}_0 \ln z + i\sum_{n\in \mathbb{Z}/\{ 0 \}} \frac{1}{n} \fem{j}_{n}z^{-n} 
\end{equation}
with,
\begin{equation}
    [\fem{i}_n, \fem{j}_m] = n\delta_{n+m,0}\delta_{ij}
\end{equation}
The corresponding Klein factors $F_j$ $F^\dagger_j$ have the following \textit{anti-commutation} relations,
\begin{equation}
    \begin{split}
        \{ F_i, F_j \} =& 0 \quad (i \neq j)\\
        \{ F_i, F^\dagger_j \} =& 2\delta_{ij} \\
    \end{split}
\end{equation}
and the following commutation relations with the field modes,
\begin{equation}
    [\fem{i}_n, F_j] = \delta_{n,0}\delta_{ij}F_j
\end{equation}
As for the single boson field case, the vacuum state, $\ket{0}$, is the unique state defined by, $\fem{j}_n\ket{0} = 0$ for $n \geq 0$. The Hilbert space is then generated by applying polynomials in $\fem{j}_n$, $F_j$ and $F^\dagger_j$ on $\ket{0}$. The energy-momentum tensor is simply a sum of the individual energy-momentum tensors of the two chiral boson species.

Finally, the vertex operators are defined by,
\begin{equation}
    \begin{split}
        V_j(z) =& F_j :e^{i\fe{j}(z)}: \\
        V^\dagger_j(z) =& F_j^\dagger :e^{-i\fe{j}(z)}: \\
    \end{split}
\end{equation}
where the modes $V_{j,k} = \oint \frac{dz}{2\pi i}z^{k - 1/2}V_j(z)$ and $V^\dagger_{j,k} = \oint \frac{dz}{2\pi i}z^{k - 1/2}V^\dagger_j(z)$ have the anti-commutation relations,
\begin{equation}
    \begin{split}
        \{ V_{i,k}, V_{j,l} \} =& 0 \\
        \{V_{i,k}, V^\dagger_{j, l} \} =& \delta_{ij}\delta_{k+l,0} \\
    \end{split}
\end{equation}
Furthermore, the Hilbert space can also be generated by polynomials in $V_{i,k}$ and $V^\dagger_{i,k}$ applied on $\ket{0}$. Hence, this system is also equivalent to two independent species of complex fermions. 

To generate the ground state wave function, the $\Omega(z)$ operator will now have to have some anti-holomorphic dependence, $\Omega(z,\bar{z})$. We consider the operator,
\begin{equation}
    \Omega(z, \bar{z}) = V_1(z) + \bar{z} V_2(z) 
\end{equation}
In the following, we  only consider the case when $N$ is odd (so that the lowest angular momentum $\nu = 2$ state is unique). Now consider the wave function,
\begin{equation} \label{Eq:nu2GroundGen}
    \Psi_{\bra{0}}(\mathbf{z}) = \bra{0}(F_2^\dagger)^{(N+1)/2}(F_1^\dagger)^{(N-1)/2} \prod_{i=1}^N \Omega(z_i, \bar{z}_i) \ket{0}
\end{equation}
By expanding out the $\Omega(z_i,\bar{z}_i)$, this correlation function can be expressed as a sum of correlation functions of products of $V_1(z)$ and $V_2(z)$. As the $U(1)$ charge of each boson field is separately conserved, the only terms that will contribute are those with $(N-1)/2$ $V_1$'s and $(N+1)/2$ $V_2$'s. Let $P(n)$ be a permutation of $\{ 1, 2, \dots, N \}$ such that $P(i) < P(j)$ for $0<i<j\leq q$ or $q < i < j \leq N$ with $q$ being a fixed integer ($0 < q \leq N$). Let the set of such permutations be denoted by $\mathcal{P}_{N,q}$. Further letting $C(N) = (F_2^\dagger)^{N_1}(F_1^\dagger)^{N_2}$ and $N_1 = (N-1)/2$, $N_2=(N+1)/2$, we can express our wave function as,
\begin{equation}
    \begin{split}
        \Psi_{\bra{0}}(\mathbf{z},\bar{\mathbf{z}}) =& \sum_{P \in \mathcal{P}_{N, N_1}} \text{sgn}(P) \bra{0}C(N) \prod_{i=1}^{N_1} V_1(z_{P(i)}) \\ 
        &\times \prod_{j=N_1 + 1}^N \bar{z}_{P(j)} V_2(z_{P(j)}) \ket{0} \\
        =& \sum_{P \in \mathcal{P}_{N, N_1}} \text{sgn}(P) 
        \begin{vmatrix}
            z_{P(1)}^{N_1 - 1} & z_{P(2)}^{N_1-1} & \cdots & z_{P(N_1)}^{N_1 - 1} \\
            \vdots & \vdots & \ddots & \vdots \\
            z_{P(1)} & z_{P(2)} & \cdots & z_{P(N_1)} \\
            1 & 1 & \cdots & 1 \\
        \end{vmatrix} \\
        &\times 
        \begin{vmatrix}
            \bar{z}_{P(N_1 + 1)}z_{P(N_1 + 1)}^{N_2 - 1} & \cdots & \bar{z}_{P(N)}z_{P(N)}^{N_2 - 1} \\
            \vdots & \ddots & \vdots \\
            \bar{z}_{P(N_1 + 1)}z_{P(N_1 + 1)} & \cdots & \bar{z}_{P(N)}z_{P(N)} \\
            \bar{z}_{P(N_1 + 1)} & \cdots & \bar{z}_{P(N)} \\
        \end{vmatrix} \\
    \end{split}
\end{equation}
where sgn$(P)$ appears because the $V_j(z)$ anti-commute. One can interpret this as a sum over wave functions involving two particle species, where we sum over all the possible ways of allocating the particles into these two species with $N_1$ particles in species $1$ and $N_2$ particles in species $2$. These permutations $P(n)$ are in one-to-one correspondence with each allocation. When we move onto the edge state map we  see that the number of particles in these species can be varied.

This sum, over permutations, of products of Slater determinants can be simplified to one Slater determinant, with the resulting wave function given by,
\begin{equation}
    \Psi_{\bra{0}}(\mathbf{z},\bar{\mathbf{z}}) =
    \begin{vmatrix}
        z_1^{N_1 - 1} & z_2^{N_1 - 1} & \cdots & z_N^{N_1 - 1} \\
        \vdots & \vdots & \ddots & \vdots \\
        z_1 & z_2 & \cdots & z_N \\
        1 & 1 & \cdots & 1 \\
        \bar{z}_1z_1^{N_2 - 1} & \bar{z}_2z_2^{N_2 - 1} & \cdots & \bar{z}_Nz_N^{N_2 - 1} \\
        \vdots & \vdots & \ddots & \vdots \\
        \bar{z}_1 z_1 & \bar{z}_2 z_2 & \cdots & \bar{z}_N z_N \\
        \bar{z}_1 & \bar{z}_2 & \cdots & \bar{z}_N \\
    \end{vmatrix}
\end{equation}
This is the non-interacting $\nu = 2$ ground state wave function of $N$ particles (with $N$ odd). Hence, we have expressed this wave function using a CFT correlation function without any explicit anti-symmetrization in Eq. \ref{Eq:nu2GroundGen}, from the fact that $V_j(z)$ anti-commute with each other, which follows from the spin-statistics theorem as these fields have half-integral conformal spin. 

The associated chiral algebra of this state is that generated by repeated operator product expansions of $V_j(z)$ and $V^\dagger_j(z)$. Note that now the chiral algebra is generated by four fields rather than just $\Omega(z)$ and $\Omega^\dagger(z)$, which was discussed earlier in this section. As with $\nu = 1$, there is only one irreducible representation of this chiral algebra (i.e. the free complex fermion representation), which reflects the fact that the $\nu = 2$ state has no topologically non-trivial excitations.

We now consider the edge state mapping. Assume $\bra{v}$ to be a state that generates the Slater determinant,
\begin{equation}
    \Psi_{\bra{v}}(\mathbf{z},\bar{\mathbf{z}}) =
    \begin{vmatrix}
        z_1^{m^{(1)}_1} & z_2^{m^{(1)}_1} & \cdots & z_N^{m^{(1)}_1} \\
        z_1^{m^{(1)}_2} & z_2^{m^{(1)}_2} & \cdots & z_N^{m^{(1)}_2} \\
        \vdots & \vdots & \ddots & \vdots \\
        z_1^{m^{(1)}_{N_1}} & z_2^{m^{(1)}_{N_1}} & \cdots & z_N^{m^{(1)}_{N_1}} \\
        \bar{z}_1z_1^{m^{(2)}_1} & \bar{z}_2 z_2^{m^{(2)}_1} & \cdots & \bar{z}_N z_N^{m^{(2)}_1} \\
        \bar{z}_1z_1^{m^{(2)}_2} & \bar{z}_2 z_2^{m^{(2)}_2} & \cdots & \bar{z}_N z_N^{m^{(2)}_2} \\
        \vdots & \vdots & \ddots & \vdots \\
        \bar{z}_1z_1^{m^{(2)}_{N_2}} & \bar{z}_2 z_2^{m^{(2)}_{N_2}} & \cdots & \bar{z}_N z_N^{m^{(2)}_{N_2}} \\
    \end{vmatrix}
\end{equation}
By $U(1)$ charge conservation the state $\bra{v}V_{2,k}$, with $N_2 - k - 1/2 = m^{(2)}_j$, must generate a $N-1$ particle wave function with,
\begin{equation}
    \begin{split}
        &\Psi_{\bra{v}V_{2,k}}(\mathbf{z},\bar{\mathbf{z}}) \\
        =& \bra{v}V_{2,k}C(N) \prod_{i=1}^{N-1}\Omega(z_i,\bar{z}_i)\ket{0} \\
        =& \sum_{P \in \mathcal{P}_{N - 1, N_1}} \text{sgn}(P) \bra{v}V_{2,k}C(N) \prod_{i=1}^{N_1} V_1(z_{P(i)}) \\ 
        &\times \prod_{l=N_1 - 1}^{N-1} \bar{z}_{P(l)} V_2(z_{P(l)}) \ket{0} \\
        =& (-1)^{j-1} \sum_{P \in \mathcal{P}_{N + 1, N_1}}
        \begin{vmatrix}
            z_{P(1)}^{m^{(1)}_{1}} & \cdots & z_{P(N_1)}^{m^{(1)}_1} \\
            \vdots & \ddots & \vdots \\
            z_{P(1)}^{m^{(1)}_{N_1}} & \cdots & z_{P(N_1)}^{m^{(1)}_{N_1}} \\
        \end{vmatrix} \\
        &\times 
        \begin{vmatrix}
            \bar{z}_{P(N_1+1)}z_{P(N_1 + 1)}^{m^{(2)}_1} & \cdots & \bar{z}_{P(N-1)} z_{P(N-1)}^{m^{(2)}_1} \\
            \vdots & \ddots & \vdots \\
            \bar{z}_{P(N_1+1)}z_{P(N_1 + 1)}^{m^{(2)}_{j-1}} & \cdots & \bar{z}_{P(N-1)}z_{P(N-1)}^{m^{(2)}_{j-1}} \\
            \bar{z}_{P(N_1+1)}z_{P(N_1 + 1)}^{m^{(2)}_{j+1}} & \cdots & \bar{z}_{P(N-1)}z_{P(N-1)}^{m^{(2)}_{j+1}} \\
            \vdots & \ddots & \vdots \\
            \bar{z}_{P(N_1+1)}z_{P(N_1 + 1)}^{m^{(2)}_{N_2}} & \cdots & \bar{z}_{P(N-1)} z_{P(N-1)}^{m^{(2)}_{N_2}} \\
        \end{vmatrix} \\
        =& (-1)^{j-1} 
        \begin{vmatrix}
            z_{1}^{m^{(1)}_{1}} & \cdots & z_{N}^{m^{(1)}_1} \\
            \vdots & \ddots & \vdots \\
            z_{1}^{m^{(1)}_{N_1}} & \cdots & z_{N}^{m^{(1)}_{N_1}} \\
            \bar{z}_1z_1^{m^{(2)}_1} & \cdots & \bar{z}_{N-1} z_{N-1}^{m^{(2)}_1} \\
            \vdots & \ddots & \vdots \\
            \bar{z}_{1}z_{1}^{m^{(2)}_{j-1}} & \cdots & \bar{z}_{N-1}z_{N-1}^{m^{(2)}_{j-1}} \\
            \bar{z}_{1}z_{1}^{m^{(2)}_{j+1}} & \cdots & \bar{z}_{N-1}z_{N-1}^{m^{(2)}_{j+1}} \\
            \vdots & \ddots & \vdots \\
            \bar{z}_{1}z_{1}^{m^{(2)}_{N_2}} & \cdots & \bar{z}_{N-1} z_{N-1}^{m^{(2)}_{N_2}} \\
        \end{vmatrix}
    \end{split}
\end{equation}
Clearly, if $N_2 - k - 1/2 \notin \{ m^{(2)}_1, m^{(2)}_2, \dots , m^{(2)}_{N_2} \}$ then $\Psi_{\bra{v}V_{2,k}}(\mathbf{z},\bar{\mathbf{z}}) = 0$. 

By the same argument, we also have,
\begin{equation}
    \Psi_{\bra{v}V^\dagger_{2,k}}(\mathbf{z},\bar{\mathbf{z}}) = 
    \begin{vmatrix}
        z_1^{m^{(1)}_1} & \cdots & z_N^{m^{(1)}_1} \\
        z_1^{m^{(1)}_2} & \cdots & z_N^{m^{(1)}_2} \\
        \vdots & \vdots & \vdots \\
        z_1^{m^{(1)}_{N_1}} & \cdots & z_N^{m^{(1)}_{N_1}} \\
        \bar{z}_1z_1^{N_2 + k - 1/2} & \cdots & \bar{z}_N z_N^{N_2 + k - 1/2} \\
        \bar{z}_1z_1^{m^{(2)}_1} & \cdots & \bar{z}_N z_N^{m^{(2)}_1} \\
        \bar{z}_1z_1^{m^{(2)}_2} & \cdots & \bar{z}_N z_N^{m^{(2)}_2} \\
        \vdots & \vdots & \ddots & \vdots \\
        \bar{z}_1z_1^{m^{(2)}_{N_2}} & \cdots & \bar{z}_N z_N^{m^{(2)}_{N_2}} \\
    \end{vmatrix}
\end{equation}
One can also easily find analogous expressions for $\Psi_{\bra{v}V_{1,k}}(\mathbf{z},\bar{\mathbf{z}})$ and $\Psi_{\bra{v}V^\dagger_{1,k}}(\mathbf{z},\bar{\mathbf{z}})$. 

We summarize this edge state mapping in second quantised notation as,
\begin{equation}
    \begin{split}
        \ket{\Psi_{\bra{v}V_{i,k}}}\rangle =& (-1)^{N - N_i} \Tilde{d}_{i,N_1 - k - 1/2} \ket{\Psi_{\bra{v}}}\rangle \\
        \ket{\Psi_{\bra{v}V^\dagger_{i,k}}}\rangle =& (-1)^{N - N_i} \Tilde{c}^\dagger_{i,N_1 + k - 1/2} \ket{\Psi_{\bra{v}}}\rangle \\
    \end{split}
\end{equation}
where $\Tilde{c}^\dagger_{1,m}$ and $\Tilde{c}^\dagger_{2,m}$ are the creation operators for orbitals $z^m$ and $\bar{z}z^{m+1}$ respectively, and $\Tilde{d}^\dagger_{i,m}$ are creation operators for single particle orbitals such that $\{ \Tilde{d}_{i,m}, \Tilde{c}^\dagger_{j,m'} \} = \delta_{ij}\delta_{mm'}$. These operator mappings are strictly only valid for $N_i - k - 1/2 \geq 0$ for $V_{i,k}$ and $N_i + k - 1/2 \geq 0$ for $V^\dagger_{i,k}$.

In Appendix \ref{Sec:nu2InnerProd} we show what form the inner products of the CFT generated edge state inner products take for large $N$. To express these inner products in a form that generalises to other states, we first define the currents, $J^3(z) = [i\partial\fe{2}(z) - i\partial\fe{1}(z)]/2$, $J^+(z) = V_2(z) V^\dagger_1(z)$ and $J^-(z) = V_1(z)V^\dagger_2(z)$. These fields have the OPEs,
\begin{equation}
    \begin{split}
        J^3(z)J^3(w) &\sim \frac{1/2}{(z-w)^2} \\
        J^3(z)J^\pm(w) &\sim \pm \frac{J^\pm (w)}{z - w} \\
        J^+(z)J^-(w) &\sim \frac{1}{(z-w)^2} + \frac{2J^3(w)}{z-w} \\
    \end{split}
\end{equation}
which is precisely the OPEs for $\widehat{\mathfrak{su}}(2)_1$ WZW model currents. Finally, we also define $\Phi(z) = [\fe{1}+\fe{2}]/\sqrt{2}$ with it's corresponding modes $\Tilde{a}_n$. Using these currents the edge state inner products can be expressed as,
\begin{equation}
    \langle \braket{\Psi_{\bra{w}} | \Psi_{\bra{v}} } \rangle/Z_N = \bra{v}R^{2L_0}e^{S}\ket{w}
\end{equation}
where the radius of the droplet is $R = \sqrt{2N_1}$ and,
\begin{equation} \label{Eq:nu2innerprodaction}
    \begin{split}
        S =& \sqrt{2}[N_1\ln N_1 - N_1 + (1/2)\ln N_1 + \ln(2\pi\sqrt{2\pi})]\Tilde{a}_0 \\
        &+ 3\ln (2N_1) \bigg [ J^3_0 - \frac{J^1_0}{N_1} \bigg ]
    \end{split}
\end{equation}
where, $J^1(z) = (J^+(z) + J^-(z))/2$. One should note that the currents $i\partial\Phi(z)$, $J^3(z)$ and $J^\pm(z)$ form a basis of the space of neutral fields with scaling dimension one in this theory. Thus, we can see that for $\nu = 2$ we have other divergent terms in $S$ that are zero modes of these extra currents, where we cannot cancel these terms with a simple $\Omega \rightarrow \sqrt{\lambda_N}\Omega$. As will be seen later in this work, this is a generic feature.

Finally, in Appendix \ref{Sec:nu2Angular} we show that the angular momentum operator maps over to the CFT as,
\begin{equation}
    \sum_i z_i\partial_i - \bar{z}_i\overline{\partial}_i \rightarrow [L_0 + (2N_1 - 1)\Tilde{a}_0/\sqrt{2} + N_1(N_1 - 1) - 1]
\end{equation}
As for the $\nu = 1$ case, we can see that all CFT states with the same $\Tilde{a}_0$ (which map to states of a fixed number of particles) and $L_0$ eigenvalues will map to wave functions with the same angular momentum.

\subsubsection{$\nu = n$}
We can now easily generalise from the $\nu = 2$ case. To generate the ground state we  require $n$ independent free chiral boson fields $\fe{i}(z)$ along with their Klein factors $F_i$. The corresponding vertex operators are defined as $V_j(z) = F_j:e^{i\fe{j}(z)}:$ with the generating $\Omega$ operator given by,
\begin{equation}
    \Omega(z, \bar{z}) = \sum_{j=1}^n \bar{z}^{j-1}V_j(z)
\end{equation}
Let $N_1$ be a positive integer. Then for systems of $N = nN_1 + n(n-1)/2$ particles, we can show, using a simple extension of the method used for $\nu = 2$ that,
\begin{equation}
    \begin{split}
        \Psi_{\bra{0}}(\mathbf{z},\bar{\mathbf{z}}) =& \bra{0}(F_n^\dagger)^{N_1 + n - 1} (F_{n-1}^\dagger)^{N_1 + n - 2} \dots (F_1^\dagger)^{N_1} \\
        &\times \prod_{i=1}^N \Omega(z_i,\bar{z}_i) \ket{0} \\
        =&
        \begin{vmatrix}
            z_1^{N_1 - 1} & \cdots & z_N^{N_1 - 1} \\
            \vdots & \ddots & \vdots \\
            z_1 & \cdots & z_N \\
            1 & \cdots & 1 \\
            \bar{z_1}z_1^{N_1} & \cdots & \bar{z}_N z_N^{N_1} \\
            \vdots & \ddots & \vdots \\
            \bar{z_1}z_1 & \cdots & \bar{z}_Nz_N \\
            \bar{z_1} & \cdots & \bar{z}_N \\
             \vdots & \ddots & \vdots \\
             \bar{z_1}^{n-1}z_1^{N_1+n-2} & \cdots & \bar{z}^{n-1}_N z_N^{N_1+n-2} \\
            \vdots & \ddots & \vdots \\
            \bar{z_1}^{n-1}z_1 & \cdots & \bar{z}^{n-1}_Nz_N \\
            \bar{z_1}^{n-1} & \cdots & \bar{z}^{n-1}_N \\
        \end{vmatrix}
    \end{split}
\end{equation}
Thus we can generate the non-interacting $\nu = n$ ground state wave function. 

For the edge state mapping, one can show, again using a simple generalisation of the $\nu = 2$ case that,
\begin{equation}
    \begin{split}
        \ket{\Psi_{\bra{v}V_{i,k}}}\rangle =& (-1)^{N - N_i} \Tilde{d}_{i,N_1 - k - 1/2} \ket{\Psi_{\bra{v}}}\rangle \\
        \ket{\Psi_{\bra{v}V^\dagger_{i,k}}}\rangle =& (-1)^{N - N_i} \Tilde{c}^\dagger_{i,N_1 + k - 1/2} \ket{\Psi_{\bra{v}}}\rangle \\
    \end{split}
\end{equation}
where $\Tilde{c}^\dagger_{j,m}$ is the creation operator for the orbital $\bar{z}^{j-1}z^{m+j-1}$ and $\Tilde{d}_{j,m}^\dagger$ is the creation operator for a single particle orbital such that $\{ \Tilde{d}_{i,m}, \Tilde{c}_{j,m'}^\dagger \} = \delta_{ij}\delta_{mm'}$. These operator mappings are strictly only valid, once again, for $N_i - k - 1/2 \geq 0$ for $V_{i,k}$ and $N_i + k - 1/2 \geq 0$ for $V^\dagger_{i,k}$. Hence, the edge state mapping will generate the entire edge space for general $\nu = n$ with states of the form $\bra{0}($ product of $V_{i,k}$'s and $V^\dagger_{i,k}$'s $)$ generating the Slater determinant wave functions. In principle, one can calculate the large $N$ edge state inner products, however, this will not be discussed here.

Finally, the angular momentum operator maps over as,
\begin{equation}
    \begin{split}
        \sum_i z_i\partial_i - \bar{z}_i\overline{\partial}_i \rightarrow & L_0 + \frac{2N_1 - 1}{2}\big ( \sum_{i=1}^n \fem{i}_0 \big ) \\
        & + \frac{nN_1(N_1 - 1)}{2} - \frac{n(n-1)}{2}
    \end{split}
\end{equation}

\section{CFT construction of trial wave functions and edge state mappings} \label{Sec:CFTPartonBig}
In Sec. \ref{Sec:CFTMthods} we discussed how previous works had written certain FQHE trial wave functions using a correlation function of a CFT \textit{defined} by some chiral algebra $\mathcal{A}$, which was denoted $\CFT{\mathcal{A}}$. Assuming the generalised screening hypothesis, $\CFT{\mathcal{A}}$ contains the information on the corresponding wave functions' topological properties. For any such wave function, one would never work directly with $\CFT{\mathcal{A}}$. Instead, one would \textit{embed} $\CFT{\mathcal{A}}$ in another CFT for which we know how to calculate the correlation functions, and then, the correlation functions of $\CFT{\mathcal{A}}$ would be computed using this other CFT. It should be emphasised that throughout this section we are assuming that for all the chiral algebras $\mathcal{A}$ considered that $\CFT{\mathcal{A}}$ exists. This is a mathematical question well beyond the scope of this paper, which we elaborate on further in Sec. \ref{Sec:PartonGenWFCon}.

In the following section, we  show how both unprojected and projected composite fermion and parton ground and edge state trial wave functions, as defined in Sec. \ref{Sec:CFandPartonWF}, can be written as CFT correlation functions without explicit anti-symmetrization. In each case, we  first detail the CFT that will actually be used to compute the correlation functions, which we denote $\CFT{L}$. The CFTs that are used are directly motivated by either the composite fermion or parton edge theories. Within each $\CFT{L}$ we  define a sub-chiral algebra $\mathcal{A}$ which is generated by the possible ``electron'' operators, and their conjugates, in the given edge theory. This gives an embedding of $\CFT{\mathcal{A}}$ in $\CFT{L}$. We then demonstrate how $\CFT{L}$ can be used to generate the corresponding wave function, where it will then be demonstrated that the generating correlation function is actually a correlation function of $\CFT{\mathcal{A}}$, which generally follows from the fact that the wave function is generated by these ``electron'' operators. This is important when formulating a generalised screening hypothesis, as there generally exist operators of $\CFT{L}$ which do not change the wave function when inserted into the generating correlation function. The degrees of freedom corresponding to such fields would then remain gapless in the perturbed field theory used to formulate the generalised screening hypothesis, which would complicate matters. 

We then go on to discuss the edge-state mapping, in each case. These edge-state mappings can always be understood in two ways: a linear map from $\CFT{L}$ to the space of wave functions or a linear map from $\CFT{\mathcal{A}}$ to the space of wave functions. The map from $\CFT{L}$ will be such that the states orthogonal to the space of $\CFT{\mathcal{A}}$ states embedded in $\CFT{L}$, will map to zero. We also demonstrate, in certain specific cases, that the existence of these states that map to zero can be shown without any reference to $\CFT{\mathcal{A}}$ and can be understood as a ``gauge'' symmetry that naturally arises in the CFT formalism, which is the same ``gauge'' symmetry imposed on the usual corresponding edge theory.

Importantly, we also show for each edge state map how one can determine the angular momentum of the wave function a given CFT state will map to. This then allows us to give rigorous upper bounds for edge state counting in terms of state counting in the corresponding $\CFT{\mathcal{A}}$.

A more detailed breakdown of the following section is as follows.

We give two detailed examples of these CFT constructions. The first is the $\nu = 2/5$ composite fermion state (see Ssc. \ref{Sec:CFWFCon}), where the construction used here is directly inspired by that of Kvorning \cite{Kvorning2013}. The second example is the $\nu = 1$ bosonic $\phi_2^2$ parton state (see Sec. \ref{Sec:PartonWFCon}). The symmetric parton states of the form $\phi^m_n$ are also discussed in Sec. \ref{Sec:PartonWFCon}, where it is shown that the corresponding ground and edge state wave functions can be expressed using the $\hat{\mathfrak{u}}(1) \oplus \widehat{\mathfrak{su}}(n)_m$ WZW model. 

The general construction for the chiral composite fermion states is detailed in Sec. \ref{Sec:CFGenCaseCon}. In Sec. \ref{Sec:PartonGenWFCon} the general chiral parton case is discussed where it is shown that all chiral LLL projected and unprojected parton ground and edge state trial wave functions can be generated using CFT correlation functions with each state having a corresponding chiral algebra $\mathcal{A}$. Finally, in Sec. \ref{Sec:CFGenCaseCon} we discuss the general structure that the parton wave functions point towards for generating trial ground and edge states using CFT, which potentially could be used to generate new trial wave functions beyond the parton states.

\subsection{Composite fermions} \label{Sec:CFWFCon}

\subsubsection{$\nu = 2/5$ example: The CFT} \label{Sec:CFCFT}
We begin with three chiral boson fields on the complex plane, $\fef{1}(z)$, $\fef{2}(z)$ and $\phi(z)$, with the following mode expansions,
\begin{equation}
    \begin{split}
        \fef{j}(z) =& -i\fefm{j}_0 \ln z + \sum_{n \neq 0} \frac{1}{n}\fefm{j}_n z^{-n} \\
        \phi(z) =& -i\fmp_0 \ln z + \sum_{n \neq 0} \frac{1}{n} \fmp_n z^{-n} \\
    \end{split}
\end{equation}
where,
\begin{equation}
    \begin{split}
        [\fefm{i}_n, \fefm{j}_m] =& n \delta_{ij} \delta_{n+m,0} \\
        [\fmp_n, \fmp_m] =& n \delta_{n+m,0} \\
    \end{split}
\end{equation}
with all other commutation relations being trivial.

The $\fef{i}(z)$ are the same chiral bosons used to construct the $\nu = 2$ wave functions in Sec. \ref{Sec:integerQHCFT}, with their corresponding Klein factors denoted by $\Tilde{F}_i$. The $\phi(z)$ is different as its $\fmp_0$ eigenvalues are quantized to be multiples of $\sqrt{2}$ (as opposed to one) with its corresponding Klein factor denoted by $F_\phi$, with the following \textit{commutation} relations,
\begin{equation} \label{Eq:KleinCom}
    \begin{split}
        [F_\phi, \Tilde{F}_i] &= 0 \\
        [\fefm{i}_n, F_\phi] &= 0 \\
        [\fmp_n, F_\phi] &= \sqrt{2}\delta_{n,0}F_\phi \\
    \end{split}
\end{equation} 
One should note that as $\fmp_0$ eigenvalues are quantised to multiples of $\sqrt{2}$ it's corresponding vertex operator $V_\phi(z) = F_\phi : e^{i\sqrt{2}\phi(z)} :$ will have scaling dimension one and so must be a \textit{bosonic} field. This is why $F_\phi$ commutes with $\Tilde{F}_i$. The corresponding vertex operators of the $\fef{i}(z)$ fields will be writen as $\Tilde{V}_j(z) \equiv \Tilde{F}_j:e^{i\fef{j}(z)}:$.

The \textit{vacuum} state, $\ket{0}$ is defined by, $\fefm{i}_n \ket{0} = \fmp_n \ket{0} = 0$ for $n \geq 0$. All states of the Hilbert space, $\mathcal{H}_{\text{CFT}}$, can be expressed as polynomials in $\fefm{i}_{-n}$, $\fmp_{-n}$, $\Tilde{F}_i$, $\Tilde{F}^\dagger_i$, $F_\phi$ and $F^\dagger_\phi$ acting on the vacuum sate (where we only need to use the field modes with $n > 0$). This then defines the $\CFT{L}$ that will later be used for the $\nu = 2/5$ wave function.

In terms of a possible edge theory for the $\nu = 2/5$ state, one can think of $\fef{i}(z)$ as the bosonized composite fermions and $\phi(z)$ as the edge mode resulting from the flux attaching Chern-Simons theory of the bulk, where $i\sqrt{2}\partial \phi(z)$ represents the density of flux quanta on the edge. From this perspective, we know that once flux attachment has been enforced the only allowed fluctuations will be in the fields,
\begin{equation}
    \begin{split}
        \fe{1}(z) = \sqrt{2}\phi(z) + \fef{1}(z) \\
        \fe{2}(z) = \sqrt{2}\phi(z) + \fef{2}(z) \\
    \end{split}
\end{equation}
whose modes we  denote as $\fem{i}_n$.
One can easily check that $[\fem{i}_n, \fem{j}_m] = nK_{ij}\delta_{n+m, 0}$, where $K_{ij}$ is the $K$-matrix of the $\nu = 2/5$ state, with $K_{ij}$ given by,
\begin{equation}
    K = 
    \begin{pmatrix}
        3 & 2 \\
        2 & 3 \\
    \end{pmatrix}
\end{equation}. 

The corresponding Klein factors for these fields are,
\begin{equation}
    F_1 \equiv F_\phi \Tilde{F}_1 \quad F_2 \equiv F_\phi \Tilde{F}_2
\end{equation}
Hence, as an edge theory, the space of physical states, $\mathcal{H}_{\text{phys}} \subset \mathcal{H}_{\text{CFT}}$, is generated by polynomials in $F_i$, $F_i^\dagger$ and the modes of $\fe{i}(z)$, applied on the vacuum state. 

Within the physical CFT, the fields that create and annihilate composite fermions are the vertex operators given by,
\begin{equation}
    \begin{split}
        V_j(z) &\equiv F_j: e^{i\fe{j}(z)} : \\
        V_j^\dagger(z) &\equiv F_j^\dagger : e^{-i\fe{j}(z)} : \\
    \end{split}
\end{equation}
The operator product expansion (OPE) of these fields is given by,
\begin{equation}
    V^\dagger_j(z)V_j(w) \sim \frac{1}{(z - w)^{K_{jj}}} + \frac{i\partial \fe{j}(w)}{(z - w)^{K_{jj}-1}} + \dots
\end{equation}
It follows from the OPE that we can generate all the modes of $i\partial \fe{j}(z)$ through anti-commutations between the modes of $V^\dagger_j(z)$ and $V_j(w)$. In addition, it can also be shown that a given polynomial in $F_j$ and $F^\dagger_j$ applied on the vacuum is equivalent to some polynomial in the modes of $V_j(z)$ and $V^\dagger_j(z)$ applied on the vacuum. We thus see that all the states of $\mathcal{H}_{\text{phys}}$ can be generated by polynomials in the modes of these fields. 

We then identify the chiral algebra generated by repeated OPEs of $V_j(z)$ and $V_j^\dagger(z)$ as the chiral algebra $\mathcal{A}$ corresponding to this $\nu = 2/5$ state. The physical edge Hilbert space $\mathcal{H}_{\text{phys}}$ then forms the vacuum representation of $\mathcal{A}$.

Another way of defining $\mathcal{H}_{\text{phys}}$ is through a \textit{gauge-invarience} condition. Let $\mathcal{J}(z) = \sqrt{\frac{2}{5}}[i\partial\fef{1}(z) + i\partial\fef{2}(z)] - \frac{1}{\sqrt{5}}i\partial\phi(z)$. We can then define $\mathcal{H}_{\text{phys}}$ as the states invariant under transformations generated by $\mathcal{J}(z)$, $\ket{v} \in \mathcal{H}_{\text{\text{phys}}} \Leftrightarrow \mathcal{J}_n \ket{v} = 0$ for $n \geq 0$. As we can write the modes $\fefm{i}_m$ and $\fem{\phi}_n$ in terms of $\fem{i}_n$ and $\mathcal{J}_n$, we can choose a basis of $\mathcal{H}_{\text{CFT}}$ with elements of the form $\prod_{n_i}\fem{1}_{-n_i}\prod_{n_j}\fem{2}_{-n_j}\prod_{n_k}\mathcal{J}_{-n_k} \Tilde{F_1}^{p_1}\Tilde{F_2}^{p_2}F_\phi^{p_\phi}\ket{0}$ with $n_i,n_j,n_k > 0$ and $p_1, p_2, p_\phi \in \mathbb{Z}$.
The condition $\mathcal{J}_0\ket{v} = 0$ implies that $\ket{v}$ must be a linear combination of basis elements with $p_1 + p_2 = p_\phi$. Next, the condition $\mathcal{J}_n\ket{v} = 0$ for $n \geq 0$ implies that $\ket{v}$ must be a linear combination of basis elements of the form $\prod_{n_i}\fem{1}_{-n_i}\prod_{n_j}\fem{2}_{-n_j} \Tilde{F_1}^{p_1}\Tilde{F_2}^{p_2}F_\phi^{p_1 + p_2}\ket{0}$, which are themselves gauge invariant and so must span $\mathcal{H}_{\text{phys}}$. Hence, this gauge invariance definition of $\mathcal{H}_{\text{phys}}$ is equivalent to our earlier definition.

Within an edge theory for these composite fermions there should be operators that generate $SU(2)$ rotations of the composite fermion orbitals (i.e. $\psi_n^i \rightarrow U_{i1}\psi_n^1 + U_{i2}\psi_n^2$ for $U \in SU(2)$). One then expects there to be $SU(2)$ currents within the edge theory. To this end, consider the fields,
\begin{equation} \label{Eq:CFSU2Fields}
    \begin{split}
        J^3(z) &\equiv \frac{i\partial\fe{2}(z) - i\partial\fe{1}(z)}{2} \\
        J^\pm(z) &\equiv  (F_2 F^\dagger_1)^{\pm 1} : e^{\pm i (\fe{2}(z) - \fe{1}(z))} : \\
        \Phi(z) &\equiv \frac{1}{\sqrt{10}} ( \fe{1}(z) + \fe{2}(z) ) 
    \end{split}
\end{equation}
We write the modes of $i\partial\Phi(z)$ as $a_n$. The OPEs of these fields are given by,
\begin{equation}
    \begin{split}
        J^3(z)J^3(w) &\sim \frac{1/2}{(z-w)^2} \\
        J^3(z)J^\pm(w) &\sim \pm \frac{J^\pm (w)}{z - w} \\
        J^+(z)J^-(w) &\sim \frac{1}{(z-w)^2} + \frac{2J^3(w)}{z-w} \\
        i\partial\Phi(z) i \partial \Phi(w) &\sim \frac{1}{(z-w)^2} \\
    \end{split}
\end{equation}
with all other OPEs between these fields being regular. Thus, in a similar way to the $\nu = 2$ case in Sec. \ref{Sec:integerQHCFT}, we then see that this theory contains a $\hat{\mathfrak{u}}(1) \oplus \widehat{\mathfrak{su}}(2)_1$ Kac-Moody algebra, with $i\partial\Phi(z)$ representing the $\hat{\mathfrak{u}}(1)$ current and the $J$'s being the $\widehat{\mathfrak{su}}(2)_1$ currents. From its definition, the $U(1)$ charge of $\Phi$ is related to the total number of composite fermions by, $($number of composite fermions$) = \sqrt{\nu}\times(U(1)$ charge of $\Phi)$. In other words, $\sqrt{\nu}i\partial\Phi(z)$ would represent the density of composite fermions as an edge theory. We also have that, $J^i(z)V_j(w) \sim (1/2)\sum_k \sigma^i_{p(j),p(k)} V_k(w)/(z-w)$, where $\sigma^i$ are the usual Pauli spin matrices and $p$ is a permutation with $p(1) = 2$ and $p(2) = 1$. Hence, the $V_j(z)$'s transform as a spin $1/2$ representation under these $SU(2)$ transformations. This can be interpreted as the $J$'s generating $SU(2)$ transformations of the composite fermion orbitals. The construction used here is the free field representation of the $SU(2)_1$ WZW model \cite{DiFrancesco1997, kass_affine_1990}. In terms of the $\hat{\mathfrak{u}}(1)\oplus \widehat{\mathfrak{su}}(2)_1$ WZW model, we can represent the fields $V_j(z)$ as $V_1(z) = :e^{i\sqrt{5/2}\Phi(z)}:\phi_{1/2,-1/2}(z)$ and $V_2(z) = :e^{i\sqrt{5/2}\Phi(z)}:\phi_{1/2,1/2}(z)$, where $\phi_{1/2,m}(z)$ are the spin-$\frac{1}{2}$ $\widehat{\mathfrak{su}}(2)_1$ WZW primary fields. To simplify notation the Klein factor for $\Phi(z)$ is implicit. We would like to emphasise that this is one way of representing the chiral algebra $\mathcal{A}$ and $\CFT{\mathcal{A}}$ is not strictly equivalent to $\hat{\mathfrak{u}}(1) \oplus \widehat{\mathfrak{su}}(2)_1$ WZW model.

The space of states with a fixed number of composite fermions, $N_{CF}$, can be generated by sums of states of the form, $($polynomial in modes of $\fe{j}(z))\times($product of $(F_2 F^\dagger_1)^{\pm 1})\times (F_2F_1)^{N_{CF}-\delta_{N_{CF}}}F_2^{\delta_{N_{CF}}}\ket{0}$, where $\delta_{N_{CF}} = N_{CF} (\text{mod }2)$. One can show that the product of $(F_2 F^\dagger_1)^{\pm 1}$ can be replaced by a product of the modes of $J^\pm(z)$. In addition, by their definitions the polynomial in the modes of $\fe{j}(z)$ can be replaced by a polynomial in the modes of $J^3(z)$ and $\Phi(z)$. Consequently, the space of states of fixed $N_{CF}$ forms an irreducible representation of the $\hat{\mathfrak{u}}(1) \oplus \widehat{\mathfrak{su}}(2)_1$ algebra. For an even $N_{CF}$ there is only one state with the lowest $L_0$ eigenvalue, and hence this space of states will give a spin-$0$ representation of $\widehat{\mathfrak{su}}(2)_1$. For an odd $N_{CF}$ there are two states with the lowest $L_0$ eigenvalue and so these spaces of fixed $N_{CF}$ will form a spin-$\frac{1}{2}$ representation of $\widehat{\mathfrak{su}}(2)_1$. Viewing $\mathcal{H}_{\text{phys}}$ from the perspective of this algebra is useful in that it allows us to more easily understand the effect of redefining the composite fermion orbitals via some $SU(2)$ rotation. This is the main idea behind Witten's non-Abelian bosonization \cite{witten_non-abelian_1984}.

\subsubsection{$\nu = 2/5$ example: ground state wave function} \label{Sec:CFWFCFTCon}
We  now show how the $\nu = 2/5$ ground state trial wave function can be generated using the above CFT without explicit anti-symmetrization. The unprojected composite fermion wave function will be considered first and then the projected one. In the following, we  only consider the case where the ground state trial wave functions contain an odd number, $N$, of fermions. 

We first define the background charge operator as $C(N) = (F_2^\dagger)^{N_2}(F_1^\dagger)^{N_1}$, where $N_1 = (N-1)/2$ and $N_2 = (N+1)/2$. We then define the generating $\Omega$ operator as,
\begin{equation}
    \Omega(z,\bar{z}) = V_1(z) + \bar{z}V_2(z)
\end{equation}
We now consider the wave functions,
\begin{equation}
    \begin{split}
        \Psi_{\bra{0}}(\mathbf{z},\bar{\mathbf{z}}) =& \bra{0}C(N)\prod_{i = 1}^N \Omega(z_i,\bar{z}_i)\ket{0} \\
        =& \bra{0}C(N) \prod_{i = 1}^N (\Tilde{V}_1(z_i) + \bar{z}_i\Tilde{V}_2(z_i))V_\phi(z_i) \ket{0} \\
        &= \bra{0}(\Tilde{F}_2^\dagger)^{N_2} (\Tilde{F}_2^\dagger)^{N_2}  \prod_{i = 1}^N (\Tilde{V}_1(z_i) + \bar{z}_i\Tilde{V}_2(z_i)) \ket{0} \\
        &\times \bra{0} (F^\dagger_\phi)^N \prod_{j=1}^N V_\phi(z_j) \ket{0}
    \end{split}
\end{equation}
where the correlation functions factorise as the $\Tilde{V}_j(z)$ are independent of $V_\phi(z)$. From Sec. \ref{Sec:integerQHCFT} we know that the first factor will generate the $\nu = 2$ ground state. The second factor will generate the $\nu = 1/2$ bosonic Laughlin wave function. Hence, the wave function is given by,
\begin{equation}
    \begin{split}
        \Psi_{\bra{0}}(\mathbf{z},\bar{\mathbf{z}}) =&
        \begin{vmatrix}
            z_1^{N_1 - 1} & z_2^{N_1 - 1} & \cdots & z_N^{N_1 - 1} \\
            \vdots & \vdots & \ddots & \vdots \\
            z_1 & z_2 & \cdots & z_N \\
            1 & 1 & \cdots & 1 \\
            \bar{z}_1z_1^{N_2 - 1} & \bar{z}_2z_2^{N_2 - 1} & \cdots & \bar{z}_Nz_N^{N_2 - 1} \\
            \vdots & \vdots & \ddots & \vdots \\
            \bar{z}_1 z_1 & \bar{z}_2 z_2 & \cdots & \bar{z}_N z_N \\
            \bar{z}_1 & \bar{z}_2 & \cdots & \bar{z}_N \\
        \end{vmatrix} \\
        &\times \prod_{i<j}^N(z_i - z_j)^2
    \end{split}
\end{equation}
which is the unprojected $\nu = 2/5$ composite fermion trial ground state wave function.

To project this wave function to the lowest Landau level we move the $\bar{z}$'s to the left and replace them with $2\partial$. Hence, to generate the projected wave function we simply replace $\Omega(z,\bar{z})$ with $\Omega(z) = V_1(z) + 2\partial V_2(z)$. In full we have,
\begin{equation}
    \begin{split}
        P_{LLL} \Psi_{\bra{0}}(\mathbf{z},\bar{\mathbf{z}}) =& \bra{0}C(N) \prod_{i = 1}^N \Omega(z_i) \ket{0} \\
        =& \bra{0}C(N) \prod_{i = 1}^N( V_1(z_i) + 2\partial_i V_2(z_i)) \ket{0} \\
    \end{split}
\end{equation}

We  now show that these correlation functions can be entirely computed within the vacuum representation of the chiral algebra $\mathcal{A}$ generated by $V_i(z)$ and $V^\dagger_i(z)$. First of all as $[C(N), \mathcal{J}(z)] = 0$, then for any $\ket{v} \in \mathcal{H}_{\text{phys}}$ we have $C(N)\ket{v} \in \mathcal{H}_{\text{phys}}$. Hence, as $\mathcal{H}_{\text{phys}}$ is entirely equivalent to the vacuum representation of $\mathcal{A}$, $C(N)$ can be redefined as purely an operator within this vacuum representation. Thus, the state $\ket{x} \equiv C(N) \prod_{i=1}^N(V_1(z_i) + \bar{z}_iV_2(z_i))\ket{0}$ belongs to this vacuum representation. The final correlation function can be compactly written as $\braket{0|x}$ and as $\ket{0}$ belongs to the vacuum representation this inner product can be entirely computed within the vacuum representation of $\mathcal{A}$. As we have seen here, however, it can be easier to compute the correlation function via embedding in a ``larger'' CFT. 

Finally, we would like to point out, from the discussion of Sec. \ref{Sec:CFCFT}, we can see that, in principle, we could have generated this wave function with the $\hat{\mathfrak{u}}(1) \oplus \widehat{\mathfrak{su}}(2)_1$ WZW model. This is merely another way of computing the correlation functions of the chiral algebra $\mathcal{A}$.

\subsubsection{$\nu = 2/5$ example: CFT to edge state mapping} \label{Sec:CFEdgeMapping}
Throughout this section, we  focus only on the unprojected wave functions. For any unprojected edge state that is generated by the following mapping, we can generate its projected wave function by simple replacement $\Omega(z,\bar{z}) \rightarrow \Omega(z) = V_1(z) + 2\partial V_2(z)$.

The edge state mapping is defined in second quantisation notation as,
\begin{equation} \label{Eq:CFEdgeStateMap}
    \ket{\Psi_{\bra{v}}}\rangle \equiv \bra{v} C(N) e^{\int d^2z e^{-|z|^2/4} \Omega(z, \bar{z})\otimes c^\dagger(z,\bar{z}) } \ket{0}\otimes\ket{0}\rangle
\end{equation}
When $\bra{v}$ has a definite $U(1)$ charge, $q$, with respect to the current $i\partial\Phi(z)$, this mapping will produce a state with a definite number of particles as the CFT correlation function must be $U(1)$ charge neutral. The resulting state will have $N + \sqrt{\nu}q$ particles with the wave function,
\begin{equation}
    \Psi_{\bra{v}}(\mathbf{z},\bar{\mathbf{z}}) = \bra{v}C(N)\prod_{i=1}^{N + \sqrt{\nu}q}\Omega(z_i,\bar{z}_i) \ket{0}
\end{equation}

As $C(N)\prod_{i=1}^{N + \sqrt{\nu}q}\Omega(z_i,\bar{z}_i) \ket{0} \in \mathcal{H}_{\text{phys}}$, all the states in the orthogonal complement of $\mathcal{H}_{\text{phys}}$ must map to zero. Hence, if we restrict this map to $\mathcal{H}_{\text{phys}}$ we do not change the image of the resulting map. In other words, we can really think of this edge state mapping as being from the vacuum representation of $\mathcal{A}$ (i.e. $\mathcal{H}_{\text{phys}}$) to the space of wave functions. It is, however, instructive to also consider this as a map from $\mathcal{H}_{\text{CFT}}$ to the space of wave functions, as w now do.

From the discussion of Sec. \ref{Sec:integerQHCFT} and from the fact that any correlation function can be factorised in terms of $\Tilde{V}_i(z)$ and $V_\phi(z)$, all states of the form 
$\psi_{\text{edge}}(\mathbf{z},\bar{\mathbf{z}}) \prod_{i<j}^{N + n}(z_i - z_j)^2$
%$\psi_{\text{edge}}(\mathbf{z},\bar{\mathbf{z}}) \prod_{i<j}^{N + n}(z_I - z_j)^2$
, where $\psi_{\text{edge}}$ is a Slater determinant in the $z^m$ and $\bar{z}z^m$ orbitals, can be generated from CFT states of the form $\bra{v} = \bra{0}(F^\dagger_\phi)^n \prod_{i=1}^2[ \prod_{k^{(i)}_j}V^\dagger_{i, k^{(i)}_j} \prod_{l^{(i)}_j} V_{i,l^{(i)}_j} ]$, where the number of $V^\dagger$'s minus the number of $V$'s must equal $n$ and all $k^{(i)}_j,l^{(i)}_j > 0$. Thus, this image of this map must contain the composite fermion edge space defined in Sec. \ref{Sec:CFandPartonWF}. 

We  now show that the entire image of this map is in fact the composite fermion edge space. First, we note that from standard constructive bosonization (see Sec. \ref{Sec:integerQHCFT}) there must exist an invertible basis transform from the states of the form $\bra{v} = \bra{0}(F^\dagger_\phi)^n \prod_{i=1}^2[ \prod_{k^{(i)}_j}V^\dagger_{i, k^{(i)}_j} \prod_{l^{(i)}_j} V_{i,l^{(i)}_j} ]$, with all $k^{(i)}_j,l^{(i)}_j > 0$, to the states of the form $\bra{0} (F_2^\dagger)^{n_2}(F_1^\dagger)^{n_1} \prod_{i=1}^2[\prod_{m^{(i)}_j}\fefm{i}_{m^{(i)}_j}]$, with all $m^{(i)}_j > 0$. Such states can also be expressed as,
\begin{equation*}
    \begin{split}
        &\bra{0} (F_2^\dagger)^{n_2}(F_1^\dagger)^{n_1} \\
        &\times [\prod_{m^{(1)}_j}( \alpha \fem{1}_{m^{(1)}_j} + \beta \fem{2}_{m^{(1)}_j} + \gamma \mathcal{J}_{m^{(1)}_j} )] \\
        &\times [\prod_{m^{(2)}_j}( \alpha \fem{2}_{m^{(2)}_j} + \beta \fem{1}_{m^{(2)}_j} + \gamma \mathcal{J}_{m^{(2)}_j} )] \\
    \end{split}
\end{equation*}
where $\alpha, \beta, \gamma \in \mathbb{R}$ with $\alpha \neq \beta$. As $[\mathcal{J}_n, \Omega(z, \bar{z})] = 0$ and $\mathcal{J}_n \ket{0} = 0$, the above state must map to the same wave function as,
\begin{equation} \label{Eq:guageBasis}
    \begin{split}
        &\bra{0} (F_2^\dagger)^{n_2}(F_1^\dagger)^{n_1} \\
        &\times [\prod_{m^{(1)}_j}( \alpha \fem{1}_{m^{(1)}_j} + \beta \fem{2}_{m^{(1)}_j})] \\
        &\times [\prod_{m^{(2)}_j}( \alpha \fem{2}_{m^{(2)}_j} + \beta \fem{1}_{m^{(2)}_j})] \\
    \end{split}
\end{equation}
In addition, as $\alpha \neq \beta$ there must exist an invertable basis transformation between the modes $(\alpha \fem{1}_m + \beta \fem{2}_m), (\alpha \fem{2}_m + \beta \fem{1}_m)$ and $\fem{1}_m, \fem{2}_m$. Hence, the states in Eq. \ref{Eq:guageBasis} must form a basis of $\mathcal{H}_{\text{phys}}$. This then implies that the image of the edge state map is, in fact, equivalent to the composite fermion edge space.

The fact that $[\mathcal{J}(w), \Omega(z, \bar{z})] = 0$ and $\mathcal{J}_n\ket{0} = 0$ for $n > 0$, can be interpreted as an emergent gauge redundancy in that the correlation function that generates a given wave function must be invariant under any transformations generated by this current $\mathcal{J}(z)$. As we have seen above, only mapping from the gauge invariant states, $\mathcal{H}_{\text{phys}}$, is sufficient to reproduce the entire image of the map. As discussed in Sec. \ref{Sec:CFCFT}, the gauge invariance condition that can be used to define $\mathcal{H}_{\text{phys}}$ is equivalent to flux attachment (in the edge theory). Hence, in some sense, one can think of the gauge redundancy of the edge state map as enforcing flux attachment.  

Finally, as the angular momentum is the sum of the angular momenta of the $\nu = 2$ and the Jastrow factor components and using the fact that in $\Hp$ $\sqrt{2/5}a_0 = (\fefm{1}_0 + \fefm{2}_0) = \sqrt{1/p}\fem{\phi}_0$, the angular momentum operator maps over to the CFT as,
\begin{equation} \label{Eq:nu25Angular}
    \begin{split}
        &\sum_{i = 1}^N (z_i \partial_i - \bar{z}_i\partial_i)\Psi_{\bra{v}}(\mathbf{z},\bar{\mathbf{z}}) \\
        &\rightarrow L_0 + \bigg [ \frac{2N_1-1}{2} + 2N - 1 \bigg ]\sqrt{\frac{2}{5}}a_0 \\
        & + N_1(N_1-1) - 1 + N(N-1)
    \end{split}
\end{equation}
As for the integer quantum Hall cases, the angular momentum mapping implies that for a fixed number of particles the edge state mapping will map states with the same $L_0$ eigenvalue to wave functions with the same angular momentum.

\subsubsection{General case} \label{Sec:CFGenCaseCon}
The discussion of Sec. \ref{Sec:CFCFT}, \ref{Sec:CFWFCFTCon} and \ref{Sec:CFEdgeMapping} can be straightforwardly generalised to generate the $\nu = \frac{n}{np + 1}$ composite fermion wave function, with $n, p \in \mathbb{Z}^+$, where even and odd $p$ corresponds to fermionic and bosonic composite fermion states respectively. One expects to use the CFT involving $n$ independent chiral boson fields, $\fef{i}(z)$, with compactification radius one and corresponding Klein factors $\Tilde{F}_i$, along with another independent chiral boson field $\phi(z)$ with $\fmp_0$ eigenvalues quantised to multiplies of $\sqrt{p}$ and corresponding Klein factor $F_\phi$ with $[\fem{\phi}_n, F_\phi] = \sqrt{p}\delta_{n,0}F_\phi$. The corresponding vertex operators are defined by $\Tilde{V}_j = \Tilde{F}_j :e^{i\fef{j}(z)}:$ and $V_\phi(z) = F_\phi :e^{i\sqrt{p}\phi(z)}:$.

One can think of this CFT as a possible edge theory for the corresponding composite fermion state, with $\fef{j}(z)$ being the bosonized $j^{\text{th}}$ Lambda level and $\sqrt{p}i\partial\phi(z)$ representing the density of flux quanta on the edge. Of course, a proper composite fermion edge theory must have flux attachment, with the space of flux attached states denoted by $\mathcal{H}_{\text{phys}}$. This space can be defined as the ``gauge'' invariant states, from the transformations generated by the current $\mathcal{J}(z) = \sqrt{p/(np + 1)}(\sum_{j=1}^n i\partial\fef{j}(z)) - \sqrt{np + 1}^{-1}i\partial\phi(z)$, $\ket{v} \in \mathcal{H}_{\text{phys}} \Leftrightarrow \mathcal{J}_n\ket{v} = 0$ for $n \geq 0$. By a simple extension of argument for $\nu = 2/5$, $\mathcal{H}_{\text{phys}}$ can be generated by polynomials in the modes of $V_j(z) \equiv \Tilde{V}_j(z)V_{\phi}(z)$ and their conjugates applied on the vacuum state $\ket{0}$. The chiral algebra $\mathcal{A}$ generated by repeated OPEs of $V_j(z)$ and $V^\dagger_j(z)$ is then identified with the corresponding composite fermion state. Consequently, $\mathcal{H}_{\text{phys}}$ must form the vacuum representation of $\mathcal{A}$.

This edge theory also contains the currents $\Phi(z) \equiv \sqrt{n(np + 1)}^{-1}(\sum_{j=1}^n \fef{j}(z) + \sqrt{p}\phi(z))$ and $J^a(z) = \sum_{ij}t^a_{ij}:V_i(z)V^\dagger_j(z):$, where $t^a$ are the generators of $SU(n)$. These fields have the OPEs of the currents of the $\hat{\mathfrak{u}}(1) \oplus \widehat{\mathfrak{su}}(n)_1$ WZW model. Any space of states with a fixed number of composite fermions in $\mathcal{H}_{\text{phys}}$ can be generated by polynomials in the modes of the fields $J^a(z)$ and $i\partial\Phi(z)$ applied on some state within that subspace, with this result following from a simple extension of non-Abelian bosonization. Furthermore, the chiral algebra $\mathcal{A}$ can be represented by fields of the $\hat{\mathfrak{u}}(1) \oplus \widehat{\mathfrak{su}}(n)_1$ WZW model, with $V_j(z) = :e^{i\Phi(z)/\sqrt{\nu}}:\phi_j(z)$ where $\phi_j(z)$ are WZW primaries corresponding the fundamental representation of $SU(n)$. 

One can define a background charge operator as $C(N) = C_n(N) (F^\dagger_\phi)^N$, where $C_n(N)$ is the background charge used to generate the $\nu = n$ integer quantum Hall wave function using the $\Tilde{V}_j(z)$ fields, which is given in Sec. \ref{Sec:integerQHCFT}. We have $[C(N),\mathcal{J}(z)] = 0$, so $\mathcal{H}_{\text{phys}}$ is an invariant subspace under the action of the background charge operator, which means $C(N)$ can be defined purely as an operator that acts within the vacuum representation of $\mathcal{A}$. The generating $\Omega$ for the unprojected wave function is given by,
\begin{equation}
    \Omega(z, \bar{z}) = \sum_{j=1}^n \bar{z}^{j-1} V_j(z)
\end{equation}
where the ground state wave function is given by the usual form $\Psi_{\bra{0}}(\mathbf{z},\bar{\mathbf{z}}) = \bra{0}C(N) \prod_{i=1}^N \Omega(z_i, \bar{z}_i)\ket{0}$. As we can also write $\Omega(z, \bar{z}) = [\sum_{j=1}^n \bar{z}^{j-1} \Tilde{V}_j(z)]V_\phi(z)$ this correlation function will factorize into a correlation function that generates the $\nu = n$ non-interacting integer quantum Hall ground state and another which generates the $\nu = 1/p$ Laughlin trial wave function. Hence, $\Psi_{\bra{0}}(\mathbf{z},\bar{\mathbf{z}}) = \Phi_n(\mathbf{z},\bar{\mathbf{z}})\prod_{i<j}(z_i-z_j)^p$, where $\Phi_n(\mathbf{z},\bar{\mathbf{z}})$ is the non-interacting $\nu = n$ integer quantum Hall ground state. Thus, we can generate the general unprojected composite fermion ground trial wave function. To generate the projected composite fermion ground state wave function one can simply use the replacement $ \Omega(z, \bar{z}) \rightarrow \Omega(z) = \sum_{j=1}^n (2\partial)^{j-1} V_j(z)$. 

As for the $\nu = 2/5$ case, the fact that $C(N) \prod_{i=1}^N \Omega(z_i, \bar{z}_i)\ket{0} \in \mathcal{H}_{\text{phys}}$ implies that the generating correlation function $\bra{0}C(N) \prod_{i=1}^N \Omega(z_i, \bar{z}_i)\ket{0}$ can be computed entirely within the vacuum representation of $\mathcal{A}$.

The edge state mapping is given by the obvious generalisation of Eq. \ref{Eq:CFEdgeStateMap}. From the discussion of Sec. \ref{Sec:integerQHCFT} and from the factorisation of the resulting correlation function, all composite fermion edge states can be generated by CFT states of the form $\bra{v} = \bra{0}(F^\dagger_\phi)^m \prod_{i=1}^n[ \prod_{k^{(i)}_j}V^\dagger_{i, k^{(i)}_j} \prod_{l^{(i)}_j} V_{i,l^{(i)}_j} ]$. Hence, the space of composite fermion edge state trial wave functions generally belongs to the image of the edge state map. Furthermore, as $[\mathcal{J}_n, \Omega(z, \bar{z})] = 0$ and $\mathcal{J}_n \ket{0} = 0$ for $n \geq 0$, the edge state map is invariant under transformations generated by $\mathcal{J}(z)$. By a simple generalisation of the argument given for $\nu = 2/5$, this implies that the image of the edge state map is equivalent to the composite fermion edge space. 

Finally, by a simple generalisation of the calculation for the $\nu = 2/5$ case, the angular momentum operator can be mapped to the CFT as,
\begin{equation}
    \begin{split}
        \sum_i z_i\partial_i - \bar{z}_i\overline{\partial}_i \rightarrow & L_0 + \big [ \frac{(2N_1-1)}{2} + \frac{p(2N-1)}{2} \big ]\sqrt{\nu}a_0 \\
        & + \frac{pN(N-1)}{2} + \frac{nN_1(N_1 - 1)}{2} \\ 
        & - \frac{n(n-1)}{2} \\ 
    \end{split}
\end{equation}
Once again, the main point that is emphasised by this mapping is that the edge state counting at fixed angular momentum, and a fixed number of particles, has a rigorous upper bound given by the number of linearly independent $\Hp$ states with the corresponding $a_0$ and $L_0$ eigenvalues.

\subsection{Parton states} \label{Sec:PartonWFCon}

\subsubsection{$\phi_2^2$ example: the CFT} \label{Sec:partonCFT}
Let $\fef{ij}(z)$ denote four independent chiral boson fields ($i,j=1,2$) with corresponding Klein factors $\Tilde{F}_{ij}$, where each boson field has compactification radius $1$ (i.e. $[\fefm{ij}_0 ,\Tilde{F}_{kl}] = \delta_{ik}\delta_{jl}\Tilde{F}_{ij}$). The Hilbert space of this theory, $\mathcal{H}_{CFT}$, is generated by polynomials of the field modes and Klein factors (both $\Tilde{F}_{ij}$ and $\Tilde{F}^\dagger_{ij}$) applied on the vacuum state, $\ket{0}$. We also denote the corresponding vertex operators as, $V_{jk}(z) \equiv \Tilde{F}_{jk} : e^{i\fef{jk}(z)} :$ and $V^\dagger_{jk}(z) \equiv \Tilde{F}^\dagger_{jk} : e^{-i\fef{jk}(z)} :$. Polynomials in the modes of these vertex operators applied on the vacuum state can also generate the Hilbert space.

As a possible edge theory for this parton state, we  take the first index $i$ in $V_{ij}(z)$ to denote the parton species and $j$ as a Landau level index. Within this theory, there are two actions of an $SU(2)$ algebra: the first rotates the Landau orbitals, $V_{ij} \rightarrow \sum_k U_{jk}V_{ik}$ and the second rotates between the parton species $V_{ij} \rightarrow \sum_k V_{ik}W_{kj}$ with $U, W \in SU(2)$. Furthermore, we have the usual $U(1)$ transformations $V_{jk} \rightarrow e^{i\alpha}V_{jk}$. As for the composite fermion case, these transformations have corresponding currents,
\begin{equation} \label{Eq:partonSU2Currents}
    \begin{split}
        \mathcal{J}^a(z) &\equiv \sum_{ijk} t^a_{ij} : V_{ik}(z) V^\dagger_{jk}(z): \\
        J^a(z) &\equiv \sum_{ijk} t^a_{ij} : V_{ki}(z) V^\dagger_{kj}(z): \\
        i\partial\Phi(z) &\equiv \frac{1}{2} \sum_{jk} i\partial\fef{jk}(z) \\ 
    \end{split}
\end{equation}
where $t^a \equiv \Tilde{\sigma}^a/2$, $\Tilde{\sigma}^a_{ij} = \sigma^a_{p(i),p(j)}$, $\sigma^a$ are the Pauli spin matrices and $p$ is a permutation with $p(1) = 2$ and $p(2) = 1$. We denote the modes of $\mathcal{J}^a(z)$, $J^a(z)$, and $i\partial\Phi(z)$ as $\mathcal{J}^a_n$, $J^a_n$ and $a_n$ respectively. We use these $\Tilde{\sigma}^a$ so that higher parton Landau levels correspond to higher $J^3_0$ eigenvalues. The OPEs of these currents are given by,
\begin{equation}
    \begin{split}
        \mathcal{J}^a(z)\mathcal{J}^b(w) &\sim \frac{\delta_{ab}}{(z - w)^2} + \frac{if^{abc}\mathcal{J}^c(w)}{z - w} \\
        J^a(z)J^b(w) &\sim \frac{\delta_{ab}}{(z-w)^2} + \frac{if^{abc}J^c(w)}{z - w} \\
        i\partial \Phi(z) i\partial \Phi(w) &\sim \frac{1}{(z - w)^2} \\
    \end{split}
\end{equation}
where $f^{abc}$ are the structure constants of $\mathfrak{su}(2)$ and all other OPEs between these currents are regular. It then follows that the modes of these currents must form a $\hat{\mathfrak{u}}(1) \oplus \widehat{\mathfrak{su}}(2)_2 \oplus \widehat{\mathfrak{su}}(2)_2$ Kac-Moody algebra. Furthermore, it was demonstrated by Affleck \cite{affleck_exact_1986} that the energy-momentum tensor $T(z)$ can be expressed as a quadratic form of these currents and so this is a conformal embedding of $\hat{\mathfrak{u}}(1) \oplus \widehat{\mathfrak{su}}(2)_2 \oplus \widehat{\mathfrak{su}}(2)_2$ within $\mathcal{H}_{\text{CFT}}$. This then allows us to express $L_0$ as $L_0 = L^{\hat{\mathfrak{u}}(1)}_0 + L^{\widehat{\mathfrak{su}}(2)_2}_0 + \Tilde{L}^{\widehat{\mathfrak{su}}(2)_2}_0$, where $L^{\hat{\mathfrak{u}}(1)}_0$, $L^{\widehat{\mathfrak{su}}(2)_2}_0$ and $\Tilde{L}^{\widehat{\mathfrak{su}}(2)_2}_0$ are the zeroth modes of the energy-momentum tensors of the $i\partial\Phi(z)$, $J^a(z)$ and $\mathcal{J}^a(z)$ currents respectively. 

As argued by Wen \cite{Wen1991a}, on a physical edge we should consider the transformations induced by $\mathcal{J}^a(z)$ as \textit{gauge} transformations. Thus, we  define the space of physical states $\mathcal{H}_{\text{phys}}$ to be the space of gauge invariant states, $\ket{v} \in \mathcal{H}_{\text{phys}} \Leftrightarrow \mathcal{J}^a_n \ket{v} = 0$ for $n \geq 0$. To understand the structure of $\mathcal{H}_{\text{phys}}$ we must first decompose $\mathcal{H}_{\text{CFT}}$ into the irreducible representations of the Kac-Moody algebra of these currents.

Let $\mathcal{H}_{N_p}$ be the space of states with $N_p$ total partons. As the currents defined above do not change the number of partons when applied to a given state, $\mathcal{H}_{N_p}$ must decompose into irreducible representations of the $\hat{\mathfrak{u}}(1) \oplus \widehat{\mathfrak{su}}(2)_2 \oplus \widehat{\mathfrak{su}}(2)_2$ algebra, $\mathcal{H}_{N_p} = \bigoplus_\lambda \mathcal{M}_\lambda$. Each $\mathcal{M}_\lambda$ must be a highest weight irreducible representation, as the eigenvalues of $L_0$ are bounded from below. Hence, within each $\mathcal{M}_\lambda$ there is a space of states with basis $\ket{\lambda; i}$ such that $\mathcal{J}^a_n\ket{\lambda; i} = J^a_n\ket{\lambda; i} = a_n\ket{\lambda; i} = 0$ for $n > 0$. Such states are referred to as WZW primaries. Moreover, for a given $\mathcal{M}_\lambda$ the states $\ket{\lambda; i}$ must all be eigenstates of $L_0$ each with the same eigenvalue, with all other states in $\mathcal{M}_\lambda$ being expressible as polynomials in the modes of these currents applied to these WZW primaries. Hence, we can understand how $\mathcal{H}_{N_p}$ is decomposed by understanding the space of WZW primaries within $\mathcal{H}_{N_p}$. We can see that the gauge invariance condition $\mathcal{J}^a_n \ket{v} = 0$ for $n \geq 0$, already implies that $\ket{v}$ must be a WZW primary of the $\mathcal{J}^a(z)$ currents. Moreover, $\ket{v}$ must be a spin-$0$ as $\mathcal{J}^a_0\ket{v} = 0$. Hence the only $\mathcal{M}_\lambda$ that contain gauge invariant states are those whose primary state is spin-$0$ with respect to $\mathcal{J}^a(z)$. The only states that are gauge invariant in $\mathcal{M}_\lambda$ are those that are expressed as polynomials only in $J^a_n$ and $a_n$ applied on the primary state of $\mathcal{M}_\lambda$.

In Appendix \ref{Sec:22CFTDetails} we show that $\Hp$ has the decomposition $\mathcal{H}_{\text{phys}} = \bigoplus_{N_p = 2n, n\in \mathbb{Z}} \mathcal{M}_{\lambda(N_p)}$, where $\mathcal{M}_{\lambda(N_p)}$ are highest weight irreducible representations of the modes $a_n$ and $J^a_n$. When $N_p$ is and is not a multiple of 4, $\mathcal{M}_{\lambda(N_p)}$ has primary states that are spin-$0$ and spin-$1$ in $J^a_0$ respectively. The $U(1)$ charge of all states in $\mathcal{M}_{\lambda(N_p)}$ is $N_p/2$.

Following Wen \cite{wen_theory_1992}, the space of possible operators that create or annihilate the underlying bosons at the edge have a basis,
\begin{equation}
    \begin{split}
        V_1(z) &\equiv V_{12}(z)V_{22}(z) \\
        V_0(z) &\equiv [V_{12}(z)V_{21}(z) + V_{11}(z)V_{22}(z)]/\sqrt{2} \\
        V_{-1}(z) &\equiv V_{11}(z)V_{21}(z) \\
    \end{split}
\end{equation}
where it should be noted that these operators are all gauge invariant.

It is also shown in Appendix \ref{Sec:22CFTDetails} that all the states of $\mathcal{H}_{\text{phys}}$ can be generated by repeated application of the modes of $V_m(z)$ and $V^\dagger_m(z)$ on $\ket{0}$. We then identify the chiral algebra $\mathcal{A}$ generated by repeated OPEs of $V_m(z)$ and $V^\dagger_m(z)$, with this parton state.

The chiral algebra $\mathcal{A}$ can also be represented by fields from the $\hat{\mathfrak{u}}(1) \oplus \widehat{\mathfrak{su}}(2)_2$ WZW model. Under this representation we can express the fields $V_m(z)$ as $V_m(z) = :e^{i\Phi(z)}:\phi_{1,m}(z)$, where $\phi_{1,m}(z)$ are the spin-$1$ $\widehat{\mathfrak{su}}(2)_2$ WZW primary fields. The spin-$1$ representation of $\mathfrak{su}(2)$ corresponds to the totally symmetric rank two tensor representation of $\mathfrak{su}(2)$. 

\subsubsection{$\phi_2^2$ example: ground state wave function}
We  now show how the CFT defined above can be used to generate the $\nu = 1$, $\phi_2^2$ parton ground state trial wave function. As this is the square of the $\nu = 2$ ground state, we  focus on the case where the number of bosons $N$ is odd. We first demonstrate this for the unprojected wave function and then show how to generate the projected wave function.

We first define the background charge operator as $C(N) = (\Tilde{F}^\dagger_{22})^{N_2}(\Tilde{F}^\dagger_{21})^{N_1}(\Tilde{F}^\dagger_{12})^{N_2}(\Tilde{F}^\dagger_{11})^{N_1}$, with $N_1 = (N-1)/2$ and $N_2 = (N+1)/2$. Then the generating $\Omega$ operator is defined by,
\begin{equation}
    \Omega(z, \bar{z}) = \prod_{i=1}^2 ( V_{i1}(z) + \bar{z}V_{i2}(z) )
\end{equation}
Now consider the wave function,
\begin{equation}
    \begin{split}
        &\Psi_{\bra{0}}(\mathbf{z},\bar{\mathbf{z}}) \\
        =& \bra{0} C(N) \prod_{i=1}^N \Omega(z_i,\bar{z}_i) \ket{0} \\
        =& (-1)^{N(N-1)/2} \bra{0} \\
        &\times(\Tilde{F}^\dagger_{12})^{N_2}(\Tilde{F}^\dagger_{11})^{N_1} \prod_{i=1}^N (V_{11}(z_i) + \bar{z}_iV_{12}(z_i)) \\
        &\times(\Tilde{F}^\dagger_{22})^{N_2}(\Tilde{F}^\dagger_{21})^{N_1} \prod_{i=1}^N (V_{21}(z_i) + \bar{z}_iV_{22}(z_i)) \\
        &\times \ket{0} \\
        =& (-1)^{N(N-1)/2} \\
        &\times \bra{0}(\Tilde{F}^\dagger_{12})^{N_2}(\Tilde{F}^\dagger_{11})^{N_1} \prod_{i=1}^N (V_{11}(z_i) + \bar{z}_iV_{12}(z_i)) \ket{0} \\
        &\times \bra{0} (\Tilde{F}^\dagger_{22})^{N_2}(\Tilde{F}^\dagger_{21})^{N_1} \prod_{i=1}^N (V_{21}(z_i) + \bar{z}_iV_{22}(z_i)) \ket{0} \\
    \end{split}
\end{equation}
where the $(-1)^{N(N-1)/2}$ appears from rearranging the fermionic $(V_{i1}(z) + \bar{z}V_{i2}(z))$ operators, and the correlation function factorises in the last line as $V_{1i}(z)$ are independent of $V_{2i}(z)$. From Sec. \ref{Sec:integerQHCFT}, we can clearly see that these two factors are in fact the non-interacting $\nu = 2$ ground state wave functions. Hence, this wave function is given by,
\begin{equation}
    \begin{split}
        &\Psi_{\bra{0}}(\mathbf{z},\bar{\mathbf{z}}) = \\ 
        &  (-1)^{\frac{N(N-1)}{2}}
        \begin{vmatrix}
            z_1^{N_1 - 1} & z_2^{N_1 - 1} & \cdots & z_N^{N_1 - 1} \\
            \vdots & \vdots & \ddots & \vdots \\
            z_1 & z_2 & \cdots & z_N \\
            1 & 1 & \cdots & 1 \\
            \bar{z}_1z_1^{N_2 - 1} & \bar{z}_2z_2^{N_2 - 1} & \cdots & \bar{z}_Nz_N^{N_2 - 1} \\
            \vdots & \vdots & \ddots & \vdots \\
            \bar{z}_1 z_1 & \bar{z}_2 z_2 & \cdots & \bar{z}_N z_N \\
            \bar{z}_1 & \bar{z}_2 & \cdots & \bar{z}_N \\
        \end{vmatrix}^2
    \end{split}
\end{equation}
which is the $\phi_2^2$ parton ground state trial wave function. 

To generate the projected ground state wave function, we first note that the $\Omega$ operator can be expressed as,
\begin{equation} \label{Eq:expandOmega}
    \begin{split}
        \Omega(z, \bar{z}) =& (V_{11}(z) + \bar{z}V_{12}(z))(V_{21}(z) + \bar{z}V_{22}(z)) \\
        =& V_{-1}(z) + \sqrt{{2}}\bar{z}V_0(z) + \bar{z}^2V_1(z) \\
    \end{split}
\end{equation}
Thus, the projected wave function will be generated by,
\begin{equation} \label{Eq:PartonOmegaProj}
    \Omega(z) = V_{-1}(z) + 2\sqrt{2}\partial V_0(z) + (2\partial)^2V_1(z)
\end{equation}
with,
\begin{equation}
    P_{LLL} \Psi_{\bra{0}}(\mathbf{z},\bar{\mathbf{z}}) = \bra{0}C(N)\prod_{i=1}^N \Omega(z_i)\ket{0}
\end{equation}

From Eq. \ref{Eq:expandOmega} one can see that $\Omega$ is expressed entirely in terms of the generators of the chiral algebra $\mathcal{A}$ corresponding to this state. As $[\mathcal{J}^a(z), C(N)] = 0$, for any state $\ket{v} \in \mathcal{H}_{\text{phys}}$ we must have $C(N)\ket{v} \in \mathcal{H}_{\text{phys}}$. Hence, we can define $C(N)$ within $\mathcal{H}_{\text{phys}}$, which is the vacuum representation of the chiral algebra $\mathcal{A}$. So the state $\ket{x} \equiv C(N) \prod_{i=1}^N \Omega(z_i,\bar{z}_i)\ket{0}$ must belong to $\mathcal{H}_{\text{phys}}$. Thus, the correlation function that defines the wave function $\braket{0|x}$ must be computable entirely in the vacuum representation of $\mathcal{A}$. However, in a similar way to the composite fermion case in Sec. \ref{Sec:CFWFCFTCon}, it is far more efficient to compute the correlation functions using the ``larger'' CFT.

\subsubsection{$\phi_2^2$ example: CFT to edge state mapping} \label{Sec:partonEdgeMap}
Inline with our presentation for the composite fermion example in Sec. \ref{Sec:CFEdgeMapping}, we  discuss this mapping for the unprojected wave functions. Once again, the projected edge state mapping can be easily understood from the following discussion by the $\Omega$ replacement of Eq. \ref{Eq:PartonOmegaProj}.

The edge state mapping is defined in second quantized notation as,
\begin{equation} \label{Eq:partonEdgeMap}
    \ket{\Psi_{\bra{v}}}\rangle \equiv \bra{v} C(N) e^{\int d^2z e^{-|z|^2/4} \Omega(z, \bar{z})\otimes c^\dagger(z,\bar{z}) } \ket{0}\otimes\ket{0}\rangle
\end{equation}
with states $\bra{v}$ having a definite $U(1)$ charge $q$, with respect to the current $i\partial\Phi(z)$, generating an $N+\sqrt{\nu}q$ particle state with a wave function,
\begin{equation}
    \Psi_{\bra{v}}(\mathbf{z},\bar{\mathbf{z}}) = \bra{v}C(N) \prod_{i=1}^{N+\sqrt{\nu}q}\Omega(z_i,\bar{z}_i) \ket{0}
\end{equation}
As $C(N) \prod_{i=1}^{N+\sqrt{\nu}q}\Omega(z_i,\bar{z}_i) \ket{0} \in \mathcal{H}_{\text{phys}}$ we must have that all states in the orthogonal complement of $\mathcal{H}_{\text{phys}}$ must map to zero. Hence, restricting this map to the vacuum representation of $\mathcal{A}$ (i.e. $\mathcal{H}_{\text{phys}}$) does not change the image of the map. So, one can also think of this map as being purely from this vacuum rep to the space of wave functions.

To understand the image of this map, however, it is easier to consider it over the full $\mathcal{H}_{\text{CFT}}$. From Sec. \ref{Sec:integerQHCFT} we know how to generate all the $\nu = 2$ edge state wave functions. Furthermore, from the independence of the fields $V_{1i}(z)$ from $V_{2i}(z)$, we know that up to a minus sign we can factorise any correlation function which can be expressed as a product of operators of these two parton species. Hence, any wave function of the form $\Phi_{1,\text{edge}}(\mathbf{z},\bar{\mathbf{z}})\Phi_{2,\text{edge}}(\mathbf{z},\bar{\mathbf{z}})$, where $\Phi_{i,\text{edge}}(\mathbf{z},\bar{\mathbf{z}})$ are Slater determinants of the orbitals $z^m$ and $\bar{z}z^m$, can be generated by states of the form,
\begin{equation}
    \bra{0}\prod_{i,j=1}^2[ \prod_{m^{(ij)}_l}V^\dagger_{ij, m^{(ij)}_l} \prod_{n^{(ij)}_l} V_{ij,n^{(ij)}_l} ]
\end{equation}
The image of this edge state mapping must then be equivalent to the parton edge space defined in \ref{Sec:CFandPartonWF}. 

As the $V_{ij}(z)$ operators transform as spin-$\frac{1}{2}$ WZW primaries for transformation generated by $\mathcal{J}^a(z)$, with the transformation only acting on the species index, it can be easily shown that $[\mathcal{J}^a_n, \Omega(z,\bar{z})] = 0$. We also have that $\mathcal{J}^a_n\ket{0} = 0$ for $n \geq 0$. It then follows that the correlation functions of the edge state mapping must be invariant under transformations generated by $\mathcal{J}^a(z)$. As for the composite fermion case, this can be interpreted as a gauge redundancy, where we know that we only need the gauge invariant states ($\mathcal{H}_{\text{phys}}$) to reproduce the image of the edge state map. 

As the image of this map has an over complete basis which are products of $\nu = 2$ edge wave functions, we must have that the angular momentum operator when mapped to the CFT is simply the sum of the $\nu = 2$ CFT angular momentum operators from each parton species. The $\nu = 2$ result in Sec. \ref{Sec:integerQHCFT} then gives,
\begin{equation}
    \begin{split}
        \sum_i z_i\partial_i - \bar{z}_i\overline{\partial}_i \rightarrow& L_0 + \frac{2N_1 - 1}{2} \bigg (\sum_{i,j = 1}^2 \fem{ij}_0 \bigg ) \\
        &+ 2(N_1(N_1 - 1) - 1) \\
        =&  L_0 + (2N_1 - 1)a_0 + 2(N_1(N_1 - 1) - 1)
    \end{split}
\end{equation}
As for all other cases considered so far, we see that states of the CFT with the same $L_0$ eigenvalues will map to wave functions with the same angular momentum. As we know that just mapping from $\mathcal{H}_{\text{phys}}$ can generate the full image of the map, we can then give rigorous upper bounds on the dimension of each angular momentum eigenspace within the parton edge space. 

\subsubsection{The $\phi_n^m$ symmetric parton states} \label{Sec:symmetricPartons}
Another example of particular interest is that of the symmetric parton states that take the form $\phi_n^m$. We only consider the case where $N = nN_1 + n(n-1)/2$ for some positive integer $N_1$. This is so that the lowest angular momentum $\nu = n$ integer quantum Hall state of $N$ particles is unique, which implies that there is a unique ``densest'' parton ground trial wave function. To generate this wave function we simply need $m$ copies of the CFT that generates the $\nu = n$ wave function. That is, $nm$ copies of chiral boson CFT (with compactification radius one) with each vertex operator denoted by $V_{ij}(z)$, where $i = 1, 2, \dots, m$ is the species index and $j = 1,2,\dots n$ is the ``Landau level'' index. The generating $\Omega$ is simply given as $\Omega(z,\bar{z}) = \prod_{i=1}^m(\sum_{j=1}^n \bar{z}^{j-1}V_{ij}(z))$, with the wave function being expressed as,
\begin{equation}
    \Psi_{\bra{0}}(\mathbf{z},\bar{\mathbf{z}}) = \bra{0}C(N)\prod_{i=1}^N \Omega(z_i,\bar{z_i})\ket{0}
\end{equation}
where $C(N) = \prod_{i=1}^m C_i(N)$ with $C_i(N)$ being the background charge operator used to generate the $\nu = n$ ground state with the $i^{\text{th}}$ chiral boson species. By the factorisation of the resulting correlation function, the wave function can be expressed as the $\nu = n$ integer quantum Hall ground state raised to the $m^{\text{th}}$ power thus giving the desired parton trial ground state.  

One can expand $\Omega$ as,
\begin{equation}
    \Omega(z,\bar{z}) = \sum_{l=m}^{mn} \bar{z}^{l-m}V_l(z)
\end{equation}
where
\begin{equation}
    V_l(z) = \sum_{i_1, i_2, \dots, i_m = 1}^n \delta_{\sum_{j=1}^m i_m, l} V_{1i_1}(z)V_{2i_2}(z)\dots V_{mi_m}(z)
\end{equation}
The chiral algebra we associate with this state, $\mathcal{A}(n)_m$, is generated by repeated OPEs of $V_l(z)$ and their conjugates. Let $\Hp$ be the space of states generated by the modes of $V_l(z)$ and $V^\dagger_l(z)$ applied on the vacuum state, which will then form the vacuum representation of $\mathcal{A}(n)_m$. From the general result of Appendix \ref{Sec:PartonBackgroundCharge} it follows that if $\ket{v} \in \Hp$ then $C(N)\ket{v} \in \Hp$ (i.e. $\Hp$ is an invariant subspace of $C(N)$). This then implies that $C(N)\prod_{i=1}^N \Omega(z_i,\bar{z_i})\ket{0} \in \Hp$. As $\ket{0} \in \Hp$ (by definition) it follows that the correlation function $\bra{0}C(N)\prod_{i=1}^N \Omega(z_i,\bar{z_i})\ket{0}$ can be computed entirely within the vacuum representation of $\mathcal{A}(n)_m$ (as the correlation function is simply an inner product within $\Hp$). 

In Appendix \ref{Sec:partonAlgebra} we show that all the $U(1)$ neutral fields of $\mathcal{A}(n)_m$ form the vacuum representation for the $\hat{\mathfrak{u}}(1)\oplus\widehat{\mathfrak{su}}(n)_m$ WZW current algebra. This is what one would expect based on previous work by Wen \cite{Wen1991a}. Consequently, the $V_l(z)$ operators can be represented by the $U(1)\otimes SU(n)_m$ WZW model with the form $V_l(z) = :e^{i\sqrt{\frac{m}{n}}\Phi(z)}: \phi_l(z)$ where $\phi_l(z)$ are $SU(n)_m$ WZW primaries corresponding to the totally symmetric rank $m$ tensor representation of $SU(n)$. Thus, the $\phi_n^m$ parton ground state trial wave function can be expressed using the $U(1)\otimes SU(n)_m$ WZW model conformal blocks.

Another way of showing correlation functions of $V_l(z)$ are the same as those of $:e^{i\sqrt{\frac{m}{n}}\Phi(z)}: \phi_l(z)$, for the $\phi_n^m$ case, is by using the general results of Refs. \cite{nakanishi_level-rank_1992, naculich_duality_1990-1, naculich_duality_1990}. We first note that within the CFT of $nm$ chiral bosons, which we  denote $\CFT{L}$ for the remainder of this section, we can define the currents,
\begin{equation} \label{Eq:partonCurrents}
    \begin{split}
        \mathcal{J}^a(z) &\equiv \sum_{ijk} t^a_{ij} : V_{ik}(z) V^\dagger_{jk}(z): \\
        J^a(z) &\equiv \sum_{ijk} T^a_{ij} : V_{ki}(z) V^\dagger_{kj}(z): \\
        i\partial\Phi(z) &\equiv \frac{1}{\sqrt{mn}} \sum_{jk} i\partial\fef{jk}(z) \\ 
    \end{split}
\end{equation}
where $t^a$ are the generaters of $SU(m)$ and $T^a$ are the generators of $SU(n)$. As demonstrated in Ref. \cite{affleck_exact_1986} these $J^a(z)$ and $\mathcal{J}^a(z)$ form an $\widehat{\mathfrak{su}}(n)_m$ and $\widehat{\mathfrak{su}}(m)_n$ current algebra respectively, with $\mathcal{J}^a i\partial\Phi(w) \sim 0$, $\mathcal{J}^a(z)J^b(w) \sim 0$ and $J^a(z)i\partial\Phi(w) \sim 0$. We can thus organise the fields of $\CFT{L}$ into representations of the $\hat{\mathfrak{u}}(1)\oplus \widehat{\mathfrak{su}}(n)_m \oplus \widehat{\mathfrak{su}}(m)_n$ current algebra. Let $\phi_{q,\lambda, \lambda'}(z)$ be a field which is a WZW primary relative to the currents $i\partial\Phi(z)$, $J^a(z)$ and $\mathcal{J}^a(z)$, which has a $U(1)$ charge $q$, transforms under the action of $J^a(z)$ as a field in the $\mathfrak{su}(n)$ representation labelled by $\lambda$ and transforms under the action of $\mathcal{J}^a(z)$ as a field in the $\mathfrak{su}(m)$ representation labelled by $\lambda'$. As shown discussed in Refs. \cite{nakanishi_level-rank_1992, naculich_duality_1990-1, naculich_duality_1990}, the correlation function $\bra{0}\prod_{i}\phi_{q_i,\lambda_i,\lambda_i'}\ket{0}$ has the decomposition,
\begin{equation} \label{Eq:generalCorrDecomp}
    \begin{split}
        \bra{0}\prod_{i}\phi_{q_i,\lambda_i,\lambda_i'}(z_i)\ket{0} = & \mathcal{F}^{\hat{\mathfrak{u}}(1)}(z_1, z_2, \dots) \\
        &\times \sum_{ab} \bigg [ C_{ab} \mathcal{F}_a^{\widehat{\mathfrak{su}}(n)_m}(z_1,z_2,\dots) \\
        &\times \mathcal{F}_b^{\widehat{\mathfrak{su}}(m)_n}(z_1,z_2,\dots) \bigg ]
    \end{split}
\end{equation}
where $C_{ab}$ are constants, $\mathcal{F}^{\hat{\mathfrak{u}}(1)}(z_1, z_2, \dots)$ is a conformal block of the $U(1)$ vertex operators $\bra{0} \prod_j :e^{iq_j\Phi(z_j)}: \ket{0}$, $\mathcal{F}_a^{\widehat{\mathfrak{su}}(n)_m}(z_1,z_2,\dots)$ is a conformal block of the $\widehat{\mathfrak{su}}(n)_m$ WZW model related to the correlation function $\bra{0}\prod_i \phi_{\lambda_i}(z_i,\bar{z}_i)\ket{0}$, and $\mathcal{F}_b^{\widehat{\mathfrak{su}}(m)_n}(z_1,z_2,\dots)$ is a conformal block of the $\widehat{\mathfrak{su}}(m)_n$ WZW model related to the correlation function $\bra{0}\prod_i \phi_{\lambda'_i}(z_i,\bar{z}_i)\ket{0}$. The $V_l(z)$ field has a $U(1)$ charge $q=\sqrt{m/n}$ and is a WZW primary relative to the $i\partial\Phi(z)$, $J^a(z)$ and $\mathcal{J}^a(z)$, where under the action of $J^a(z)$ it transforms according to the $\mathfrak{su}(n)$ representation formed by totally symmetric rank $m$ tensors, which we denote $\lambda_m$, and under the action of $\mathcal{J}^a(z)$, $V_l(z)$ transforms as the identity field $\mathbf{1}(z)$. The conjugate fields $V^\dagger_l(z)$ are also WZW primaries relative to these currents, where they have $U(1)$ charge $q=-\sqrt{m/n}$, transform under the action of $J^a(z)$ according to the representation of $\mathfrak{su}(m)$ which is the conjugate of $\lambda_m$ ($\lambda_m^\dagger$), and transform under the action of $\mathcal{J}^a(z)$ as the identity field $\mathbf{1}(z)$. Let $\phi_l(z)$ and $\phi^\dagger_l(z)$ be fields of the $\widehat{\mathfrak{su}}(n)_m$ WZW model that transform under the action of $J^a(z)$ the same way as $V_l(z)$ and $V^\dagger_l(z)$ respectively. These $\phi_l(z)$ and $\phi^\dagger_l(z)$ are simple currents, which is to say their fusion rule with any other field can only have one result. Hence, any correlation function involving just the $\phi_l$ and $\phi^\dagger_l$ fields must have \textit{only one} corresponding conformal block. Clearly, any correlation function of $\mathbf{1}(z)$ in the $\widehat{\mathfrak{su}}(m)_n$ WZW has only one corresponding conformal block which is trivially a constant $\braket{ \prod_{i}\mathbf{1}(z_i) } = 1$. From Eq. \ref{Eq:generalCorrDecomp}, we then have,
\begin{equation}
    \begin{split}
        \braket{\prod_jV_{l_j}(z_j) \prod_{k} V^\dagger_{l_k}(w_k)} =& \langle \prod_j:e^{i\sqrt{\frac{m}{n}}\Phi(z_j)}: \\ 
        & \times \prod_k :e^{-i\sqrt{\frac{m}{n}}\Phi(w_k) }: \rangle \\
        &\times \braket{\prod_j \phi_{l_j}(z_j)\prod_k\phi^\dagger_{l_k}(w_k) } \\
    \end{split}
\end{equation}
Hence, any correlation functions involving only the $V_l(z)$ and $V^\dagger_l(z)$ are equivelant to the correlation functions of the fields $:e^{i\sqrt{\frac{m}{n}}\Phi(z)}:\phi_l(z)$ and $:e^{-i\sqrt{\frac{m}{n}}\Phi(z)}:\phi^\dagger_l(z)$ in the $\hat{\mathfrak{u}}(1)\otimes \widehat{\mathfrak{su}}(n)_m$ WZW model. This implies that the chiral algebra generated by repeated OPEs of $V_l(z)$ and $V^\dagger_l(z)$ is \textit{equivelant} to the chiral algebra generated by repeated OPEs of $:e^{i\sqrt{\frac{m}{n}}\Phi(z)}:\phi_l(z)$ and $:e^{-i\sqrt{\frac{m}{n}}\Phi(z)}:\phi^\dagger_l(z)$.

Let us now consider the edge state map which can be expressed in second quantisation in the usual form,
\begin{equation}
    \ket{\Psi_{\bra{v}}}\rangle \equiv \bra{v} C(N) e^{\int d^2z e^{-|z|^2/4} \Omega(z, \bar{z})\otimes c^\dagger(z,\bar{z}) } \ket{0}\otimes\ket{0}\rangle
\end{equation}
Once again, if $\bra{v}$ has a definite amount of total $U(1)$ charge $q$ the resulting wave function will have a definite number of particles $N + q\sqrt{\nu}$ with the resulting wave function being,
\begin{equation}
    \Psi_{\bra{v}}(\mathbf{z},\bar{\mathbf{z}}) = \bra{v}C(N)\prod_{i=1}^{N + q\sqrt{\nu}}\Omega(z_i,\bar{z}_i) \ket{0}
\end{equation}
Now consider the state $\bra{v} = \bra{v_1}\otimes\bra{v_2}\otimes \dots \otimes \bra{v_m} $ where $\bra{v_i}$ is a state of the $i^{\text{th}}$ parton species. From the factorisation of the resulting correlation function, the wave function $\Psi_{\bra{v}}(\mathbf{z},\bar{\mathbf{z}})$ must be expressible as a product of $m$ $\nu = n$ integer quantum Hall edge states. Hence, the image of this edge state map is spanned by wave functions that are expressible as products of $m$ wave functions which are all $\nu = n$ integer quantum Hall wave functions.

From the definition of the edge state map, we must have that all the states that belong to the orthogonal complement of $\Hp$ must map to zero. Hence, if we restrict the edge state map to $\Hp$ we do not alter the image of the map. In other words, as for all other cases presented, the edge state map can be considered a map from the vacuum representation of $\mathcal{A}(n)_m$ to the space of wave functions.

Finally, the angular momentum operator can be expressed in the CFT as a sum of the angular momentum operators of the $\nu = n$ integer quantum Hall edge state map for each parton species. This mapping of the angular momentum operator can then be expressed as,
\begin{equation}
    \begin{split}
        \sum_i z_i\partial_i - \bar{z}_i\overline{\partial}_i \rightarrow & L_0 + \frac{\sqrt{nm}(2N_1 - 1)}{2}a_0 \\
        &+ m\frac{nN_1(N_1 - 1)}{2} - m\frac{n(n-1)}{2}
    \end{split}
\end{equation}
where $a_0$ is the zeroth mode of $i\partial\Phi(z)$ and is proportional to the total number of particles added to the edge. Thus, for a fixed number of particles and angular momentum, in the actual wave function, the edge state counting must have a rigorous upper bound given by the state counting in $\Hp$ for the corresponding $L_0$ and $a_0$ eigenvalues.

\subsubsection{General case} \label{Sec:PartonGenWFCon}
The above discussion can be expanded to general parton states by the following observation. Suppose we know how to generate some trial wave functions $\Psi_1$ and $\Psi_2$ using the operators $\Omega_1(z,\bar{z})$ and $\Omega_2(z,\bar{z})$ from $\CFT{1}$ and $\CFT{2}$ respectively. We can then generate the trial wave function $\Psi_1\Psi_2$ using the CFT $\CFT{2}\otimes\CFT{1}$ as follows. Let $C(N) = C_1(N)C_2(N)$, where $C_i(N)$ is the background charge operator used to generate $\Psi_i$, and $\Omega(z, \bar{z}) = \Omega_1(z,\bar{z})\Omega_2(z,\bar{z})$. We then have,
\begin{equation} \label{Eq:prodWFCorr}
    \begin{split}
        \Psi_{\bra{0}}(\mathbf{z},\bar{\mathbf{z}}) =& \bra{0}C(N)\prod_{i=1}^N\Omega(z_i,\bar{z}_i)\ket{0} \\
        \propto & \bra{0}\bigg [ C_2(N) \prod_{i=1}^N\Omega_2(z_i,\bar{z}_i) \bigg ] \ket{0} \\
        &\times \bra{0} \bigg [ C_1(N) \prod_{j=1}^N\Omega_1(z_j,\bar{z}_j) \bigg ] \ket{0} \\
        &= \Psi_1(\mathbf{z},\bar{\mathbf{z}})\Psi_2(\mathbf{z},\bar{\mathbf{z}}) \\
    \end{split}
\end{equation}
where $\propto$ appears as, depending on the conformal spin of the $\Omega_i$'s, we may obtain an additional minus one factor from rearranging the $\Omega_i$'s. Thus, simply by the fact the resulting correlation function will factorise, we can generate $\Psi_1\Psi_2$ from a CFT correlation function. 

One can expand $\Omega(z,\bar{z}) = \sum_j \bar{z}^j\phi_j(z)$ where each $\phi_j(z)$ can be expressed in terms of $\CFT{1}\otimes \CFT{2}$. We identify the chiral algebra $\mathcal{A}$ generated by repeated OPEs with $\phi_j(z)$ and their conjugates, with the new wave function $\Psi_1\Psi_2$. The subspace of $\CFT{1}\otimes\CFT{2}$ that is generated by the modes of the fields in $\mathcal{A}$, we  denote $\mathcal{H}_{\text{phys}}$. This $\mathcal{H}_{\text{phys}}$ forms the vacuum representation of $\mathcal{A}$. In Appendix \ref{Sec:PartonBackgroundCharge}, we show that $\mathcal{H}_{\text{phys}}$ is an invariant subspace of $C(N)$ provided the ``densest'' (or lowest angular momentum) parton trial wave function is unique at the fixed number of particles $N$. Hence, the correlation function that generates $\Psi_1\Psi_2$ can be computed entirely in the vacuum representation of $\mathcal{A}$. For the remainder of this section, we  assume that the ``densest'' parton trial wave function at the fixed number of particles $N$ is unique. We briefly consider the case when this is not so at the end of the section. 

An important point in formulating a CFT from $\mathcal{A}$ is whether it contains an energy-momentum tensor $T(z)$. We also show in Appendix \ref{Sec:PartonBackgroundCharge} that $\mathcal{A}$ has an energy-momentum tensor. A more general proof of this can be seen in Ref. \cite{carpi_vertex_2018}. This is a necessary condition for $\CFT{\mathcal{A}}$ to exist. We further emphasise that we are assuming that a CFT with the chiral algebra $\mathcal{A}$ exists, preferably satisfying the axioms set out by Moore and Seiberg for rational conformal field theory \cite{Moore1989}. There do exist certain results in the mathematical literature on vertex operator algebras in relation to this question \cite{zhu_modular_1996}, however it is not clear if these results apply in all cases considered in this paper. 

The edge state map is defined by the obvious generalization of Eq. \ref{Eq:partonEdgeMap}. Consider the state $\ket{v} \in \CFT{1}\otimes\CFT{2}$ with, $\ket{v} = \ket{v_1}\otimes \ket{v_2}$ and $\ket{v_i}\in \CFT{i}$. By the factorization of the resulting correlation function we must have that, $\Psi_{\bra{v_1}\otimes\bra{v_2}}(\mathbf{z},\bar{\mathbf{z}}) \propto \Psi_{\bra{v_1}}^{(1)}(\mathbf{z},\bar{\mathbf{z}}) \Psi_{\bra{v_2}}^{(2)}(\mathbf{z},\bar{\mathbf{z}})$, where $\Psi_{\bra{v_i}}^{(i)}(\mathbf{z},\bar{\mathbf{z}})$ is the wave function resulting from the edge state map of $\CFT{i}$ (i.e. the edge state map associated with $\Psi_i$). Thus, the image of the edge state map must be spanned by wave functions which are products of wave functions from the images of the edge state maps of $\CFT{1}$ and $\CFT{2}$. Furthermore, we know, from the discussion of Appendix \ref{Sec:PartonBackgroundCharge}, that $C(N)\prod_{i=1}^N\Omega(z_i,\bar{z}_i)\ket{0} \in \mathcal{H}_{\text{phys}}$, which implies all states of the orthogonal complement of $\mathcal{H}_{\text{phys}}$ must map to zero. Hence, restricting this map to the states of the vacuum representation of $\mathcal{A}$ must preserve the image of the map. Thus, one can view the map as being from the vacuum representation of $\mathcal{A}$ to the space of wave functions. 

In Appendix \ref{Sec:AMParton} show that, for these cases where the ``densest'' state is unique for the fixed number of particles $N$ and when restricting the edge state map to be from $\Hp$, the angular momentum operator can be mapped over to the CFT, as,
\begin{equation}
    \sum_i z_i\partial_i - \bar{z}_i\overline{\partial}_i \rightarrow L_0 + v(N)a_0 + u(N)
\end{equation}
where $L_0$ is the zeroth Virasoro mode, $a_0$ is the total $U(1)$ charge operator which is proportional to the number of particles added to the edge, and $v(N)$ and $u(N)$ are some real-valued functions. Thus, for a fixed number of particles and angular momentum, in the actual wave function, the edge state counting must have a rigorous upper bound given by the state counting in the vacuum representation of $\mathcal{A}$ for the corresponding $L_0$ and $a_0$ eigenvalues.

In Appendix \ref{Sec:221Example} we give the $\phi_2^2\phi_1$ state as an example of this process.

Finally, let us now briefly consider the case where the ``densest'' parton trial ground state at the fixed number of particles $N$ is not unique. More precisely this occurs when there does not exist a number of particles $N$ such that each integer quantum Hall component of the parton state has a unique lowest angular momentum state at the given number of particles $N$. An example of this is the $\phi_4\phi_2$ state. For $\nu = 2$ there is only a unique lowest angular momentum state when the number of particles is odd (where for $N$ even we get two lowest angular momentum states). For $\nu = 4$ the lowest angular momentum state is unique when $N$ can be expressed as $N = 4N_1 + 6$ with $N_1$ being some non-negative integer. Thus, in this case, there is no number of particles $N$ where both the $\nu = 2$ and the $\nu = 4$ components of this parton state have unique lowest angular momentum states.

In such cases, it is not clear if the background charge operator can be defined in a simple way. One can however proceed without a background charge operator. Suppose we can write a ground state wave function in the form $\bra{N} \prod_i \Omega(z_i,\bar{z}_i) \ket{0}$ for some state $\bra{N}$ (which can clearly be done for all cases that do have a well-defined background charge operator), where we can then, once again, expand $\Omega(z,\bar{z}) = \sum_j \bar{z}^{j-1}\phi_j(z)$. The correlation function $\bra{N} \prod_i \Omega(z_i,\bar{z}_i) \ket{0}$ can then be expressed within the vacuum representation of the chiral algebra $\mathcal{A}$ generated by $\phi_j(z)$ and their conjugates. One can then define the edge state map as before simply with the background charge operator omitted, which will be a linear map from the vacuum representation of $\mathcal{A}$ to the space of wave functions. One can then reapply all the inductive steps discussed here to show this structure exists for all chiral parton states even when the background charge operator is perhaps less well defined. It should be noted, however, that in these more general cases, the lack of a background charge operator makes edge state counting less straightforward.

\subsection{General structure} \label{Sec:CFTGenStructure}
The discussions of Sec. \ref{Sec:CFWFCFTCon} and Sec. \ref{Sec:PartonWFCon} point towards a general structure for constructing wave functions from CFT. Suppose we wish to construct a wave function using some chiral algebra $\mathcal{A}$ that is generated by repeated OPEs of $n$ fields $\phi_l(z)$ and their conjugates $\phi^\dagger_l(z)$, with $l=0,1,\dots n-1$. We can then define the operator,
\begin{equation}
    \Omega(z,\bar{z}) = \sum_{l=0}^{n-1} \bar{z}^l\phi_l(z)
\end{equation}
along with an appropriately chosen background charge operator $C(N)$, to generate the trial wave function,
\begin{equation} \label{Eq:generalCFTWFGen}
    \Psi_{\bra{0}}(\mathbf{z},\bar{\mathbf{z}}) = \bra{0}C(N) \prod_{i=1}^N\Omega(z,\bar{z}) \ket{0}
\end{equation}
To generate the projected wave function one can then simply use the generating operator $\Omega(z) = \sum_{l=0}^{n-1} (2\partial)^l \phi_l(z)$.

One can then construct edge-state trial wave functions using the general edge-state mapping,
\begin{equation} \label{Eq:CFEdgeStateMapGeneral}
    \ket{\Psi_{\bra{v}}}\rangle \equiv \bra{v} C(N) e^{\int d^2z e^{-|z|^2/4} \Omega(z, \bar{z})\otimes c^\dagger(z,\bar{z}) } \ket{0}\otimes\ket{0}\rangle
\end{equation}
This gives a linear map from the vacuum representation of $\mathcal{A}$ to the space of wave functions.

In all the cases described in Sec. \ref{Sec:CFWFCon} and for the symmetric parton wave functions in Sec. \ref{Sec:PartonWFCon}, we showed the chiral algebra $\mathcal{A}$ could be \textit{represented} using a CFT of the form $\CFT{U(1)}\otimes\CFT{\chi}$, where $\CFT{U(1)}$ is the chiral boson CFT and $\CFT{\chi}$ is another CFT which can be referred to as the ``statistics'' sector. With this representation the fields $\phi_l(z)$ take the form,
\begin{equation} \label{Eq:generalGenerator}
    \phi_l(z) = :e^{i\varphi(z)/\sqrt{\nu}}:\chi_l(z)
\end{equation}
where $\varphi(z)$ is the $U(1)$ chiral boson, $\chi_l(z)$ are primaries of $\CFT{\chi}$ all with the same scaling dimension, and $\nu$ is the filling fraction of the resulting trial wave function. This construction can then be seen as a natural generalisation of formalism discussed in Sec. \ref{Sec:CFTMthods}.

Importantly, the $\chi_l(z)$ are \textit{simple currents} of $\CFT{\chi}$. A simple current is a field whose fusion with any other field can only have one result. As shown by Schoutens and Wen, the conformal blocks from various simple current algebras can be used to model the statistics of a large class of anyon theories \cite{schoutens_simple-current_2016} (also see Ref. \cite{fukusumi2023fermionic} and references therein for further discussion of these simple current constructions in the FQHE context).

Of course, this is not the only way of generating a trial wave function, and edge states, from the given chiral algebra $\mathcal{A}$. For example, we could instead use $\Omega(z,\bar{z}) = \sum_{l=0}^{n-1} (a^\dagger)^l\phi_l(z)$, where $a^\dagger = (\bar{z} - 2\partial)/\sqrt{2}$. This construction has the added feature that the index $l$ in $\phi_l(z)$ is directly related to the Landau level index, which could be of some use in the description of multilayer systems (such as multilayer graphene \cite{faugno_non-abelian_2020, faugno_unconventional_2021, wu_non-abelian_2017, kim_even_2019, timmel_non-abelian_2023}) where the layer index is taken as a pseudo-Landau level. Defining $\Omega(z,\bar{z})$ this way does, however, have the disadvantage that projecting the wave function will in general alter the topological order of the wave function. This is because it would only be $\phi_0(z)$ that would generate the projected wave function and, thus, one would expect the corresponding topological order would be encoded in the chiral algebra generated only by $\phi_0(z)$ and $\phi^\dagger_0(z)$. In fact, provided the background charge operator is chosen such that the unprojected ground state wave function has a definite number of particles in each Landau level, then clearly the projection would simply give a zero wave function in general (which would be the case for all the $\phi_n^m$ parton states for example). Numerical computation of such unprojected wave functions may also be inefficient due to the extra derivatives in their definition, although these unprojected wave functions may be amenable to certain MPS methods.

\section{Generalised screening and the inner product action} \label{Sec:GenScreening}
We  now formulate the generalised screening hypothesis for trial wave functions constructed using some chiral algebra $\mathcal{A}$ according to the general structure outlined in Sec. \ref{Sec:CFTGenStructure}. We then go on in Sec. \ref{Sec:CFTInnerProducts} to discuss the form edge state inner products of parton states take given generalised screening. Finally, in Sec. \ref{Sec:RSES} we  then show how the arguments of DRR \cite{Dubail2012} can be easily extended to understand the structure of the RSES of these CFT-constructed wave functions.

Throughout this section, we  only consider unprojected wave functions. The discussion of this section does not rely on these wave functions being unprojected. One can formulate a generalised screening hypothesis and repeat the same arguments given here for the projected wave functions by a simple replacement $\Omega(z, \bar{z}) \rightarrow \Omega(z) = \sum_{l=0}^{n-1} (2\partial)^l \phi_l(z)$

\subsection{Generalised screening} \label{Sec:GenScreeningGen}

\begin{figure}
    \centering
    \includegraphics[scale=0.24]{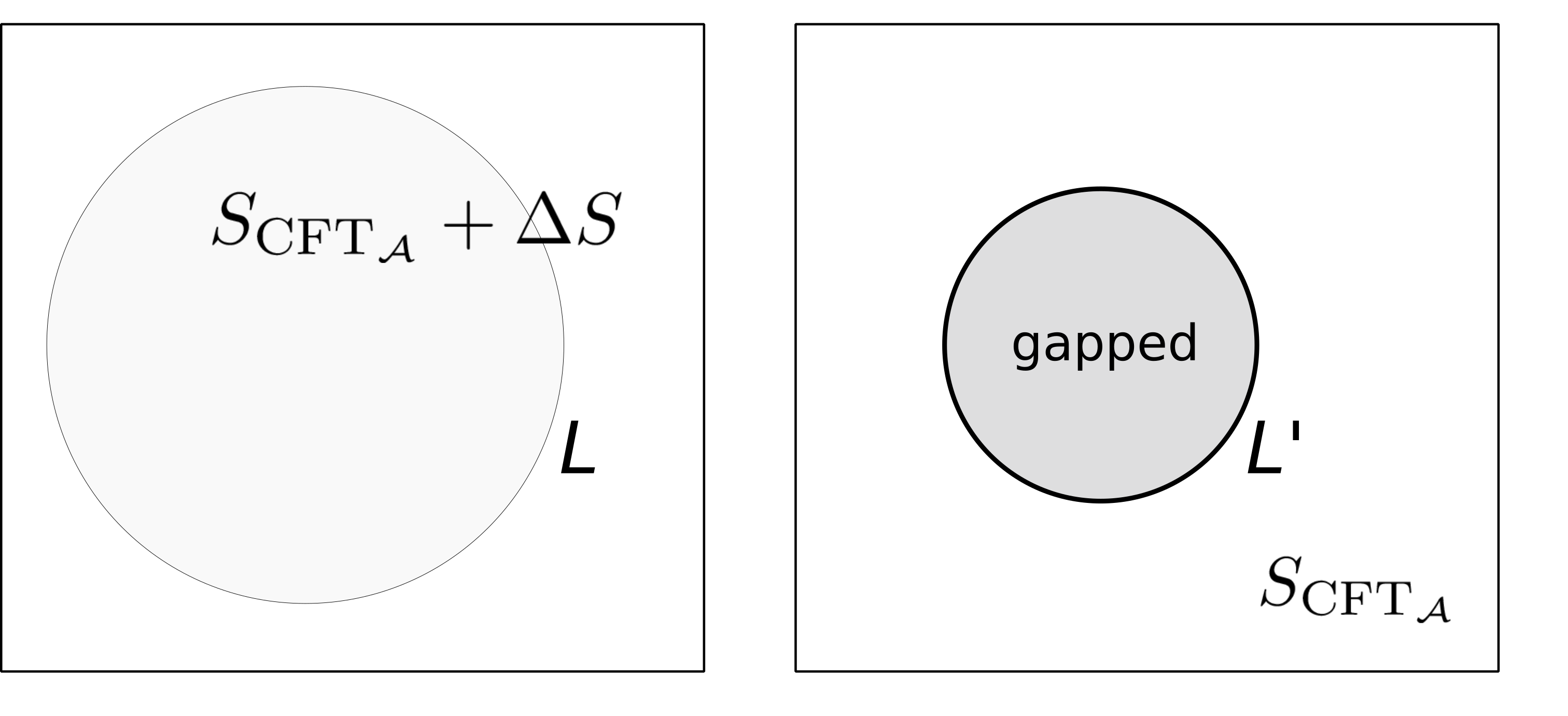}
    \caption{Generalised screening is when the given ``perturbed'' CFT under RG flows to a gapped field theory inside the droplet and back to the unperturbed field theory outside the droplet. Here $L$ and $L'$ denote the lengths of the droplet before and after a certain amount of RG flow.}
    \label{fig:genralisedScreening}
\end{figure}

Consider now a ground state trial wave function generated by the general construction of Sec. \ref{Sec:CFTGenStructure}. The full CFT defined by this chiral algebra $\mathcal{A}$, which we denote $\CFT{\mathcal{A}}$, will also contain the anti-chiral copy of this algebra $\overline{\mathcal{A}}$, where these two algebras are independent (i.e. correlation functions of just the fields of $\mathcal{A}$ and $\overline{\mathcal{A}}$ will factorise into a correlation function of just the $\mathcal{A}$ operators times a correlation function of just the $\overline{\mathcal{A}}$ operators). The matching of $\phi_j(z) \in \mathcal{A}$ with its anti-chiral copy $\bar{\phi}_j(\bar{z}) \in \overline{\mathcal{A}}$ is such that a correlation function of a number of $\phi_j(z)$'s is the complex conjugate of the correlation function of the corresponding $\bar{\phi}_j(\bar{z})$'s. Let the anti-chiral version of the generating operator $\Omega(z,\bar{z})$ be written as $\overline{\Omega}(z,\bar{z}) = \sum_l z^l \bar{\phi}_l(\bar{z})$. Further, let $\overline{C}(N)$ be the anti-chiral copy of the background charge operator $C(N)$. The complex conjugate of the ground state wave function can be expressed as $\overline{\Psi_{\bra{0}}(\mathbf{z},\bar{\mathbf{z}})} = \bra{0}\overline{C}(N)\prod_{i=1}^N\overline{\Omega}(z_i,\bar{z}_i)\ket{0}$. One can then repeat the calculation shown in Sec. \ref{Sec:CFTMthods} to show that the norm of the ground state wave function is given by,
\begin{equation} \label{Eq:partitionFunction}
    \begin{split} 
        Z_N \equiv& \langle \braket{\Psi_{\bra{0}} | \Psi_{\bra{0}}} \rangle \\
        =& \braket{ \overline{C}(N) C(N) e^{\int D^2z \overline{\Omega}(z,\bar{z}) \Omega(z, \bar{z}) } }
    \end{split}
\end{equation}
where it should be noted that when the $\phi_l(z)$ have half integral conformal dimension this is true up to an $N$-dependant -1 factor that can appear from rearranging the $\phi_l$ and $\bar{\phi}_l$ (as mentioned before under Eq. \ref{Eq:partitionFunctionBackgroundSection}). We  omit such factors for clarity. Once again, only the term in the expansion involving $N$ insertion of $\overline{\Omega}(z,\bar{z}) \Omega(z, \bar{z})$ will actually contribute by $U(1)$ charge conservation.

We then interpret $Z_N$ as the partition function of a field theory which is a ``perturbation'' of $\CFT{\mathcal{A}}$. Of course, there is no sense in which we can say the term $\overline{\Omega}(z,\bar{z}) \Omega(z, \bar{z})$ is small and so we mean this in a rather loose sense of the word ``perturbed''. The correlation functions of this field theory are given by,
\begin{equation}
   \begin{split}
        &\braket{\phi_1(w_1,\bar{w}_1)\phi_2(w_2,\bar{w}_2)\dots}_* \\
        \equiv & \langle \overline{C}(N) C(N) \mathcal{R} \phi_1(w_1,\bar{w}_1)\phi_2(w_2,\bar{w}_2)\dots \\
        &\times e^{\int D^2z \overline{\Omega}(z,\bar{z}) \Omega(z,\bar{z})} \rangle
   \end{split}
\end{equation}
where $\mathcal{R}$ denotes radial ordering and $\phi_i(w_i,\bar{w}_i)$ are fields of $\CFT{\mathcal{A}}$. There are operators of $\CFT{\mathcal{A}}$ that will have singular OPEs with the $\phi_l(z)$ operators. Correlation functions involving these operators will then require some regularisation and subsequent renormalization, which is briefly discussed in Refs. \cite{Dubail2012, Read2009} and is of no conceptual concern here. 

As discussed in Sec. \ref{Sec:CFTMthods} and in Refs. \cite{Dubail2012}, the configurations of the $\overline{\Omega}(z,\bar{z}) \Omega(z, \bar{z})$ which are the dominant contributions to the partition function are such that the long-distance (or course-grained) density profile of these insertions are the same for each configuration, with this density profile being the same as the density of particles in the corresponding ground state trial wave function. One can use a saddle-point approximation, where we are looking for saddle points of $\ln | \Psi_{\bra{0}} ( \mathbf{z},\bar{\mathbf{z}} ) |^2$, to determine this density profile (the case of the Laughlin wave functions is discussed in Ref. \cite{cappelli_large_1993}). As the parton wave functions are constructed as products of integer quantum Hall wave functions, an obvious solution, in these cases, is a density profile of a disk with a uniform density of $\frac{\nu}{2 \pi l_B^2}$, where $\nu^{-1} = \sum_i n_i^{-1}$ and $n_i$ is the filling fraction of the $i^{\text{th}}$ integer quantum Hall factor of the trial wave function. We will not discuss the more general case here, although one would expect, based on the discussion in Ref. \cite{Dubail2012}, that when the generators of $\mathcal{A}$ take the form of Eq. \ref{Eq:generalGenerator} the resulting density profile of the trial wave function will also be a disk with uniform density $\frac{\nu}{2 \pi l_B^2}$. Thus, up to exponentailly small corrections, the $\overline{\Omega}(z,\bar{z}) \Omega(z, \bar{z})$ insertions are confined to the droplet of radius $R = \sqrt{\frac{2N}{\nu}}$. 

Let $S_{FQH}$ denote the action of the perturbed theory. This can be epxressed as $S_{FQH} = S_{\CFT{\mathcal{A}}} + \Delta S$, where $S_{\CFT{\mathcal{A}}}$ is the action of $\CFT{\mathcal{A}}$ and $\Delta S$ is the ``perturbation'' of the action. $\Delta S$ is composed of two terms: one which generates the $\overline{\Omega}(z,\bar{z}) \Omega(z, \bar{z})$ insertions and another which is a perturbation localised out at infinity that comes from the background charge term $\overline{C}(N)C(N)$. From the discussion of the previous paragraph, the term in $\Delta S$ that generates the $\overline{\Omega}(z,\bar{z}) \Omega(z, \bar{z})$ insertions is localised to the droplet (up to exponentially small corrections). 

We then take the \textit{generalised screening hypothesis} to be that under RG transformations the action inside the droplet flows to a massive infrared fixed point (see Fig. \ref{fig:genralisedScreening}). In other words, this perturbed field theory has short-range correlations inside the droplet. Outside the droplet, the action is that of $S_{\CFT{\mathcal{A}}}$ plus the background charge term at infinity (up to exponentially small corrections) and, hence, the action outside the droplet should be invariant under any RG transformation. 

We now wish to discuss a slightly different formulation that will be useful when discussing edge state inner products in the Sec. \ref{Sec:CFTInnerProducts}. In the formulation above we have placed the background charge operators at the left of the correlation function and so they are equivalent to some field being placed at infinity. As discussed in Appendix \ref{Sec:PartonBackgroundCharge} the operator $C(N)$ has the property $C(N)\phi_l(z)C^\dagger(N) \propto z^{k(N) + l}\phi_l(z)$, with $k(N) \in \mathbb{Z}$. One can use this property to move the background charge to the right of the correlation function and, thus, we can express $Z_N$ as,
\begin{equation} \label{Eq:partitionFuncModForm}
    Z_N = \braket{ e^{\int D^2z \overline{\Tilde{\Omega}}(z,\bar{z}) \Tilde{\Omega}(z, \bar{z}) } \overline{C}(N)C(N)}
\end{equation}
where
\begin{equation}
    \overline{\Tilde{\Omega}}(z,\bar{z}) \Tilde{\Omega}(z, \bar{z}) = \overline{C}(N)C(N)\overline{\Omega}(z,\bar{z}) \Omega(z, \bar{z})C^\dagger(N)\overline{C}^\dagger(N)
\end{equation}
Having the background charge operator at the right of the correlation function is equivalent to some local field placed at $z = 0$. Simply moving the background charge like this will not affect the configurations of the $\overline{\Tilde{\Omega}}(z,\bar{z}) \Tilde{\Omega}(z, \bar{z})$ which are the dominant contributions to the partition function. Hence, the $\overline{\Tilde{\Omega}}(z,\bar{z}) \Tilde{\Omega}(z, \bar{z})$ should still be confined to the droplet of radius $R$. We now have a perturbed field theory with an action $\Tilde{S}_{FQH} = S_{\CFT{A}} + \Delta \Tilde{S}$, where $\Delta \Tilde{S}$ is entirely localised to the droplet (up to exponentially small corrections). Moreover, as the transformation of any field $\phi(z,\bar{z})$ from the background charge $\overline{C}(N)C(N)\phi(z,\bar{z})C^\dagger(N)\overline{C}^\dagger(N)$ is expressible as a sum of local fields at the same position as $\phi(z,\bar{z})$ (with coefficients that may depend on this position), we must have that if the field theory with action $S_{FQH}$ has short-range correlations in the droplet, then the field theory with action $\Tilde{S}_{FQH}$ must also have short-range correlations in the droplet. Hence, if generalised screening holds for the field theory with the action $S_{FQH}$ then it must hold for the field theory with the action $\Tilde{S}_{FQH}$.

\subsection{Edge-state inner products} \label{Sec:CFTInnerProducts}
We  now generalise the result of DRR \cite{Dubail2012} to the wave functions considered here. That is, we now discuss what form edge-state inner products of parton states take given generalised screening. We follow a line of argument similar to DRR. Throughout this section, we use concepts and techniques from boundary critical phenomena \cite{diehl_theory_1997, Cardy2015} and boundary conformal field theory (bCFT) \cite{Cardy2004, Andrei2020, Ishibashi1988, Cardy1989, behrend_boundary_2000, li_boundary_2022}. Other than the use of these standard methods, the following argument relies on an additional assumption, which was eluded to in the introduction. This assumption is that the matrix of Eq. \ref{Eq:generalBoundaryCon} is invertible.

In this section, we  apply these arguments to the parton states only where there always exists a unique ``densest'' (i.e. lowest angular momentum for a given number of particles) trial wave function. In these cases, the resulting boundary critical problem has full rotational invariance. For other parton states, one can still apply the arguments we use here. However, care must be taken when considering what form the fixed point boundary condition and boundary action will take for the resulting boundary-critical problem (due to the loss of full rotational invariance).

In what follows we will only be interested in inner products that correspond to the ``scaling region''. Roughly speaking, these are inner products between edge states where the fluctuation of the radius of the droplet $\delta R$ for either state is such that $\frac{\delta R}{R} \ll 1$. For a more precise definition of this, we refer the reader to Sec. III.C of \cite{Dubail2012}.

Firstly, we note that we can use the anti-chiral $\overline{\Omega}(z,\bar{z})$ to generate the complex conjugate of the wave function $\Psi_{\bra{v}}(\mathbf{z},\bar{\mathbf{z}})$ (with this wave function being defined by the general edge-state mapping of Eq. \ref{Eq:CFEdgeStateMapGeneral}),
\begin{equation}
    \overline{\Psi_{\bra{v}}(\mathbf{z},\bar{\mathbf{z}})} = \overline{\bra{v}}\overline{C}(N) \prod_{i=1}^N \overline{\Omega}(z_i,\bar{z}_i)\ket{0}
\end{equation}
where $\overline{\bra{v}}$ is the anti-chiral copy of $\bra{v}$. The edge-state inner products can then be expressed as,
\begin{equation}
    \begin{split}
        \braket{ \braket{ \Psi_{\bra{w}} | \Psi_{\bra{v}} } } =& \overline{\bra{w}}\bra{v} e^{\int D^2z \overline{\Tilde{\Omega}}(z,\bar{z})\Tilde{\Omega}(z, \bar{z}) } \\
        &\times \overline{C}(N)C(N) \ket{0} \\
    \end{split}
\end{equation}
As the state $\overline{\bra{w}}\bra{v}$ can be written as a polynomial in the modes of $\phi_l(z)$, $\phi^\dagger_l(z)$, $\bar{\phi}_l(\bar{z})$ and $\bar{\phi}^\dagger_l(\bar{z})$ applied on $\bra{0}$, this inner product can be expressed as contour integrals of correlation functions of this perturbed field theory, with the partition function of Eq. \ref{Eq:partitionFuncModForm}, where these contours are \textit{outside} the droplet. 

Thus, to understand these edge-state inner products we need to understand the structure of correlation functions of this field theory with the action $\Tilde{S}_{FQH}$, where all field insertions are outside the droplet. To this end, one can, in principle, ``integrate out'' the droplet. We are then left with a field theory that lives on the complex plane with a disk of radius $R$ cut out of it. Assuming short-range correlations inside the droplet, the action of this new field theory $\hat{S}_{FQH}$ can be written as $\hat{S}_{FQH} = S_{\CFT{\mathcal{A}}} + S_b(R)$, where $S_{\CFT{\mathcal{A}}}$ is the action of $\CFT{\mathcal{A}}$ outside the droplet and $S_b(N)$ is a \textit{boundary action} that is localised on the edge of the droplet (i.e. an integral of local operators along the droplet). We note that as the perturbation in the partition function of Eq. \ref{Eq:partitionFuncModForm} has explicit radial dependence one would expect that the boundary action $S_b(N)$ has some dependence on the number of particles in the droplet (as $R \propto \sqrt{N}$). 

If the number of particles in the droplet $N$ is large, then, in principle, we can perform an RG transformation of $\hat{S}_{FQH}$. Such an RG procedure would shrink the radius of the droplet edge, however, if $N$ is large enough then one can perform a sufficient amount of RG flow while keeping the radius of the droplet edge much larger than the magnetic length. Under the usual assumptions in the study of boundary critical phenomena, one expects that under RG flow the action outside the droplet, $S_{\CFT{A}}$, will remain invariant, as this is the action of a CFT, and only the boundary action (on the edge of the droplet), $S_b(N)$, will change. After a sufficient amount of RG flow, one would expect $S_b(N)$ to be close to some \textit{fixed point} boundary action $S_b^*(N)$, $S_b(N) \rightarrow S_b^*(N)$. As $S_b(N)$ may have some $N$ dependence the fixed point boundary action that it flows towards $S_b^*(N)$ may also have some $N$ dependence. Thus, edge-state inner products which correspond to long wavelength modes of the correlation functions of this field theory should be accurately described by the fixed point boundary action $S^*_b(N)$.

In bCFT, these fixed point boundary actions are described by \textit{fixed point boundary conditions}\cite{Cardy1989, Ishibashi1988, behrend_boundary_2000, li_boundary_2022}. Such boundary conditions typically take the form where some field of $\mathcal{A}$ inserted at the boundary, with a scaling dimension $h$, can be replaced by some linear combination of fields of $\overline{\mathcal{A}}$ with the same scaling dimension $h$ (inside correlation functions). Thus, one would expect $S^*_b(N)$ to be described by a boundary condition (on the edge of the droplet) of the form,
\begin{equation} \label{Eq:generalBoundaryCon}
    \phi^\dagger_l(z) = \bigg ( \frac{\bar{z}}{z} \bigg )^h \sum_{l'} M(N)_{ll'}\bar{\phi}_{l'}(\bar{z})
\end{equation}
where, by $U(1)$ charge conservation, the $\phi^\dagger_l(z)$ can only be replaced with $\bar{\phi}_{l'}(\bar{z})$, $h$ is the scaling dimension of $\phi^\dagger_l(z)$ and $M(N)$ is a matrix that may depend on the number of particles in the droplet. The $(\bar{z}/z)^h$ factor appears as $\phi^\dagger_l(z)$ has the opposite conformal spin of $\bar{\phi}_{l'}(\bar{z})$. We assume that $M(N)_{ll'}$ is invertible, which is the additional assumption eluded to in the abstract and introduction of this paper.

One can argue more directly, although not rigorously, that this boundary condition should occur. As we discussed in Sec. \ref{Sec:CFTMthods}, the $\overline{\Tilde{\Omega}}(z,\bar{z}) \Tilde{\Omega}(z, \bar{z})$ insertions behave analogously to a screening plasma. Let the ``electric'' charge of $\overline{\Tilde{\Omega}}(z,\bar{z}) \Tilde{\Omega}(z, \bar{z})$ be one, which then gives the electric charge of $\phi_l(z)$ and $\phi^\dagger_l(z)$ to be $1/2$ and $-1/2$ respectively. Now consider a correlation function with $\phi^\dagger_l(z)$ inserted very close (i.e. on the order of a magnetic length) to the droplet edge with all other field insertions of the correlation function outside the droplet. By the screening property, the configuration of the $\overline{\Tilde{\Omega}}(z,\bar{z}) \Tilde{\Omega}(z, \bar{z})$ insertions that are the dominant contributions to the correlation function, are such that their density profile is uniform over the droplet except near $\phi^\dagger_l(z)$ where there must be a $1/2$ charge excess to screen the charge of $\phi^\dagger_l(z)$. In any such configuration there must be a $\overline{\Tilde{\Omega}}(z,\bar{z}) \Tilde{\Omega}(z, \bar{z})$ insertion which is closest to $\phi^\dagger_l(z)$. One can then take an OPE with this $\overline{\Tilde{\Omega}}(z,\bar{z}) \Tilde{\Omega}(z, \bar{z})$ insertion and $\phi^\dagger_l(z)$, where one expects the most singular of which to be the dominant contribution to the long wavelength properties of this correlation function. This leaves some linear combination of the $\bar{\phi}_l(\bar{z})$ where $\phi^\dagger_l(z)$ was and a configuration of the $\overline{\Tilde{\Omega}}(z,\bar{z}) \Tilde{\Omega}(z, \bar{z})$ insertions with a uniform density throughout the droplet except at the location where $\phi^\dagger_l(z)$ was where there is now a $-1/2$ charge deficit. Repeating this process for all other contributing $\overline{\Tilde{\Omega}}(z,\bar{z}) \Tilde{\Omega}(z, \bar{z})$ insertion configurations and then averaging, we then have, on long length scales, a $\phi^\dagger_l(z)$ insertion at the droplet edge can be replaced with some linear combination of $\bar{\phi}_{l'}(\bar{z})$ insertions at the same location. This then leads us to the boundary condition of Eq. \ref{Eq:generalBoundaryCon}. 

Among these possible boundary conditions is one that takes the form,
\begin{equation} \label{Eq:SpecialBoundaryCon}
    \phi^\dagger_l(z) = \bigg ( \frac{\bar{z}}{z} \bigg )^h \bar{\phi}_l(\bar{z})
\end{equation}
on the droplet edge. Let $\braket{\dots}_{1}$ denote the correlation function of the boundary CFT which is $\CFT{\mathcal{A}}$ on the complex plane with a disk of radius $R$, centred at the origin, removed, with this boundary condition at the edge of the disk. Such correlation functions can be computed by inner products in $\CFT{\mathcal{A}}$ on the full complex plane. That is the correlation function of some field insertions, $\phi_i(z_i)$ and $\bar{\phi}_j(\bar{z}_j)$, can be expressed as, $\braket{\prod_i \phi_i(z_i) \prod_j \bar{\phi}_j(\bar{z}_j) }_{1} = \bra{0} \prod_i \phi_i(z_i) \prod_j \bar{\phi}_j(\bar{z}_j) R^{L_0 + \bar{L}_0} \ket{B}$, for some state $\ket{B}$. We have included a factor of $R^{L_0 + \bar{L}_0}$ so that $\ket{B}$ is a state that satisfies the condition $[\phi^\dagger_l(z) - ( \bar{z}/z )^h \bar{\phi}_l(\bar{z})]\ket{B} = 0$ where $|z|=1$ (which enforces the required boundary condition for the $\braket{ \dots }_{1}$ correlation functions). Written in terms of the modes of the fields this reads,
\begin{equation} \label{Eq:boundaryConMode}
    [\phi^\dagger_{l,n} - \bar{\phi}_{l,-n}]\ket{B} = 0
\end{equation}
As we require $\braket{1}_{1} = 1$, one also has $\bra{0}\ket{B} = 1$. Furthermore, as we are only interested in correlation functions of fields of $\mathcal{A}$ and $\overline{\mathcal{A}}$, one can take $\ket{B} \in \mathcal{H}_0 \otimes \overline{\mathcal{H}}_0$, where $\mathcal{H}_0$ is the vacuum representation of $\mathcal{A}$ which is irreducible. One can show, using Schur's lemma, that the state $\ket{B}$ is completely determined by the condition of Eq. \ref{Eq:boundaryConMode} and $\bra{0}\ket{B} = 1$ (which is a standard result in bCFT). Such a state $\ket{B}$ is referred to as an Ishibashi state \cite{Ishibashi1988}. It can also be shown from these conditions that the state $\ket{B}$ has the property $\overline{\bra{w}}\bra{v} \ket{B} = \braket{v | w}$\footnote{As pointed out in Ref. \cite{Dubail2012} this can be seen directly by expressing the states as modes of the $\phi_l$ fields applied on the vacuum. For example, consider the case of just a chiral boson where the modes of the field have the boundary condition $a_n \ket{B} = \bar{a}_{-n}\ket{B}$. Now compute $\bra{0}a_1a_2\bar{a}_1\bar{a}_2 \ket{B} = \bra{0}a_1a_2\bar{a}_1a_{-2} \ket{B} = 2 \bra{0}a_1\bar{a}_{1}\ket{B} = 2 \bra{0}a_1a_{-1}\ket{B} = 2\braket{0|B} = 2 = \bra{0}a_1a_2a_{-1}a_{-2}\ket{0}$. One can generalize this approach to see that in the case where the $\phi_l$ have half integral conformal dimension, which implies that the modes of the $\phi_l$ and $\bar{\phi}_l$ must anti-commute, one can generally obtain an overall minus sign that depends on the number of $\phi_l$ modes in either of the states $\ket{v}$ or $\ket{w}$.}. Note that when the $\phi_l$'s have half integral conformal spin $\overline{\bra{w}}\bra{v} \ket{B} = \braket{v | w}$ is true up to an overall minus sign that depends on the $U(1)$ charges of $\ket{v}$ and $\ket{w}$.  Finally, as every field of $\mathcal{A}$ is generated by repeated OPEs of $\phi_l(z)$ and $\phi^\dagger_l(z)$, the boundary condition of Eq. \ref{Eq:SpecialBoundaryCon} implies that any field of $\mathcal{A}$ inserted at the boundary can be replaced by some linear combination of insertions of fields of $\overline{\mathcal{A}}$ at the same point. We can then take $\mathcal{A}$ as the set of \textit{boundary operators}\footnote{Strictly speaking, we can have boundary operators which belong to other representations of $\mathcal{A}$, however, in the present case we are only interested in correlation functions in this bCFT that only involve fields of $\mathcal{A}$ and $\overline{\mathcal{A}}$.}. It can also be shown that $[L_n - \bar{L}_{-n}]\ket{B} = 0$, which means the boundary condition of Eq. \ref{Eq:SpecialBoundaryCon} is \textit{conformal}.

We can use the $\braket{...}_1$ theory to describe the correlation functions of the theory on the complex plane with the disk of radius $R$ removed around the origin with the boundary action $S_b^*(N)$ on the edge of the disk, which enforces the boundary condition of Eq. \ref{Eq:generalBoundaryCon}. Let the correlation functions of this theory be denoted by $\braket{...}_{2,N}$. Now define the operator $\hat{M}$ that only acts on the chiral sector by $\phi^\dagger_l(z)\hat{M} = \hat{M}\sum_{l'} M(N)_{ll'}\phi^\dagger_{l'}(z)$, $\phi_l(z)\hat{M} = \hat{M}\sum_{l'}[M(N)^{-1}]^*_{ll'}\phi_{l'}(z)$ (with $*$ denoting the complex conjugate) and $\hat{M}\ket{0} = \ket{0}$. Using this the correlation functions of the $\braket{...}_{2,N}$ theory can be expressed as $\braket{\prod_i \phi_i(z_i) \prod_j \bar{\phi}_j(\bar{z}_j) }_{2,N} = \bra{0} \prod_i \phi_i(z_i) \prod_j \bar{\phi}_j(\bar{z}_j) \hat{M} R^{L_0 + \bar{L}_0} \ket{B}$, where one can check that the defining properties of $\hat{M}$ reproduces the boundary condition of Eq. \ref{Eq:generalBoundaryCon}.

We can normalise the $\phi_l$ fields such that $\phi_l(z)\phi_{l'}^\dagger(w) = \delta_{ll'}(z-w)^{-2h} + \dots$, where $h$ is the scaling dimension of $\phi_l(z)$. With this normalisation we then have that for large-$N$, $\braket{\braket{\Psi_{\bra{0}\phi^\dagger_{l,h}} | \Psi_{\bra{0}\phi^\dagger_{l',h}} }} \propto M(N)_{ll'}$, which implies that $M(N)$ must be a Hermitian positive semi-definite matrix. Furthermore, as we have already assumed $M(N)$ is invertible, it follows that $M(N)$ must a Hermitian positive definite matrix. This then implies that we can express $\hat{M}$ as $\hat{M} = e^{\delta S_b^*(N)}$, where $\delta S_b^*(N)$ is another operator that only acts on the chiral sector. As $\hat{M}$ implements a local scale-invariant boundary condition at the edge of the disk, we must have that, by analytic continuation from the boundary, for any $\phi(z) \in \mathcal{A}$, $\hat{M}^{-1}\phi(z)\hat{M}$ must be expressible as a sum of fields from $\mathcal{A}$ at the same location $z$ all having the same scaling dimension as $\phi(z)$. From this, it follows that $\delta S_b^*(N)$ must be a conserved charge of $\CFT{\mathcal{A}}$. Hence, $\delta S_b^*(N)$ must be expressable as a one-dimensional integral of local fields in $\mathcal{A}$, as is implied by Noether's theorem. 

We can now express the correlation functions of the $\braket{...}_{2,N}$ theory as $\braket{\prod_i \phi_i(z_i) \prod_j \bar{\phi}_j(\bar{z}_j) }_{2,N} = \braket{\prod_i \phi_i(z_i) \prod_j \bar{\phi}_j(\bar{z}_j) e^{\delta S_b^*(N)}}_1$, where in this expression the integral in $\delta S_b^*(N)$ can be taken to be along the edge of the droplet. Hence, the $\braket{...}_{2,N}$ theory can be interpreted as a local boundary perturbation of the $\braket{...}_1$ theory. Thus, these correlation functions can also be expressed as $\braket{\prod_i \phi_i(z_i) \prod_j \bar{\phi}_j(\bar{z}_j) }_{2,N} = \bra{0} \prod_i \phi_i(z_i) \prod_j \bar{\phi}_j(\bar{z}_j) e^{\delta S^*_b(N)} R^{L_0 + \bar{L}_0} \ket{B}$. 

The operators in $\delta S_b^*(N)$ must be $U(1)$ neutral, by $U(1)$ charge conservation, and must have scaling dimension one, for this to be a marginal \textit{boundary} perturbation. Thus, $\delta S^*_b(N) = \sum_a \oint_{|z|=R} \frac{dz}{2\pi i} f_a(N; \theta) J^a(z) $, where $J^a(z)$ form a basis of $U(1)$ neutral fields of $\mathcal{A}$ with scaling dimension one, $\theta$ is the usual polar coordinate around the edge of the disk and $f_a(N; \theta)$ are some complex-valued functions. By rotational invariance, however, $f_a(N; \theta)$ must take the form $f_a(N; \theta) = f_a(N)$. Hence, we are left with $\delta S_b^*(N) = \sum_a f_a(N) \oint_{|z|=R} \frac{dz}{2\pi i} J^a(z) = \sum_a f_a(N) J^a_0$. This then implies that $[\delta S_b^*(N), L_n] = 0$, as $\delta S_b^*(N)$ is composed of zero modes of scaling dimension one fields, which means the boundary condition for the $\braket{\dots}_{2, N}$ theory is conformal. As the $\phi_l(z)$ do not generally form a basis of the space of fields with the same scaling dimension and $U(1)$ charge of the $\phi_l(z)$ fields, to obtain a boundary condition of the form of Eq. \ref{Eq:generalBoundaryCon} one may need to impose some constraints on $f_a(N)$ such that $[\delta S^*_b(N), \phi^\dagger_l(z)]$ is expressable as a sum of $\phi^\dagger_{l'}(z)$ fields. Such constraints will not be required for the cases considered later in this paper.

Thus, assuming generalised screening, one expects the inner products that correspond to long wavelength modes of correlation functions of the field theory with the action $\hat{{S}}_{FQH}$, for large $N$, to take the form,
\begin{equation}
    \begin{split}
        \frac{\braket{ \braket{ \Psi_{\bra{w}} | \Psi_{\bra{v}} } }}{Z_N} \approx &\overline{\bra{w}}\bra{v}e^{\delta S_b^*(N)}R^{L_0 + \bar{L}_0}\ket{B} \\
        =& \bra{v} e^{\delta S_b^*(N)}R^{2L_0} \ket{w} \\
        =& \bra{v} R^{2L_0} e^{\delta S_b^*(N)} \ket{w}
    \end{split}
\end{equation}
where in going from line one to line two one should recall $\delta S_b^*(N)$ only contains operators of $\mathcal{A}$ and in going from line two to line three one should also recall $[\delta S_b^*(N), L_0] = 0$. Note that in the case when $\phi_l$ have half integral conformal dimension the minus signs that can appear in $\overline{\bra{w}}\bra{v} \ket{B} = \braket{v | w}$ cancels the other minus signs that can appear as mentioned under Eq. \ref{Eq:partitionFunctionBackgroundSection} and Eq. \ref{Eq:partitionFunction}.  From the discussion of Sec. \ref{Sec:integerQHCFT}, one can easily see that the edge state inner products in the $\nu=1$ and $\nu=2$ cases take this form for large $N$, where the $f_a(N)$ terms diverge as $N$ increases. By modifying the edge state map by the replacement $\bra{v} \rightarrow \bra{v}e^{-\frac{\delta S_b^*(N)}{2}}R^{-L_0}$ then we have, for large $N$, 
\begin{equation}
    \frac{\braket{ \braket{ \Psi_{\bra{w}e^{-\delta S_b^*(N)/2}R^{-L_0}} | \Psi_{\bra{v}e^{-\delta S_b^*(N)/2}R^{-L_0}} } }}{Z_N} \approx \braket{v | w}
\end{equation}
Hence, given generalised screening, there exists a simple modification of the edge state map which becomes an isometric isomorphism in the thermodynamic limit, where an isometric isomorphism (i.e. an invertible linear map that preserves the inner product). In the cases considered by DRR the only possible term in $\delta S_b^*(N)$ involved the zero modes of the $i\partial\Phi(z)$ current, $a_0$, which could simply be cancelled by the replacement $\Omega(z) \rightarrow \sqrt{\lambda_N}\Omega(z)$ (with an appropriately chosen $\lambda_N$). 

For intermediate system sizes $N$, one must consider the RG irrelevant terms in the boundary action $S_b(N)$. This can also be described as a perturbation of the boundary action enforcing the Eq. \ref{Eq:SpecialBoundaryCon}, which we denote $\delta S_b(N)$, that includes the exactly marginal $\delta S_b^*(N)$ term and boundary RG irrelevant terms. This can be expressed in the form, $\delta S_b(N) = \delta S_b^*(N) + \sum_{h_j > 1} \alpha_j(N) (2\pi )^{-1} \int_{|z|=R} |dz|e^{ih_j\theta } \phi_j(z)$, where $\phi_j(z)$ are fields of $\mathcal{A}$ with corresponding scaling dimension $h_j$, $\alpha(N)$ are numbers, and the $e^{ih_j\theta}$ is required by rotational invarience. In principle, the $\alpha_j(N)$ can depend on the number of particles in the droplet, because of the explicit radial dependence in the perturbation in the partition function of Eq. \ref{Eq:partitionFuncModForm}, however, previous works show that they do not, at least for the first smallest scaling dimension terms \cite{Dubail2012, Fern2018}. This then gives the following form for edge-state inner products at intermediate system sizes,
\begin{equation} \label{Eq:genInnerProdForm}
    \frac{\braket{ \braket{ \Psi_{\bra{w}} | \Psi_{\bra{v}} } }}{Z_N} = \bra{v}R^{2L_0}e^{\delta S_b(N)} \ket{v}
\end{equation}
with,
\begin{equation} \label{Eq:generalInProdAct}
    \begin{split}
        \delta S_b(N) =& \sum_a f_a(N) J^a_0 \\
        &+ \sum_{h_j > 1} \frac{\alpha_j(N)}{R^{h_j - 1}} \oint \frac{dz}{2\pi i} z^{h_j - 1} \phi_j(z) \\
    \end{split}
\end{equation}
where we have used $|dz|e^{h_j\theta} = \frac{dz}{iR^{h_j - 1}}z^{h_j - 1}$ (at $|z|=R$). We would like to emphasise to the reader that in principle $\alpha_j(N)$ can depend on $N$. From the RG analysis above, however, we do expect that $\alpha_j(N)/R^{h_j -1} \rightarrow 0$ in the thermodynamic limit in such a way that the less relevant terms fall to zero faster than the more relevant terms. When modelling the inner products this property would allow for the expansion of the inner product action to be truncated to terms below some low scaling dimension. In the numerical tests of Sec. \ref{Sec:numericalTests} we  show that the $\alpha_j$ for the $h_j = 2$ terms do not appear to have any $N$ dependence for the cases tested.

\subsection{Real-space entanglement spectra} \label{Sec:RSES}
We now will briefly discuss the structure of the real-space entanglement spectrum (RSES) for trial wave function that can be constructed from CFT in the way outlined in Sec. \ref{Sec:CFTGenStructure}. This is a rather straightforward extension of the calculation in DRR's work. We merely wish to point out that DRR's result can be extended to more general CFT constructions. For a precise definition of the RSES we refer the reader to Refs. \cite{Li2008, Dubail2012, Dubail2012a, Qi2012, Sterdyniak2012, Rodriguez2012, Rodriguez2013, Davenport2015, Amico}. 

Now consider a system of $N$ particles whose wave function can be expressed in the form given by Eq. \ref{Eq:generalCFTWFGen}. We then take a circular real-space cut centred at the origin with radius $R_c = R/\sqrt{2}$, so that the average number of particles inside the cut is $N/2$ (with $R$ again being the radius of the droplet). We have chosen this particular radius merely for simplicity of exposition and is of no conceptual significance. Let the region inside the real-space cut be $A$ and the region outside be $B$. 

We now define two edge state maps for these two subsystems. For subsystem $A$ we have,
\begin{equation} \label{Eq:CFEdgeStateMapA}
    \begin{split}
        \ket{\Psi^A_{\bra{v}}}\rangle \equiv& \bra{v} R_c^{-2L_0} C(N/2) e^{\int_A d^2z e^{-|z|^2/4} \Omega(z, \bar{z})\otimes c^\dagger(z,\bar{z}) } \\
        &\times \ket{0}\otimes\ket{0}\rangle
    \end{split}
\end{equation}
and for subsystem $B$ we have,
\begin{equation} \label{Eq:CFEdgeStateMapB}
    \begin{split}
        \ket{\Psi^B_{\ket{w}}}\rangle \equiv& \bra{0} C(N) e^{\int_B d^2z e^{-|z|^2/4} \Omega(z, \bar{z})\otimes c^\dagger(z,\bar{z}) } \\
        &\times C^\dagger(N/2) R_c^{2L_0} \ket{w}\otimes\ket{0}\rangle
    \end{split}
\end{equation}
We further assume that $N$ is such that the edge state maps for $A$ and $B$ have the property that any two CFT states with the same $L_0$ eigenvalues and $U(1)$ charge will map to wave functions with the same angular momentum (so that for a fixed particle number the $L_0$ and angular momentum operators are equivalent up to an additive constant). In the integer quantum Hall case and by extension both the symmetric parton and composite fermion cases, this is simply the condition that for $N/2$ particles confined to the lowest $n$ Landau levels the lowest angular momentum state is unique. When $N/2$ does not fulfil this condition one will need to consider the mapping of the angular momentum operator more carefully.

For subsystems $A$ and $B$ the edge state inner products can be expressed in the form, $\braket{\braket{ \Psi^A_{\bra{w}} | \Psi^A_{\bra{v}} }} = \bra{v}e^{\delta S^A_b(N/2)}\ket{w}$ and $\braket{\braket{ \Psi^B_{\ket{w}} | \Psi^B_{\ket{v}} }} = \bra{w}e^{\delta S^B_b(N/2)}\ket{v}$ respectively. We absorbed the $Z^A_N$ and $Z^B_N$ factors into $\delta S^A_b(N)$ and $\delta S^B_b(N)$ respectively (to keep the notation simple). Assuming generalised screening holds then we expect $\delta S_b^A(N)$ and $\delta S_b^B(N)$ to take the local form of Eq. \ref{Eq:generalInProdAct}.

Now define the entanglement action, $S_{ES}$, by $e^{-\frac{S_{ES}}{2}} = e^{\frac{\delta S_b^B(N)}{2}} e^{\frac{\delta S_b^A(N)}{2}}$. We write the singular value decompositon $e^{-\frac{S_{ES}}{2}} = \sum_i e^{-\xi_i} \ket{u_i}\bra{v_i}$, where $\ket{v_i}$ form an orthonormal basis and $\ket{u_i}$ form another orthonormal basis. We can then perform the following resolution of the identity,
\begin{equation}
    \begin{split}
        \mathbf{1} = & C^\dagger(N/2)R_c^{2L_0}e^{-\frac{\delta S^B_b(N)}{2}} e^{-\frac{S_{ES}}{2}} e^{-\frac{\delta S^A_b(N)}{2}} R_c^{-2L_0} C(N/2) \\
        = & \sum_i e^{-\xi_i} C^\dagger(N/2)R_c^{2L_0}e^{-\frac{\delta S^B_b(N)}{2}} \ket{u_i} \bra{v_i} \\
        & \times e^{-\frac{\delta S^A_b(N)}{2}}R_c^{-2L_0} C(N/2)
    \end{split}
\end{equation}
We can then express the ground state trial wave function in the following form,
\begin{equation}
    \begin{split}
        \ket{\Psi_{\bra{0}}}\rangle =& \bra{0} C(N) e^{\int d^2z e^{-|z|^2/4} \Omega(z, \bar{z})\otimes c^\dagger(z,\bar{z}) } \\
        &\times \ket{0}\otimes\ket{0}\rangle \\
        &= \bra{0} C(N) e^{\int_B d^2z e^{-|z|^2/4} \Omega(z, \bar{z})\otimes c^\dagger(z,\bar{z}) } \\
        &\times e^{\int_A d^2z e^{-|z|^2/4} \Omega(z, \bar{z})\otimes c^\dagger(z,\bar{z}) } \\
        &\times \ket{0}\otimes\ket{0}\rangle \\
    \end{split}
\end{equation}
Then inserting the resolution of the identity in the middle, we have,
\begin{equation}
    \begin{split}
        \ket{\Psi_{\bra{0}}}\rangle &= \sum_i e^{-\frac{\xi_i}{2}} \bigg | \Psi^B_{e^{-\frac{\delta S^B_b(N)}{2}} \ket{u_i}} \bigg \rangle \bigg \rangle \\
        &\times \bigg | \Psi^A_{\bra{v_i} e^{-\frac{\delta S^A_b(N)}{2}}} \bigg \rangle \bigg \rangle
    \end{split}
\end{equation}
This then provides a Schmidt decomposition, which implies that the set of $\xi_i$, which are the eigenvalues of $S_{ES}$, forms the entanglement spectrum. 

As both the particle number and the angular momentum are good quantum numbers in this entanglement spectrum (from the rotational invariance of the real-space cut), the states $\ket{v_i}$ must be eigenstates of the $a_0$ and $L_0$ operators. Hence, the number states with a given $L_0$ and $a_0$ in the vacuum representation of $\mathcal{A}$ must give a rigorous upper bound for the entanglement level state counting at the corresponding particle number and angular momentum sector of the entanglement spectrum. 

If the generalised screening hypothesis holds, then from $e^{-\frac{S_{ES}}{2}} = e^{\frac{\delta S_b^B(N)}{2}} e^{\frac{\delta S_b^A(N)}{2}}$ and through the Baker-Cambell-Housdorf formula one can show $S_{ES}$ will take the form of Eq. \ref{Eq:generalInProdAct}. In such an expansion we again could in principle have some $N$ dependence in the coefficients $\alpha_j(N)$, however, numerical evidence suggests they do not \cite{Henderson2021}. This form of $S_{ES}$ then implies that for states within the scaling region, the entanglement level state counting will match the state counting of the vacuum representation of $\mathcal{A}$ (i.e. the aforementioned upper-bound must now be saturated in the scaling region).

In principle then, one can use this result to explain some of the observations of Ref. \cite{anand_real-space_2022} regarding the RSES of parton wave functions. However, this will not be discussed here.

\section{Numerical Tests} \label{Sec:numericalTests}
We  now present our numerical tests of whether the edge-state inner products can be expressed in the general form given in Eq. \ref{Eq:genInnerProdForm} and Eq. \ref{Eq:generalInProdAct}, which one expects if generalised screening holds, in the case of the \textit{unprojected} $\nu = 2/5$ composite fermion state and the \textit{unprojected} $\phi_2^2$ parton state. In Sec. \ref{Sec:innerProdModels} we  first present a model of the inner product action for each state, where we only include terms up to and including scaling dimension two. Next, we present how these models can be fitted to the actual inner products. Finally, in Sec. \ref{Sec:numResults} the result of this fitting procedure and how the fitted model parameters scale with the system size, where we find the edge inner-products are consistent with generalised screening. 

\subsection{The models} \label{Sec:innerProdModels}
The following models for the inner product action only contain terms which are integrals of fields up to scaling dimension two, for simplicity. In each case, there are, in principle, many fields from the corresponding chiral algebra at scaling dimension two that could appear. The models we present here contain far fewer fields than otherwise could be included, as we \textit{emprically} find them to be sufficient to model the inner product action. One may be able to constrain the possible inner product action using translational symmetry of the droplet \cite{Dubail2012, Fern2018} and thereby give an explanation as to why these simpler models work. However, we will not pursue this here. 

For the $\nu = 2/5$ composite fermion case the model inner product action is given by,
\begin{equation} \label{Eq:CFInnerProdModel}
   \begin{split}
        \delta \hat{S}_b = & \alpha J^3_0 + \beta J^1_0 \\
        & + \oint \frac{dz}{2\pi i} z [\gamma :(J^3(z))^2: + \sqrt{2}\delta J^3(z)i\partial\Phi(z)] 
   \end{split}
\end{equation}
where the $\widehat{\mathfrak{su}}(2)_1$ currents, $J^a(z)$, are defined in Eq. \ref{Eq:CFSU2Fields}, and $\alpha, \beta$, $\gamma$ and $\delta$ are model parameters that require fitting. The zero mode of $i\partial\Phi(z)$, $a_0$, is not included as we only fit the inner product action using inner products for a fixed number of particles and so the presence of such a term would not be detectable. From the result for $\nu = 2$ in Eq. \ref{Eq:nu2innerprodaction}, we expect that $\alpha \sim \ln N$ and $\beta \sim (\ln N)/N$. Furthermore, from previous works \cite{Fern2018, Dubail2012} we expect $\gamma, \delta \sim 1/\sqrt{N}$ (i.e. under the assumption that the $\alpha_i(N)$ of Eq. \ref{Eq:generalInProdAct} do not depend on $N$).

As mentioned at the start of this section, terms in the model of Eq. \ref{Eq:CFInnerProdModel} are not all the possible terms that could appear up to scaling dimension 2. At scaling dimension one the other possible term that could have been there is the zero mode of $J^2(z)$. At scaling dimension two the other possible terms, which produce linearly independent terms when integrated, are $:(i\partial\Phi(z))^2:$, $:J^3(z)J^1(z):$, $:J^3(z)J^2(z):$, $i\partial\Phi(z)J^2(z)$ and $i\partial\Phi(z)J^1(z)$. We find that not including these terms in the model reproduces the numerically estimated inner-product action matrix elements with sufficient accuracy (after fitting).

For the $\phi^2_2$ state the model inner product action is,
\begin{equation} \label{Eq:partInnProdModel}
    \begin{split}
         \delta \hat{S}_b = & \alpha J^3_0 + \beta J^1_0 \\
        &+ \oint \frac{dz}{2\pi i} z \bigg [ \frac{\gamma}{2} :(J^3(z))^2: \\ 
        &+  \frac{\delta}{2} [ :(J^2(z))^2: + :(J^3(z))^2: ] + \epsilon J^3(z) i\partial\Phi(z) \bigg ]
    \end{split}  
\end{equation}
where the $J^a(z)$ and $\Phi(z)$ fields are defined in Eq. \ref{Eq:partonSU2Currents}. As for the composite fermion case we do not include the zeroth mode of $i\partial\Phi(z)$ and we expect $\alpha \sim \ln N$, $\beta \sim (\ln N)/N$ and $\gamma, \delta, \epsilon \sim 1/\sqrt{N}$. 

As for the $\nu = 2/5$ there are other terms that could appear in this model. At scaling dimension one the other possible term that could have been there is the zero mode of $J^2(z)$. At scaling dimension two the other possible terms, which produce linearly independent terms when integrated, are $:(i\partial\Phi(z))^2:$, $:J^3(z)J^1(z):$, $:J^3(z)J^2(z):$, $i\partial\Phi(z)J^2(z)$ and $i\partial\Phi(z)J^1(z)$. Again we find that not including these terms in the model reproduces the numerically estimated inner-product action matrix elements with sufficient accuracy (after fitting).

\subsection{Fitting procedure} \label{Sec:fitting}
We first detail the general fitting procedure for some CFT to edge state mapping. Let $\ket{M;i}$ be an orthonormal basis of states in the given CFT which have $L_0$ eigenvalue $M$ and are all $U(1)$ neutral (i.e. $a_0 \ket{M;i} = 0$) so that these states map to edge state wave functions of $N$ particles. We assume that the background charge operator, $C(N)$, has been chosen such that these the wave functions $\ket{\Psi_{\bra{M;i}}}\rangle$ all have angular momentum $M$ \textit{relative} to the ground state trial wave function $\ket{\Psi_{\bra{0}}}\rangle$. This then gives $\braket{\braket{ \Psi_{\bra{M;i}} | \Psi_{\bra{M';j}} }} = \bra{M';j}R^{2L_0}e^{\delta S_b}\ket{M;i} = 0$ for $M \neq M'$. Thus, $\bra{M';j}\delta S_b\ket{M;i} = 0$ for $M \neq M'$, which means the operator $\delta S_b$ must be a block diagonal matrix relative to the $\ket{M;j}$ basis. We then define the series of matrices $G(M)_{ij} \equiv \braket{\braket{ \Psi_{\bra{M;j}} | \Psi_{\bra{M;i}} }}/R^{2M} = \bra{M;i}e^{\delta S_b}\ket{M;j}$. We then see that the log of these matrices must be given by $[\ln G(M)]_{ij} = \bra{M;i}\delta S_b\ket{M;j}$. Hence, if one can compute $G(M)_{ij}$ then the matrix elements of $\delta S_b$ can be determined.

To fit the model entanglement action we first compute the $G(M)_{ij}$ using Monte Carlo integration and then the log of these matrices is taken to give estimates for the matrix elements of $\delta S_b$. The parameters of the model inner product action are then determined by minimizing the sum of squared differences between the model and estimated matrix elements of $\delta S_b$, over the $M = 1,2$ subspaces where each matrix element is equally weighted. 

For the $\nu = 2/5$ composite fermion state the chosen basis takes the form $\ket{\lambda_1;\lambda_2;p} \equiv \prod_{n_1 \in \lambda_1}a_{-n_1} \prod_{n_2 \in \lambda_2} J^3_{n_2} (F_2F^\dagger_1)^p\ket{0}/\sqrt{\mathcal{N}}$, where $\lambda_i$ are partitions and $\mathcal{N}$ is a normalisation factor. From the gauge invarience of the edge state map the $\ket{\lambda_1;\lambda_2;p}$ state will map to the same edge state as $\prod_{n_1 \in \lambda_1}\sqrt{5/2}(\fefm{1}_{-n_1} + \fefm{2}_{-n_1}) \prod_{n_2 \in \lambda_2} \sqrt{2}^{-1} (\fefm{2}_{n_2} - \fefm{1}_{n_2}) (F_2F^\dagger_1)^p\ket{0}/\sqrt{\mathcal{N}}$. Using the usual bosonisation relations, which we discuss in Sec. \ref{Sec:integerQHCFT}, such states can be expressed as polynomials in the modes of the $\Tilde{V}_j(z)$ fields applied on the vacuum, which then ultimately allows for the $\Psi_{\bra{\lambda_1;\lambda_2;p}}(\mathbf{z},\bar{\mathbf{z}})$ wave functions to be expressed as a sum of Slater determinants of the $z^m$ and $\bar{z}z^m$ orbitals, times the flux attaching Jastrow factor. By expressing the $\Psi_{\bra{\lambda_1;\lambda_2;p}}(\mathbf{z},\bar{\mathbf{z}})$ wave function in this form the $G(M)$ matrices can then be computed using Monte Carlo integration. To find the matrix elements of the model $\delta S_b$ we first note that $J^3_0$ and $([J^3]^2)_0 \equiv \oint \frac{dz}{2\pi i} z :(J^3(z))^2:$ are diagonal in the $\ket{\lambda_1;\lambda_2;p}$ basis with $J^3_0\ket{\lambda_1;\lambda_2;p} = p\ket{\lambda_1;\lambda_2;p}$ and $([J^3]^2)_0\ket{\lambda_1;\lambda_2;p} = [ p^2 + (\sum_{n\in\lambda_2} n) ]\ket{\lambda_1;\lambda_2;p}$. The matrix elements of $J^1_0$ can be computed by first noting that $J^1_0 = (J^+_0 + J^-_0)/2$. Then define the chiral boson field $\eta(z) \equiv -i\sqrt{2}\ln z J^3_0 + i\sum_n J^3_n \frac{z^{-n}}{n}$, which can be used to express the $J^\pm(z)$ fields as $J^\pm(z) = (F_2F^\dagger_1)^{\pm 1} : e^{\pm i\sqrt{2}\eta(z)} :$ (i.e. the free field representation of $\widehat{\mathfrak{su}}(2)_1$ \cite{DiFrancesco1997}). The matrix elements of $J^\pm_0$ operators are then equivalent to the matrix elements of modes of chiral vertex operators of a chiral boson. We discuss the computation of such matrix elements in Appendix \ref{Sec:vertexOpMatrixEle}.

For the $\phi_2^2$ parton case the basis we used has a somewhat more involved definition. The space of states in the vacuum representation of $\mathcal{A}(2)_2$ (i.e. the chiral algebra corresponding to this state) that are $U(1)$ charge neutral form the vacuum representation of $\hat{\mathfrak{u}}(1) \oplus \widehat{\mathfrak{su}}(2)_2$. The $\hat{\mathfrak{u}}(1)$ part is generated by the modes of $i\partial\Phi(z)$ and the $\widehat{\mathfrak{su}}(2)_2$ part can be \textit{represented} using a Majorana field $\psi(z)$ and a chiral boson $\varphi(z)$ with compactification radius one and corresponding Klein factor denoted by $F_\varphi$ (where this Klein factor should anti-commute with the Majorana field) \cite{gepner_modular_1987}. We discuss this in more detail in Appendix \ref{Sec:SU2Free} and will give a summary here. We denote the chosen basis for the system containing the $\Phi(z)$, $\psi(z)$ and $\varphi(z)$ as $\ket{\lambda_1;\mu, \lambda_2;p} = \prod_{n_1 \in \lambda_1}a_{n_1}\prod_{n_3\in \lambda_3}\fem{\varphi}_{-n_3}F_{\varphi}^p\prod_{n_2\in \mu} \psi_{-\frac{n_2}{2}}\ket{0}/\sqrt{\mathcal{N}}$, where $\fem{\varphi}_n$ are the modes of $\varphi(z)$, $\lambda_i$ are partitions, $\mu$ is a partition with no repeated elements and where all elements are odd, $p \in \mathbb{Z}$, and $\mathcal{N}$ is used to normalise the state. The vacuum representation of $\hat{\mathfrak{u}}(1) \oplus \widehat{\mathfrak{su}}(2)_2$ is spanned by the basis elements $\ket{\lambda_1;\mu;\lambda_2;p}$, such that the parity of the number of elements in $\mu$ is equal to the parity of $p$ (i.e. if $p$ is odd $\mu$ will have an odd number of elements and if $p$ is even then $\mu$ must have an even number of elements). We further detail how this basis of the $\hat{\mathfrak{u}}(1) \oplus \widehat{\mathfrak{su}}(2)_2$ can be mapped back to states of $\mathcal{H}_{\text{CFT}}$ (defined in Sec. \ref{Sec:partonCFT}) with the given states being expressed as modes of the $\Tilde{V}_{ij}(z)$ and $\Tilde{V}^\dagger_{ij}(z)$ fields applied on the vacuum. Expressing the basis this way allows the corresponding edge states, $\ket{\Psi_{\bra{\lambda_1;\mu;\lambda_2;p}}}\rangle$, under the edge state map, to be expressed as a sum of products of Slater determinants of the $z^m$ and $\bar{z}z^m$ orbitals from the discussion of Sec. \ref{Sec:partonEdgeMap}. By expressing $\ket{\Psi_{\bra{\lambda_1;\mu;\lambda_2;p}}}\rangle$ this way we then computed the $G(M)$ matrices using Monte Carlo integration. The computation of the matrix elements of the model inner product action is also discussed in Appendix \ref{Sec:SU2Free}. 

\subsection{Results} \label{Sec:numResults}
\subsubsection{$\nu = 2/5$ composite fermion}
\begin{figure*}[ht!]
    \centering
    \includegraphics{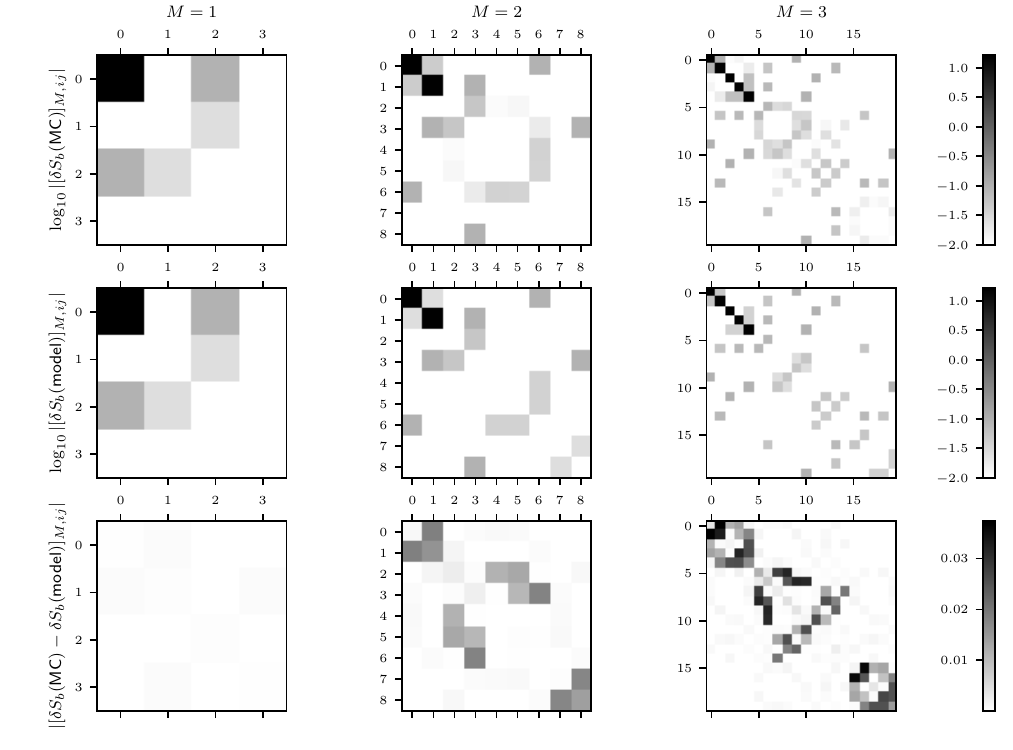}
    \caption{Comparison between the Monte Carlo (MC) estimated and model (Eq. \ref{Eq:CFInnerProdModel}) inner product action matrix elements, for the case of $\nu = 2/5$ composite fermion state where the ground state wave function contains $N = 55$ fermions. We use the shorthand notation $[\delta S_b]_{M,ij} \equiv \bra{M;i}\delta S_b \ket{M; j}$, where the basis reference for the index $i$ can be found in Appendix \ref{Sec:furtherMatElFit}. Note that when presenting the $\log_{10}$ of the absolute value of the matrix elements, we have used a cutoff where any matrix elements whose absolute value is below $10^{-2}$ have been replaced by $10^{-2}$, for clarity. No cuttof has been used for the matrix element errors, $|[\delta S_b(\text{MC}) - \delta S_b(\text{Model}) ]_{M,ij}|$. Some sample matrix elements can be found in Appendix \ref{Sec:furtherMatElFit}.}
    \label{fig:fittingTestCF}
\end{figure*}

Fig. \ref{fig:fittingTestCF} shows a comparison of the resulting fitted model (Eq. \ref{Eq:CFInnerProdModel}) inner product action matrix elements and the Monte Carlo (MC) estimated inner product action matrix elements for the $U(1)$ neutral CFT states with $L_0$ eigenvalues $M = 1,2,3$, in the case where the ground state trial wave function contains $N = 55$ fermions. We use the shorthand notation $[\delta S_b]_{M,ij} \equiv \bra{M;i}\delta S_b \ket{M;j}$ for the inner product action matrix elements, where the basis reference for the index $i$ is given in Tables \ref{tab:CFBasisRef1}, \ref{tab:CFBasisRef2} and \ref{tab:CFBasisRef3} in Appendix \ref{Sec:furtherMatElFit} for $M = 1,2,3$ respectively. The first two rows show a colour map of the $\log_{10}$ of the absolute value of the MC matrix elements and fitted model matrix elements, where we have introduced a cutoff such that any matrix element with an absolute value below $10^{-2}$ is replaced  with $10^{-2}$, for clarity. The third row shows a colour map of the absolute values of the errors of the matrix elements (i.e. the difference between the MC and model matrix elements), where no cutoff is used. We find the model matrix elements to be in good agreement with the MC matrix elements with errors $\lessapprox 3\times 10^{-2}$.

\begin{figure}[h!]
    \centering
    \includegraphics{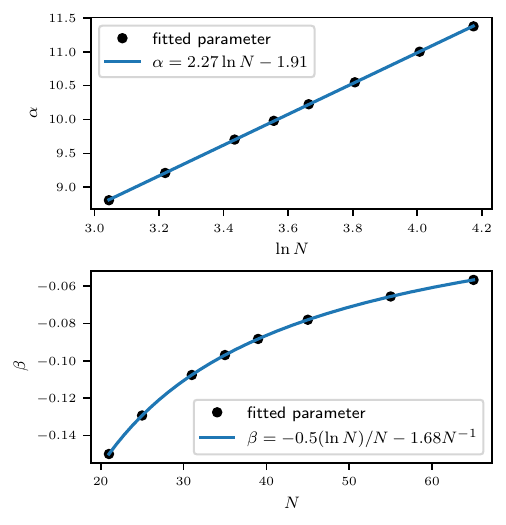}
    \caption{Shows the fitted $\nu = 2/5$ composite fermion inner product action model parameters $\alpha$ and $\beta$, of Eq. \ref{Eq:CFInnerProdModel}, for various system sizes $N$, along with a fit to a particular functional form for the $N$ dependence of each parameter.}
    \label{fig:CFParamScaling1}
\end{figure}

Fig. \ref{fig:CFParamScaling1} shows the fitted $\alpha$ and $\beta$ parameters, model of Eq. \ref{Eq:CFInnerProdModel}, for various system sizes $N$, along with a fit to a particular functional form for the $N$ dependence of each parameter. By extrapolating the fitted functions from Fig. \ref{fig:CFParamScaling1}, we expect that for large $N$ $\alpha \approx 2.3\ln N$ and $\beta \approx -0.5(\ln N)/N$. This is the same functional forms for the corresponding $\alpha$ and $\beta$ in the result of $\nu = 2$ in Eq. \ref{Eq:nu2innerprodaction}. 

\begin{figure}[h!]
    \centering
    \includegraphics{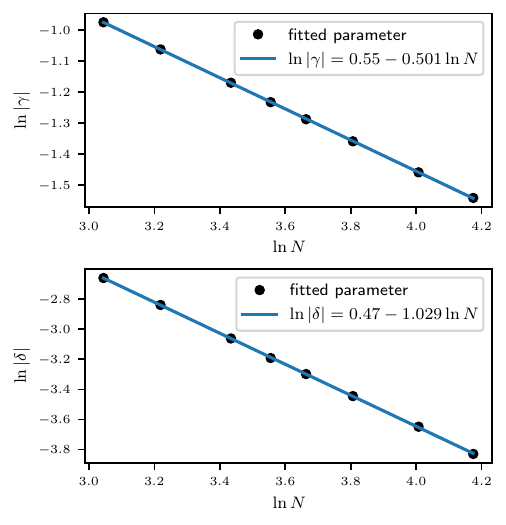}
    \caption{Shows the fitted $\nu = 2/5$ composite fermion inner product action model parameters $\gamma$ and $\delta$, of Eq. \ref{Eq:CFInnerProdModel}, for various system sizes $N$, along with a fit to a particular functional form for the $N$ dependence of each parameter.}
    \label{fig:CFParamScaling2}
\end{figure}

Fig. \ref{fig:CFParamScaling2} shows the fitted $\gamma$ and $\delta$ parameters, for various system sizes $N$, where the $N$ dependence of each parameter was fit to the functional form $\sim N^{-a}$ with $a \in \mathbb{R}$. We can then see that $\gamma$ can be fit very well to the form $\gamma \sim \frac{1}{\sqrt{N}}$, as expected. The $\delta$ parameter, however, has a fitted $N$ dependence of the form $\delta \sim \frac{1}{N}$. In Eq. \ref{Eq:CFInnerProdModel} $\delta$ is the coefficient of the term $\oint \frac{dz}{2 \pi i}z J^3(z) i\partial\Phi(z)$. By integration by parts, we can see that $\oint \frac{dz}{2 \pi i}z J^3(z) i\partial\Phi(z) = - \frac{1}{2} \oint \frac{dz}{2 \pi i}z^2 \partial(J^3(z) i\partial\Phi(z))$. Thus, we can get the same term from the integral of a scaling dimension $3$ operator \cite{Fern2018, Fern2018a}, for which we expect the leading $N$ dependence to be $\sim \frac{1}{N}$ \footnote{This is assuming that the $\alpha(N)$ of Eq. \ref{Eq:generalInProdAct} have no $N$ dependence.}. We can then interpret $\delta$ as being the coefficient of this scaling dimension $3$ term with the scaling dimension $2$ term being absent.

\subsubsection{$\phi_2^2$ parton}
\begin{figure*}[ht!]
    \centering
    \includegraphics{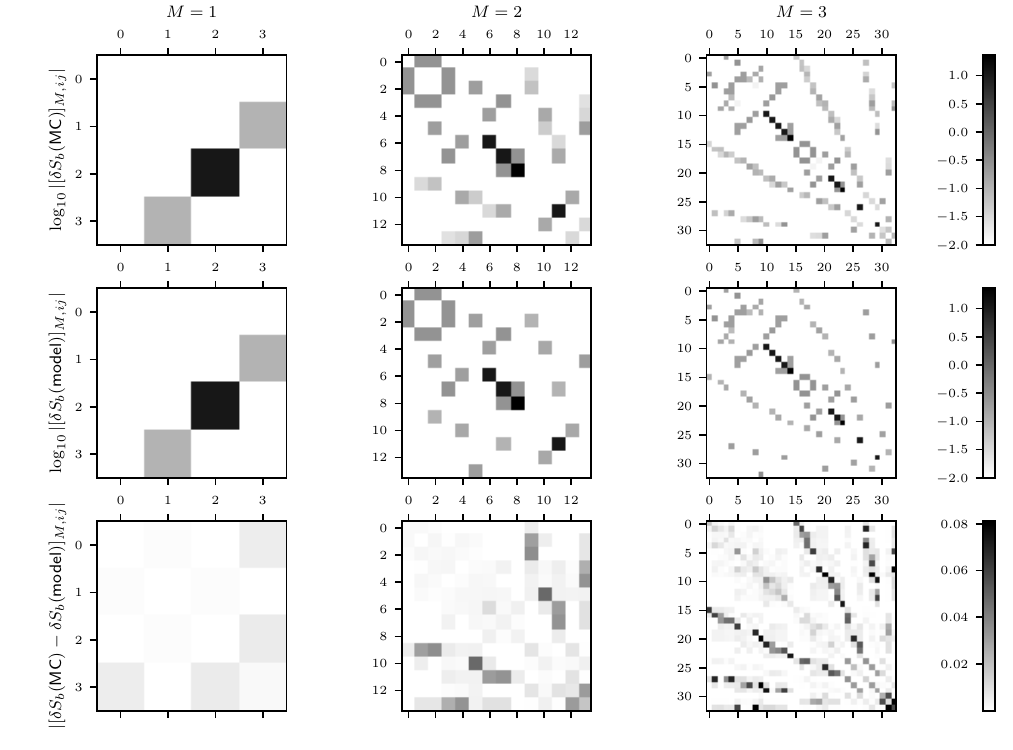}
    \caption{Comparison between the Monte Carlo (MC) estimated and model (Eq. \ref{Eq:partInnProdModel}) inner product action matrix elements, for the case of $\phi_2^2$ parton state where the ground state wave function contains $N = 31$ bosons. We use the shorthand notation $[\delta S_b]_{M,ij} \equiv \bra{M;i}\delta S_b \ket{M; j}$, where the basis reference for the index $i$ can be found in Appendix \ref{Sec:furtherMatElFit}. Note that when presenting the $\log_{10}$ of the absolute value of the matrix elements, we have used a cutoff where any matrix elements whose absolute value is below $10^{-2}$ have been replaced by $10^{-2}$, for clarity. No cutoff has been used for the matrix element errors, $|[\delta S_b(\text{MC}) - \delta S_b(\text{Model}) ]_{M,ij}|$. Some sample matrix elements can be found in Appendix \ref{Sec:furtherMatElFit}.}
    \label{fig:fittingTestParton}
\end{figure*}

Fig. \ref{fig:fittingTestParton} shows the same comparison between the MC estimated and model inner product action (Eq. \ref{Eq:partInnProdModel}) matrix elements that was done for the composite fermion case, but now for the $\phi_2^2$ parton case where the ground state trial wave function contains $N = 31$ bosons. We find the MC estimated matrix elements to be in good agreement with the fitted model matrix elements with errors $\lessapprox 8 \times 10^{-2}$.

\begin{figure}[h!]
    \centering
    \includegraphics{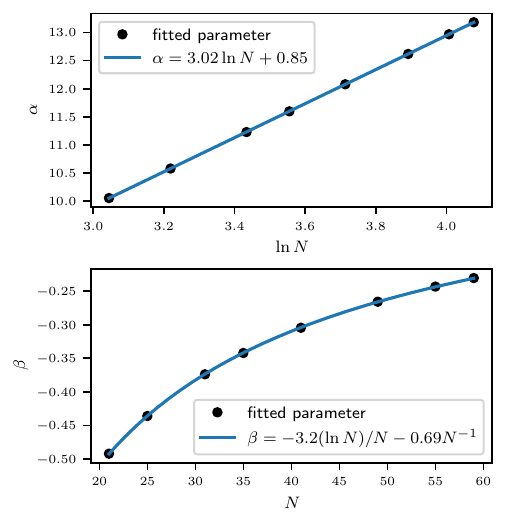}
    \caption{Shows the fitted $\phi_2^2$ parton inner product action model parameters $\alpha$ and $\beta$, of Eq. \ref{Eq:partInnProdModel}, for various system sizes $N$, along with a fit to a particular functional form for the $N$ dependence of each parameter.}
    \label{fig:PartonParamScaling1}
\end{figure}

Fig. \ref{fig:PartonParamScaling1} shows the fitted $\alpha$ and $\beta$ parameters, model of Eq. \ref{Eq:partInnProdModel}, for various system sizes $N$, along with a fit to a particular functional form for the $N$ dependence of each parameter. Based on the functional form these parameters can be fit to, we expect that for large $N$, $\alpha \approx 3\ln N$ and $\beta \approx -3.2(\ln N)/N$. These are the same functional forms as the result for $\nu = 2$, in Eq. \ref{Eq:nu2innerprodaction}, where, interestingly, the coefficients of the $\ln N$ and $(\ln N)/N$ terms are almost the same as those of Eq. \ref{Eq:nu2innerprodaction}.  

\begin{figure}[h!]
    \centering
    \includegraphics{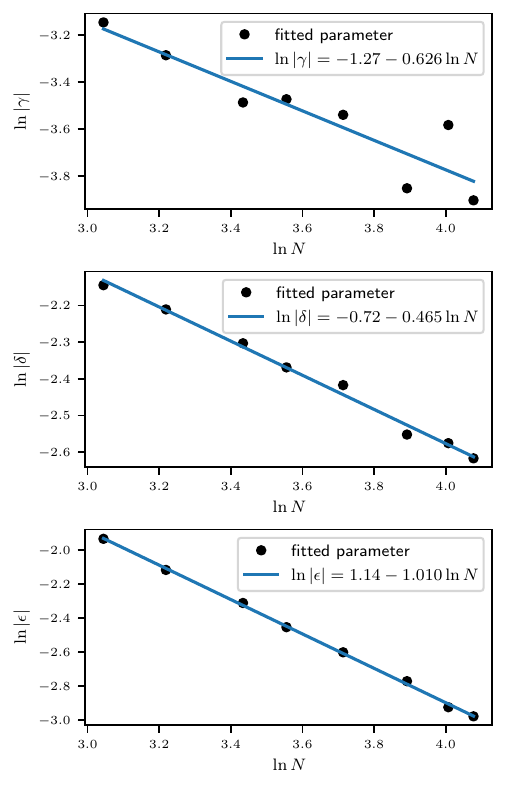}
    \caption{Shows the fitted $\phi_2^2$ parton inner product action model parameters $\gamma$, $\delta$ and $\epsilon$, of Eq. \ref{Eq:partInnProdModel}, for various system sizes $N$, along with a fit to a particular functional form for the $N$ dependence of each parameter.}
    \label{fig:PartonParamScaling2}
\end{figure}

Fig. \ref{fig:PartonParamScaling2} shows the fitted $\gamma$, $\delta$ and $\epsilon$ parameters, for various system sizes $N$, where the $N$ dependence of each parameter was fit to the functional form $\sim N^{-a}$ with $a \in \mathbb{R}$. Despite significant noise compared with the composite fermion case, we can then see that both $\gamma$ and $\delta$ roughly have an $N$ dependence of the form $\gamma, \delta \sim \frac{1}{\sqrt{N}}$, as expected. The $\epsilon$ parameter has an $N$ dependence of the form $\epsilon \sim \frac{1}{N}$, which is the same $N$ dependence for the corresponding term in the composite fermion case. As discussed above for the composite fermion case, this is consistent with interpreting $\epsilon$ as the coefficient of an integral of a scaling dimension $3$ operator, which is related to the original term in Eq. \ref{Eq:partInnProdModel} through integration by parts. 

\section{Quasi-particles and braiding from the generalised screening hypotheis} \label{Sec:QPBraiding}
As summarised in Sec. \ref{Sec:CFTMthods}, previous works have shown that if a wave function can be written as a correlation function of a particular $\CFT{\mathcal{A}}$, then there exist quasi-particle excitations of this wave function corresponding to representations of the chiral algebra $\mathcal{A}$, whose braiding statistics are given by monodromy of conformal blocks of $\CFT{\mathcal{A}}$ provided the generalised screening hypothesis holds and assuming there exits a trapping Hamiltonian for which these quasi-particle states have the lowest energy (see Sec. \ref{Sec:CFTMthods} and Ref. \cite{Read2009} for further details). In the previous works discussed in Sec. \ref{Sec:CFTMthods} the chiral algebra $\mathcal{A}$ was generated by a $\Omega(z)$ and $\Omega^\dagger(z)$. We  now show how this can be generalised to the cases considered in this work. We first discuss the general quasi-particle construction in Sec. \ref{Sec:qpgenStruc}, and how this can be applied to cases where $\mathcal{A}$ can be represented using a simple current algebra, as detailed in Sec. \ref{Sec:CFTGenStructure}, where we  conclude that the symmetric parton wave functions $\phi^m_n$ can host quasi-particles whose braiding statistics can be described by the monodromy of conformal blocks of the $\widehat{\mathfrak{su}}(n)_m$ WZW model, provided the generalised screening hypothesis holds and assuming the existence of the aforementioned quasi-particle trapping Hamiltonian. Then in Sec. \ref{Sec:su2Series} we  consider the $\phi_2^k$ series of parton wave functions, where we  show that the quasi-particles of these states are related to those of the Read-Rezayi series wave functions. We will not explicitly discuss the quasi-particles of the composite fermion wave functions here as this has been detailed at length in other works (see Ref. \cite{Hansson2017} and references therein). Throughout this section, we  only discuss the unprojected wave functions, with corresponding results for the projected wave functions being obtained by the replacement $\bar{z} \rightarrow 2\partial$. 

\subsection{General construction} \label{Sec:qpgenStruc}
Now consider a ground state trial wave function which is generated by an $\Omega(z,\bar{z}) = \sum_l \bar{z}^l\phi_l(z)$ operator belonging to a given $\CFT{\mathcal{A}}$. Let $\phi_j(z)$ be a field which belongs to the $\mathcal{H}_j$ representation of the chiral algebra $\mathcal{A}$. Schematically, quasi-particle states are generated by these fields by inserting them into the generating correlation function $\Psi(\mathbf{w};\mathbf{z},\bar{\mathbf{z}}) \sim \bra{0} C(N) \prod_{i=1}^m \phi_{j_i}(w_i) \prod_{k=1}^{N-q\sqrt{\nu}}\Omega(z_k,\bar{z}_k)\ket{0}$, where $\mathbf{w} = w_1,w_2,\dots,w_m$ and $q$ is the total $U(1)$ charge of the fields $\phi_{j_i}(w_i)$ (recall that the correlation function must be $U(1)$ neutral). Just as in Sec. \ref{Sec:CFTMthods}, this expression is only well defined if the way in which the fields $\phi_{j_i}(w_i)$ fuse to the identity field (or vacuum) is specified. For a given set of $\phi_{j_i}(w_i)$ there are multiple possible wave functions. More formally, these are expressed as,
\begin{equation}
    \Psi_a(\mathbf{w};\mathbf{z},\bar{\mathbf{z}}) = \mathcal{F}_a(\mathbf{w};\mathbf{z},\bar{\mathbf{z}})
\end{equation}
where $\mathcal{F}_a(\mathbf{w};\mathbf{z},\bar{\mathbf{z}})$ are the conformal blocks corresponding to the correlation function of $\CFT{\mathcal{A}}$ given by,
\begin{equation}
    \begin{split}
        &\bra{0} C(N) \overline{C}(N) \prod_{i=1}^m \phi_{j_i}(w_i,\bar{w}_i) \prod_{k=1}^{N-q\sqrt{\nu}}\Omega(z_k,\bar{z}_k)\overline{\Omega}(z_i,\bar{z}_i)\ket{0} \\
        &= \sum_a |\mathcal{F}_a(\mathbf{w}; \mathbf{z},\bar{\mathbf{z}})|^2
    \end{split}
\end{equation}
with $a$ labeling the way in which the fields $\phi_{j_i}(w_i)$ fuse to the identity. As mentioned in Sec. \ref{Sec:rationalCFT} the $\mathcal{F}_a(\mathbf{w};\mathbf{z},\bar{\mathbf{z}})$ are in one-to-one correspondence with the conformal blocks of a correlation function just involving the $\phi_{j_i}(w)$ fields, where $\mathcal{F}_a(\mathbf{w};\mathbf{z},\bar{\mathbf{z}})$ has the same monodromy properties as its corresponding conformal block. Note that the $\mathcal{F}_a(\mathbf{w};\mathbf{z},\bar{\mathbf{z}})$ functions will have some $\bar{z}$ dependence, which arises from expressing $\Omega$ as $\Omega(z,\bar{z}) = \sum_l \bar{z}^l\phi_l(z)$. For the $\mathcal{F}_a(\mathbf{w};\mathbf{z},\bar{\mathbf{z}})$ to be valid wave functions we must choose the fields from $\mathcal{H}_{j_i}$ such that the OPE of $\phi_{j_i}(w)$ and the $\phi_l(z)$ operators have no singular terms $\phi_{j_i}(w)\phi_l(z) \sim 0$. We will not elaborate on whether this can be done in general. However, we  show, later in this section, this can be done in the case when $\mathcal{A}$ can be represented by a simple current algebra (see Sec. \ref{Sec:CFTGenStructure}).  Note that as the $\phi_l(z)$ operators belong to the chiral algebra $\mathcal{A}$ the resulting $\Psi_a(\mathbf{w};\mathbf{z},\bar{\mathbf{z}})$ will be single-valued in the $z$ coordinates as the $\phi_l(z)$ will have trivial monodromy with the $\phi_{j_i}(w_i)$ fields.

If generalised screening holds, then the resulting short-range correlations inside the droplet will imply that the excitations created by the $\phi_{j}(w)$ operators will be ``point like'' on length scales much larger than the magnetic length. Furthermore, the argument of Sec. \ref{Sec:CFTMthods} can be reapplied to show that the charge of the excitation created by $\phi_j(w)$ is equal to $-q_j\sqrt{\nu}$, where $q_j$ is the $U(1)$ charge of $\phi_j(w)$. 

By construction, the $\mathcal{F}_a(\mathbf{w};\mathbf{z},\bar{\mathbf{z}})$ are holomorphic in the $w_i$ coordinates. We can then reapply the arguments of Read \cite{Read2009} to show that if generalised screening holds, and assuming a quasi-particle trapping Hamiltonian exists, the transformation of the wave function $\Psi_a(\mathbf{w};\mathbf{z},\bar{\mathbf{z}})$ after adiabatically moving the quasi-particles along some path $w_i(\tau)$ is equivalent, up to the usual magnetic Berry phase and the unphysical time dependant phase factor, to the monodromy transformation of $\mathcal{F}_a(\mathbf{w};\mathbf{z},\bar{\mathbf{z}})$ along this path $w_i(\tau)$. We assume the existence of these quasi-particle trapping Hamiltonians for the remainder of this section. We would like to emphasise that if $\phi_j(z)$ and $\phi'_j(z)$ belong to the same representation of $\mathcal{A}$ then a conformal block involving $\phi_j(z)$ will have the same monodromy properties as the conformal block with $\phi_j(z)$ replaced by $\phi'_j(z)$. Thus, the possible anyon types of the quasi-particles described by these conformal blocks must correspond to the irreducible representations of $\mathcal{A}$.

\subsection{Simple current algebra constructions} \label{Sec:simpleCurrentBraid}
Here we describe how this formalism can be applied to the case, discussed in Sec. \ref{Sec:CFTGenStructure}, when the $\phi_l(z)$ can be represented by fields from a CFT of the form $\CFT{U(1)}\otimes \CFT{\chi}$ with $\phi_l(z) = :e^{i\Phi(z)/\sqrt{\nu}}:\chi_l(z)$ where $\chi_l(z)$ are \textit{simple currents} of $\CFT{\chi}$. 

A simple current $\chi$ of a given $\CFT{\chi}$ is a field whose fusion with any other field must only have one result. That is for any primary field $\xi$ of $\CFT{\chi}$ (i.e. primary with respect to the chiral algebra of $\CFT{\chi}$) the fusion with $\chi$ takes the form $\chi \times \xi = \xi^{(1)}$ where $\xi^{(1)}$ is another primary of $\CFT{\chi}$. Note that regardless of which $\chi_l$ is fused with $\xi$ the result must be a primary field that is in the same representation, of the chiral algebra of $\CFT{\chi}$, as $\xi^{(1)}$. The \textit{orbit} of $\xi$ are the set of fields produced by repeated fusion with $\chi$. We use $\xi^{(n)}$ to denote the result of $n$ fusions with $\xi$ (i.e. $\xi^{(n+1)} = \chi \times \xi^{(n)}$). Assuming $\CFT{\chi}$ is rational, which we take as given, this orbit must be finite. Let $n_\xi$ be the smallest positive number such that $\xi^{(n_\xi)} = \xi$. The size of the orbit is then given by $n_\xi$. Note that the size of an orbit may be one in some cases (i.e. $\xi^{(1)} = \xi$).

As shown by Schellekens et al. in Ref. \cite{schellekens_extended_1989} the OPE of the $\chi_l$ simple currents must take the form,
\begin{equation}
    \chi_l(z)\chi_{l'}(w) \propto (z-w)^{-r/n_\chi}\chi^{(1)}(w) + \dots
\end{equation}
where $r$ is an integer known as the monodromy parameter and $n_\chi$ is the length of the orbit of $\chi$. As $\phi_l(z)$ forms a well-defined chiral algebra, we must have that the OPE $\phi_l(z)\phi_{l'}(w)$ can only contain integer powers of $(z-w)$. Furthermore, as $\phi_l(z)$ is used to generate a many-body wave function, we must also require that the OPE $\phi_l(z)\phi_{l'}(w)$ contains only non-negative integer powers of $(z-w)$. This OPE takes the form $\phi_l(z)\phi_{l'}(w) = :e^{i\Phi(z)/\sqrt{\nu}}:\chi_l(z) :e^{i\Phi(w)/\sqrt{\nu}}:\chi_{l'}(w) \propto (z-w)^{(\nu^{-1} - r/n_\chi)} :e^{i2\Phi(w)/\sqrt{\nu}}:\chi^{(1)}(w) + \dots$. Thus, $\nu^{-1} = m + r/n_\chi$ where $m$ is a non-negative integer.

It is further shown by Schellekens et al. that the OPE of $\xi$ and $\chi$ must take the form,
\begin{equation}
    \chi_l(z)\xi(w) \propto (z-w)^{-t_\xi /n_\chi}\xi^{(1)}(w) + \dots
\end{equation}
where $t_\xi$ is an integer, and that there exists a conserved charge $Q$ which takes the form $Q(\xi^{(n)}) = t_\xi/n_\chi + nr/n_\chi \pmod{1}$. The existence of this charge $Q$ then gives, $Q(\xi^{(n_\xi)}) = Q(\xi)$, which implies that $n_\xi r/n_\chi \in \mathbb{Z}$.

The fields of $\CFT{\chi}$ can be used to create representations of the chiral algebra $\mathcal{A}$, which is generated by $\phi_l(z)$ and $\phi^\dagger_l(z)$. Consider the field $\phi(z) = :e^{iq\Phi(z)}:\xi(z)$, where $q$ is chosen such that the OPE of $\phi(z)$ with $\phi_l(w)$ only contains integer powers of $(z-w)$. A representation of $\mathcal{A}$ can then be formed by the space of states generated by polynomials in the modes of $\phi_l(z)$ and $\phi^\dagger_l(z)$ applied on the state $\phi(0)\ket{0}$. We refer to the representation of $\mathcal{A}$ constructed in this way as $\chi$-reps. Now consider the OPE,
\begin{equation}
    \begin{split}
        &:e^{i\Phi(z)/\sqrt{\nu}}:\chi_l(z) e^{iq\phi(w)}\xi(w) \\
        &\propto (z-w)^{q\sqrt{\nu}^{-1} - t_\xi/n_\chi} :e^{i(q + \sqrt{\nu}^{-1})\Phi(w)}:\xi^{(1)}(w) + \dots
    \end{split}
\end{equation}
For this to contain only integer powers of $(z-w)$ we then require $q_p = \sqrt{\nu}(p + t_\xi/n_\chi )$ with $p \in \mathbb{Z}$. We can label these irreducible representations by $(\xi, p)$, where we  denote the actual representation by $\mathcal{H}_{\xi,p}$. From this OPE we can see that the state $:e^{i(q + \sqrt{\nu}^{-1})\Phi(0)}:\xi^{(1)}(0)\ket{0}$ must belong to this representation and, by induction, so too must the states $:e^{i(q + n\sqrt{\nu}^{-1})\Phi(0)}:\xi^{(n)}(0)\ket{0}$. Hence, the field $:e^{i(q_p + xn_\xi \sqrt{\nu}^{-1})\Phi(z)}:\xi(z)$ will generate the same representation where $x \in \mathbb{Z}$. Thus, $\mathcal{H}_{\xi,(p + xn_\xi \nu^{-1})} = \mathcal{H}_{\xi, p}$. From the above OPE we can then see that there must exist such an $x$ so that $:e^{i(q_p + xn_\xi \sqrt{\nu}^{-1})\Phi(z)}:\xi(z) :e^{i\Phi(w)/\sqrt{\nu}}: \chi_l(w) \sim 0$ (i.e. the OPE contains no singular terms). Such fields are then suitable representatives from this representation to generate quasi-particle wave functions. We can then choose $q_p$ such that the OPE $:e^{iq_p\Phi(z)}:\xi(z) :e^{i\Phi(w)/\sqrt{\nu}}:\chi_l(w)$ contains only non-negative integer powers of $(z-w)$, wherein making such an additional restriction does not miss any $\chi$-reps. In fact, such a restriction simply amounts to choosing a $p 
\geq 0$. The primary fields of $\CFT{\chi}$ can be organised into the distinct orbits generated by $\chi$. For the $j^{\text{th}}$ orbit one can choose a particular $\xi_j$, belonging to that orbit, as its representative. The distinct $\chi$-reps can then be labelled by an orbit $j$ and an integer $p$ such that $\mathcal{H}_{j,p} \equiv \mathcal{H}_{\xi_j,p}$. For a given orbit $j$ the distinct representations are given by $0, 1, 2, \dots, m_j - 1$ where $m_j = n_{\xi_j}\nu^{-1}$.

The fusion rules of these representations are given by $U(1)$ charge conservation and the fusion rules of $\CFT{\chi}$.

These $\phi_{j,p}(z) = :e^{iq_p\Phi(z)}:\xi_j(z)$ fields can then be used to generate quasi-particle wave functions. Such conformal blocks decompose into a product of a $\CFT{U(1)}$ conformal block times a $\CFT{\chi}$ conformal block. Thus, up to an Abelian phase given by the $U(1)$ factor, the braiding properties of the resulting anyons will be given by monodromy properties of the $\CFT{\chi}$ conformal blocks, provided generalised screening holds and that there exists a quasi-particle trapping Hamiltonian. 

As discussed in Sec. \ref{Sec:PartonGenWFCon} and in Appendix \ref{Sec:partonAlgebra}, the symmetric parton wave functions, $\phi^m_n$, have a chiral algebra $\mathcal{A}(n)_m$ which can be represented using this simple current construction with $\CFT{\chi} = \CFT{\widehat{\mathfrak{su}}(n)_m}$, where $\CFT{\widehat{\mathfrak{su}}(n)_m}$ is the $\widehat{\mathfrak{su}}(n)_m$ WZW model. In these cases, the fields $\chi_l$ correspond to the WZW primaries which transform as the rank $m$ symmetric tensor representation of $SU(n)$. One can compute the various orbits generated by the $\chi_l(z)$ using the method given in Ref. \cite{schellekens_extended_1989}. We can then generate quasi-particle wave functions using the $:e^{iq_p\Phi(z)}:\phi_\lambda(z)$ fields (with $p \geq 0$), where $\phi_\lambda$ is the WZW primary which corresponds to the highest weight $\lambda$. As for the general case, the resulting conformal block quasi-particle wave functions will decompose into a $U(1)$ conformal block and an $\widehat{\mathfrak{su}}(n)_m$ WZW conformal block. Hence, if generalised screening holds and there exists a quasi-particle trapping Hamiltonian, the symmetric parton trial wave function $\phi^m_n$ can host quasi-particles whose braiding properties are described by the conformal blocks of the $\widehat{\mathfrak{su}}(n)_m$ WZW model, up to an Abelian phase. This is precisely what one would expect based on Wen's arguments \cite{Wen1991a}. 

The monodromy properties of the $\widehat{\mathfrak{su}}(n)_m$ conformal blocks are described by the braid group representations given by the corresponding quantum group, which is discussed in Ref. \cite{slingerland_quantum_2001}.

\subsection{The $\phi_2^k$ series} \label{Sec:su2Series}
The $\phi^k_2$ parton trial wave function can be expressed as a correlation function of the $\mathcal{A}(2)_k$ chiral algebra, as detailed in Sec. \ref{Sec:PartonGenWFCon}. In Appendix \ref{Sec:A2kReps} we show how the representations of $\mathcal{A}(2)_k$ can be determined using elementary methods, under certain assumptions which we  take as given for the remainder of this section. Under these assumptions (detailed in Appendix \ref{Sec:A2kReps}), these representations are uniquely labelled by a pair $(j,p)$, where $j$ can take values $0, \frac{1}{2}, 1, \frac{3}{2} , \dots, \frac{k}{4}$ for $k$ even and $0, \frac{1}{2}, 1, \frac{3}{2} , \dots, \frac{k-1}{4}$ for $k$ odd. For a given $j$, $p$ can take values $p = 0,1,2,\dots, k - 1$ for $j\neq k/4$ and $p = 0,1,2,\dots, k/2-1$ for $j = k/4$. 

Every $\mathcal{H}_{j,p}$ representation can be constructed from the $\hat{\mathfrak{u}}(1)\oplus \widehat{\mathfrak{su}}(2)_k$ WZW model. First, let $\phi_{j,m}(z)$ be the spin-$j$ WZW primary with $J^3_0$ eigenvalue $m$ and is normalised such that we have the following two-point function $\braket{\phi_{j,m}(z)\phi_{j,-m}(w)} = (z-w)^{-2h}$ (with $h$ being the conformal dimension of $\phi_{j,m}(z)$). Under this representation we can express $\Omega(z,\bar{z}) = \sum_{l=0}^k \bar{z}^l V_l(z)$ with $V_l(z) = \sqrt{\binom{k}{l}}:e^{i\sqrt{k/2}\Phi(z)}:\phi_{k/2, -k/2 + l}(z)$. The $\mathcal{H}_{j,p}$ representation is then constructed by applying polynomials in the modes of $V_l(z)$ and $V^\dagger_l(z)$ on the state $\zeta_{j,p}(0)\ket{0}$, where the field $\zeta_{j,p}(z)$ is given by. 
\begin{equation}
    \zeta_{j,p}(z) = :e^{iq_{j,p}\Phi(z)}:\phi_{j,j}(z)
\end{equation}
with,
\begin{equation}
    q_{j,p} = \sqrt{\frac{2}{k}}(p + j)
\end{equation}
The fusion rules for these representations are inherited from the $\widehat{\mathfrak{su}}(2)_k$ fusion rules \cite{gepner_string_1986, zamolodchikov_operator_1986}. These fusion rules are detailed in Appendix \ref{Sec:A2kFusionRules}.

By construction $\zeta_{j,p}(z)V_l(w) \sim 0$. Hence, the $\zeta_{j,p}(z)$ fields can be used to generate quasi-particle wave functions with the resulting conformal block wave function taking the form $\Psi_a(\mathbf{w};\mathbf{z},\bar{\mathbf{z}}) \sim \bra{0}C(N) \prod_i \zeta_{j_i,p_i}(w_i) \prod_r \Omega(z_r,\bar{z}_r)\ket{0}$. The conformal block wave functions $\Psi_a(\mathbf{w};\mathbf{z},\bar{\mathbf{z}})$ will then decompose into a product of a $\hat{\mathfrak{u}}(1)$ conformal block and an $\widehat{\mathfrak{su}}(2)_k$ conformal block. Assuming generalised screening, the braiding properties of the $\zeta_{j,p}$ quasi-particles will be given by the monodromy properties of these conformal block wave functions, up to the magnetic Berry phase as usual. In other words, if the $\zeta_{j_i,p_i}(w_i)$ quasi-particles are adiabatically moved along a braid $w_i(\tau)$ then the wave function will transform as $\Psi_a \rightarrow  \sum_b B_{ba} \Psi_b$ where the matrix $B$ is given by $e^{i\theta}B_{\widehat{\mathfrak{su}}(2)_k}$ (up to the magnetic Berry phase) where $B_{\widehat{\mathfrak{su}}(2)_k}$ is the monodromy matrix for the $\widehat{\mathfrak{su}}(2)_k$ conformal block and $e^{i\theta}$ is a phase given by the monodromy of the $\hat{\mathfrak{u}}(1)$ conformal block. The relevant data to compute the $B_{\widehat{\mathfrak{su}}(2)_k}$ matrices (i.e. the $F$ and $R$ matrices) can be found in Ref. \cite{bonderson_non-abelian_2007}. The monodromy properties of the $\widehat{\mathfrak{su}}(2)_k$ conformal blocks were first investigated by Tsuchiya and Kanie \cite{tsuchiya_vertex_1987}, where they found such properties are related to the Jones polynomial.

As shown and detailed in Refs. \cite{zamolodchikov_nonlocal_1985, gepner_modular_1987} the $\widehat{\mathfrak{su}}(2)_k$ fields can be represented using fields of a CFT of the form $\CFT{\hat{\mathfrak{u}}(1)}\otimes \CFT{\psi_k}$ where $\CFT{\psi_k}$ is the $Z_k$ parafermion theory (of Ref. \cite{zamolodchikov_nonlocal_1985}). With this representation the fields $\phi_{j,m}(z)$ can be expressed as,
\begin{equation} \label{Eq:parafermionDecomp}
    \phi_{j,m}(z) = \Phi^{2j}_{2m}(z) :e^{im\sqrt{\frac{2}{k}}\varphi(z)}:
\end{equation}
where $\Phi^l_{2m}(z)$ are the $Z_k$ parafermion primary fields and $\varphi(z)$ is the chiral boson of $\CFT{\hat{\mathfrak{u}}(1)}$, with $\braket{\varphi(z)\varphi(w)} = -\ln (z-w)$. Thus, the $\Psi_a(\mathbf{w};\mathbf{z},\bar{\mathbf{z}})$ quasi-particle wave functions decompose further into a product of two $\hat{\mathfrak{u}}(1)$ factors times a conformal block of $\CFT{\psi_k}$. Hence, up to Abelian phases, the braiding properties of the $\zeta_{j,p}$ quasi-particles are the same as the braiding properties of the quasi-particles of the Read-Rezayi series \cite{Read1999}. We would like to emphasise this is only up to Abelian phases, which means the $\zeta_{j,p}$ anyons are \textit{not equivelant} to the corresponding anyons of the Read-Rezayi series, in that their corresponding braiding matrices differ by an overall phase factor which is determined by the $\hat{\mathfrak{u}}(1)$ factors of the conformal block wave functions.

It should also be noted that the Read-Rezayi states can be directly understood in terms of the more general projective parton construction, where the electron operator is not written as just a product of parton operators but as a sum of products of parton operators \cite{barkeshli2010effective, barkeshli2011bilayer}. Roughly speaking, these projective constructions ``mod out'' the extra chiral boson field that can be seen in Eq. \ref{Eq:parafermionDecomp}. It is this removal of this chiral boson degree of freedom that alters the phase factors that occur when braiding the various anyonic excitations of these states. More generally, there should also exist a similar relationship between the symmetric parton states $\phi_n^m$ and the $SU(N-1)$ singlet states of Ref. \cite{fuji2017non}, which should be given by the relationship between the $SU(N)$ WZW models and Gepner's generalised parafermion theories \cite{gepner1987new}. 

\subsubsection{Example: $\phi^2_2$}
The $\mathcal{A}(2)_2$ algebra has three irreducible representations which are the vacuum $\mathbf{1} \equiv \zeta_{0,0}$ and two other fields denoted by $\zeta_1 \equiv \zeta_{\frac{1}{2}, 0}$ and $\zeta_2 \equiv \zeta_{0,1}$. The identity field has the usual fusion rule with all other fields $\mathbf{1} \times \zeta_i = \zeta_i$, with the remaining fusion rules being,
\begin{equation}
    \begin{split}
        \zeta_2 \times \zeta_2 = & \mathbf{1} \\
        \zeta_1 \times \zeta_2 = & \zeta_1 \\
        \zeta_1 \times \zeta_1 = & \mathbf{1} + \zeta_2 \\
    \end{split}
\end{equation}
which are the Ising anyon fusion rules. In this case, the $\widehat{\mathfrak{su}}(2)_2$ fields can be represented by the fields of a chiral boson theory $\varphi(z)$ combined with the fields of the $Z_2$ parafermion. The $Z_2$ parafermion theory is equivalent to the Ising CFT which includes a Majorana field $\psi(z)$ and the spin field $\sigma(z)$. With this representation we have $\Omega(z,\bar{z}) = :e^{i\Phi(z)}:( :e^{-i\varphi(z)}: + \sqrt{2}\bar{z}\psi(z) + \bar{z}^2:e^{i\varphi(z)}: )$, $\zeta_2(z) = :e^{i\Phi(z)}:$ and $\zeta_1(z) = :e^{i\frac{\Phi(z)}{2}}: :e^{i\frac{\varphi(z)}{2}}:\sigma(z)$. Notice that the $\zeta_2$ quasi-particle is a fermion, which we  still treat as topologically non-trivial as the underlying particles of the $\phi_2^2$ wave function are bosons. Now consider the wave functions of four $\zeta_1$ particles $\Psi_a(w_1,\dots,w_4; \mathbf{z},\bar{\mathbf{z}}) \sim \bra{0}C(N) \prod_{i=1}^4 \zeta_1(w_i) \prod_{r=1}^{N-2}\Omega(z_r,\bar{z}_r)\ket{0}$. From the representation in terms of the Ising CFT, we can immediately see that there are two linearly independent conformal block wave functions, $\Psi_1$ and $\Psi_2$, which can be factorised as $\Psi_i = \mathcal{F}_\Phi \mathcal{F}_\varphi \mathcal{F}_{\sigma,i}$, where $\mathcal{F}_\Phi$, $\mathcal{F}_\varphi$ and $\mathcal{F}_{\sigma, i}$ are the conformal blocks of the $\Phi$ and $\varphi$ and $\sigma$ fields respectively. The basis of $\sigma$ conformal blocks can be chosen such that the monodromy transformation of braiding quasi-particle $3$ around quasi-particle $2$ is given by $\mathcal{F}_{\sigma, 1} \rightarrow e^{\frac{2\pi i}{8}}\mathcal{F}_{\sigma, 2}$ and $\mathcal{F}_{\sigma, 2} \rightarrow e^{\frac{2\pi i}{8}}\mathcal{F}_{\sigma, 1}$ \cite{Moore1991}. The monodromy transformation of braiding quasi-particle $3$ around quasi-particle $2$ for the $\mathcal{F}_{\Phi} \rightarrow e^{\frac{2\pi i}{4}}\mathcal{F}_{\Phi}$ and $\mathcal{F}_{\varphi} \rightarrow e^{\frac{2\pi i}{4}}\mathcal{F}_{\varphi}$. Hence, the full monodromy transformation of the quasi-particle wave functions is given by,
\begin{equation}
    \begin{pmatrix}
        \Psi_1 \\
        \Psi_2 \\
    \end{pmatrix}
    \rightarrow
    e^{\frac{5\pi i}{4}}
    \begin{pmatrix}
        0 & 1 \\
        1 & 0 \\
    \end{pmatrix}
    \begin{pmatrix}
        \Psi_1 \\
        \Psi_2 \\
    \end{pmatrix}
\end{equation}
which, assuming generalised screening and that there exits a quasi-particle trapping Hamiltonian, is the same transformation induced by adiabatically braiding quasi-particle $3$ around quasi-particle $2$, up to the area-dependent magnetic Berry phase and the unphysical time dependent phase factor. In the case of the $\nu = 1$ bosonic Pfaffian state, the corresponding braiding of four $\sigma$ particles will produce the same transformation, except with the upfront phase factor being $e^{\frac{3\pi i}{4}}$ instead of $e^{\frac{5\pi i}{4}}$. The difference between these upfront phase factors then implies the anyons in the two cases are \textit{not equivalent}.

\subsubsection{Example: $\phi_2^3$}
The $\mathcal{A}(2)_3$ algebra has six irreducible representations, with the corresponding fields being given by $\zeta_p \equiv \zeta_{0,p}$ for $p = 0,1,2$ and $\tau_p \equiv \zeta_{\frac{1}{2},p}$ for $p=0,1,2$. Note that $\zeta_0$ corresponds to the identity field $\zeta_0 = \mathbf{1}$. These have the following fusion rules, where for simplicity of notation we take $\zeta_{p+3} = \zeta_p$ and $\tau_{p+3} = \tau_p$,
\begin{equation}
    \begin{split}
        \zeta_{p_1}\times \zeta_{p_2} = & \zeta_{(p_1+p_2)} \\
        \zeta_{p_1} \times \tau_{p_2} = & \tau_{(p_1+p_2)} \\
        \tau_{p_1} \times \tau_{p_2} = & \zeta_{(p_1+p_2+1)} + \tau_{(p_1 + p_2 + 2)} \\
    \end{split}
\end{equation}
The third fusion rule is a modified form of the Fibonacci anyon fusions rules \cite{bonderson_non-abelian_2007}. The $\phi_2^3$ state has been detailed elsewhere \cite{ahari_partons_2022}, where the quasi-particle braiding properties were computed using alternative methods.

\subsubsection{Example: $\phi_2^4$}
We now give the $\phi_2^4$ state as a final example. The $\mathcal{A}(2)_4$ algebra has ten irreducible representations given by $\zeta_{0,p}$ for $p=0,1,2,3$, $\zeta_{\frac{1}{2}, p}$ for $p=0,1,2,3$ and $\zeta_{1,p}$ for $p = 0,1$. To express the fusion rules we extend the labelling system so that $\zeta_{0,p+4} = \zeta_{0,p}$, $\zeta_{\frac{1}{2}, p + 4} = \zeta_{\frac{1}{2}, p}$ and $\zeta_{1,p+2} = \zeta_{1,p}$. We can then write the fusion rules as,
\begin{equation}
    \begin{split}
        \zeta_{0,p_1} \times \zeta_{j,p_2} = & \zeta_{j,p_1+p_2} \\
        \zeta_{\frac{1}{2},p_1} \times \zeta_{\frac{1}{2}, p_2} = & \zeta_{0,p_1+p_2+1} + \zeta_{1,p_1+p_2} \\
        \zeta_{\frac{1}{2}, p_1} \times \zeta_{1,p_2} = & \zeta_{\frac{1}{2}, p_1+p_2 + 1} + \zeta_{\frac{1}{2}, p_1+p_2-1} \\
        \zeta_{1,p_1} \times \zeta_{1,p_2} = & \zeta_{0,p_1+p_2 + 2} + \zeta_{1,p_1+p_2+1} + \zeta_{0,p_1+p_2} \\ 
    \end{split}
\end{equation}
Now consider the wave functions of two $\zeta_{\frac{1}{2}, 0}$ particles and two $\zeta_{\frac{1}{2}, 1}$,
\begin{equation}
    \begin{split}
        &\Psi_a(w_1,\dots,w_4; \mathbf{z},\bar{\mathbf{z}}) \sim \\
        & \bra{0}C(N) \zeta_{\frac{1}{2},1}(w_1) \zeta_{\frac{1}{2},0}(w_2) \zeta_{\frac{1}{2},0}(w_3) \zeta_{\frac{1}{2},1}(w_4) \\
        &\times \prod_{r=1}^{N-2}\Omega(z_r,\bar{z}_r)\ket{0}
    \end{split}
\end{equation}
From the above fusions rules, we can see there must be two linearly independent wave functions for a given set of $w_i$ positions, $\Psi_1$ and $\Psi_2$. Furthermore, as $\zeta_{\frac{1}{2},0}(w) = :e^{i\Phi(w)/(2\sqrt{2})}:\phi_{\frac{1}{2},\frac{1}{2}}(w)$ and $\zeta_{\frac{1}{2},1}(w) = :e^{i3\Phi(w)/(2\sqrt{2})}:\phi_{\frac{1}{2},\frac{1}{2}}(w)$, these wave functions must factorise as $\Psi_i = \mathcal{F}_{\Phi}\mathcal{F}_{\widehat{\mathfrak{su}}(2)_4, i}$ where $\mathcal{F}_\Phi$ is the conformal block of the $\Phi$ fields and $\mathcal{F}_{\widehat{\mathfrak{su}}(2)_4,i}$ are the conformal blocks of the $\widehat{\mathfrak{su}}(2)_4$ fields. From the discussion of Refs. \cite{slingerland_quantum_2001, fern_how_2017}, we can see there must exist a basis of the $\widehat{\mathfrak{su}}(2)_4$ conformal blocks such that under the monodromy transformation of \textbf{swapping} quasi-particle $2$ with quasi-particle $3$ in the clockwise direction the conformal blocks transform as $\mathcal{F}_{\widehat{\mathfrak{su}}(2)_4,1} \rightarrow e^{\frac{i\pi}{12}}[ (e^{\frac{i\pi}{6}}/\sqrt{3})\mathcal{F}_{\widehat{\mathfrak{su}}(2)_4,1} - (e^{\frac{-i\pi}{6}}\sqrt{2/3}) \mathcal{F}_{\widehat{\mathfrak{su}}(2)_4,2}]$ and $\mathcal{F}_{\widehat{\mathfrak{su}}(2)_4,2} \rightarrow e^{\frac{i\pi}{12}}[ -(e^{\frac{-i\pi}{6}}\sqrt{2/3})\mathcal{F}_{\widehat{\mathfrak{su}}(2)_4,1} + (e^{\frac{i\pi}{2}}/\sqrt{3}) \mathcal{F}_{\widehat{\mathfrak{su}}(2)_4,2}]$. Also under this monodromy transformation, we have $\mathcal{F}_\Phi \rightarrow e^{\frac{\pi i}{8}}\mathcal{F}_\Phi$. Hence, under this monodromy transformation, the quasi-particle wave functions transform as,
\begin{equation}
    \begin{pmatrix}
        \Psi_1 \\
        \Psi_2 \\
    \end{pmatrix}
    \rightarrow
    e^{\frac{5\pi i}{24}}
    \begin{pmatrix}
        \frac{e^{\frac{i\pi}{6}}}{\sqrt{3}} & -e^{\frac{-i\pi}{6}}\sqrt{\frac{2}{3}} \\
        -e^{\frac{-i\pi}{6}}\sqrt{\frac{2}{3}} & \frac{e^{\frac{i\pi}{2}}}{\sqrt{3}} \\
    \end{pmatrix}
    \begin{pmatrix}
        \Psi_1 \\
        \Psi_2 \\
    \end{pmatrix}
\end{equation}
which, assuming generalised screening and the existence of a quasi-particle trapping Hamiltonian, is the same transformation of the wave functions induced by adiabatically swapping quasi-particle $2$ and quasi-particle $3$ in the clockwise direction, up to the magnetic Berry phase and the unphysical time dependent phase factor. As shown by Fern et al. \cite{fern_how_2017} the braid group representations formed by the monodromy transformation of $\widehat{\mathfrak{su}}(2)_4$ conformal blocks can be constructed using the $Z_3$ parafermion operator algebra (of the Fradkin-Kadanoff-Fendley \cite{fendley_free_2014} type and not the Zamalodchikov-Fateev type \cite{zamolodchikov_nonlocal_1985}).

\section{Conclusion}
\subsection{Summary}
In this work, it has been demonstrated, by first considering how IQH ground and edge state wave functions can be expressed using CFT, that all chiral parton LLL projected and unprojected ground and edge state trial wave functions, in the planar geometry, (as defined in Sec. \ref{Sec:CFandPartonWF}) can be expressed using CFT correlation functions where to each parton state we can associate a chiral algebra $\mathcal{A}$ such that the CFT defined by $\mathcal{A}$, $\CFT{\mathcal{A}}$, is, in some sense, the ``smallest'' CFT than can generate all the ground and edge state trial wave functions of the corresponding parton state, where we are assuming there exists a CFT with the chiral algebra $\mathcal{A}$. We then formulated a field-theoretic generalisation of Laughlin's plasma analogy, known as generalised screening \cite{Read2009, Dubail2012}, where if this holds for a given parton state then various topological properties of the state can be directly related to properties of the corresponding chiral algebra $\mathcal{A}$. In particular, if generalised screening holds, along with one other mild assumption (see Sec. \ref{Sec:CFTInnerProducts}), then for parton states where the densest trial wave function is unique the edge state trial wave function state counting and the entanglement level counting in the real-space entanglement spectrum of a given parton wave function can be directly related to the state counting in the vacuum representation of the corresponding chiral algebra $\mathcal{A}$. We have further discussed how one can use the conformal blocks of $\CFT{\mathcal{A}}$ to generate quasi-particle trial wave functions for the given parton state, where if generalised screening holds the adiabatic braiding statistics of the quasi-particles is given, up to the magnetic Berry phase, by the monodromy properties of the corresponding conformal block. This then allows one to relate the various possible anyonic types of the given parton state, that can be generated by these conformal blocks, to the irreducible representations of the corresponding chiral algebra $\mathcal{A}$.

More specifically, we gave two detailed examples of how these ground and edge state trial wave functions can be generated using CFT in the case of the $\nu = 2/5$ composite fermion state and the $\nu = 1$ bosonic $\phi_2^2$ parton state. We further discussed how all chiral composite fermion states (i.e. those with no reverse flux attachment or negative effective magnetic field) trial ground and edge state wave functions, as defined in Sec. \ref{Sec:CFandPartonWF}, can be constructed from CFT correlation functions \textit{without} any explicit anti-symmetrization or symmetrization of the correlation functions. The symmetric parton states of the form $\phi_n^m$ were also discussed where it was shown that the corresponding ground and edge state wave functions could be expressed using the conformal blocks of the $\hat{\mathfrak{u}}(1) \oplus \widehat{\mathfrak{su}}(n)_m$ WZW model. This had the consequence that, even without generalised screening holding, the state counting of the edge state trial wave functions have rigorous upper bounds given by the state counting in the $\hat{\mathfrak{u}}(1) \oplus \widehat{\mathfrak{su}}(n)_m$ WZW model.

We further discussed how given we know how to generate two wave functions $\Psi_1$ and $\Psi_2$ using CFT we can then generate the product wave function $\Psi_1\Psi_2$ using CFT where the chiral algebra corresponding to $\Psi_1\Psi_2$ can be understood in terms of the chiral algebras corresponding to $\Psi_1$ and $\Psi_2$. As we discussed how all IQH ground and edge state wave functions can be generated using CFT, it is precisely this inductive step that shows that all chiral parton ground and edge state trial wave functions can be expressed using CFT correlation functions. The general construction for generating trial wave functions using CFT that these parton states point towards was then discussed. In short, the general construction allows one to generate trial wave functions for a rational CFT whose chiral algebra $\mathcal{A}$ is generated by multiple fields $\phi_l(z)$ and their conjugates $\phi_l^\dagger(z)$. This includes the chiral algebras $\mathcal{A}$ that can be understood as simple current constructions, see Ref. \cite{schoutens_simple-current_2016}, where the simple current algebras (not to be confused with the chiral algebra) are generated by one simple current representation. 

The generalised screening hypothesis was also detailed for these CFT constructions. Provided this holds, we were then able to map the problem of computing edge state trial wave function inner products to a boundary critical problem, as done in Ref. \cite{Dubail2012}, where, under an additional mild assumption, edge state inner products can be expressed as the matrix elements of an exponentiated CFT operator, called the inner product action, that takes the form of a sum of integrals of local operators, where we gave this argument in full for parton states where the lowest angular momentum trial wave function (i.e. densest wave function), for a fixed number of particles, is unique. This then generalises the result of Ref. \cite{Dubail2012}. The additional mild assumption that was required was that the matrices that define the fixed point boundary conditions seen in the boundary critical problems are invertible. Notably, using standard renormalisation group methods it could be argued that the scaling of the coefficients of the various terms in the inner product action scale with the system size in such a way that in the thermodynamic limit, there exists a simple modification of edge state map such that it preserves the inner product in the thermodynamic limit. This then allows one to equate edge state counting with state counting in the vacuum representation of the corresponding chiral algebra. It was then shown that the calculations of Ref. \cite{Dubail2012} can be easily extended to these CFT constructions where, assuming generalised screening and the additional assumption, the real-space entanglement spectrum of a given ground state trial wave function can be expressed as the spectrum of a CFT operator known as the entanglement action which takes the same form as the inner product action. This implied that in the thermodynamic limit, the entanglement level counting will match the state counting in the vacuum representation of the corresponding chiral algebra.

Explicit numerical tests of this edge state inner product result were presented in the case of the unprojected $\nu = 2/5$ composite fermion state and the unprojected bosonic $\nu = 1$ $\phi_2^2$ parton state. In each case, by estimating the edge state inner products using Monte-Carlo methods, we were able to obtain estimates for matrix elements of the inner product actions. These were then fitted to model inner product actions that take the local form implied by generalised screening. It was found that the model inner product action matrix elements could be fit very well to the numerically estimated matrix elements, where the scaling of the model parameters with the system size were also found to be consistent with generalised screening.

In addition, the general construction, for a state with a corresponding chiral algebra $\mathcal{A}$, of quasi-particle trial wave functions in terms of conformal blocks of $\CFT{\mathcal{A}}$ was detailed. We pointed out that the arguments of Read \cite{Read2009} still applied to these CFT constructions, which implies that if generalised screening holds, and that there exists a quasi-particle trapping Hamiltonian, the adiabatic braiding statistics of the quasi-particle wave functions are given by the monodromy properties of the corresponding conformal block, up to the usual magnetic berry phase. It was then shown for chiral algebras that can be understood by a simple current construction how certain representations can be found in terms of a chiral boson CFT and the CFT of the simple currents. In these cases, the conformal blocks can then be expressed as a product of a chiral boson conformal block and a conformal block of the simple current CFT. More specifically, as the chiral algebras of the symmetric parton states $\phi_n^m$ can be understood as simple current constructions, this implied that there exist quasi-particle trial wave functions of these states that can be expressed as a product of a chiral boson conformal block and a conformal block of the $\widehat{\mathfrak{su}}(n)_m$ WZW model. Thus, if generalised screening holds and there exits a corresponding quasi-particle trapping Hamiltonian, the adiabatic braiding of certain quasi-particles of the $\phi_n^m$ states can be directly related to the monodromy properties of the conformal blocks of the $\widehat{\mathfrak{su}}(n)_m$ WZW models. Finally, we then considered the $\phi_2^k$ series in detail, where the relation to the Read-Rezayi series was discussed and several examples of computing quasi-particle adiabatic statistics were given.

\subsection{Outlook}
It should be emphasised that the work presented here only considered the chiral parton states (i.e. those that do not contain a complex conjugated IQH wave function). It is not at all obvious if the formalism presented here can be extended to understand non-chiral states, with the main obstacle being that one would need to understand what mathematical object can replace the chiral algebra as the encoder of the topological data of the corresponding state. Some progress has already been made in this direction in the case of the CFT-generated hierarchy wave functions \cite{Suorsa2011a, Suorsa2011, Hansson2017}.

Furthermore, we have only considered these wave functions in the planar geometry. As a matter of completeness, it would be interesting to understand how these constructions can be extended to other geometries such as the sphere, cylinder and torus. In particular, if this can be well understood for the cylinder geometry, then one may be able to extend the methods of Refs. \cite{Zaletel2012, estienne_matrix_2013, estienne_fractional_2013} to obtain analytically computable arbitrarily precise matrix product state representations of the parton wave functions using their corresponding CFTs. Importantly, this could allow for direct tests of the implications of generalised screening in these trial wave functions \cite{wu_braiding_2014}.

Although we have shown that there corresponds a chiral algebra to each parton state which encodes the state's topological properties, we have not detailed this correspondence for all parton states. It would then be interesting to mathematically classify the chiral algebras that correspond to parton wave functions, which would allow one to determine precisely what topological orders can be described by chiral parton trial wave functions. Indeed it would also be interesting to understand if for each such chiral algebra there exists mathematically well-defined rational CFT with that chiral algebra.

We have also only considered parton states where the electron operator is written as a product of parton operators (see Sec. \ref{Sec:CFandPartonWF}). There exists a more general projective parton construction where the electron operator is expressed as a sum of products of parton operators, which has been useful in understanding many FQHE states that can be expressed this way \cite{wen_projective_1999, barkeshli2010effective, barkeshli2011bilayer}. In principle, one should be able to extend the formalism presented here to include these projective constructions, where the resulting parton trial wave function does not entirely reside within the lowest Landau level.

Finally, we also do not discuss the issue of special parent Hamiltonians, with the parton trial wave function being exact zero energy states, in this work. Using a formalism which is a generalisation of the root partition description of fractional quantum Hall states centred around the so-called ``entangled Pauli principle'', it has been shown that such parent Hamiltonian do indeed exist for certain unprojected parton states \cite{ahari_partons_2022, bandyopadhyay_entangled_2018, Bandyopadhyay2020}. As well as allowing one to find special parent Hamiltonians, this formalism also allows for the extraction of various topological properties such as edge state counting, and quasi-particle fusion and adiabatic statistics. It would, perhaps, therefore be enlightening to understand the connection between this formalism and the CFT formalism presented in this work.

\begin{acknowledgments}
GJH would like to thank B. Yang, Y. Fukusumi, G. Ji, Y. Wang, Z. Nussinov, A. Seidel, M. Barkeshli and M. Yutushui for useful comments, feedback and typo corrections. GJH would also like to thank B. Yang's research group at Nanyang Technological University Singapore for their kind hospitality during his visit, where some of this work was completed.
GJS would like to thank TIFR, Mumbai and JQI and  CMTC, University of Maryland for their hospitality during the completion of this work. We thank National Supercomputing Mission, Government of India for providing computing resources of `PARAM Brahma' at IISER Pune.  SHS and GJH were partially funded by EPSRC grant EP/S020527/1.   Statement of compliance with EPSRC policy framework  on research data: This publication is theoretical work
that does not require supporting research data.
\end{acknowledgments}

\appendix

\section{Unitary irreducible representations of the Moore-Seiberg algebra} \label{Sec:LaughlinReps}
We demonstrate how the unitary representations of the chiral algebra of Sec. \ref{Sec:LaghlinExample} can be deduced by elementary methods, under certain assumptions. Note that this is not fully mathematically rigorous and we shall point out the assumption we make. A fully mathematically rigorous exploration of these representations would require a more precise definition of these algebras and what constitutes a representation. We do not require much information about this algebra to deduce these representations. We only require Eq. \ref{Eq:LaughlinAlgebra}, the fact that $\mathcal{A}$ contains both the Heisenberg algebra, generated by $a_n$, and the Virasoro algebra, generated by the modes $L_n$ where $L_0 = \frac{(a_0)^2}{2} + \sum_{n>0} a_{-n}a_n$, and the following further commutation relations,
\begin{equation} \label{Eq:usefulCom}
    \begin{split}
        [a_n, \Omega_k] =& \sqrt{m}\Omega_{k + n} \\
        [a_n, \Omega^\dagger_k] =& -\sqrt{m}\Omega^\dagger_{k + n} \\
        [L_0, \Omega_k] =& -k\Omega_k \\
        [L_0, \Omega^\dagger_k] =& -k\Omega^\dagger_k
    \end{split}
\end{equation}
Furthermore, we  use the fact that all the modes of fields in $\mathcal{A}$ can be expressed in terms of the modes $\Omega_k$ and $\Omega^\dagger_k$, as all the fields of $\mathcal{A}$ are generated by OPEs of $\Omega(z)$ and $\Omega^\dagger(z)$.

Consider now some unitary representation of $\mathcal{A}$, $\mathcal{H}_j$, where the eigenvalues of $L_0$ are bounded from below (or highest weight representations in other words). By unitary we mean that the inner-product in $\mathcal{H}_j$ is such that the modes of $\mathcal{A}$ acting on this Hilbert space have Hermitian conjugates given by $(\Omega_k)^\dagger = \Omega^\dagger_{-k}$ and $(a_n)^\dagger = a_{-n}$. 

$\mathcal{A}$ contains the Heisenberg algebra. \textit{We assume that $\mathcal{H}_j$ decomposes into irreducible representations (irr. rep.) thereof.} Any irr. rep. of the Heisenberg algebra can be defined by a vector $\ket{q}$ such that $a_n\ket{q} = 0$ for $n > 0$ and $a_0\ket{q} = q\ket{q}$, $q \in \mathbb{R}$, with all states of the rep. being linear combinations of $\prod_{n_i} a_{-n_i} \ket{q}$ where $n_i > 0$. We can thus understand $\mathcal{H}_j$ through what $\ket{q}$ vectors it contains. The Hilbert space of the irr. rep. of the Heisenberg algebra defined by vector $\ket{q}$ will be denoted $\mathcal{H}_q$ 

We  now show that for a given $q$ such that there is at least one $\ket{q} \in \mathcal{H}_j$ then the space of states satisfying $a_n\ket{q} = 0$ for $n > 0$ and $a_0\ket{q} = q\ket{q}$ (with $q$ fixed) is one dimensional. In other words, when decomposing $\mathcal{H}_j$ into irr. reps. of the Heisenberg algebra, then the multiplicity of any irr. rep. that does appear is one, $\mathcal{H}_j = \oplus_{\text{possible } q} \mathcal{H}_q$.

If this is not the case, then there must exist two states, $\ket{q;1}$ and $\ket{q;2}$, such that, $\braket{q;1|q;2} = 0$ and $\braket{q;1|q;1} = \braket{q;2|q;2} = 1$. One can easily show that the space of states of the form (polynomial in $\Omega_k$ and $\Omega^\dagger_k$)$\ket{q;1}$ must be an invariant subspace of the action of $\mathcal{A}$ on $\mathcal{H}_j$. As $\mathcal{H}_j$ is an irr. rep. of $\mathcal{A}$, this space of states must, in fact, be $\mathcal{H}_j$. Thus, we must have $\ket{q;2} =$(some polynomial in $\Omega_k$ and $\Omega^\dagger_k$)$\ket{q;1}$. Let $X(p)$ denote a generic term that could be in this polynomial which is a product of $p$ $\Omega_k$'s and $p$ $\Omega^\dagger_k$'s, where there must be the same number of $\Omega_k$'s and $\Omega^\dagger_k$'s as $\ket{q;1}$ and $\ket{q;2}$ have the same $a_0$ eigenvalues. 

We  now show, inductively, that $\bra{q;2}X(p)\ket{q;1} = 0$ for all $X(p)$. First, consider $\Omega_k\ket{q;1}$. If this is non-zero then by the commutation relations of Eq. \ref{Eq:usefulCom}, it must be an eigenvector of $a_0$ with eigenvalue $q + \sqrt{m}$ and it must be an eigenvector of $L_0$ with eigenvalue $\frac{q^2}{2} - k$. For an eigenvector of $a_0$, with some eigenvalue $Q$, that is also an eigenvector of $L_0$, with eigenvalue $M$, we must have $M \geq \frac{Q^2}{2}$ (as $L_0 = \frac{(a_0)^2}{2} + \sum_{n>0} a_{-n}a_n$). Thus, if $\Omega_k\ket{q;1} \neq 0 \Rightarrow \frac{(q + \sqrt{m})^2}{2} \leq \frac{q^2}{2} - k \Rightarrow k \leq -q\sqrt{m} - \frac{m}{2}$. Similarly, if $\bra{q;2}\Omega_k \neq 0 \Rightarrow \frac{(q - \sqrt{m})^2}{2} \leq \frac{q^2}{2} + k \Rightarrow k \geq -q\sqrt{m} + \frac{m}{2}$. As both inequalities cannot be satisfied, $\Omega_k\ket{q;1} \neq 0 \Rightarrow \bra{q;2}\Omega_k = 0$. By similar reasoning, if $\Omega_k^\dagger\ket{q;1} \neq 0 \Rightarrow \bra{q;2}\Omega^\dagger_k = 0$. Now assume $\bra{q;2}X(p)\ket{q;1} = 0$ for all $X(p)$ with $p \leq P$. Now take some $X(P+1)$ and let the right most mode appearing in it be $x$ (i.e. $x$ is either some $\Omega_k$ or some $\Omega_k^\dagger$). If $x\ket{q;1} = 0$ then $\bra{q;2}X(P+1)\ket{q;1} = 0$. If $x\ket{q;1} \neq 0$ then we must have $\bra{q;2}x = 0$. Let $Y$ a the product of modes in $X(P+1)$ that does not include $x$, so $X(P+1) = Yx$. Now we can write $\bra{q;2}X(P+1)\ket{q;1} = \bra{q;2}\{ Y, x \} \ket{q;1}$. The anti-commutator $\{ Y, x \}$ must be a sum of terms where each term is a product of modes and only one anti-commutator between $x$ and another mode of opposite charge. As given by Eq. \ref{Eq:LaughlinAlgebra}, such anti-commutators are normal ordered polynomials in $a_n$. By using Eq. \ref{Eq:usefulCom}, thee $a_n$ can be moved around so that $\{ Y, x \}$ is a sum of terms of the form $[\prod_{n_i}a_{-n_i}]X(P)\prod_{n_j}a_{n_j}$ where $n_i, n_j \geq 0$ (this may include some terms that involve no $a_n$ at all). Recalling that for $n > 0$ $\bra{q;2}a_{-n} = 0$ and $a_n\ket{q;1} = 0$, it follows that $\bra{q;2}X(P+1)\ket{q;1} = \bra{q;2}\{ Y, x \} \ket{q;1}$ is expresable as a sum of term of the form $\bra{q;2}X(P)\ket{q;1}$ and thus, $\bra{q;2}X(P+1)\ket{q;1} = 0$. The base case is trivial as $\bra{q;2}X(0)\ket{q;1} = \braket{q;2|q;1} = 0$. Hence, $\bra{q;2}X(p)\ket{q;1} = 0$ for all $X(p)$.

From this we now must have $\bra{q;2}($polynomial in $\Omega_k$ and $\Omega_k^\dagger) \ket{q;1} = 0$ for all such for all such $($polynomial in $\Omega_k$ and $\Omega_k^\dagger)$. However, we argued earlier that $\ket{q; 2} = ($some polynomial in $\Omega_k$ and $\Omega_k^\dagger)\ket{q;1}$ and, thus, $\braket{q;2|q;2} = 0$, which is a contradiction. We can then conclude for any irr. rep. of the Heisenberg algebra in $\mathcal{H}_j$ has multiplicity one.

Now given some $\mathcal{H}_j$ we  figure out which $q$ can appear in it. First, if $\ket{q} \in \mathcal{H}_j$ then there must exist some $k$ such that $\Omega_k \ket{q} \neq 0$ and some $l$ such that $\Omega_l^\dagger\ket{q} \neq 0$, otherwise this $\mathcal{H}_j$ irr. rep. would be trivial. Thus, if $\ket{q} \in \mathcal{H}_j$ and $\ket{q'} \in \mathcal{H}_j$, then there is some $n \in \mathbb{Z}$ such that $q' = q + n\sqrt{m}$. So the possible $U(1)$ charges in $\mathcal{H}_j$ forms a one-dimensional lattice with lattice spacing $\sqrt{m}$. Hence, we can label each irr. rep. by a charge $q_* \geq 0$ which is the smallest such positive charge in the given irr. rep., $\mathcal{H}_{q_*}$.

Now we consider the state $\ket{q_* + \sqrt{m}}$, which has $L_0$ eigenvalue $\frac{(q_* + \sqrt{m})^2}{2}$. By the same arguments as before, $\ket{q_* + \sqrt{m}} = ($some polynomial in $\Omega_l$ and $\Omega_l^\dagger) \ket{q_*}$. Hence, there exists some odd integer $k$ such that $\frac{(q_* + \sqrt{m})}{2} = \frac{q_*^2}{2} + \frac{k}{2} \Rightarrow q_* = \frac{k - m}{2\sqrt{m}}$.

In conclusion, the allowed $q_*$ are $q_* \in \{0, 1/\sqrt{m}, 2/\sqrt{m}, \dots, (m - 1)/\sqrt{m}\}$. We have thus found all the possible irr. reps. of this algebra, assuming that each irr. rep. must decompose into irr. reps. of the Heisenberg algebra. In each case, $\mathcal{H}_{q_*}$, one can take the state $\ket{q_*}$ is the primary state, where the $\mathcal{H}_{q_*}$ is generated by polynomials in the modes $\Omega_k$ and $\Omega_k^\dagger$ acting on this state. 

\section{IQH edge state inner products: further details}
\subsection{$\nu = 1$} \label{Sec:nu1InnerProd}
This simple mapping, given in Eqs. \ref{Eq:VMapn1} and Eqs. \ref{Eq:VdMapn1}, from $V^\dagger_k$ and $V_k$ to $c_m^\dagger$ and $c_m$ allows us to express edge state inner-products in the CFT as,
\begin{equation}
    \langle \braket{ \Psi_{\bra{w}} | \Psi_{\bra{v}} } \rangle / Z_N = \bra{v}e^\mathcal{S} \ket{w}
\end{equation}
where,
\begin{equation}
    \begin{split}
        \mathcal{S} =& \sum_{k > 0} \ln [ \mathcal{N}(N+k-1/2) ] V_{-k}V^\dagger_k \\
        &- \sum_{0<k\leq N-1/2}\ln [\mathcal{N}(N - k - 1/2)] V^\dagger_{-k} V_k \\
        =& \sum_{k = -N + 1/2}^\infty \ln[\mathcal{N}(N+k-1/2)] : V_{-k}V^\dagger_k :
    \end{split}
\end{equation}
For large $N$ and $|(k-1/2)/N| \ll 1$ we can use the Stirling series to expand $\ln [ \mathcal{N}(N+k-1/2) ] \approx N\ln N - N + (1/2)\ln N + (3/2)\ln(2\pi) + 1/(12N) + (k-1/2)\ln 2N  + [(k-1/2)^2 + (k-1/2)]/(2N)$. Then we note the following, $\sum_k (k-1/2) :V_{-k}V^\dagger_k: = \oint \frac{dz}{2\pi i} z :(\partial V(z))V^\dagger(z): = \oint \frac{dz}{2\pi i} z [i\partial^2 \varphi(z) + :(i\partial\varphi(z))^2:]/2 = -a_0/2 + L_0$ (where we use fermionic normal ordering for products of $V_k$ and $V^\dagger_k$ operators), and $\sum_k (k-1/2)^2:V_{-k}V^\dagger_k: = \oint \frac{dz}{2\pi i} z^2 [:(\partial^2 V(z))V^\dagger(z): + z^{-1}:(\partial V(z))V^\dagger(z):] = \oint \frac{dz}{2\pi i} z^2 [:(i\partial\varphi(z))^3: + (3/2) \partial :(i\partial\varphi(z))^2: + i\partial^3\varphi(z)]/3 + L_0 -a_0/2 = \oint \frac{dz}{2\pi i} z^2 :(i\partial\varphi(z))^3:/3 - L_0 + (1/6)a_0$. These relations can be derived by using $V(z)V^\dagger(w) - 1/(z-w) = \sum_{n=1}^\infty (n!)^{-1}(z-w)^{n-1}):e^{-i\varphi(w)}\partial^n e^{i\varphi(w)}:$. We can then write the inner products in the following form,
\begin{equation}
    \langle \braket{ \Psi_{\bra{w}} | \Psi_{\bra{v}} } \rangle / Z_N = \bra{v}R^{2L_0}e^S \ket{w}
\end{equation}
where $R = \sqrt{2N}$ (i.e. the radius of the droplet) and,
\begin{equation}
    \begin{split}
        S =& (N\ln N - N + \ln[2\pi\sqrt{\pi}] - 1/(12N))a_0 \\
        &+ \frac{1}{6N}  \oint \frac{dz}{2\pi i} z^2 :(i\partial\varphi(z))^3: 
    \end{split}
\end{equation}

\subsection{$\nu = 2$} \label{Sec:nu2InnerProd}
To express edge state inner products in the CFT, we first define $R = \sqrt{2N_1}$,  $M(m)_{ij} = \{ \Tilde{c}_{i,m}, \Tilde{c}^\dagger_{j,m} \}$ (i.e. $M(m)_{ij}$ is the inner-product matrix of the orbitals $z^m$ and $\bar{z}z^{m+1}$). Then the inner products take the usual form $\langle \braket{\Psi_{\bra{w}} | \Psi_{\bra{v}} } \rangle/Z_N = \bra{v}e^{\mathcal{S}}\ket{w}$, with,
\begin{equation}
    \mathcal{S} = \sum_k \sum_{ij} (-1)^{i+j} [\ln M( N_1 + k - 1/2 )]_{ij} :V_{i,-k}V^\dagger_{j,k}:
\end{equation}
with the sum of $k$ having the obvious restriction.

By explicit diagonalization and keeping terms up to and including $1/R$, for large $N$ we find,
\begin{equation}
    \begin{split}
        \ln M(N_1 + k - 1/2) \approx& 6\ln R
        \begin{pmatrix}
            -\frac{1}{2} & \frac{1}{R^2} \\
            \frac{1}{R^2} & \frac{1}{2} \\
        \end{pmatrix} \\
        &+ [N_1\ln N_1 - N_1 + \ln N_1 \\
        &+ \ln(4\pi\sqrt{\pi}) \\
        &+ (k-1/2)\ln 2N_1] \mathbf{1}_{2\times 2}
    \end{split}
\end{equation}
where $\mathbf{1}_{2\times 2}$ is the two by two identity matrix.

Using the $\widehat{\mathfrak{su}}(2)_1$ defined in Sec. \ref{Sec:nu2CFT}, we can now express the inner products as $\langle \braket{\Psi_{\bra{w}} | \Psi_{\bra{v}} } \rangle/Z_N = \bra{v}R^{2L_0}e^{S}\ket{w}$, where,
\begin{equation}
    \begin{split}
        S =& \sqrt{2}[N_1\ln N_1 - N_1 + (1/2)\ln N_1 + \ln(2\pi\sqrt{2\pi})]\Tilde{a}_0 \\
        &+ 3\ln (2N_1) \bigg [ J^3_0 - \frac{J^1_0}{N_1} \bigg ]
    \end{split}
\end{equation}
where, $J^1(z) = (J^+(z) + J^-(z))/2$.

\section{Angular momentum calculations}
\subsection{$\nu = 1$} \label{Sec:nu1Angular}
Given that $L_0\ket{0} = 0$ and $[L_0, V(z)] = V(z)/2 + z\partial V(z)$, then,
\begin{equation}
    \begin{split}
        \sum_{i=1}^N z_i\partial_i \Psi_{\bra{v}}(\mathbf{z},\bar{\mathbf{z}}) =& \bra{v}(F^\dagger)^N(L_0 - a_0/2)\prod_{i=1}^N V(z_i) \ket{0} \\
        =& \bra{v}(L_0 + (2N-1)a_0/2 + N(N-1)/2)\\
        &\times(F^\dagger)^N \prod_{i=1}^N V(z_i) \ket{0} \\
        =& \Psi_{\bra{v}(L_0 + (2N-1)a_0/2 + N(N-1)/2)}(\mathbf{z})
    \end{split}
\end{equation}
Hence, we can see that the angular momentum operator maps to the CFT as $\sum_i z_i \partial_i \rightarrow L_0 + (2N-1)a_0/2 + N(N-1)/2$. Other operator mappings can be found in \cite{Fern2018}.

\subsection{$\nu = 2$} \label{Sec:nu2Angular}
The angular momentum operator $\sum_i z_i\partial_i - \bar{z}_i\overline{\partial}_i$ can be mapped over using the same method for the $\nu = 1$ case to give,
\begin{equation}
    \begin{split}
        [\sum_{i=1}^N z_i\partial_i - \bar{z}_i\overline{\partial}_i] \Psi_{\bra{v}}(\mathbf{z},\bar{\mathbf{z}}) =& \bra{v}(F^\dagger_2)^{N_2} (F_1^\dagger)^{N_1} \\
        &\times (L_0 - (1/2)\fem{1} - (3/2)\fem{2}_0) \\
        &\times \prod_{i=1}^N\Omega(z,\bar{z}) \ket{0} \\
        =& \bra{v}[L_0 + (2N_1 - 1)\Tilde{a}_0/\sqrt{2} \\
        &+ N_1(N_1 - 1) - 1] \\
        &\times (F^\dagger_2)^{N_2} (F_1^\dagger)^{N_1} \prod_{i=1}^N \Omega(z,\bar{z}) \ket{0}
    \end{split}
\end{equation}

\section{Further details for $\phi_2^2$ CFT} \label{Sec:22CFTDetails}
Now let us find the decomposition of $\Hp$ into the irreducible representations of $\hat{\mathfrak{u}}(1) \oplus \widehat{\mathfrak{su}}(2)_2$. As the primaries of the gauge invarient $\mathcal{M}_\lambda$ are spin-$0$ with repsect to $\mathcal{J}^a(z)$, we must have $\Tilde{L}^{\widehat{\mathfrak{su}}(2)_2}_0 \ket{\lambda;i} = 0$. So, the $L_0$ eigenvalue of $\ket{\lambda;i}$ must take the form, $h_\lambda = h^{\hat{\mathfrak{u}}(1)} + h^{\widehat{\mathfrak{su}}(2)_2}$ where $h^{\hat{\mathfrak{u}}(1)}$ and $h^{\mathfrak{su}(2)_2}$ are the eigenvalues of $L_0^{\hat{\mathfrak{u}}(1)}$ and $L_0^{\mathfrak{su}(2)_2}$ respectively. Furthermore, as $\ket{\lambda; i}$ is a WZW primary of $i\partial \Phi(z)$, we must have $L_0^{\hat{\mathfrak{u}}(1)}\ket{\lambda; i} = \frac{(a_0)^2}{2}\ket{\lambda ; i} = h^{\hat{\mathfrak{u}}(1)}\ket{\lambda ; i}$. For all states $\ket{v} \in \mathcal{H}_{N_p}$ we have $a_0\ket{v} = (N_p/2)\ket{v}$. Hence, $h^{\hat{\mathfrak{u}}(1)} = (N_p)^2/8$. As the number of partons must be exactly double the number of underlying bosons, $N_p$ must be an even number. For a spin-$j$ primary of $J^a(z)$ we have $h^{\widehat{\mathfrak{su}}(2)_2} = j(j+1)/4$, where the allowed $j$ are $0,\frac{1}{2}$ and $1$ \cite{DiFrancesco1997}. One can also directly check that for even $N_p$ all $L_0$ eigenvalues with $\mathcal{H}_{N_p}$ must be integers.

Hence, when $N_p$ is a multiple of 4,  $h^{\hat{\mathfrak{u}}(1)}$ must be an even positive integer and we can only have spin-$0$ primary states of $J^a(z)$, with such a primary having $h_\lambda = (N_p)^2/8$. Within $\mathcal{H}_{N_p}$ there is only one state with this $L_0$ eigenvalue and it is the unique state with the lowest $L_0$ eigenvalue in $\mathcal{H}_{N_p}$. So when $N_p$ is a multiple of four there is only one gauge invariant $\mathcal{M}_\lambda$ and so all gauge invariant states of $\mathcal{H}_{N_p}$ can be expressed as polynomials of the modes of $i\partial\Phi(z)$ and $J^a(z)$ applied on the state with lowest $L_0$ eigenvalue in $\mathcal{H}_{N_p}$.

When $N_p$ is not a multiple of $4$ then $h^{\hat{\mathfrak{u}}(1)}$ must be half-integral. Hence, the only allowed gauge invariant $\mathcal{M}_\lambda$ must be such that its primary states are spin-$1$ with respect to $J^a(z)$. In this case, the gauge-invariant primaries of $\mathcal{H}_{N_p}$ now must have $h_\lambda = (N_p)^2/8 + 1/2$, which corresponds to lowest eigenvalue of $L_0$ within $\mathcal{H}_{N_p}$. Within $\mathcal{H}_{N_p}$ the space of states with the lowest $L_0$ eigenvalue can be decomposed into representations of the $J^a_0$ and $\mathcal{J}^a_0$ as $1\otimes 0 \oplus 0 \otimes 1$. Only the states in $1 \otimes 0$, which are spin-$1$ with respect to $J^a_0$ can be gauge invariant. Hence, there is again only one gauge invariant $\mathcal{M}_\lambda$ with all gauge invariant states being expressed as polynomials in $a_n$ and $J^a_n$ applied on the states of the $1 \otimes 0$ representation within the space of states with lowest $L_0$ eigenvalue. 

Now we  show that the modes of $V_m(z)$ and $V_m^\dagger(z)$ (see  \footnote{We have $V^\dagger_1(z) = V^\dagger_{22}(z)V^\dagger_{12}(z)$, $V^\dagger_0(z) = [V^\dagger_{21}(z)V^\dagger_{12}(z) + V^\dagger_{22}(z)V^\dagger_{11}(z)]/\sqrt{2}$ and $V^\dagger_{-1}(z) = V^\dagger_{21}(z)V^\dagger_{11}(z)$.} for a defintion of $V^\dagger_m(z)$) fields can be used to generate $\Hp$. These fields have the following OPEs,
\begin{equation}
    \begin{split}
        V_{\pm 1}(z)V^\dagger_{\pm 1}(w) \sim & \frac{1}{(z-w)^2} + \frac{i\partial\Phi(w) \pm J^3(w)}{z-w} \\
        V_{\pm 1}(z)V^\dagger_0(w) \sim & \pm \frac{J^{\pm}(w)/\sqrt{2}}{z-w} \\
    \end{split}
\end{equation}
From these OPEs it follows that we can express the modes $a_n$ and $J^a_n$ in terms of the modes of $V_m(z)$ and $V^\dagger_m(z)$. Hence, in any $\mathcal{M}_{\lambda(N_p)}$, the states can be generated by the application of the modes of $V_m(z)$ and $V^\dagger_m(z)$ on the primary states. When $N_p$ is a multiple of 4, we only need to consider one primary state of $\mathcal{M}_{\lambda(N_p)}$, which we denote $\ket{\lambda(N_p); 0}$. Now consider the states $V_{m, -(N_p/2 + 1)}\ket{\lambda(N_p); 0}$. If non-zero, these states have a $U(1)$ charge corresponding to the states with $N_p + 2$ partons and hence belong to $\mathcal{H}_{N_p+2}$. Furthermore, they have an $L_0$ eigenvalue which is the lowest of $\mathcal{H}_{N_p+2}$. We also have that $V^\dagger_{m, N_p/2 + 1}\ket{\lambda(N_p); 0} = 0$, as if it was non-zero it would have an $L_0$ eigenvalue lower than any state with that $U(1)$ charge and so we would have a contradiction. It then follows $\bra{\lambda(N_p);0} V^\dagger_{m, N_p/2 + 1} V_{m, -(N_p/2+1) }\ket{\lambda(N_p); 0} = \bra{\lambda(N_p);0} [V^\dagger_{m, N_p/2 + 1}, V_{m, -(N_p/2+1) } ]\ket{\lambda(N_p); 0} = \bra{\lambda(N_p);0} ( N_p/2 + 1 - a_0 - mJ^3_0 )\ket{\lambda(N_p); 0} = \bra{\lambda(N_p);0} ( N_p/2 + 1 - N_p/2 )\ket{\lambda(N_p); 0} = \braket{\lambda(N_p);0 | \lambda(N_p);0} > 0$. Thus, the states $V_{m,N_p/2 + 1}\ket{\lambda(N_p); 0}$ are non-zero gauge invariant, linearly independent (as they different $J^3_0$ eigenvalues), and have $L_0$ eigenvalue corresponding to the lowest within $\mathcal{H}_{N_p + 2}$. Hence, these states must correspond to the three primaries $\ket{\lambda(N_p); m}$ of $\mathcal{M}_{\lambda(N_p + 2)}$. Similarly, it can be shown that when $N_p$ is a multiple of 4 the primaries $\ket{\lambda(N_p - 2);m}$ can be generated from $\ket{\lambda(N_p);0}$ by applying mode of $V^\dagger_m(z)$. It can also be shown (by the same method) that for $N_p$ not a multiple of 4, the primaries $\ket{\lambda(N_p + 2);0}$ and $\ket{\lambda(N_p - 2);0}$ can be generated by applying the modes of $V_m(z)$ and $V^\dagger_m(z)$ on $\ket{\lambda(N_p); m'}$ respectively. Hence, all primaries $\ket{\lambda(N_p);i}$ can be generated by repeated application of the modes of $V_m(z)$ and $V^\dagger_m(z)$ on $\ket{0}$. 

\section{$\mathcal{A}(n)_m$ algebras} \label{Sec:partonAlgebra}
We first detail the general structure of the $\mathcal{A}(n)_m$ algebras as defined in Sec. \ref{Sec:PartonGenWFCon}. Then we  discuss the representations of $\mathcal{A}(2)_k$.

\subsection{General structure} \label{Sec:partonAlgebraGenStructure}
As given in Sec. \ref{Sec:symmetricPartons}, $\mathcal{A}(n)_m$ is defined as being a sub-chiral algebra of a CFT of $nm$ chiral bosons with vertex operators $V_{ij}(z)$ where $i = 1, 2, \dots, m$ and $j = 1, 2, \dots, n$. The generating fields of $\mathcal{A}(n)_m$ are labeled by $l = m, m+1, \dots, nm$, where,
\begin{equation}
    V_l(z) = \sum_{i_1, i_2, \dots i_m}^n \delta_{(\sum_{j=1}^m) i_j, l}  V_{1i_1}(z)V_{2i_2}(z) \dots V_{mi_m}(z)
\end{equation}
The space of states generated by the modes of $V_l(z)$ and $V^\dagger_l(z)$ applied on the vacuum state will be denoted as $\mathcal{H}_{\text{phys}}$ (as usual). By operator-state correspondence, each state in $\mathcal{H}_{\text{phys}}$ is in one-to-one correspondence with fields of $\mathcal{A}(n)_m$. We denote the full space of states of the multiple chiral boson theory as $\mathcal{H}_{F}$. 

Within $\mathcal{H}_F$ we have the currents defined in Eq. \ref{Eq:partonCurrents}. In the notation used in this section, $t^a$ are indexed by two numbers $a = kl$ with each in the range $1,2,\dots, n$. For $k\neq l$ we have $t^{kl}_{ij} = \delta_{ik}\delta_{lj}$. The generators $t^{ll}$ only exist for $l = 1, 2, \dots , n - 1$ with $t^{ll}_{ij} = \delta_{ij}(\delta_{li} - \delta_{ni})$ being the convention used here. As further demonstrated in Ref. \cite{affleck_exact_1986} this is a conformal embedding of these current algebras in that the energy-momentum tensor of $\mathcal{H}_F$ is a sum of the Suguwara energy-momentum tensors of each set of currents $i\partial\Phi(z)$, $J^a(z)$ and $\mathcal{J}^a(z)$. That is $T(z) = T^{\hat{\mathfrak{u}}(1)}(z) + T^{\widehat{\mathfrak{su}}(n)_m}(z) + T^{\widehat{\mathfrak{su}}(m)_n}(z)$. 

As for the case discussed in Sec. \ref{Sec:partonCFT}, we define the space of gauge invarient states $\mathcal{H}_G$ by $\ket{v} \in \mathcal{H}_G \Leftrightarrow \mathcal{J}^a_q\ket{v} = 0$ for $q \geq 0$. One can easily verify that $[\mathcal{J}^a_q, V_l(z)] = 0$ ($q\in \mathbb{Z}$). Hence, we must have $\mathcal{H}_{\text{phys}} \subset \mathcal{H}_G$ (as $\ket{0} \in \mathcal{H}_G$). 

Now consider the space of $U(1)$ neutral states $\mathcal{H}_0$ (i.e. those states with $a_0\ket{v} = 0$ with $a_0$ being the zeroth mode of $i\partial\Phi(z)$). This space must be an invariant subspace under the action of the modes $a_n$, $J^a_n$ and $\mathcal{J}^a_n$. Thus, $\mathcal{H}_0$ must decompose into a direct sum of irreducible representations of the $\hat{\mathfrak{u}}(1)\oplus\widehat{\mathfrak{su}}(n)_m\oplus\widehat{\mathfrak{su}}(m)_n$ algebra, $\mathcal{H}_0 = \bigoplus_\lambda \mathcal{M}_\lambda$. The only $\mathcal{M}_\lambda$ that can contain gauge invariant states are those which take the form $\mathcal{M}_\lambda = \mathcal{M'}_{\lambda'} \otimes \mathcal{V}$ where $\mathcal{M'}_{\lambda'}$ is an irreducible representation of $\hat{\mathfrak{u}}(1)\oplus \widehat{\mathfrak{su}}(n)_m$ and $\mathcal{V}$ is the vacuum representation of $\widehat{\mathfrak{su}}(m)_n$. As given by the branching rules given in Ref. \cite{nakanishi_level-rank_1992, DiFrancesco1997} the only such $\mathcal{M}_\lambda$ is such that $\mathcal{M'}_{\lambda'} = \mathcal{V}'$ where $\mathcal{V}'$ is the vacuum representation of $\hat{\mathfrak{u}}(1)\otimes\widehat{\mathfrak{su}}(n)_m$. Furthermore, these branching rules indicate that $\mathcal{V}'\otimes\mathcal{V}$ has multiplicity one in $\mathcal{H}_0$. This implies that all the gauge invariant states of $\mathcal{H}_0$ can be generated by applying the modes of the $i\partial\Phi(z)$ and $J^a(z)$ on the vacuum.

Now consider the OPEs,
\begin{equation}
    \begin{split}
        V_l(z) V^\dagger_l(w) =& \frac{p_m(l)}{(z-w)^m} \\
        &+ \frac{ \sum_{j=1}^n p_{m-1}(l-j) \sum_{k=1}^m i\partial \Tilde{\varphi}^{(kj)}(w)}{(z-w)^{m-1}} \\ 
        &+ \dots
    \end{split}
\end{equation}
where $p_m(l)$ is an integer which is the number column vectors of $m$ integers such that each member of the set is in the range $1, 2, \dots, n$ and their sum is $l$ (note this can be zero if no such sets exist). By first considering $l=nm$ and then $l=nm - 1$ one can easily show that by suitable linear combinations of these OPEs we can generate the currents $i\partial\Phi(z)$ and $J^{ii}(z)$. We also have the OPEs,
\begin{equation}
    V_{nm}(z)V^\dagger_{nm - i}(w) \propto \frac{J^{ni}(w)}{(z-w)^{m-1}} + \dots
\end{equation}
with $i = 1, 2, \dots, n - 1$ and,
\begin{equation}
    V_{nm-i}(z)V^\dagger_{nm}(w) \propto \frac{J^{in}(w)}{(z-w)^{m-1}} + \dots
\end{equation}
Furthermore, we also have $[J^{in}_0, J^{nj}_0] = J^{ij}_0$ with $i \neq j$. Thus, all the currents $i\partial\Phi(z)$ and $J^a(z)$ must belong to $\mathcal{A}(n)_m$. As $\mathcal{H}_{\text{phys}} \subset \mathcal{H}_G$, this implies that all gauge invarient states of $\mathcal{H}_0$ is the same space of states as all the $U(1)$ netural guage invarient states of $\mathcal{H}_{\text{phys}}$, as $\mathcal{H}_{\text{phys}} \subset \mathcal{H}_G$. Thus, by operator state correspondence, all the $U(1)$ neutral fields of $\mathcal{A}(n)_m$ form the chiral algebra of the $\hat{\mathfrak{u}}(1) \oplus \widehat{\mathfrak{su}}(n)_m$ WZW model. 

When $m$ is even, this implies the commutation relations,
\begin{equation}
    \begin{split}
        [V_{l,r}, V^\dagger_{k, s}] =& \delta_{s+r, 0}\delta_{lk} p(l)_m \bigg [ \prod_{j=1}^{m-1}\frac{r + m/2 - j}{j} \bigg ] \\
        & + ( \text{normal ordered polynomial in } \\
        & a_q \text{ and } J^a_q ) \\
    \end{split}
\end{equation}
where $q$ is meant to represent some integer. For odd $m$ we have anti-commutators of the form,
\begin{equation}
    \begin{split}
        \{ V_{l,r}, V^\dagger_{k, s} \} =& \delta_{s+r, 0}\delta_{lk} p(l)_m \bigg [ \prod_{j=1}^{m-1}\frac{r + m/2 - j}{j} \bigg ] \\
        & + ( \text{normal ordered polynomial in } \\
        & a_q \text{ and } J^a_q ) \\
    \end{split}
\end{equation}
\cite{schellekens_extended_1989}

We now comment, very briefly, on the representation theory of general $\mathcal{A}(n)_m$ algebras. We first note that the $V_l(z)$ fields are WZW primaries with respect to the $J^a(z)$ fields corresponding to the totally symmetric rank $m$ tensor representation. Such WZW primaries have fusion rules corresponding to \textit{simple currents} \cite{schellekens_extended_1989}. A simple current $V$ has a fusion rule $V\times \phi_{\lambda_1} = \phi_{\lambda_2}$ where $\phi_{\lambda_1}$ and $\phi_{\lambda_2}$ are two WZW primaries (i.e. the fusion of $J$ with any other field has only one possible outcome). The \textit{orbit} of a field $\phi_\lambda$ are all the WZW primaries generated by repeated fusion with $V$. Any irreducible representation of $\mathcal{A}(n)_m$, $\mathcal{H}_j$ will decompose into irreducible representations of $\hat{\mathfrak{u}}(1)\otimes \widehat{\mathfrak{su}}(n)_m$ $\mathcal{H}_j = \bigoplus_{q,\lambda}\mathcal{M}_{q,\lambda}$ where $q$ is the $U(1)$ charge and $\lambda$ labels an $\widehat{\mathfrak{su}}(n)_m$ representation. $\mathcal{H}_j$ must be built up by applying the modes of $V_l(z)$ and $V^\dagger_l(z)$ onto some state belonging to one of the $\mathcal{M}_{q,\lambda}$. Clearly then, the only possible other $\mathcal{M}_{q',\lambda'}$ must be such that $q'$ differs from $q$ by an integer number of the $U(1)$ charge of $V_l(z)$ and $\lambda'$ must belong to the orbit of $\lambda$. Thus, the possible irreducible representations of $\mathcal{A}(n)_m$ can be labelled by orbits of this simple current in the set of possible $\lambda$. This labelling is not one-to-one.

One can always construct representations of $\mathcal{A}(n)_m$ using the $\hat{\mathfrak{u}}(1)\oplus \widehat{\mathfrak{su}}(n)_m$ WZW model from the space of states that are generated by repeated application of the modes of $:e^{i\sqrt{\frac{m}{n}}\Phi(z)}:\phi_l(z)$ and $:e^{-i\sqrt{\frac{m}{n}}\Phi(z)}:\phi^\dagger_l(z)$ on the state $:e^{iq\Phi(0)}:\phi_\lambda(0)\ket{0}$, where $\phi_\lambda(z)$ is an $\widehat{\mathfrak{su}}(n)_m$ WZW primary and $q$ is chosen such that the OPE $V_l(z)e^{iq\Phi(w)}\phi_\lambda(w)$ only contains integer powers of $(z-w)$. These representations have the property that each $\mathcal{M}_{q',\lambda'}$ has multiplicity one and for a given $q'$ there is only one $\mathcal{M}_{q',\lambda'}$ in the decomposition of the irreducible representation (i.e. if $\mathcal{M}_{q',\lambda_1} \subset \mathcal{H}_j$ and $\mathcal{M}_{q',\lambda_2} \subset \mathcal{H}_j$ then $\lambda_1=\lambda_2$). One can show that the conformal blocks involving primaries of these representations directly correspond to the conformal blocks of $\hat{\mathfrak{u}}(1)\otimes \widehat{\mathfrak{su}}(n)_m$. In other words, the conformal blocks of these representations are equivalent to the conformal blocks of the fields $:e^{iq\Phi(z)}:\phi_\lambda(z)$, with such conformal blocks transforming among themselves under monodromy transformations. This can be shown by considering the possible chiral vertex operators \cite{tsuchiya_vertex_1987, Moore1988, Moore1989} that generate the conformal blocks of $\mathcal{A}(n)_m$, and from $U(1)$ charge conservation. 

\subsection{Representations of $\mathcal{A}(2)_k$} \label{Sec:A2kReps}
In the case of $\mathcal{A}(2)_k$, $V_l(z)$ transforms as a WZW primary under the action of $J^a(z)$ corresponding to the totally symmetric rank $k$ tensor representation of $\mathfrak{su}(2)$. The possible representations of $\widehat{\mathfrak{su}}(2)_k$ are labelled by the usual $j$ with the possible values being $j = 0,1/2,1,\dots, k/2$ with corresponding scaling dimension $h_j = j(j+1)/(k+2)$ \cite{DiFrancesco1997}. The totally symmetric rank $k$ tensor representation corresponds to $j=k/2$ and has the fusion rules $\phi_j\times \phi_{k/2} = \phi_{k/2 - j}$. 

We  now show how the irreducible representations can be determined assuming that each irreducible representation can be decomposed into irreducible representations of the $\widehat{\mathfrak{su}}(2)_k$ algebra. As for the chiral algebra associated with the Laughlin wave functions the discussion here is not fully mathematically rigorous.

Let $\mathcal{H}_a$ denote an irreducible representation of $\mathcal{A}(2)_k$, where the labelling system $a$ is yet to be determined. Let $\mathcal{M}_{q,j}$ be an irreducible representation of $\hat{\mathfrak{u}}(1)\oplus \widehat{\mathfrak{su}}(2)_k$, where $q$ is the $U(1)$ charge and $j$ labels the $\widehat{\mathfrak{su}}(2)_k$ representation. Now by the discussion in the previous section, for any $\mathcal{H}_a$ we know there must exist a $j$ such that $\mathcal{H}_a = [\oplus_{\text{possible } q_1} \mathcal{M}_{q_1, j} ] \oplus [\oplus_{\text{possible } q_2} \mathcal{M}_{q_2, k/2-j} ]$. Let $\ket{q,j}$ denote the primary of $\mathcal{M}_{q,j}$. We know $V_{l,x}\ket{q,j}$ must have $L_0$ eigenvalue $\frac{q^2}{2} + \frac{j(j+1)}{k + 2} - x$. At the same time $V_{l,x}\ket{q,j} \in \mathcal{M}_{q + \sqrt{k/2}, k/2 - j}$ and so if $V_{l,x}\ket{q,j} \neq 0$ we must have $\frac{(q + \sqrt{k/2})^2}{2} + \frac{(k-2j)(k-2j + 2)}{4(k+2)} \leq \frac{q^2}{2} + \frac{j(j+1)}{k + 2} - x$, as all states of $\mathcal{M}_{q + \sqrt{k/2}, k/2 - j}$ must have an $L_0$ eigenvalue larger than or equal to the eigenvalue of $\ket{q + \sqrt{k/2}, k/2 - j}$. This constraint simplifies to, $x \leq -q\sqrt{k/2} - \frac{k-2j}{2}$. Similarly, for $\bra{q, j}V_{l,x} \neq 0$ then $x \geq -q\sqrt{k/2} + \frac{k - 2j}{2}$. Also for either $V^\dagger_{l,x}\ket{q,j} \neq 0$ or $\bra{q,j}V^\dagger_{l,x} \neq 0$ we must have $x \geq q\sqrt{k/2} - \frac{k - 2j}{2}$ and $x \geq q\sqrt{k/2} + \frac{k-2j}{2}$ respectively. Hence, we can reuse the argument from Sec. \ref{Sec:LaughlinReps} to show that any $\mathcal{M}_{q,j}$ can appear at most once in the decomposition of $\mathcal{H}_a$ into irreducible representations of the $\hat{\mathfrak{u}}(1)\oplus \widehat{\mathfrak{su}}(2)_k$ algebra. 

Furthermore, as we must be able to generate state $\ket{q-\sqrt{k/2}, k/2 - j}$ from state $\ket{q, j}$ by applying modes of the $V_l(z)$ and $V^\dagger_l(z)$ fields (as this is an irreducible representation), there must exist an integer $x$ such that, $\frac{x}{2} = q\sqrt{k/2} - \frac{k-2j}{2}$, where if $k$ is odd or even $x$ is odd or even respectively. Hence, $q = \frac{x + k - 2j}{\sqrt{2k}}$. As $x+k$ must be even this can be expressed more compactly as $q = \sqrt{\frac{2}{k}}(p + j)$ where $p \in \mathbb{Z}$. For a $\mathcal{M}_{q_1,j} \subset \mathcal{H}_a$ and $\mathcal{M}_{q_2, j} \subset \mathcal{H}_a$ then $q_1 = q_2 + \kappa \sqrt{2k}$ with $\kappa \in \mathbb{Z}$. This follows as we must be able to obtain a state of $\mathcal{M}_{q_2, j}$ from a state in $\mathcal{M}_{q_1,j}$ by an even number of applications of modes of $V_l(z)$ or $V^\dagger_l(z)$. Similarly, if $\mathcal{M}_{q,j}\subset \mathcal{H}_a$ then $\mathcal{M}_{q+\sqrt{k/2}, k/2 - j} \subset \mathcal{H}_a$. Hence, we must have the decomposition,
\begin{equation}
    \mathcal{H}_{j,x} = \bigoplus_{\kappa \in \mathbb{Z}} [ \mathcal{M}_{q_{j,p,0}(\kappa), j} \oplus \mathcal{M}_{q_{j,p,1}(\kappa),k/2 - j} ]
\end{equation}
where we now can adopt the labelling system $a = (j,p)$ and,
\begin{equation}
    q_{j,p,r}(\kappa) = \sqrt{\frac{2}{k}}( p + \kappa k + r(k/2) + j) \\
\end{equation}
So we have the irreducible representations $\mathcal{H}_{j,p}$. with the inequivelent ones labelled by $j = 0, 1/2, 1, \dots, k/4$ for even $k$ and $j = 0, 1/2, 1, \dots, (k-1)/4$ for odd $k$, with $p = 0, 1, 2, \dots, k - 1$ for $j\neq k/4$ and $p = 0,1,2,\dots,k/2 - 1$ for $j=k/4$. Thus, for a given $j$ there are $k$ possible inequivalent representations with one exception when $k$ is even: there are only $k/2$ representations with $j = k/4$.

In each case, we can construct the $\mathcal{H}_{j,p}$ using the $U(1)\otimes SU(2)_k$ WZW model by applying modes of the $V_l(z)$ and $V^\dagger_l(z)$ fields on the state $:e^{iq_p\Phi(0)}:\phi_j(0)\ket{0}$ where $\Phi(z)$ is the $U(1)$ field, $\phi_j(z)$ is a spin-$j$ $SU(2)_k$ WZW primary field $q_p = \sqrt{\frac{2}{k}}(p + j)$, and $V_l(z)$ is represented by $V_l(z) = \sqrt{\binom{k}{l}}e^{i\sqrt{k/2}\Phi(z)}\phi_{k/2, -k/2 + l}(z)$ with $\phi_{k/2,m}(z)$ is the spin-$k/2$ WZW primary with $J^3_0$ eigenvalue $m$. From the discussion at the end of Sec. \ref{Sec:partonAlgebraGenStructure}, we can then see that all the conformal blocks of $\CFT{\mathcal{A}(2)_k}$ must correspond to conformal blocks of the $\hat{\mathfrak{u}}(1)\oplus\widehat{\mathfrak{su}}(2)_k$ WZW model. Where we emphasise that the irreducible representation of this algebra can always be decomposed into irreducible representation of $\widehat{\mathfrak{su}}(2)_k$.

\subsection{Fusion rules of $\mathcal{A}(2)_k$} \label{Sec:A2kFusionRules}
We now give the fusion rules for $\mathcal{A}(2)_k$, whose representations correspond to the fields $\zeta_{j,p}$ as detailed in Sec. \ref{Sec:su2Series}. So that the fusion rules can be expressed in a compact way we  extend the labelling system $(j,p)$ such that $\zeta_{j,p + nk} = \zeta_{j,p}$ for $n \in \mathbb{Z}$ and $\zeta_{k/2 - j,p + 2j} = \zeta_{j,p}$. Note that this then gives $\zeta_{k/4,p + k/2} = \zeta_{k/4,p}$. The fusions rules can then be expressed as,
\begin{equation}
    \zeta_{j_1, p_1} \times \zeta_{j_2, p_2} = \sum_{\substack{j = |j_1 - j_2| \\ (j_1 + j_2 - j) = 0 \pmod{1}}}^{\min (j_1+j_2, k-j_1-j_2)} \zeta_{j, p_1+p_2+j_1+j_2-j}
\end{equation}

\section{Further discussion for general parton construction} \label{Sec:PartonBackgroundCharge}

\subsection{The background charge operator $C(N)$}
We now show inductively that the correlation function of Eq. \ref{Eq:prodWFCorr} can be computed in the vacuum representation of $\mathcal{A}$, which was defined in Sec. \ref{Sec:PartonGenWFCon}. To do this we merely need to show that $C(N)$ can be defined purely as an operator in the vacuum representation of $\mathcal{A}$, which we  denote $\mathcal{H}_{\text{phys}}$. The following arguments apply when the lowest angular momentum parton state for a fixed number of particles $N$ is unique. We assume that the base cases are integer quantum Hall wave functions (i.e. we are only taking products of parton wave function). It is trivially the case that the integer quantum Hall wave functions are generated by a chiral algebra as identified above (as can be seen in Sec. \ref{Sec:integerQHCFT}). The inductive assumptions, which can be easily shown for the integer quantum Hall base cases, are as follows. First, we assume $\Omega_i(z, \bar{z}) = \sum_{j=0}^{n_i} \bar{z}^j\phi^{(i)}_j(z)$ with $n_i \in \mathbb{Z}^+$, with $\mathcal{A}^{(i)}$ denoting the chiral algebra generated by $\phi^{(i)}_j(z)$ and their conjugates, with the vacuum representation $\mathcal{H}^{(i)}_{\text{phys}}$. It is assumed that $C_i(N)$ is defined within $\mathcal{H}^{(i)}_{\text{phys}}$, so that the generating correlation function of $\Psi_i$ can be computed in $\mathcal{H}^{(i)}_{\text{phys}}$, with the property $C_i(N)\phi^{(i)}_j(z)C_i(N)^\dagger \propto z^{k_i(N) + j}\phi^{(i)}_j(z) \rightarrow C_i(N)\phi^{(i)}_{j,m}C(N)^\dagger \propto \phi^{(i)}_{j,k_i(N) +j + m}$, where $k_i(N) \in \mathbb{Z}$. There also exits three Hermitian operators $a^{(i)}_0$, $L^{(i)}_0$, and $J^{(i)}_0$ within $\mathcal{H}^{(i)}_{\text{phys}}$ that commute with each other, with the properties, $[a^{(i)}_0, \phi^{(i)}_j(z)] = q_i \phi^{(i)}_j(z)$ ($q_i \in \mathbb{R}$), $[L^{(i)}_0, \phi^{(i)}_j(z)] = h_i \phi^{(i)}_j(z) + z\partial\phi^{(i)}_j(z)$ ($h_i\in \mathbb R, h_i>0$), $[J^{(i)}_0, \phi^{(i)}_j(z)] = j\phi^{(i)}_j(z)$, and $a^{(i)}_0\ket{0} = L^{(i)}_0\ket{0} = J^{(i)}_0\ket{0} = 0$. Finally, we assume $C_i(N)^\dagger\ket{0}$ is a simultaneous eigenstate of $a^{(i)}_0$, $L^{(i)}_0$, and $J^{(i)}_0$, such that it is the unique state with the lowest $L^{(i)}_0 - J^{(i)}_0$ eigenvalue amounst all other states with the same $a^{(i)}_0$ eigenvalue. 

By explicitly expanding $\Omega(z,\bar{z})$ we get $\phi_j(z) = \sum_{j_1j_2}\delta_{j_1+j_2, j}\phi^{(1)}_{j_1}(z)\phi^{(2)}_{j_2}(z)$. It then follows that, $C(N)\phi_j(z)C(N)^\dagger \propto z^{k_1(N) + k_2(N) + j}\phi_j(z)$. Within $\CFT{1}\otimes\CFT{2}$. we then define the operators $a_0 = a^{(1)}_0 + a^{(2)}_0$, $L_0 = L^{(1)}_0 + L^{(2)}_0$ and $J_0 = J^{(1)}_0 + J^{(2)}_0$. Clearly these have the desired properties, $[a_0, \phi_j(z)] = (q_1 + q_2) \phi_j(z)$, $[L_0, \phi_j(z)] = (h_1 + h_2) \phi_j(z) + z\partial\phi_j(z)$, $[J_0, \phi_j(z)] = j\phi_j(z)$, and $a_0\ket{0} = L_0\ket{0} = J_0\ket{0} = 0$. Hence, $\mathcal{H}_{\text{phys}}$ must be an invarient subspace of the operators $a_0$, $L_0$ and $J_0$. Furthermore, $[a^{(i)}_0, \phi_j(z)] = q_i \phi_j(z)$, so $\mathcal{H}_{\text{phys}}$ is an invariant subspace of $a^{(i)}_0$ and, hence, within $\mathcal{H}_{\text{phys}}$ $a^{(1)}_0 \propto a^{(2)}_0$. Clearly, $C(N)^\dagger\ket{0}$ must be the unique state in $\CFT{1}\otimes\CFT{2}$, with the property that it is a simultaneous eigenstate of $a^{(i)}_0$, $L_0$ and $J_0$ with the property that it has the lowest $L_0-J_0$ eigenvalue of all states with the same $a^{(i)}_0$ eigenvalues. As, $\bra{0}C(N)\prod_{i=1}^N\Omega(z_i,\bar{z}_i)\ket{0} \neq 0$ then $P C(N)^\dagger \ket{0} \neq 0$ where $P$ is the projector onto $\mathcal{H}_{\text{phys}}$. Thus, $C(N)^\dagger\ket{0} = \ket{x} + \ket{y}$ where $\ket{x} \in \mathcal{H}_{\text{phys}}$ and $P\ket{y} = 0$, with both $\ket{x}$ and $\ket{y}$ having the same eigenvalues of $a^{(i)}_0$, $L_0$ and $J_0$. Given the uniqueness properties given above, it must be the case that $\ket{y} = 0$. So, $C(N)^\dagger\ket{0} \in \mathcal{H}_{\text{phys}}$. So $C(N)$ is well defined within $\mathcal{H}_{\text{phys}}$ and so we can write the wave function $\Psi_1\Psi_2$ using the vacuum representation of $\mathcal{A}$. As $a^{(1)} \propto a^{(2)}$ in $\mathcal{H}_{\text{phys}}$ then $C(N)^\dagger\ket{0}$ must the must be the unique state in $\mathcal{H}_{\text{phys}}$, with the property that it is a simulaneous eigenstate of $a_0$, $L_0$ and $J_0$ with the property that it has the lowest $L_0-J_0$ eigenvalue of all states with the same $a_0$ eigenvalue.

\subsection{Angular momentum mapping} \label{Sec:AMParton}
We  now show what form the angular momentum operator generally takes in the CFT provided the lowest angular momentum parton ground state trial wave function is unique for a fixed number of particles $N$.

We again consider the product wave function $\Psi_1\Psi_2$. We then assume that the edge state mapping for $\Psi_i$ is such that the angular momentum operator can be expressed in the CFT as $L_0^{(i)} + v_i(N)a^{(i)}_0 + u_i(N)$ for some real-valued functions $v_i(N)$ and $u_i(N)$. This is true for the integer quantum Hall base cases so long as $N$ is such that the lowest angular momentum state is unique at this fixed number of particles. 

For the product wave function $\Psi_1\Psi_2$ the angular momentum operator can mapped to the CFT simply as $L_0^{(1)} + L_0^{(2)} + v_1(N)a^{(1)}_0 + v_2(N)a^{(2)}_0 + u_1(N) + u_2(N)$. Then recall that within $\Hp$ $a^{(1)}_0 \propto a^{(2)}_0$. Thus within $\Hp$ the angular momentum operator maps over as $L_0 + v(N)a_0 + u(N)$ for some real-valued functions $v(N)$ and $u(N)$. 

\subsection{Energy-momentum tensor of $\mathcal{A}$}
We  now show that $\mathcal{A}$ contains an energy-momentum tensor, which is required for the construction of $\CFT{\mathcal{A}}$. Let $T(z)$ denote the energy-momentum tensor of $\CFT{1}\otimes \CFT{2}$ (which is simply the sum of the energy-momentum tensors of $\CFT{1}$ and $\CFT{2}$). Its corresponding state is $\ket{T} \equiv T(0)\ket{0}$. Once again, let $P$ be the projector onto $\mathcal{H}_{\text{phys}}$. We can then define a new field $\Tilde{T}(z)$ from state-operator correspondence $\Tilde{T}(0)\ket{0} \equiv P\ket{T}$. As $P\ket{T} \in \mathcal{H}_{\text{phys}}$ we must have $\Tilde{T}(z) \in \mathcal{A}$. 

Let $\phi(z)$ be any field of $\CFT{1}\otimes \CFT{2}$, $\phi(z) \in \CFT{1}\otimes\CFT{2}$. We then define $(L_n\phi)(z) \equiv \oint_{z}\frac{dw}{2\pi i} (w-z)^{n+1} T(w)\phi(z)$. This allows one to express the OPE $T(z)\phi(w)$ as,
\begin{equation}
    T(z)\phi(w) = \sum_n \frac{(L_n\phi)(w)}{(z-w)^{n+2}}
\end{equation}
The corresponding state of $(L_n\phi)(z)$ is $L_n\ket{\phi}$ where $L_n$ are the usual Virasoro generators, $L_n = \oint \frac{dz}{2\pi i}z^{n+1}T(z)$. For the field $\Tilde{T}(z)$ we also define $(\Tilde{L}_n\phi)(z) \equiv \oint_{z}\frac{dw}{2\pi i} (w-z)^{n+1} \Tilde{T}(w)\phi(z)$, where $(\Tilde{L}_n\phi)(0)\ket{0} = \Tilde{L}_n\ket{\phi}$, with $\Tilde{L}_n = \oint \frac{dz}{2 \pi i }z^{n+1} \Tilde{T}(z)$. As for $T(z)$, we can express the OPE $\Tilde{T}(z)\phi(w)$ as,
\begin{equation}
    \Tilde{T}(z)\phi(w) = \sum_n \frac{(\Tilde{L}_n\phi)(w)}{(z-w)^{n+2}}
\end{equation}

We now show that for any field $\phi(z)$ in the chiral algebra $\mathcal{A}$, $\phi(z) \in \mathcal{A}$, we have $(L_n\phi)(z) = (\Tilde{L}_n\phi)(z)$ for $n \geq -1$. Firstly, we note that the fields that generate $\mathcal{A}$, $\phi_j(z)$, are primary fields with some scaling dimension $h$, and so we must have the OPE,
\begin{equation}
    T(z) \phi_j(w) \sim \frac{h\phi_j(w)}{(z-w)^2} + \frac{\partial\phi_j(z)}{z-w}
\end{equation}
This OPE can also be expressed as,
\begin{equation}
    \phi_j(z)T(w) \sim \frac{h\phi_j(w)}{(z-w)^2} + \frac{(h-1)\partial\phi_j(w)}{z-w}
\end{equation}
We can further convert this OPE to vector form using state-operator correspondence,
\begin{equation}
    \phi_j(z)\ket{T} \sim z^{-2}h\ket{\phi_j} + z^{-1}(h-1)\ket{\partial\phi_j}
\end{equation}
We know that $\ket{\partial\phi_j} \in \mathcal{H}_{\text{phys}}$ as $\phi_{j, -h-1}\ket{0} = \oint \frac{dz}{2\pi i} z^{-2}\phi_j(z)\ket{0} = \partial\phi_j(0)\ket{0} = \ket{\partial\phi_j}$. Furthermore, let $\mathcal{H}^\perp_{\text{phys}}$ be the orthogonal complement of $\mathcal{H}_{\text{phys}}$. From the definition of $\mathcal{H}_{\text{phys}}$, if $\ket{v} \in \mathcal{H}_{\text{phys}}$ then $\phi^\dagger_{j,-n}\ket{v} \in \mathcal{H}_{\text{phys}}$. So if $\ket{w} \in \mathcal{H}^\perp_{\text{phys}}$ then for any $\ket{v} \in \mathcal{H}_{\text{phys}}$ we have $\bra{v}\phi_{j,n}\ket{w} = 0$ as $\phi^\dagger_{j,-n}\ket{v} \in \mathcal{H}_{\text{phys}}$ and, hence, $\phi_{j,n}\ket{w} \in \mathcal{H}^\perp_{\text{phys}}$. We then have that $[\phi_j(z), P] = 0$. It then follows that,
\begin{equation}
    \begin{split}
        \phi_j(z) P \ket{T} =& P\phi_j(z)\ket{T} \\
        \sim & P( z^{-2}h\ket{\phi_j} + z^{-1}(h-1)\ket{\partial\phi_j} ) \\
        =& z^{-2}h\ket{\phi_j} + z^{-1}(h-1)\ket{\partial\phi_j}
    \end{split}
\end{equation}
In OPE form we have,
\begin{equation}
    \Tilde{T}(z) \phi_j(w) \sim \frac{h\phi_j(w)}{(z-w)^2} + \frac{\partial\phi_j(z)}{z-w}
\end{equation}
Hence, $[L_n - \Tilde{L}_n, \phi(z)] = 0$ for $n \geq -1$. We also have for $n \geq -1$ $\Tilde{L}_n\ket{0} = \oint\frac{dz}{2\pi i}z^{n+1}\Tilde{T}(z)\ket{0} = 0$ and $L_n\ket{0} = 0$. Thus, $(L_n - \Tilde{L}_n)\ket{0} = 0$ for $n \geq -1$. As $\ket{v} \in \mathcal{H}_{\text{phys}}$ must be a polynomial in the modes of $\phi_j(z)$ and $\phi^\dagger_j(z)$ applied on the vacuum $\ket{0}$, we must also have that $(L_n - \Tilde{L}_n)\ket{v} = 0$ for $n \geq -1$. Thus, for any $\phi(z) \in \mathcal{A}$ $(L_n\phi)(z) = (\Tilde{L}_n\phi)(z)$ for $n \geq -1$. This then implies that for $\phi \in \mathcal{A}$ the singular terms in the OPE $T(z)\phi(w)$ are the same as the singular terms in the OPE $\Tilde{T}(z)\phi(w)$. We then also have the property that for $\ket{v} \in \mathcal{H}_{\text{phys}}$ then $L_n\ket{v} \in \mathcal{H}_{\text{phys}}$ for $n\geq -1$. A special case of this is $L_1$, where if $\ket{w} \in \mathcal{H}^\perp_{\text{phys}}$ and $\ket{v} \in \mathcal{H}_{\text{phys}}$ then $\bra{v}L_{1}\ket{w} = 0$ as $L_{-1}\ket{v} \in \mathcal{H}_{\text{phys}}$ and so $L_1\ket{w} \in \mathcal{H}^\perp_{\text{phys}}$. This then implies that $[L_1, P] = 0$. 

Now as $[L_0, \phi_{j,n}] = -n\phi_{j,n}$ we must have that $\mathcal{H}_{\text{phys}}$ is an invarient subspace of $L_0$. Also, as $L_0$ is Hermitian, $\mathcal{H}^\perp_{\text{phys}}$ is also an invarient subspacde of $L_0$. Hence, $[L_0, P] = 0$. So $L_0P\ket{T} = 2P\ket{T}$, which implies $\Tilde{T}(z)$ must be a field of scaling dimension two. Thus, $(L_0\Tilde{T})(z) = 2\Tilde{T}(z)$ and $(L_{-1}\Tilde{T})(z) = \partial\Tilde{T}(z)$. Also, from the standard OPE of $T(z)T(w)$ we know that $L_1\ket{T} = 0$. Thus, $L_1P\ket{T} = PL_1\ket{T} = 0 \Rightarrow (L_1\Tilde{T})(z) = 0$. $L_2P\ket{T}$ must be a state with $L_0$ eigenvalue zero. Only the vacuum state $\ket{0}$ has $L_0$ eigenvalue zero. It then follows $(L_2\Tilde{T})(z) = (\Tilde{c}/2)\mathbf{1}(z)$, where $\Tilde{c}$ is some constant. For $n > 2$, $(L_n\Tilde{T})(z) = 0$ as there are no fields with negative scaling dimension (as $\CFT{1}\otimes\CFT{2}$ is assumed to be unitary). The OPE $T(z)\Tilde{T}(w)$ must then be,
\begin{equation}
    T(z)\Tilde{T}(w) \sim \frac{\Tilde{c}/2}{(z-w)^4} + \frac{2\Tilde{T}(w)}{(z-w)^2} + \frac{\partial\Tilde{T}(w)}{(z-w)}
\end{equation}
As argued earlier, $T(z)\Tilde{T}(w)$ must have the same singular terms as $\Tilde{T}(z)\Tilde{T}(w)$ because $\Tilde{T}(z) \in \mathcal{A}$. Thus,
\begin{equation} \label{Eq:newTOPE}
    \Tilde{T}(z)\Tilde{T}(w) \sim \frac{\Tilde{c}/2}{(z-w)^4} + \frac{2\Tilde{T}(w)}{(z-w)^2} + \frac{\partial\Tilde{T}(w)}{(z-w)}
\end{equation}

For any operator $\phi(z)$ of $\CFT{1}\otimes\CFT{2}$, we have that the Hermitian conjugate of $\phi(z)$ is given by $[\phi(z)]^\dagger = \bar{z}^{-2h}\phi^\dagger(\bar{z}^{-1})$, where $\phi^\dagger(z)$ is the \textit{conjugate} of $\phi(z)$ and $h$ is the scaling dimension of $\phi(z)$. We can then define an anti-unitary operator $\mathcal{C}$ which acts as $\mathcal{C}\ket{\phi} = \ket{\phi^\dagger}$. One can think of $\mathcal{C}$ is the usual ``charge'' conjugation operator in quantum field theory. By taking the Hermitian conjugate of the product $\phi_1(z)\phi_2(w)$, it can be seen that the fields in the OPE of $\phi_1^\dagger(z)\phi_2^\dagger(w)$ are simply the conjugates of the fields that appear in the OPE $\phi_1(z)\phi_2(w)$. As $\phi_j^\dagger(z) \in \mathcal{A}$ and $\phi_j(z) \in \mathcal{A}$, it inductively follows that for any $\phi(z) \in \mathcal{A}$ then $\phi^\dagger(z) \in \mathcal{A}$. In vector form, if $\ket{v} \in \mathcal{H}_{\text{phys}}$ then $\mathcal{C} \ket{v} \in \mathcal{H}_{\text{phys}}$. As $\mathcal{C}$ is an anti-unitary operator, it also follows that if $\ket{w} \in \mathcal{H}^\perp_{\text{phys}}$ then $\mathcal{C}\ket{w} \in \mathcal{H}^\perp_{\text{phys}}$. So, $\mathcal{C}P = P\mathcal{C}$. We then have $\mathcal{C}P\ket{T} = P\mathcal{C}\ket{T} = P\ket{T}$ (as $T(z)$ is self-conjugate). Hence, $\Tilde{T}(z)$ is self-conjugate $\Tilde{T}^\dagger(z) = \Tilde{T}(z)$. We then have the property $(\Tilde{L}_n)^\dagger = \Tilde{L}_{-n}$. 

From the OPE of Eq. \ref{Eq:newTOPE} it follows that $\Tilde{L}_n$ must form a Virasoro algebra with central charge $\Tilde{c}$. Furthermore, as $\bra{0}\Tilde{L}_2\Tilde{L}_{-2}\ket{0} \geq 0$ and $\bra{0}\Tilde{L}_2\Tilde{L}_{-2}\ket{0} = \Tilde{c}/2$ then $\Tilde{c} \geq 0$. The field $\Tilde{T}(z)$ can be taken as a candidate energy-momentum tensor for $\mathcal{A}$.

Finally, we show that $\Tilde{T}(z)$ is unique. Suppose we have another candidate energy-momentum tensor $\hat{T}(z)$. We must require it to have the OPE $\hat{T}(z)\phi_j(w) \sim h\phi_j(w)/(z-w)^2 + \partial\phi_j(w)/(z-w)$. Clearly then, the modes of $\hat{T}$ must be such that $\Tilde{L}_n - \hat{L}_n$ must commute with all the modes of $\phi_j(z)$ and $\phi_j^\dagger(z)$. We must further require that $\hat{L}_n\ket{0} = 0$ for $n \geq -1$. So $(\Tilde{L}_n - \hat{L}_n)\ket{0} = 0$ for $n \geq -1$ and $\bra{0}(\Tilde{L}_n - \hat{L}_n) = 0$ for $n \leq 1$. As any two states, $\ket{a}$ and $\ket{b}$, of $\mathcal{H}_{\text{phys}}$ are polynomials in the modes of $\phi_j(z)$ and $\phi^\dagger_j(z)$ applied on $\ket{0}$, it then follows that $\bra{a}(\Tilde{L}_n - \hat{L}_n)\ket{b} = 0$ for all $n$. Hence, for any $\ket{v} \in \mathcal{H}_{\text{phys}}$ we must have $(\Tilde{L}_n - \hat{L}_n)\ket{v} = 0$ for all $n$. Thus, within $\mathcal{H}_{\text{phys}}$, $\Tilde{T}(z) = \hat{T}(z)$. $\Tilde{T}(z)$ is then the \textit{unique} candidate energy-momentum tensor. 

In conclusion, we have found an energy-momentum tensor for $\mathcal{A}$ with the unique property that any Virasoro primary field of $\CFT{1}\otimes\CFT{2}$ that is in $\mathcal{A}$ is also a Virasoro primary relative to $\Tilde{T}(z)$ with the same scaling dimension.

\subsection{The $\phi_2^2\phi_1$ example} \label{Sec:221Example}
As another example to illustrate the general process of generating a product wave function outlined in Sec. \ref{Sec:PartonGenWFCon}, let us briefly consider the $\phi_2^2\phi_1$ state (or the $221$ state in another notation). This state is at filling fraction $\nu = 1/2$ and was, at one point, considered as a possible candidate wave function for $\nu = 5/2$ \cite{jain_thirty_2020}. We already know how to produce the $\phi_2^2$ from Sec. \ref{Sec:PartonWFCon} and we know how to construct the $\nu = 1$ wave function from Sec. \ref{Sec:integerQHCFT}. Let $\mathcal{A}_1$ be the chiral algebra corresponding to $\phi_2^2$ with the wave function being generated by a correlation function of $\CFT{\mathcal{A}_1}$. Let $\CFT{2}$ be the CFT that generates the $\nu = 1$ wave function where we  use the same notation from Sec. \ref{Sec:integerQHCFT}. We can then generate the $\phi_2^2\phi_1$ wave function in $\CFT{\mathcal{A}_1}\otimes \CFT{2}$ with $\Omega(z, \bar{z}) = (V_{-1}(z) + \bar{z}\sqrt{2}V_{0}(z) + \bar{z}^2V_{1}(z))V(z)$. The chiral algebra we associate $\mathcal{A}$ with this state is that generated by repeated OPEs of $\Tilde{V}_m(z) \equiv V_{m}(z)V(z)$ and their conjugates. Let $\mathcal{H}_{\text{phys}}$ be space of states in $\CFT{\mathcal{A}_1}\otimes\CFT{2}$ that is generated by applying the modes of $\Tilde{V}_m(z)$ and $\Tilde{V}^\dagger_m(z)$ on $\ket{0}$. Now define the current $\mathcal{J}(z) = (i\partial\Phi(z) - i\partial\varphi(z))/\sqrt{2}$ and the space of gauge invarient states $\mathcal{H}_G$ by $\ket{v} \in \mathcal{H}_G \Leftrightarrow \mathcal{J}_n\ket{v} = 0$ for $n>0$. As $[\mathcal{J}(z), \Tilde{V}_m(z)] = 0$ we must have $\mathcal{H}_{\text{phys}} \subset \mathcal{H}_G$. Further define $\Tilde{\Phi}(z) = (\Phi(z) + \varphi(z))/\sqrt{2}$. One can easily show that the space of states with a fixed number of partons and fermions of $\CFT{2}$ can be generated by the action of modes of $i\partial\Tilde{\Phi}(z), J^a(z)$ and $\mathcal{J}(z)$ on a state with which has the lowest $L_0$ eigenvalue of that space. The only such spaces that can contain gauge invariant states are those where the number of partons is double the number of fermions. The states with the lowest $L_0$ eigenvalues in such spaces can be generated by modes of $\Tilde{V}_m(z)$ and their conjugates. One can also explicitly check that the fields $\Tilde{\Phi}(z)$ and $J^a(z)$ can be generated by OPEs of $\Tilde{V}_m(z)$ and $\Tilde{V}^\dagger_m(z)$. It then follows that all gauge invariant states can be generated by modes of the fields of $\mathcal{A}$, which implies $\mathcal{H}_G = \mathcal{H}_{\text{phys}}$. By state operator correspondence, all neutral fields of $\mathcal{A}$ correspond to fields that are descendants of the identity in the $\hat{\mathfrak{u}}(1)\oplus \widehat{\mathfrak{su}}(2)_2$ WZW model. One should note, however, that the compactification radius of $\Phi(z)$ is different from $\Tilde{\Phi}(z)$. For a fixed number of particles the edge state counting of $\phi_2^2\phi_1$ must have upper bounds given by the state counting of the $\hat{\mathfrak{u}}(1)\oplus\widehat{\mathfrak{su}}(2)_2$ WZW model.

\section{Matrix elements of modes of chiral boson vertex operators} \label{Sec:vertexOpMatrixEle}
Consider the chiral boson $\varphi(z)$ with a corresponding Klein factor $F$ such that $[a_0, F] = \sqrt{r} F$, where $r \in \mathbb{Z}$ and $r > 0$. We denote the basis of its Hilbert space by $\ket{\lambda; m} \equiv \prod_{n\in \lambda}a_{-n}F^m\ket{0}/\sqrt{\mathcal{N}}$, where $m$ is an integer, $\lambda$ is a partition and $\mathcal{N}$ is used to normalise the state.

The matrix elements we are interested in take the form,
\begin{equation}
    \bra{\lambda'; m'} F^l:e^{il\sqrt{r} \varphi}: \ket{\lambda; m}
\end{equation}
where $l \in \mathbb{Z}$. This can be split up into several independent sectors. From the $F$, $a_0$ sector we simply get a $\delta_{m'-m,l}z^{lrm}$ factor. Further, all momentum modes are decoupled and we simply need to find matrix elements of the form, where we  let $\sigma \equiv l\sqrt{r}$,
\begin{equation}
    \begin{split}
        &\bra{0}(a_n)^q e^{ \frac{\sigma}{n} a_{-n}z^n} e^{\frac{-\sigma}{n}a_{n} z^{-n}} (a_{-n})^p \ket{0} \\
        &= \bigg ( \frac{\partial}{\partial y} \bigg )^q_{x,y = 0} \bigg ( \frac{\partial}{\partial x} \bigg )^q_{x,y = 0} \bra{0}e^{ya_n} e^{ \frac{\sigma}{n} a_{-n}z^n} e^{-\frac{\sigma}{n}a_{n} z^{-n}} \\
        & \times e^{xa_{-n}} \ket{0}
    \end{split}
\end{equation}
We can then calculate,
\begin{equation}
    \begin{split}
        &\bra{0}e^{ya_n} e^{ \frac{\sigma}{n} a_{-n}z^n} e^{-\frac{\sigma}{n}a_{n} z^{-n}} e^{xa_{-n}} \ket{0} \\
        &= \bra{0}e^{ya_n} e^{xa_{-n}} \ket{0} e^{y\sigma z^n - x \sigma z^{-n}} \\
        &= e^{y\sigma z^n + nxy - x \sigma z^{-n}} \\
    \end{split}
\end{equation}

Then,
\begin{equation}
    \begin{split}
        &\bigg ( \frac{\partial}{\partial y} \bigg )^q_{x,y = 0} \bigg ( \frac{\partial}{\partial x} \bigg )^q_{x,y = 0} e^{y\sigma z^n + nxy - x \sigma z^{-n}} \\
        &=  \bigg ( \frac{\partial}{\partial y} \bigg )^q_{x,y = 0} (yn - \sigma z^{-n})^p e^{y\sigma z^n + nxy - x \sigma z^{-n}} \\
        &= \sum_{i=0}^{\min(p,q)} \binom{q}{i} n^i \frac{p!}{(p-i)!} ( - \sigma z^{-n})^{p-i} (\sigma z^n)^{q-i} \\
        &= z^{(q-p)n} (-1)^p \sigma^{q+p} \sum_{i=0}^{\min(p,q)} \frac{q!p!(-n)^i}{\sigma^{2i}i!(q-i)!(p-i)!} \\
    \end{split}
\end{equation}

Now let $p_n(\lambda)$ be the multiplicity of $n$ in the partition $\lambda$ and let $f(\lambda)$ be the \textit{set} of distinc elements in $\lambda$. We can then express the full matrix element as,
\begin{equation}
    \begin{split}
        & \bra{\lambda'; m'} F^l:e^{il\sqrt{r} \varphi}: \ket{\lambda; m} \\
        &= \delta_{m'-m,l}z^{lrm} \\
        & \times \prod_{n \in f(\lambda')\cup f(\lambda)} \bigg [ z^{(p_n(\lambda')-p_n(\lambda))n} (-1)^{p_n(\lambda)} (l\sqrt{r})^{p_n(\lambda')+p_n(\lambda)} \\
        & \times \sum_{j=0}^{\min(p_n(\lambda),p_n(\lambda'))} \frac{(p_n(\lambda'))!(p_n(\lambda))!(-n)^j}{(l\sqrt{r})^{2j}j!(p_n(\lambda')-j)!(p_n(\lambda)-j)!} \bigg ]
    \end{split}
\end{equation}

\section{Free field representation of $\widehat{\mathfrak{su}}(2)_2$ currents and parton mapping} \label{Sec:SU2Free}
We first detail how a combination of a Majorana field $\psi(z)$ and a free chiral boson $\varphi(z)$, with compactification radius one and corresponding Klein factor $F_\varphi$, can be used to generate the vacuum representation of the $\widehat{\mathfrak{su}}(2)_2$ Kac-Moody algebra. The computation of the matrix elements of the model inner product action of Eq. \ref{Eq:partInnProdModel}, using this representation, will then be discussed. It will then be detailed how this vacuum representation can be mapped back to the one occurring in $\mathcal{H}_{\text{CFT}}$ (defined in Sec. \ref{Sec:partonCFT}).

The $\psi(z)$ field has conformal dimension of $1/2$ and has the following OPE,
\begin{equation}
    \psi(z)\psi(w) = \frac{1}{z-w} + (z-w)2T^{\psi}(w) + \dots
\end{equation}
where $T^\psi (z)$ is the energy-momentum tensor of this Majorana degree of freedom given by,
\begin{equation}
    T^\psi (z) = -\frac{1}{2}: \psi(z) \partial \psi(z) :
\end{equation}
The above OPE implies the following anti-commutation relations,
\begin{equation}
    \{ \psi_n, \psi_m \} = \delta_{n+m,0}
\end{equation}
where $n$ and $m$ are half-integral $n,m \in \mathbb{Z} + \frac{1}{2}$.

We then define the usual vertex operator for $\varphi(z)$ as $V_\varphi (z) \equiv F_\varphi : e^{i\varphi(z)} :$. Let the Hilbert space of this combined system of $\psi(z)$ and $\varphi(z)$ be $\mathcal{H}_{\psi,\varphi}$. This hilbert space can be spanned by an orthonormal basis of the form $\ket{\mu;\lambda;p} = \prod_{n_2\in \lambda}\fem{\varphi}_{-n_2} F_\varphi^p \prod_{n_1\in \mu} \psi_{-\frac{n_1}{2}} \ket{0}/\sqrt{\mathcal{N}}$, where $\lambda$ is a partition, $\mu$ is a partition with no repeated elements and with all elements being odd, $p \in \mathbb{Z}$, $\fem{\varphi}_n$ are the modes of $i\partial\varphi(z)$, and $\mathcal{N}$ is used to normalise the state.  

Now consider the following currents,
\begin{equation} \label{Eq:su2_2MajPhi}
    \begin{split}
        J^+(z) \equiv & \sqrt{2}\psi(z) V_\varphi(z) \\
        J^-(z) \equiv & \sqrt{2}V_\varphi^\dagger(z) \psi(z) \\
        J^3(z) \equiv & i\partial \varphi(z) \\
    \end{split}
\end{equation}
From the usual OPEs $i\partial\varphi(z)V_\varphi(w) \sim V_\varphi(w)/(z-w)$, $i\partial\varphi(z) V_\varphi^\dagger(w) \sim - V_\varphi^\dagger(w)/(z-w)$ and $i\partial\varphi(z)i\partial\varphi(w) \sim 1/(z-w)^2$, we have the OPEs,
\begin{equation}
    \begin{split}
        J^3(z)J^\pm(w) \sim & \pm \frac{J^\pm(w)}{z-w} \\
        J^3(z)J^3(w) \sim & \frac{1}{(z-w)^2} \\
    \end{split}
\end{equation}
We also have the OPE,
\begin{equation}
    \begin{split}
        J^+(z)J^-(w) = & 2\psi(z)V_\varphi (z) V_\varphi^\dagger(w)\psi(w) \\
        = & 2\psi(z)\psi(w) V_\varphi (z) V_\varphi^\dagger(w) \\
        \sim & \frac{2}{z - w} \bigg [ \frac{1}{(z-w)} + i\partial\varphi(w) \bigg ] \\
        = & \frac{2}{(z-w)^2} + \frac{2 J^3(w)}{z - w}
    \end{split}
\end{equation}
Thus, the $J^3(z)$ and $J^\pm(z)$ currents have the OPEs corresponding to the $\widehat{\mathfrak{su}}(2)_2$ Kac-Moody algebra (i.e. their modes will form this algebra). 

Let $\mathcal{H}_{\widehat{\mathfrak{su}}(2)_2} \subset \mathcal{H}_{\psi,\varphi}$ be the Hilbert space generated by polynomials in the modes of the $J^3(z)$ and $J^\pm(z)$ fields applied on the vacuum, $\ket{0}$. By definition $\mathcal{H}_{\widehat{\mathfrak{su}}(2)_2}$ will form the vacuum representation of $\widehat{\mathfrak{su}}(2)_2$. Let $\mathcal{P}$ be an operator defined by $\mathcal{P}\ket{\mu;\lambda;p} = \ket{\mu;\lambda;p}$ if $p + $ (number of elements of $\mu$) $ = $ an even number, and $\mathcal{P}\ket{\mu;\lambda;p} = 0$ if $p + $ (number of elements of $\mu$) $ = $ an odd number. In words, $\mathcal{P}$ is the projector onto the space of states with \textit{even} fermion parity. As the $J^3(z)$ and $J^\pm(z)$ currents carry even fermion parity, the modes of these fields must commute with $\mathcal{P}$. Furthermore, as $\ket{0}$ has even fermion parity, all states of $\mathcal{H}_{\widehat{\mathfrak{su}}(2)_2}$ must have even fermion parity. We  now show that the space of states of even fermion parity is equivalent to $\mathcal{H}_{\widehat{\mathfrak{su}}(2)_2}$. 

We first note that the Hilbert space of just the Majorana field can be split up into irreducible representations of the Virasoro algebra formed by the modes, $L^\psi_n$, of $T^\psi(z)$ with central charge $c = 1/2$. It is well known that there are only three unitary irreducible representations of the Virasoro at $c = 1/2$, with the conformal dimension of the corresponding primary states being $0$, $1/2$ and $1/16$. For the Majorana system used here, with periodic boundary conditions on the complex plane, only the $0$ and $1/2$ representation can occur, as there are no states with $L_0$ eigenvalue of $1/16$ in the Hilbert space. By matching the $L_0$ eigenvalues, both of the $0$ and $1/2$ representation occur with multiplicity one with corresponding primary states $\ket{0}$ and $\psi_{-\frac{1}{2}}\ket{0}$ respectively. We can then use the following \textit{overcomplete} basis for $\mathcal{H}_{\psi,\varphi}$, with the basis elements $\ket{0;\lambda_1;\lambda_2;p} = \prod_{n_1\in \lambda_1}L^\psi_{-n_1} \prod_{n_2\in\lambda_2}\fem{\varphi}_{-n_2}F_\varphi^p\ket{0}$ and $\ket{1/2;\lambda_1;\lambda_2;p} = (-1)^p\psi_{-\frac{1}{2}}\ket{0;\lambda_1;\lambda_2;p}$, where $\lambda_i$ are partitions and $p\in\mathbb{Z}$. The space of states with even fermion parity is spanned by the basis elements $\ket{0;\lambda_1;\lambda_2;p}$ with $p$ even and $\ket{1/2;\lambda_1;\lambda_2;p}$ with $p$ odd. 

Now note that $J^+_{-1}\ket{0} = \sqrt{2}\psi_{-\frac{1}{2}}F_\varphi\ket{0}$ and $J^-_{-1}\ket{0} = \sqrt{2}F^{-1}_\varphi\psi_{-\frac{1}{2}}\ket{0}$. Hence, $\psi_{-\frac{1}{2}}F_\varphi\ket{0}, F^{-1}_\varphi\psi_{-\frac{1}{2}}\ket{0} \in \mathcal{H}_{\widehat{\mathfrak{su}}(2)_2}$. Furthermore, the OPEs,
\begin{equation} \label{Eq:JpJpOPE}
    \begin{split}
        J^+(z)J^+(w) = & -2 F_\varphi^2 : e^{2i\varphi(z)} : + \dots \\
        J^-(z)J^-(w) = & -2 F_\varphi^{-2} : e^{-2i\varphi(z)} : + \dots \\
    \end{split}
\end{equation}
imply, by induction with repeated OPE, that the $F^p_\varphi\ket{0} \in \mathcal{H}_{\widehat{\mathfrak{su}}(2)_2}$ for $p$ even. Next, we have $J^\pm_{-(1 \pm p)}F_\varphi^p\ket{0} \propto \sqrt{2}\psi_{-\frac{1}{2}}F^{p \pm 1}_\varphi\ket{0}$. It then follows that $\psi_{-\frac{1}{2}}F_\varphi^p\ket{0} \in \mathcal{H}_{\widehat{\mathfrak{su}}(2)_2}$ for $p$ odd. By definition we have $J^3_n = \fem{\varphi}_n$. Finally, including the first non-singular term in the OPE of $J^+(z)J^-(w)$ gives,
\begin{equation} \label{Eq:jpjmopemaj}
    \begin{split}
        J^+(z)J^-(w) = & 2\bigg [ \frac{1}{z-w} + (z-w)2T^\psi(w) + \dots \bigg ] \\
        & \times \bigg [ \frac{1}{z-w} + i\partial\varphi(w) \\
        &+ \frac{1}{2}(z-w)( i\partial^2\varphi(w) + :(i\partial\varphi(w))^2: ) + \dots \bigg ] \\
        = & \frac{2}{(z-w)^2} + \frac{2J^3(w)}{z - w} \\
        & + 4T^\psi(w) + (\partial J^3(w) + :(J^3(w))^2:) + \dots \\
    \end{split}
\end{equation}
which implies that the $L^\psi_n$ modes can be expressed in terms of the modes of $J^3(z)$ and $J^\pm(z)$. It then follows that the space of states of even fermion parity is equivalent to $\mathcal{H}_{\widehat{\mathfrak{su}}(2)_2}$.

Now let $\mathcal{H}_{\Phi,\psi,\varphi}$ be the Hilbert space of the combination of a chiral boson $\Phi(z)$, and the $\psi(z)$ and $\varphi(z)$ system we discussed above. Let $\mathcal{H}_{\hat{\mathfrak{u}}(1)\oplus \widehat{\mathfrak{su}}(2)_2} \subset \mathcal{H}_{\Phi,\psi,\varphi}$ be the space of states that are $U(1)$ charge natural relative to $\Phi(z)$ (i.e. $\ket{v} \in \mathcal{H}_{\hat{\mathfrak{u}}(1)\oplus \widehat{\mathfrak{su}}(2)_2} \Rightarrow a_0\ket{v}=0$) and have even fermion parity. $\mathcal{H}_{\hat{\mathfrak{u}}(1)\oplus \widehat{\mathfrak{su}}(2)_2}$ then forms the vacuum representation of the $\hat{\mathfrak{u}}(1)\oplus\widehat{\mathfrak{su}}(2)_2$ Kac-Moody algebra, with the $\hat{\mathfrak{u}}(1)$ current being $i\partial\Phi(z)$ and the $\widehat{\mathfrak{su}}(2)_2$ currents being given by Eq. \ref{Eq:su2_2MajPhi}. A basis for this space, written as $\ket{\lambda_1;\mu;\lambda_2;p}$, was given in Sec. \ref{Sec:fitting} (with $p + $ (number of elements of $\mu$) $=$ an even number). We then define the extension of the overcomplete $\ket{0;\lambda_2;\lambda_2;p}$ and $\ket{1/2;\lambda_1;\lambda_2;p}$ basis for the $\mathcal{H}_{\psi,\varphi}$ system to the $\mathcal{H}_{\Phi,\psi,\varphi}$ (where we will only be interested in the $U(1)$ neutral states) as,
\begin{equation} \label{Eq:overlcompleteBasis}
    \begin{split}
        \ket{\lambda_1: 0;\lambda_2;\lambda;p} \equiv & \prod_{n_1\in \lambda_1}a_{-n_1} \prod_{n_2\in\lambda_2} L^\psi_{-n_2}\prod_{n_3\in\lambda_3}\fem{\varphi}_{-n_3} \\
        & \times F_\varphi^p\ket{0} \\
        \ket{\lambda_1: 1/2;\lambda_2;\lambda;p} \equiv & (-1)^p \psi_{-\frac{1}{2}} \ket{\lambda_1: 0;\lambda_2;\lambda;p} \\
    \end{split}
\end{equation}

Before moving on, we  now briefly discuss how the matrix elements of the model inner product action of Eq. \ref{Eq:partInnProdModel} can be computed using this representation of $\hat{\mathfrak{u}}(1) \oplus \widehat{\mathfrak{su}}(2)_2$. We first note that $\frac{1}{2}[ : (J^1(z))^2 : + : (J^2(z))^2 : ] = T^\psi(z)$. Now let $L^\varphi_0 \equiv \frac{1}{2}\oint \frac{dz}{2\pi i} z :(J^3(z))^2: = \frac{1}{2}\oint \frac{dz}{2\pi i} z :(i\partial\varphi(z))^2:$. In the $\ket{\lambda_1;\mu;\lambda_2;p}$ basis the operators $L^\psi_0$, $L^\varphi_0$ and $J^3_0$ are diagonal with $L^\psi_0\ket{\lambda_1;\mu;\lambda_2;p} = [\sum_{n\in \mu} n/2]\ket{\lambda_1;\mu;\lambda_2;p}$, $L^\varphi_0 \ket{\lambda_1;\mu;\lambda_2;p} = [p^2/2 + (\sum_{n\in \lambda_2} n)]\ket{\lambda_1;\mu;\lambda_2;p}$ and $J^3_0 \ket{\lambda_1;\mu;\lambda_2;p} = p\ket{\lambda_1;\mu;\lambda_2;p}$. The matrix elements, in this basis, of the operator $(J^3i\partial\Phi)_0 \equiv \oint \frac{dz}{2\pi i} z J^3(z) i\partial\Phi(z)$ can be computed by expressing it as $(J^3i\partial\Phi)_0 = \sum_n \fem{\varphi}_{-n}a_n$. Finally, the matrix elements of $J^1_0$ can be computed by noting that $J^1_0 = [J^+_0 + J^-_0]/2$ with $J^+_0 = \sqrt{2}\sum_k \psi_{-k}V_{\varphi, k}$ and $J^-_0 = \sqrt{2}\sum_k V^\dagger_{\varphi, -k}\psi_k$, where the matrix elements of the $V_{\varphi, k}$ and $V^\dagger_{\varphi, k}$ modes can be computed using the result of Appendix \ref{Sec:vertexOpMatrixEle}.

Now recall the system of four chiral bosons $\fef{ij}(z)$, all with compactification radius one, and Hilbert space $\mathcal{H}_{\text{CFT}}$, which was defined in Sec. \ref{Sec:partonCFT}. Within this system we have the $\hat{\mathfrak{u}}(1) \oplus\widehat{\mathfrak{su}}(2)_2$ currents,
\begin{equation}
    \begin{split}
        i\partial\hat{\Phi}(z) = & \frac{1}{2} [i\partial\fef{11}(z) + i\partial\fef{12}(z) \\
        & + i\partial\fef{21}(z) + i\partial\fef{22}(z) ] \\
        \hat{J}^\pm(z) = & V_1^\pm(z) + V_2^\pm(z) \\
        \hat{J}^3(z) = & \frac{1}{2}[i\partial\fef{12}(z) - i\partial\fef{11}(z) \\
        & + i\partial\fef{22}(z) - i\partial\fef{21}(z)] \\
    \end{split}
\end{equation}
where,
\begin{equation}
    V^\pm_j(z) = (\Tilde{F}_{j2}  \Tilde{F}^\dagger_{j1})^{\pm 1} : e^{\pm i( \fef{j2}(z) - \fef{j1}(z) )} : 
\end{equation}

The space of states generated by applying polynomials of modes of these fields to the vacuum state of $\mathcal{H}_{\text{CFT}}$ forms the vacuum representation of the $\hat{\mathfrak{u}}(1) \oplus \widehat{\mathfrak{su}}(2)_2$ Kac-Moody algebra. We denote the vacuum representation formed this way as $\hat{\mathcal{H}}_{\hat{\mathfrak{u}}(1)\oplus\widehat{\mathfrak{su}}(2)_2}$. 

We  now construct the isomorphism that maps the $\mathcal{H}_{\hat{\mathfrak{u}}(1)\oplus\widehat{\mathfrak{su}}(2)_2}$ vacuum representation to the $\hat{\mathcal{H}}_{\hat{\mathfrak{u}}(1)\oplus\widehat{\mathfrak{su}}(2)_2}$ vacuum representation, where we  write this linear map as $\mathtt{M}: \mathcal{H}_{\hat{\mathfrak{u}}(1)\oplus\widehat{\mathfrak{su}}(2)_2} \rightarrow \hat{\mathcal{H}}_{\hat{\mathfrak{u}}(1)\oplus\widehat{\mathfrak{su}}(2)_2}$. This is the \textit{unique} linear map with the property,
\begin{equation} \label{Eq:MPropDefinition}
    \begin{split}
        \mathtt{M}a_n\ket{v} = & \hat{a}_n \mathtt{M}\ket{v} \\
        \mathtt{M}J^3_n\ket{v} = & \hat{J}^3_n \mathtt{M}\ket{v} \\
        \mathtt{M}J^\pm_n\ket{v} = & \hat{J}^\pm_n \mathtt{M}\ket{v} \\
        \mathtt{M}\ket{0} = & \ket{\hat{0}} \\
    \end{split}
\end{equation}
where $\ket{v} \in \mathcal{H}_{\hat{\mathfrak{u}}(1)\oplus\widehat{\mathfrak{su}}(2)_2}$ and $\ket{\hat{0}}$ is the vacuum state of $\hat{\mathcal{H}}_{\hat{\mathfrak{u}}(1)\oplus\widehat{\mathfrak{su}}(2)_2}$. 

By its defining property, we can immediately see that $\mathtt{M}$ must have the property,
\begin{equation} \label{Eq:varphiModeMap}
    \begin{split}
        \mathtt{M}\fem{\varphi}_n\ket{v} = & \frac{1}{2}[\fefm{12}_n - \fefm{11}_n \\
        & + \fefm{22}_n - \fefm{21}_n] \mathtt{M}\ket{v} \\
        =& \hat{J}^3_n \mathtt{M}\ket{v} \\
    \end{split} 
\end{equation}
where we one should note that the $\fefm{ij}_n$ modes can be expressed in terms of the modes of the $\Tilde{V}_{ij}(z)$ and $\Tilde{V}^\dagger_{ij}(z)$ fields using Eq. \ref{Eq:densityModeExpan}. We also have the OPEs,
\begin{equation}
    \hat{J}^\pm(z)\hat{J}^\pm(w) = 2V^\pm_1(w)V^\pm_2(w) + \dots
\end{equation}
By matching these with the OPEs of Eq. \ref{Eq:JpJpOPE}, it follows that,
\begin{equation} \label{Eq:KleinFactorMap}
    \mathtt{M}F_\varphi^{2p} \ket{v} = [-\Tilde{F}_{12}\Tilde{F}^\dagger_{11}\Tilde{F}_{22}\Tilde{F}^\dagger_{21}]^p \mathtt{M}\ket{v}
\end{equation}
where $p \in \mathbb{Z}$. Let the two fields $\phi^{(1)}$ and $\phi^{(2)}$ be given by,
\begin{equation}
    \phi^{(i)}(z) \equiv \frac{1}{\sqrt{2}}[i\partial\fef{i2} - i\partial\fef{i1}]
\end{equation}
Now consider the OPE,
\begin{equation}
    \begin{split}
        \hat{J}^+(z)\hat{J}^-(w) = & \frac{2}{(z-w)^2} + \frac{2\hat{J}^3(w)}{z-w} \\
        &+ V^+_1(w)V^-_2(w) + V^-_1(w)V^+_2(w) \\
        &+ \frac{1}{\sqrt{2}}[i\partial\phi^{(1)}(w) + i\partial\phi^{(2)}(w)] \\
        &+ : (\phi^{(1)}(w))^2 : + : (\phi^{(2)}(w))^2 : + \dots \\
        = & \frac{2}{(z-w)^2} + \frac{2\hat{J}^3(w)}{z-w} \\
        &+ V^+_1(w)V^-_2(w) + V^-_1(w)V^+_2(w) \\
        &+ \frac{1}{2}: ( \phi^{(2)}(w) - \phi^{(1)}(w) )^2 : \\
        &+ \partial\hat{J}^3(w) + : (\hat{J}^3(w))^2 : + \dots \\
    \end{split}
\end{equation}
By comparing this with the OPE of Eq. \ref{Eq:jpjmopemaj}, one can then see that the Majorana energy-momentum tensor, $T^\psi(z)$, must map over to $\hat{\mathcal{H}}_{\hat{\mathfrak{u}}(1)\oplus\widehat{\mathfrak{su}}(2)_2}$ as,
\begin{equation}
    \begin{split}
        \hat{T}^\psi(z) = & \frac{1}{4} \bigg [ V^+_1(w)V^-_2(w) + V^-_1(w)V^+_2(w) \\
        &+ \frac{1}{2}: ( \phi^{(2)}(w) - \phi^{(1)}(w) )^2 : \bigg ] \\
    \end{split}
\end{equation}
Let the modes of $\hat{T}^\psi(z)$ be $\hat{L}^\psi_n$. We can then write,
\begin{equation} \label{Eq:psiVirMap}
    \mathtt{M}L^\psi_n\ket{v} = \hat{L}^\psi_n\mathtt{M}\ket{v}
\end{equation}
Finally, we also have,
\begin{equation} \label{Eq:oddStateMap}
    \begin{split}
        \mathtt{M}\psi_{-\frac{1}{2}} F_\varphi \ket{0} = & \mathtt{M}\frac{1}{\sqrt{2}}J^+_{-1}\ket{0} \\
        = & \frac{1}{\sqrt{2}}\hat{J}^+_{-1}\ket{\hat{0}} \\
        = & \frac{1}{\sqrt{2}}[ \Tilde{V}_{12,-\frac{1}{2}}\Tilde{V}^\dagger_{11, -\frac{1}{2}} + \Tilde{V}_{22,-\frac{1}{2}}\Tilde{V}^\dagger_{21, -\frac{1}{2}} ] \ket{\hat{0}} \\
         \mathtt{M}\psi_{-\frac{1}{2}} F^\dagger_\varphi \ket{0} = & \mathtt{M}\frac{-1}{\sqrt{2}}J^-_{-1}\ket{0} \\
        = & \frac{-1}{\sqrt{2}}\hat{J}^-_{-1}\ket{\hat{0}} \\
        = & \frac{-1}{\sqrt{2}}[ \Tilde{V}_{11,-\frac{1}{2}}\Tilde{V}^\dagger_{12, -\frac{1}{2}} + \Tilde{V}_{21,-\frac{1}{2}}\Tilde{V}^\dagger_{22, -\frac{1}{2}} ] \ket{\hat{0}} \\
    \end{split}
\end{equation}

By combining Eqs. \ref{Eq:MPropDefinition}, \ref{Eq:varphiModeMap}, \ref{Eq:psiVirMap} and \ref{Eq:oddStateMap}, we find that the basis of Eq. \ref{Eq:overlcompleteBasis} maps over to $\hat{\mathcal{H}}_{\hat{\mathfrak{u}}(1)\oplus \widehat{\mathfrak{su}}(2)_2}$ as,
\begin{equation}
    \begin{split}
        \mathtt{M}\ket{\lambda_1: 0;\lambda_2;\lambda_3;p} = & \prod_{n_1\in\lambda_1}\hat{a}_{-n_1} \prod_{n_2\in \lambda_2} \hat{L}^\psi_{-n_2} \prod_{n_3\in\lambda_3} \hat{J}^3_{-n_3} \\
        & \times [-\Tilde{F}_{12}\Tilde{F}^\dagger_{11}\Tilde{F}_{22}\Tilde{F}^\dagger_{21}]^{p/2} \ket{\hat{0}}
    \end{split}
\end{equation}
for $p$ even,
\begin{equation}
    \begin{split}
        \mathtt{M}\ket{\lambda_1: 0;\lambda_2;\lambda_3;p} = & \prod_{n_1\in\lambda_1}\hat{a}_{-n_1} \prod_{n_2\in \lambda_2} \hat{L}^\psi_{-n_2} \prod_{n_3\in\lambda_3} \hat{J}^3_{-n_3} \\
        & \times [-\Tilde{F}_{12}\Tilde{F}^\dagger_{11}\Tilde{F}_{22}\Tilde{F}^\dagger_{21}]^{(p-1)/2} \\
        & \times \frac{-1}{\sqrt{2}}[ \Tilde{V}_{12,-\frac{1}{2}}\Tilde{V}^\dagger_{11, -\frac{1}{2}} + \Tilde{V}_{22,-\frac{1}{2}}\Tilde{V}^\dagger_{21, -\frac{1}{2}} ] \ket{\hat{0}} \\
    \end{split}
\end{equation}
for $p$ odd and $p>0$, and
\begin{equation}
        \begin{split}
        \mathtt{M}\ket{\lambda_1: 0;\lambda_2;\lambda_3;p} = & \prod_{n_1\in\lambda_1}\hat{a}_{-n_1} \prod_{n_2\in \lambda_2} \hat{L}^\psi_{-n_2} \prod_{n_3\in\lambda_3} \hat{J}^3_{-n_3} \\
        & \times [-\Tilde{F}_{12}\Tilde{F}^\dagger_{11}\Tilde{F}_{22}\Tilde{F}^\dagger_{21}]^{(p+1)/2} \\
        & \frac{1}{\sqrt{2}}[ \Tilde{V}_{11,-\frac{1}{2}}\Tilde{V}^\dagger_{12, -\frac{1}{2}} + \Tilde{V}_{21,-\frac{1}{2}}\Tilde{V}^\dagger_{22, -\frac{1}{2}} ] \ket{\hat{0}} \\
    \end{split}
\end{equation}
for $p$ odd and $p<0$. Note that by using Eq. \ref{Eq:densityModeExpan} along with the fact that $V^+_i(z) = \Tilde{V}_{i2} \Tilde{V}^\dagger_{i1}(z)$ and $V^-_i(z) = \Tilde{V}_{i1} \Tilde{V}^\dagger_{i2}(z)$ the modes $\hat{a}_n$, $\hat{L}_n$ and $\hat{J}^3_n$ can be expressed in terms of the modes of the $\Tilde{V}_{ij}(z)$ and $\Tilde{V}^\dagger_{ij}(z)$ fields. Furthermore, it was previously discussed in Sec. \ref{Sec:integerQHCFT} how the Klein factors acting on the vacuum can be expressed as the modes of corresponding vertex operator acting on the vacuum state. Thus, putting this altogether we can express the states $\mathtt{M}\ket{\lambda_1: 0;\lambda_2;\lambda_3;p}$ and $\mathtt{M}\ket{\lambda_1: 1/2;\lambda_2;\lambda_3;p}$ can be expressed as polynomials in the modes $\Tilde{V}_{ij,n}$ and $\Tilde{V}^\dagger_{ij,n}$ acting on $\ket{\hat{0}}$. By expressing $\mathtt{M}\ket{\lambda_1: 0;\lambda_2;\lambda_3;p}$ and $\mathtt{M}\ket{\lambda_1: 1/2;\lambda_2;\lambda_3;p}$ this way, the edge wave functions that these states map to can be straightforwardly expressed as a sum of products of Slater determinants in the orbitals $z^m$ and $\bar{z}z^m$ from the discussions of Sec. \ref{Sec:integerQHCFT} and Sec. \ref{Sec:partonEdgeMap}.

Finally, we can then map over the $\ket{\lambda_1;\mu;\lambda_2;p}$ basis by understanding how this can be expressed as linear combintations of the overlcomplete $\ket{\lambda_1: 0;\lambda_2;\lambda_3;p}$ and $\ket{\lambda_1: 1/2;\lambda_2;\lambda_3;p}$ basis. Note that the elements of both these bases are $L_0$ eigenstates, so $\ket{\lambda_1;\mu;\lambda_2;p}$ must be expressible in terms of a finite number of elements from the overcomplete basis with the same $L_0$ eigenvalue. Thus, the basis transformation can be expressed as a series of matrices, where each matrix performs the basis transform for the states of a given $L_0$ eigenvalue. To find the matrix for the $L_0$ eigenvalue $M$, which we write $Y(M)$, one can first find the matrix $X(M)$ that transforms the $\ket{\lambda_1: 0;\lambda_2;\lambda_3;p}$ and $\ket{\lambda_1: 1/2;\lambda_2;\lambda_3;p}$ states, with $L_0$ eigenvalue $M$, to the $\ket{\lambda_1;\mu;\lambda_3;p}$ states, also with $L_0$ eigenvalue $M$, by expressing the Majorana Virasoro modes as $L^\psi_n = (1/2)\sum_k (k+1/2) : \psi_{n-k}\psi_k :$. If $X(M)$ is a square matrix $Y(M)$ would then be the inverse of $X(M)$. However, as the $\ket{\lambda_1: 0;\lambda_2;\lambda_3;p}$ and $\ket{\lambda_1: 1/2;\lambda_2;\lambda_3;p}$ basis is overcomplete, the matrix $X(M)$ will represent an underdetermined system of linear equations, and so there does not exist a unique $Y(M)$. One solution, that can be implemented numerically, is to take $Y(M) = ([X(M)]^TX(M))^{-1}[X(M)]^T$ as the chosen basis transformation matrix. Once $Y(M)$ has been calculated the edge state wave functions that the $\ket{\lambda_1;\mu;\lambda_2;p}$ states map to can be expressed as sums of products of Slater determinants of the $z^m$ and $\bar{z}z^m$ orbitals. 

\section{Inner product action matrix element fitting: continued} \label{Sec:furtherMatElFit}
For the $\nu = 2/5$ composite fermion case the basis references used in Fig. \ref{fig:fittingTestCF} can be found in Tables \ref{tab:CFBasisRef1}, \ref{tab:CFBasisRef2} and \ref{tab:CFBasisRef3}, and sample matrix elements have been provided in Tables \ref{tab:CFSampleMatEl1} and \ref{tab:CFSampleMatEl2}. For the $\phi_2^2$ parton case the basis references used in Fig. \ref{fig:fittingTestParton} can be found in Tables \ref{tab:PartonBasisRef1}, \ref{tab:PartonBasisRef2} and \ref{tab:PartonBasisRef3}, and sample matrix elements have been provided in Tables \ref{tab:PartonSampleMatEl1} and \ref{tab:PartonSampleMatEl2}. Note that we have only provided sample matrix elements for the CFT states that map to edge state trial wave functions at angular momentum $\Delta M = 1,2$ above the ground state trial wave function, where only matrix elements for which the Monte Carlo estimated values have absolute values greater than 0.05 have been shown.

\begin{table}[]
    \centering
    \begin{tabular}{||c|c||}
        \hline\hline
        Index & basis element \\
        \hline\hline
        \input{CFBasisRefN55_1} \\
        \hline
    \end{tabular}
    \caption{Basis reference, used in Fig. \ref{fig:fittingTestCF}, for the $\nu = 2/5$ composite fermion CFT $U(1)$ neutral states with $L_0$ eigenvalue $M = 1$ (i.e. states that map to trial edge state wave functions with the same particle number as the ground state trial wave function and with angular momentum $M = 1$ relative to the ground state trial wave function). The corresponding basis elements are defined in Sec. \ref{Sec:fitting}.}
    \label{tab:CFBasisRef1}
\end{table}

\begin{table}[]
    \centering
    \begin{tabular}{||c|c||}
        \hline\hline
        Index & basis element \\
        \hline\hline
        \input{CFBasisRefN55_2} \\
        \hline
    \end{tabular}
    \caption{Basis reference, used in Fig. \ref{fig:fittingTestCF}, for the $\nu = 2/5$ composite fermion CFT $U(1)$ neutral states with $L_0$ eigenvalue $M = 2$ (i.e. states that map to trial edge state wave functions with the same particle number as the ground state trial wave function and with angular momentum $M = 2$ relative to the ground state trial wave function). The corresponding basis elements are defined in Sec. \ref{Sec:fitting}.}
    \label{tab:CFBasisRef2}
\end{table}

\begin{table}[]
    \centering
    \begin{tabular}{||c|c||}
        \hline\hline
        Index & basis element \\
        \hline\hline
        \input{CFBasisRefN55_3} \\
        \hline
    \end{tabular}
    \caption{Basis reference, used in Fig. \ref{fig:fittingTestCF}, for the $\nu = 2/5$ composite fermion CFT $U(1)$ neutral states with $L_0$ eigenvalue $M = 3$ (i.e. states that map to trial edge state wave functions with the same particle number as the ground state trial wave function and with angular momentum $M = 3$ relative to the ground state trial wave function). The corresponding basis elements are defined in Sec. \ref{Sec:fitting}.}
    \label{tab:CFBasisRef3}
\end{table}

\begin{table}[]
    \centering
    \begin{tabular}{||c|c|c||}
        \hline\hline
        Matrix element & MC & Model \\
        \hline\hline
        \input{CFInnerProdTableN55_1} \\
        \hline
    \end{tabular}
    \caption{Monte Carlo estimated (MC) and the fitted model (Eq. \ref{Eq:CFInnerProdModel}) matrix elements of the inner product action $\delta S_b$ for the $\nu = 2/5$ composite fermion CFT states that are $U(1)$ neutral and have $L_0$ eigenvalue $M = 1$, where we have only shown matrix elements that have an MC estimated value with an absolute value greater than $0.05$. The basis elements used are defined in Sec. \ref{Sec:fitting}}
    \label{tab:CFSampleMatEl1}
\end{table}

\begin{table}[]
    \centering
    \begin{tabular}{||c|c|c||}
        \hline\hline
        Matrix element & MC & Model \\
        \hline\hline
        \input{CFInnerProdTableN55_2} \\
        \hline
    \end{tabular}
    \caption{Monte Carlo estimated (MC) and the fitted model (Eq. \ref{Eq:CFInnerProdModel}) matrix elements of the inner product action $\delta S_b$ for the $\nu = 2/5$ composite fermion CFT states that are $U(1)$ neutral and have $L_0$ eigenvalue $M = 2$, where we have only shown matrix elements that have an MC estimated value with an absolute value greater than $0.05$. The basis elements used are defined in Sec. \ref{Sec:fitting}}
    \label{tab:CFSampleMatEl2}
\end{table}

\begin{table}[]
    \centering
    \begin{tabular}{||c|c||}
        \hline\hline
        Index & basis element \\
        \hline\hline
        \input{PartonBasisRefN31_1} \\
        \hline
    \end{tabular}
    \caption{Basis reference, used in Fig. \ref{fig:fittingTestParton}, for the $\phi_2^2$ parton CFT $U(1)$ neutral states with $L_0$ eigenvalue $M = 1$ (i.e. states that map to trial edge state wave functions with the same particle number as the ground state trial wave function and with angular momentum $M = 1$ relative to the ground state trial wave function). The corresponding basis elements are defined in Sec. \ref{Sec:fitting}.}
    \label{tab:PartonBasisRef1}
\end{table}

\begin{table}[]
    \centering
    \begin{tabular}{||c|c||}
        \hline\hline
        Index & basis element \\
        \hline\hline
        \input{PartonBasisRefN31_2} \\
        \hline
    \end{tabular}
    \caption{Basis reference, used in Fig. \ref{fig:fittingTestParton}, for the $\phi_2^2$ parton CFT $U(1)$ neutral states with $L_0$ eigenvalue $M = 2$ (i.e. states that map to trial edge state wave functions with the same particle number as the ground state trial wave function and with angular momentum $M = 2$ relative to the ground state trial wave function). The corresponding basis elements are defined in Sec. \ref{Sec:fitting}.}
    \label{tab:PartonBasisRef2}
\end{table}

\begin{table}[]
    \centering
    \begin{tabular}{||c|c||}
        \hline\hline
        Index & basis element \\
        \hline\hline
        \input{PartonBasisRefN31_3} \\
        \hline
    \end{tabular}
    \caption{Basis reference, used in Fig. \ref{fig:fittingTestParton}, for the $\phi_2^2$ parton CFT $U(1)$ neutral states with $L_0$ eigenvalue $M = 3$ (i.e. states that map to trial edge state wave functions with the same particle number as the ground state trial wave function and with angular momentum $M = 3$ relative to the ground state trial wave function). The corresponding basis elements are defined in Sec. \ref{Sec:fitting}.}
    \label{tab:PartonBasisRef3}
\end{table}

\begin{table}[]
    \centering
    \begin{tabular}{||c|c|c||}
        \hline\hline
        Matrix element & MC & Model \\
        \hline\hline
        \input{PartonInnerProdTableN31_1} \\
        \hline
    \end{tabular}
    \caption{Monte Carlo estimated (MC) and the fitted model (Eq. \ref{Eq:partInnProdModel}) matrix elements of the inner product action $\delta S_b$ for the $\phi_2^2$ parton CFT states that are $U(1)$ neutral and have $L_0$ eigenvalue $M = 1$, where we have only shown matrix elements that have an MC estimated value with an absolute value greater than $0.05$. The basis elements used are defined in Sec. \ref{Sec:fitting}.}
    \label{tab:PartonSampleMatEl1}
\end{table}

\begin{table}[]
    \centering
    \begin{tabular}{||c|c|c||}
        \hline\hline
        Matrix element & MC & Model \\
        \hline\hline
        \input{PartonInnerProdTableN31_2} \\
        \hline
    \end{tabular}
    \caption{Monte Carlo estimated (MC) and the fitted model (Eq. \ref{Eq:partInnProdModel}) matrix elements of the inner product action $\delta S_b$ for the $\phi_2^2$ parton CFT states that are $U(1)$ neutral and have $L_0$ eigenvalue $M = 2$, where we have only shown matrix elements that have an MC estimated value with an absolute value greater than $0.05$. The basis elements used are defined in Sec. \ref{Sec:fitting}.}
    \label{tab:PartonSampleMatEl2}
\end{table}

\clearpage

\bibliography{refsShort.bib}

\end{document}

%% file: CFBasisRefN55_1.tex
0 & $\ket{\emptyset;\emptyset;1}$ \\ 
1 & $\ket{1;\emptyset;0}$ \\ 
2 & $\ket{\emptyset;1;0}$ \\ 
3 & $\ket{\emptyset;\emptyset;-1}$

%% file: CFBasisRefN55_2.tex
0 & $\ket{1;\emptyset;1}$ \\ 
1 & $\ket{\emptyset;1;1}$ \\ 
2 & $\ket{2;\emptyset;0}$ \\ 
3 & $\ket{\emptyset;2;0}$ \\ 
4 & $\ket{11;\emptyset;0}$ \\ 
5 & $\ket{\emptyset;11;0}$ \\ 
6 & $\ket{1;1;0}$ \\ 
7 & $\ket{1;\emptyset;-1}$ \\ 
8 & $\ket{\emptyset;1;-1}$

%% file: CFBasisRefN55_3.tex
0 & $\ket{2;\emptyset;1}$ \\ 
1 & $\ket{\emptyset;2;1}$ \\ 
2 & $\ket{11;\emptyset;1}$ \\ 
3 & $\ket{\emptyset;11;1}$ \\ 
4 & $\ket{1;1;1}$ \\ 
5 & $\ket{3;\emptyset;0}$ \\ 
6 & $\ket{\emptyset;3;0}$ \\ 
7 & $\ket{12;\emptyset;0}$ \\ 
8 & $\ket{\emptyset;12;0}$ \\ 
9 & $\ket{2;1;0}$ \\ 
10 & $\ket{1;2;0}$ \\ 
11 & $\ket{11;1;0}$ \\ 
12 & $\ket{1;11;0}$ \\ 
13 & $\ket{\emptyset;111;0}$ \\ 
14 & $\ket{111;\emptyset;0}$ \\ 
15 & $\ket{2;\emptyset;-1}$ \\ 
16 & $\ket{\emptyset;2;-1}$ \\ 
17 & $\ket{11;\emptyset;-1}$ \\ 
18 & $\ket{\emptyset;11;-1}$ \\ 
19 & $\ket{1;1;-1}$

%% file: CFInnerProdTableN55_1.tex
$\bra{\emptyset;\emptyset;1}\delta S_b\ket{\emptyset;\emptyset;1}$ & 15.3244 & 15.3242 \\ 
$\bra{\emptyset;\emptyset;1}\delta S_b\ket{\emptyset;1;0}$ & 0.0928 & 0.0928 \\ 
$\bra{\emptyset;1;0}\delta S_b\ket{\emptyset;1;0}$ & -0.2330 & -0.2326 \\ 
$\bra{\emptyset;1;0}\delta S_b\ket{\emptyset;\emptyset;-1}$ & -0.0928 & -0.0928 \\ 
$\bra{\emptyset;\emptyset;-1}\delta S_b\ket{\emptyset;\emptyset;-1}$ & -15.7893 & -15.7894

%% file: CFInnerProdTableN55_2.tex
$\bra{1;\emptyset;1}\delta S_b\ket{1;\emptyset;1}$ & 15.3253 & 15.3242 \\ 
$\bra{1;\emptyset;1}\delta S_b\ket{1;1;0}$ & 0.0929 & 0.0928 \\ 
$\bra{\emptyset;1;1}\delta S_b\ket{\emptyset;1;1}$ & 15.0758 & 15.0916 \\ 
$\bra{\emptyset;1;1}\delta S_b\ket{\emptyset;2;0}$ & 0.0929 & 0.0928 \\ 
$\bra{2;\emptyset;0}\delta S_b\ket{\emptyset;2;0}$ & 0.0525 & 0.0521 \\ 
$\bra{\emptyset;2;0}\delta S_b\ket{\emptyset;2;0}$ & -0.4676 & -0.4652 \\ 
$\bra{\emptyset;2;0}\delta S_b\ket{\emptyset;1;-1}$ & 0.0929 & 0.0928 \\ 
$\bra{\emptyset;11;0}\delta S_b\ket{\emptyset;11;0}$ & -0.4656 & -0.4652 \\ 
$\bra{1;1;0}\delta S_b\ket{1;1;0}$ & -0.2327 & -0.2326 \\ 
$\bra{1;1;0}\delta S_b\ket{1;\emptyset;-1}$ & -0.0928 & -0.0928 \\ 
$\bra{1;\emptyset;-1}\delta S_b\ket{1;\emptyset;-1}$ & -15.7891 & -15.7894 \\ 
$\bra{\emptyset;1;-1}\delta S_b\ket{\emptyset;1;-1}$ & -16.0080 & -16.0221

%% file: PartonBasisRefN31_1.tex
0 & $\ket{\emptyset;1;\emptyset;-1}$ \\ 
1 & $\ket{\emptyset;\emptyset;1;0}$ \\ 
2 & $\ket{\emptyset;1;\emptyset;1}$ \\ 
3 & $\ket{1;\emptyset;\emptyset;0}$

%% file: PartonBasisRefN31_2.tex
0 & $\ket{\emptyset;\emptyset;\emptyset;-2}$ \\ 
1 & $\ket{\emptyset;3;\emptyset;-1}$ \\ 
2 & $\ket{\emptyset;1;1;-1}$ \\ 
3 & $\ket{\emptyset;31;\emptyset;0}$ \\ 
4 & $\ket{\emptyset;\emptyset;11;0}$ \\ 
5 & $\ket{\emptyset;\emptyset;2;0}$ \\ 
6 & $\ket{\emptyset;3;\emptyset;1}$ \\ 
7 & $\ket{\emptyset;1;1;1}$ \\ 
8 & $\ket{\emptyset;\emptyset;\emptyset;2}$ \\ 
9 & $\ket{1;1;\emptyset;-1}$ \\ 
10 & $\ket{1;\emptyset;1;0}$ \\ 
11 & $\ket{1;1;\emptyset;1}$ \\ 
12 & $\ket{11;\emptyset;\emptyset;0}$ \\ 
13 & $\ket{2;\emptyset;\emptyset;0}$

%% file: PartonBasisRefN31_3.tex
0 & $\ket{\emptyset;\emptyset;1;-2}$ \\ 
1 & $\ket{\emptyset;5;\emptyset;-1}$ \\ 
2 & $\ket{\emptyset;3;1;-1}$ \\ 
3 & $\ket{\emptyset;1;11;-1}$ \\ 
4 & $\ket{\emptyset;1;2;-1}$ \\ 
5 & $\ket{\emptyset;51;\emptyset;0}$ \\ 
6 & $\ket{\emptyset;31;1;0}$ \\ 
7 & $\ket{\emptyset;\emptyset;111;0}$ \\ 
8 & $\ket{\emptyset;\emptyset;12;0}$ \\ 
9 & $\ket{\emptyset;\emptyset;3;0}$ \\ 
10 & $\ket{\emptyset;5;\emptyset;1}$ \\ 
11 & $\ket{\emptyset;3;1;1}$ \\ 
12 & $\ket{\emptyset;1;11;1}$ \\ 
13 & $\ket{\emptyset;1;2;1}$ \\ 
14 & $\ket{\emptyset;\emptyset;1;2}$ \\ 
15 & $\ket{1;\emptyset;\emptyset;-2}$ \\ 
16 & $\ket{1;3;\emptyset;-1}$ \\ 
17 & $\ket{1;1;1;-1}$ \\ 
18 & $\ket{1;31;\emptyset;0}$ \\ 
19 & $\ket{1;\emptyset;11;0}$ \\ 
20 & $\ket{1;\emptyset;2;0}$ \\ 
21 & $\ket{1;3;\emptyset;1}$ \\ 
22 & $\ket{1;1;1;1}$ \\ 
23 & $\ket{1;\emptyset;\emptyset;2}$ \\ 
24 & $\ket{11;1;\emptyset;-1}$ \\ 
25 & $\ket{11;\emptyset;1;0}$ \\ 
26 & $\ket{11;1;\emptyset;1}$ \\ 
27 & $\ket{2;1;\emptyset;-1}$ \\ 
28 & $\ket{2;\emptyset;1;0}$ \\ 
29 & $\ket{2;1;\emptyset;1}$ \\ 
30 & $\ket{111;\emptyset;\emptyset;0}$ \\ 
31 & $\ket{12;\emptyset;\emptyset;0}$ \\ 
32 & $\ket{3;\emptyset;\emptyset;0}$

%% file: PartonInnerProdTableN31_1.tex
$\bra{\emptyset;1;\emptyset;-1}\delta S_b\ket{\emptyset;1;\emptyset;-1}$ & -11.2950 & -11.2950 \\ 
$\bra{\emptyset;1;\emptyset;-1}\delta S_b\ket{\emptyset;\emptyset;1;0}$ & -0.2631 & -0.2643 \\ 
$\bra{\emptyset;\emptyset;1;0}\delta S_b\ket{\emptyset;1;\emptyset;-1}$ & -0.2631 & -0.2643 \\ 
$\bra{\emptyset;\emptyset;1;0}\delta S_b\ket{\emptyset;\emptyset;1;0}$ & -0.0999 & -0.0999 \\ 
$\bra{\emptyset;\emptyset;1;0}\delta S_b\ket{\emptyset;1;\emptyset;1}$ & -0.2656 & -0.2643 \\ 
$\bra{\emptyset;\emptyset;1;0}\delta S_b\ket{1;\emptyset;\emptyset;0}$ & 0.0991 & 0.0991 \\ 
$\bra{\emptyset;1;\emptyset;1}\delta S_b\ket{\emptyset;\emptyset;1;0}$ & -0.2656 & -0.2643 \\ 
$\bra{\emptyset;1;\emptyset;1}\delta S_b\ket{\emptyset;1;\emptyset;1}$ & 11.1645 & 11.1645 \\ 
$\bra{1;\emptyset;\emptyset;0}\delta S_b\ket{\emptyset;\emptyset;1;0}$ & 0.0991 & 0.0991

%% file: PartonInnerProdTableN31_2.tex
$\bra{\emptyset;\emptyset;\emptyset;-2}\delta S_b\ket{\emptyset;\emptyset;\emptyset;-2}$ & -22.6614 & -22.6594 \\ 
$\bra{\emptyset;\emptyset;\emptyset;-2}\delta S_b\ket{\emptyset;3;\emptyset;-1}$ & 0.2665 & 0.2643 \\ 
$\bra{\emptyset;\emptyset;\emptyset;-2}\delta S_b\ket{\emptyset;1;1;-1}$ & 0.2654 & 0.2643 \\ 
$\bra{\emptyset;3;\emptyset;-1}\delta S_b\ket{\emptyset;\emptyset;\emptyset;-2}$ & 0.2665 & 0.2643 \\ 
$\bra{\emptyset;3;\emptyset;-1}\delta S_b\ket{\emptyset;3;\emptyset;-1}$ & -11.3254 & -11.3256 \\ 
$\bra{\emptyset;3;\emptyset;-1}\delta S_b\ket{\emptyset;31;\emptyset;0}$ & 0.2647 & 0.2643 \\ 
$\bra{\emptyset;3;\emptyset;-1}\delta S_b\ket{\emptyset;\emptyset;11;0}$ & -0.1843 & -0.1869 \\ 
$\bra{\emptyset;3;\emptyset;-1}\delta S_b\ket{\emptyset;\emptyset;2;0}$ & -0.1852 & -0.1869 \\ 
$\bra{\emptyset;1;1;-1}\delta S_b\ket{\emptyset;\emptyset;\emptyset;-2}$ & 0.2654 & 0.2643 \\ 
$\bra{\emptyset;1;1;-1}\delta S_b\ket{\emptyset;1;1;-1}$ & -11.3928 & -11.3949 \\ 
$\bra{\emptyset;1;1;-1}\delta S_b\ket{\emptyset;31;\emptyset;0}$ & 0.2669 & 0.2643 \\ 
$\bra{\emptyset;1;1;-1}\delta S_b\ket{\emptyset;\emptyset;11;0}$ & -0.1851 & -0.1869 \\ 
$\bra{\emptyset;1;1;-1}\delta S_b\ket{\emptyset;\emptyset;2;0}$ & 0.1878 & 0.1869 \\ 
$\bra{\emptyset;1;1;-1}\delta S_b\ket{1;1;\emptyset;-1}$ & 0.0633 & 0.0991 \\ 
$\bra{\emptyset;31;\emptyset;0}\delta S_b\ket{\emptyset;3;\emptyset;-1}$ & 0.2647 & 0.2643 \\ 
$\bra{\emptyset;31;\emptyset;0}\delta S_b\ket{\emptyset;1;1;-1}$ & 0.2669 & 0.2643 \\ 
$\bra{\emptyset;31;\emptyset;0}\delta S_b\ket{\emptyset;31;\emptyset;0}$ & -0.0640 & -0.0611 \\ 
$\bra{\emptyset;31;\emptyset;0}\delta S_b\ket{\emptyset;3;\emptyset;1}$ & -0.2625 & -0.2643 \\ 
$\bra{\emptyset;31;\emptyset;0}\delta S_b\ket{\emptyset;1;1;1}$ & 0.2601 & 0.2643 \\ 
$\bra{\emptyset;\emptyset;11;0}\delta S_b\ket{\emptyset;3;\emptyset;-1}$ & -0.1843 & -0.1869 \\ 
$\bra{\emptyset;\emptyset;11;0}\delta S_b\ket{\emptyset;1;1;-1}$ & -0.1851 & -0.1869 \\ 
$\bra{\emptyset;\emptyset;11;0}\delta S_b\ket{\emptyset;\emptyset;11;0}$ & -0.1987 & -0.1999 \\ 
$\bra{\emptyset;\emptyset;11;0}\delta S_b\ket{\emptyset;3;\emptyset;1}$ & 0.1891 & 0.1869 \\ 
$\bra{\emptyset;\emptyset;11;0}\delta S_b\ket{\emptyset;1;1;1}$ & -0.1869 & -0.1869 \\ 
$\bra{\emptyset;\emptyset;11;0}\delta S_b\ket{1;\emptyset;1;0}$ & 0.1383 & 0.1402 \\ 
$\bra{\emptyset;\emptyset;2;0}\delta S_b\ket{\emptyset;3;\emptyset;-1}$ & -0.1852 & -0.1869 \\ 
$\bra{\emptyset;\emptyset;2;0}\delta S_b\ket{\emptyset;1;1;-1}$ & 0.1878 & 0.1869 \\ 
$\bra{\emptyset;\emptyset;2;0}\delta S_b\ket{\emptyset;\emptyset;2;0}$ & -0.2025 & -0.1999 \\ 
$\bra{\emptyset;\emptyset;2;0}\delta S_b\ket{\emptyset;3;\emptyset;1}$ & -0.1900 & -0.1869 \\ 
$\bra{\emptyset;\emptyset;2;0}\delta S_b\ket{\emptyset;1;1;1}$ & -0.1895 & -0.1869 \\ 
$\bra{\emptyset;\emptyset;2;0}\delta S_b\ket{2;\emptyset;\emptyset;0}$ & 0.1938 & 0.1982 \\ 
$\bra{\emptyset;3;\emptyset;1}\delta S_b\ket{\emptyset;31;\emptyset;0}$ & -0.2625 & -0.2643 \\ 
$\bra{\emptyset;3;\emptyset;1}\delta S_b\ket{\emptyset;\emptyset;11;0}$ & 0.1891 & 0.1869 \\ 
$\bra{\emptyset;3;\emptyset;1}\delta S_b\ket{\emptyset;\emptyset;2;0}$ & -0.1900 & -0.1869 \\ 
$\bra{\emptyset;3;\emptyset;1}\delta S_b\ket{\emptyset;3;\emptyset;1}$ & 11.1226 & 11.1339 \\ 
$\bra{\emptyset;3;\emptyset;1}\delta S_b\ket{\emptyset;\emptyset;\emptyset;2}$ & -0.2595 & -0.2643 \\ 
$\bra{\emptyset;1;1;1}\delta S_b\ket{\emptyset;31;\emptyset;0}$ & 0.2601 & 0.2643 \\ 
$\bra{\emptyset;1;1;1}\delta S_b\ket{\emptyset;\emptyset;11;0}$ & -0.1869 & -0.1869 \\ 
$\bra{\emptyset;1;1;1}\delta S_b\ket{\emptyset;\emptyset;2;0}$ & -0.1895 & -0.1869 \\ 
$\bra{\emptyset;1;1;1}\delta S_b\ket{\emptyset;1;1;1}$ & 11.0603 & 11.0646 \\ 
$\bra{\emptyset;1;1;1}\delta S_b\ket{\emptyset;\emptyset;\emptyset;2}$ & 0.2631 & 0.2643 \\ 
$\bra{\emptyset;1;1;1}\delta S_b\ket{1;1;\emptyset;1}$ & 0.1331 & 0.0991 \\ 
$\bra{\emptyset;\emptyset;\emptyset;2}\delta S_b\ket{\emptyset;3;\emptyset;1}$ & -0.2595 & -0.2643 \\ 
$\bra{\emptyset;\emptyset;\emptyset;2}\delta S_b\ket{\emptyset;1;1;1}$ & 0.2631 & 0.2643 \\ 
$\bra{\emptyset;\emptyset;\emptyset;2}\delta S_b\ket{\emptyset;\emptyset;\emptyset;2}$ & 22.2554 & 22.2597 \\ 
$\bra{1;1;\emptyset;-1}\delta S_b\ket{\emptyset;1;1;-1}$ & 0.0633 & 0.0991 \\ 
$\bra{1;1;\emptyset;-1}\delta S_b\ket{1;1;\emptyset;-1}$ & -11.2865 & -11.2950 \\ 
$\bra{1;1;\emptyset;-1}\delta S_b\ket{1;\emptyset;1;0}$ & -0.2634 & -0.2643 \\ 
$\bra{1;\emptyset;1;0}\delta S_b\ket{\emptyset;\emptyset;11;0}$ & 0.1383 & 0.1402 \\ 
$\bra{1;\emptyset;1;0}\delta S_b\ket{1;1;\emptyset;-1}$ & -0.2634 & -0.2643 \\ 
$\bra{1;\emptyset;1;0}\delta S_b\ket{1;\emptyset;1;0}$ & -0.0956 & -0.0999 \\ 
$\bra{1;\emptyset;1;0}\delta S_b\ket{1;1;\emptyset;1}$ & -0.2657 & -0.2643 \\ 
$\bra{1;\emptyset;1;0}\delta S_b\ket{11;\emptyset;\emptyset;0}$ & 0.1381 & 0.1402 \\ 
$\bra{1;1;\emptyset;1}\delta S_b\ket{\emptyset;1;1;1}$ & 0.1331 & 0.0991 \\ 
$\bra{1;1;\emptyset;1}\delta S_b\ket{1;\emptyset;1;0}$ & -0.2657 & -0.2643 \\ 
$\bra{1;1;\emptyset;1}\delta S_b\ket{1;1;\emptyset;1}$ & 11.1640 & 11.1645 \\ 
$\bra{11;\emptyset;\emptyset;0}\delta S_b\ket{1;\emptyset;1;0}$ & 0.1381 & 0.1402 \\ 
$\bra{2;\emptyset;\emptyset;0}\delta S_b\ket{\emptyset;\emptyset;2;0}$ & 0.1938 & 0.1982

%% file: main.bbl
\begin{thebibliography}{124}%
\makeatletter
\providecommand \@ifxundefined [1]{%
 \@ifx{#1\undefined}
}%
\providecommand \@ifnum [1]{%
 \ifnum #1\expandafter \@firstoftwo
 \else \expandafter \@secondoftwo
 \fi
}%
\providecommand \@ifx [1]{%
 \ifx #1\expandafter \@firstoftwo
 \else \expandafter \@secondoftwo
 \fi
}%
\providecommand \natexlab [1]{#1}%
\providecommand \enquote  [1]{``#1''}%
\providecommand \bibnamefont  [1]{#1}%
\providecommand \bibfnamefont [1]{#1}%
\providecommand \citenamefont [1]{#1}%
\providecommand \href@noop [0]{\@secondoftwo}%
\providecommand \href [0]{\begingroup \@sanitize@url \@href}%
\providecommand \@href[1]{\@@startlink{#1}\@@href}%
\providecommand \@@href[1]{\endgroup#1\@@endlink}%
\providecommand \@sanitize@url [0]{\catcode `\\12\catcode `\$12\catcode
  `\&12\catcode `\#12\catcode `\^12\catcode `\_12\catcode `\%12\relax}%
\providecommand \@@startlink[1]{}%
\providecommand \@@endlink[0]{}%
\providecommand \url  [0]{\begingroup\@sanitize@url \@url }%
\providecommand \@url [1]{\endgroup\@href {#1}{\urlprefix }}%
\providecommand \urlprefix  [0]{URL }%
\providecommand \Eprint [0]{\href }%
\providecommand \doibase [0]{https://doi.org/}%
\providecommand \selectlanguage [0]{\@gobble}%
\providecommand \bibinfo  [0]{\@secondoftwo}%
\providecommand \bibfield  [0]{\@secondoftwo}%
\providecommand \translation [1]{[#1]}%
\providecommand \BibitemOpen [0]{}%
\providecommand \bibitemStop [0]{}%
\providecommand \bibitemNoStop [0]{.\EOS\space}%
\providecommand \EOS [0]{\spacefactor3000\relax}%
\providecommand \BibitemShut  [1]{\csname bibitem#1\endcsname}%
\let\auto@bib@innerbib\@empty
%</preamble>
\bibitem [{\citenamefont {Tsui}\ \emph {et~al.}(1982)\citenamefont {Tsui},
  \citenamefont {Stormer},\ and\ \citenamefont {Gossard}}]{Tsui}%
  \BibitemOpen
  \bibfield  {author} {\bibinfo {author} {\bibfnamefont {D.~C.}\ \bibnamefont
  {Tsui}}, \bibinfo {author} {\bibfnamefont {H.~L.}\ \bibnamefont {Stormer}},\
  and\ \bibinfo {author} {\bibfnamefont {A.~C.}\ \bibnamefont {Gossard}},\
  }\bibfield  {title} {\bibinfo {title} {Two-{Dimensional} {Magnetotransport}
  in the {Extreme} {Quantum} {Limit}},\ }\href
  {https://journals.aps.org/prl/abstract/10.1103/PhysRevLett.48.1559}
  {\bibfield  {journal} {\bibinfo  {journal} {Phys. Rev. Lett.}\ }\textbf
  {\bibinfo {volume} {48}},\ \bibinfo {pages} {1559} (\bibinfo {year}
  {1982})}\BibitemShut {NoStop}%
\bibitem [{\citenamefont {Prange}\ and\ \citenamefont
  {Girvin}(1990)}]{prange_quantum_2012}%
  \BibitemOpen
  \bibinfo {editor} {\bibfnamefont {R.~E.}\ \bibnamefont {Prange}}\ and\
  \bibinfo {editor} {\bibfnamefont {S.~M.}\ \bibnamefont {Girvin}},\ eds.,\
  \href {https://doi.org/10.1007/978-1-4612-3350-3} {\emph {\bibinfo {title}
  {The Quantum Hall Effect}}},\ Graduate Texts in Contemporary Physics\
  (\bibinfo  {publisher} {Springer New York},\ \bibinfo {year}
  {1990})\BibitemShut {NoStop}%
\bibitem [{\citenamefont {Haldane}(2017)}]{Duncan2017}%
  \BibitemOpen
  \bibfield  {author} {\bibinfo {author} {\bibfnamefont {F.~D.~M.}\
  \bibnamefont {Haldane}},\ }\bibfield  {title} {\bibinfo {title} {Nobel
  {Lecture}: {Topological} quantum matter *},\ }\href
  {https://doi.org/10.1103/RevModPhys.89.040502} {\bibfield  {journal}
  {\bibinfo  {journal} {Rev. Mod. Phys.}\ }\textbf {\bibinfo {volume} {89}},\
  \bibinfo {pages} {040502} (\bibinfo {year} {2017})}\BibitemShut {NoStop}%
\bibitem [{\citenamefont {Wen}(2017)}]{wen_colloquium_2017}%
  \BibitemOpen
  \bibfield  {author} {\bibinfo {author} {\bibfnamefont {X.~G.}\ \bibnamefont
  {Wen}},\ }\bibfield  {title} {\bibinfo {title} {\textit{{Colloquium}} : {Zoo}
  of quantum-topological phases of matter},\ }\href
  {https://link.aps.org/doi/10.1103/RevModPhys.89.041004} {\bibfield  {journal}
  {\bibinfo  {journal} {Rev. Mod. Phys.}\ }\textbf {\bibinfo {volume} {89}},\
  \bibinfo {pages} {041004} (\bibinfo {year} {2017})}\BibitemShut {NoStop}%
\bibitem [{\citenamefont {Nayak}\ \emph {et~al.}(2008)\citenamefont {Nayak},
  \citenamefont {Simon}, \citenamefont {Stern}, \citenamefont {Freedman},\ and\
  \citenamefont {Das~Sarma}}]{Nayak}%
  \BibitemOpen
  \bibfield  {author} {\bibinfo {author} {\bibfnamefont {C.}~\bibnamefont
  {Nayak}}, \bibinfo {author} {\bibfnamefont {S.~H.}\ \bibnamefont {Simon}},
  \bibinfo {author} {\bibfnamefont {A.}~\bibnamefont {Stern}}, \bibinfo
  {author} {\bibfnamefont {M.}~\bibnamefont {Freedman}},\ and\ \bibinfo
  {author} {\bibfnamefont {S.}~\bibnamefont {Das~Sarma}},\ }\bibfield  {title}
  {\bibinfo {title} {Non-{Abelian} anyons and topological quantum
  computation},\ }\href {https://doi.org/10.1103/RevModPhys.80.1083} {\bibfield
   {journal} {\bibinfo  {journal} {Rev. Mod. Phys.}\ }\textbf {\bibinfo
  {volume} {80}},\ \bibinfo {pages} {1083} (\bibinfo {year}
  {2008})}\BibitemShut {NoStop}%
\bibitem [{\citenamefont {Laughlin}(1983)}]{Laughlin1983}%
  \BibitemOpen
  \bibfield  {author} {\bibinfo {author} {\bibfnamefont {R.~B.}\ \bibnamefont
  {Laughlin}},\ }\bibfield  {title} {\bibinfo {title} {Anomalous quantum {Hall}
  effect: {An} incompressible quantum fluid with fractionally charged
  excitations},\ }\href {https://doi.org/10.1103/PhysRevLett.50.1395}
  {\bibfield  {journal} {\bibinfo  {journal} {Phys. Rev. Lett.}\ }\textbf
  {\bibinfo {volume} {50}},\ \bibinfo {pages} {1395} (\bibinfo {year}
  {1983})}\BibitemShut {NoStop}%
\bibitem [{\citenamefont {Arovas}\ \emph {et~al.}(1984)\citenamefont {Arovas},
  \citenamefont {Schrieffer},\ and\ \citenamefont
  {Wilczek}}]{arovas_fractional_1984}%
  \BibitemOpen
  \bibfield  {author} {\bibinfo {author} {\bibfnamefont {D.}~\bibnamefont
  {Arovas}}, \bibinfo {author} {\bibfnamefont {J.~R.}\ \bibnamefont
  {Schrieffer}},\ and\ \bibinfo {author} {\bibfnamefont {F.}~\bibnamefont
  {Wilczek}},\ }\bibfield  {title} {\bibinfo {title} {Fractional {Statistics}
  and the {Quantum} {Hall} {Effect}},\ }\href
  {https://link.aps.org/doi/10.1103/PhysRevLett.53.722} {\bibfield  {journal}
  {\bibinfo  {journal} {Phys. Rev. Lett.}\ }\textbf {\bibinfo {volume} {53}},\
  \bibinfo {pages} {722} (\bibinfo {year} {1984})}\BibitemShut {NoStop}%
\bibitem [{\citenamefont {Halperin}(1984)}]{Halperin1984}%
  \BibitemOpen
  \bibfield  {author} {\bibinfo {author} {\bibfnamefont {B.~I.}\ \bibnamefont
  {Halperin}},\ }\bibfield  {title} {\bibinfo {title} {Statistics of
  quasiparticles and the hierarchy of fractional quantized hall states},\
  }\href {https://doi.org/10.1103/PhysRevLett.52.1583} {\bibfield  {journal}
  {\bibinfo  {journal} {Phys. Rev. Lett.}\ }\textbf {\bibinfo {volume} {52}},\
  \bibinfo {pages} {1583} (\bibinfo {year} {1984})}\BibitemShut {NoStop}%
\bibitem [{\citenamefont {Haldane}(1985)}]{haldane_many-particle_1985}%
  \BibitemOpen
  \bibfield  {author} {\bibinfo {author} {\bibfnamefont {F.~D.~M.}\
  \bibnamefont {Haldane}},\ }\bibfield  {title} {\bibinfo {title}
  {Many-{Particle} {Translational} {Symmetries} of {Two}-{Dimensional}
  {Electrons} at {Rational} {Landau}-{Level} {Filling}},\ }\href
  {https://doi.org/10.1103/PhysRevLett.55.2095} {\bibfield  {journal} {\bibinfo
   {journal} {Phys. Rev. Lett.}\ }\textbf {\bibinfo {volume} {55}},\ \bibinfo
  {pages} {2095} (\bibinfo {year} {1985})}\BibitemShut {NoStop}%
\bibitem [{\citenamefont {Haldane}\ and\ \citenamefont
  {Rezayi}(1985)}]{haldane_periodic_1985}%
  \BibitemOpen
  \bibfield  {author} {\bibinfo {author} {\bibfnamefont {F.~D.~M.}\
  \bibnamefont {Haldane}}\ and\ \bibinfo {author} {\bibfnamefont {E.~H.}\
  \bibnamefont {Rezayi}},\ }\bibfield  {title} {\bibinfo {title} {Periodic
  {Laughlin}-{Jastrow} wave functions for the fractional quantized {Hall}
  effect},\ }\href {https://doi.org/10.1103/PhysRevB.31.2529} {\bibfield
  {journal} {\bibinfo  {journal} {Phys. Rev. B}\ }\textbf {\bibinfo {volume}
  {31}},\ \bibinfo {pages} {2529} (\bibinfo {year} {1985})}\BibitemShut
  {NoStop}%
\bibitem [{\citenamefont {Wen}\ and\ \citenamefont {Niu}(1990)}]{Wen1990a}%
  \BibitemOpen
  \bibfield  {author} {\bibinfo {author} {\bibfnamefont {X.~G.}\ \bibnamefont
  {Wen}}\ and\ \bibinfo {author} {\bibfnamefont {Q.}~\bibnamefont {Niu}},\
  }\bibfield  {title} {\bibinfo {title} {Ground-state degeneracy of the
  fractional quantum {Hall} states in the presence of a random potential and on
  high-genus {Riemann} surfaces},\ }\href
  {https://doi.org/10.1103/PhysRevB.41.9377} {\bibfield  {journal} {\bibinfo
  {journal} {Phys. Rev. B}\ }\textbf {\bibinfo {volume} {41}},\ \bibinfo
  {pages} {9377} (\bibinfo {year} {1990})}\BibitemShut {NoStop}%
\bibitem [{\citenamefont {Li}\ and\ \citenamefont {Haldane}(2008)}]{Li2008}%
  \BibitemOpen
  \bibfield  {author} {\bibinfo {author} {\bibfnamefont {H.}~\bibnamefont
  {Li}}\ and\ \bibinfo {author} {\bibfnamefont {F.~D.}\ \bibnamefont
  {Haldane}},\ }\bibfield  {title} {\bibinfo {title} {Entanglement spectrum as
  a generalization of entanglement entropy: {Identification} of topological
  order in non-{Abelian} fractional quantum hall effect states},\ }\href
  {https://doi.org/10.1103/PhysRevLett.101.010504} {\bibfield  {journal}
  {\bibinfo  {journal} {Phys. Rev. Lett.}\ }\textbf {\bibinfo {volume} {101}},\
  \bibinfo {pages} {1} (\bibinfo {year} {2008})}\BibitemShut {NoStop}%
\bibitem [{\citenamefont {L{\"a}uchli}\ \emph {et~al.}(2010)\citenamefont
  {L{\"a}uchli}, \citenamefont {Bergholtz}, \citenamefont {Suorsa},\ and\
  \citenamefont {Haque}}]{Lauchli2010}%
  \BibitemOpen
  \bibfield  {author} {\bibinfo {author} {\bibfnamefont {A.~M.}\ \bibnamefont
  {L{\"a}uchli}}, \bibinfo {author} {\bibfnamefont {E.~J.}\ \bibnamefont
  {Bergholtz}}, \bibinfo {author} {\bibfnamefont {J.}~\bibnamefont {Suorsa}},\
  and\ \bibinfo {author} {\bibfnamefont {M.}~\bibnamefont {Haque}},\ }\bibfield
   {title} {\bibinfo {title} {Disentangling entanglement spectra of fractional
  quantum hall states on torus geometries},\ }\href
  {https://doi.org/10.1103/PhysRevLett.104.156404} {\bibfield  {journal}
  {\bibinfo  {journal} {Phys. Rev. Lett.}\ }\textbf {\bibinfo {volume} {104}},\
  \bibinfo {pages} {156404} (\bibinfo {year} {2010})}\BibitemShut {NoStop}%
\bibitem [{\citenamefont {Papi{\'c}}\ \emph {et~al.}(2011)\citenamefont
  {Papi{\'c}}, \citenamefont {Bernevig},\ and\ \citenamefont
  {Regnault}}]{Papic2011}%
  \BibitemOpen
  \bibfield  {author} {\bibinfo {author} {\bibfnamefont {Z.}~\bibnamefont
  {Papi{\'c}}}, \bibinfo {author} {\bibfnamefont {B.~A.}\ \bibnamefont
  {Bernevig}},\ and\ \bibinfo {author} {\bibfnamefont {N.}~\bibnamefont
  {Regnault}},\ }\bibfield  {title} {\bibinfo {title} {Topological entanglement
  in {Abelian} and non-{Abelian} excitation eigenstates},\ }\href
  {https://doi.org/10.1103/PhysRevLett.106.056801} {\bibfield  {journal}
  {\bibinfo  {journal} {Phys. Rev. Lett.}\ }\textbf {\bibinfo {volume} {106}},\
  \bibinfo {pages} {056801} (\bibinfo {year} {2011})}\BibitemShut {NoStop}%
\bibitem [{\citenamefont {Sterdyniak}\ \emph {et~al.}(2012)\citenamefont
  {Sterdyniak}, \citenamefont {Chandran}, \citenamefont {Regnault},
  \citenamefont {Bernevig},\ and\ \citenamefont {Bonderson}}]{Sterdyniak2012}%
  \BibitemOpen
  \bibfield  {author} {\bibinfo {author} {\bibfnamefont {A.}~\bibnamefont
  {Sterdyniak}}, \bibinfo {author} {\bibfnamefont {A.}~\bibnamefont
  {Chandran}}, \bibinfo {author} {\bibfnamefont {N.}~\bibnamefont {Regnault}},
  \bibinfo {author} {\bibfnamefont {B.~A.}\ \bibnamefont {Bernevig}},\ and\
  \bibinfo {author} {\bibfnamefont {P.}~\bibnamefont {Bonderson}},\ }\bibfield
  {title} {\bibinfo {title} {Real-space entanglement spectrum of quantum {Hall}
  states},\ }\href {https://doi.org/10.1103/PhysRevB.85.125308} {\bibfield
  {journal} {\bibinfo  {journal} {Phys. Rev. B}\ }\textbf {\bibinfo {volume}
  {85}},\ \bibinfo {pages} {125308} (\bibinfo {year} {2012})}\BibitemShut
  {NoStop}%
\bibitem [{\citenamefont {Dubail}\ \emph
  {et~al.}(2012{\natexlab{a}})\citenamefont {Dubail}, \citenamefont {Read},\
  and\ \citenamefont {Rezayi}}]{Dubail2012a}%
  \BibitemOpen
  \bibfield  {author} {\bibinfo {author} {\bibfnamefont {J.}~\bibnamefont
  {Dubail}}, \bibinfo {author} {\bibfnamefont {N.}~\bibnamefont {Read}},\ and\
  \bibinfo {author} {\bibfnamefont {E.~H.}\ \bibnamefont {Rezayi}},\ }\bibfield
   {title} {\bibinfo {title} {Real-space entanglement spectrum of quantum
  {Hall} systems},\ }\href {https://doi.org/10.1103/PhysRevB.85.115321}
  {\bibfield  {journal} {\bibinfo  {journal} {Phys. Rev. B}\ }\textbf {\bibinfo
  {volume} {85}},\ \bibinfo {pages} {115321} (\bibinfo {year}
  {2012}{\natexlab{a}})}\BibitemShut {NoStop}%
\bibitem [{\citenamefont {Rodr{\'i}guez}\ \emph {et~al.}(2012)\citenamefont
  {Rodr{\'i}guez}, \citenamefont {Simon},\ and\ \citenamefont
  {Slingerland}}]{Rodriguez2012}%
  \BibitemOpen
  \bibfield  {author} {\bibinfo {author} {\bibfnamefont {I.~D.}\ \bibnamefont
  {Rodr{\'i}guez}}, \bibinfo {author} {\bibfnamefont {S.~H.}\ \bibnamefont
  {Simon}},\ and\ \bibinfo {author} {\bibfnamefont {J.~K.}\ \bibnamefont
  {Slingerland}},\ }\bibfield  {title} {\bibinfo {title} {Evaluation of {Ranks}
  of {Real} {Space} and {Particle} {Entanglement} {Spectra} for {Large}
  {Systems}},\ }\href {https://doi.org/10.1103/PhysRevLett.108.256806}
  {\bibfield  {journal} {\bibinfo  {journal} {Phys. Rev. Lett.}\ }\textbf
  {\bibinfo {volume} {108}},\ \bibinfo {pages} {256806} (\bibinfo {year}
  {2012})}\BibitemShut {NoStop}%
\bibitem [{\citenamefont {Levin}\ and\ \citenamefont {Wen}(2006)}]{Levin2006}%
  \BibitemOpen
  \bibfield  {author} {\bibinfo {author} {\bibfnamefont {M.}~\bibnamefont
  {Levin}}\ and\ \bibinfo {author} {\bibfnamefont {X.-G.}\ \bibnamefont
  {Wen}},\ }\bibfield  {title} {\bibinfo {title} {Detecting {Topological}
  {Order} in a {Ground} {State} {Wave} {Function}},\ }\href
  {https://doi.org/10.1103/PhysRevLett.96.110405} {\bibfield  {journal}
  {\bibinfo  {journal} {Phys. Rev. Lett.}\ }\textbf {\bibinfo {volume} {96}},\
  \bibinfo {pages} {110405} (\bibinfo {year} {2006})}\BibitemShut {NoStop}%
\bibitem [{\citenamefont {Kitaev}\ and\ \citenamefont
  {Preskill}(2006)}]{Kitaev2006}%
  \BibitemOpen
  \bibfield  {author} {\bibinfo {author} {\bibfnamefont {A.}~\bibnamefont
  {Kitaev}}\ and\ \bibinfo {author} {\bibfnamefont {J.}~\bibnamefont
  {Preskill}},\ }\bibfield  {title} {\bibinfo {title} {Topological
  {Entanglement} {Entropy}},\ }\href
  {https://doi.org/10.1103/PhysRevLett.96.110404} {\bibfield  {journal}
  {\bibinfo  {journal} {Phys. Rev. Lett.}\ }\textbf {\bibinfo {volume} {96}},\
  \bibinfo {pages} {110404} (\bibinfo {year} {2006})}\BibitemShut {NoStop}%
\bibitem [{\citenamefont {Chen}\ \emph {et~al.}(2010)\citenamefont {Chen},
  \citenamefont {Gu},\ and\ \citenamefont {Wen}}]{Chen}%
  \BibitemOpen
  \bibfield  {author} {\bibinfo {author} {\bibfnamefont {X.}~\bibnamefont
  {Chen}}, \bibinfo {author} {\bibfnamefont {Z.~C.}\ \bibnamefont {Gu}},\ and\
  \bibinfo {author} {\bibfnamefont {X.~G.}\ \bibnamefont {Wen}},\ }\bibfield
  {title} {\bibinfo {title} {Local unitary transformation, long-range quantum
  entanglement, wave function renormalization, and topological order},\ }\href
  {https://doi.org/10.1103/PhysRevB.82.155138} {\bibfield  {journal} {\bibinfo
  {journal} {Phys. Rev. B}\ }\textbf {\bibinfo {volume} {82}},\ \bibinfo
  {pages} {155138} (\bibinfo {year} {2010})}\BibitemShut {NoStop}%
\bibitem [{\citenamefont {Wen}(1992)}]{wen_theory_1992}%
  \BibitemOpen
  \bibfield  {author} {\bibinfo {author} {\bibfnamefont {X.~G.}\ \bibnamefont
  {Wen}},\ }\bibfield  {title} {\bibinfo {title} {Theory of the edge states in
  fractional quantum hall effects},\ }\href
  {https://www.worldscientific.com/doi/abs/10.1142/S0217979292000840}
  {\bibfield  {journal} {\bibinfo  {journal} {Int. J. Mod. Phys. B}\ }\textbf
  {\bibinfo {volume} {06}},\ \bibinfo {pages} {1711} (\bibinfo {year}
  {1992})}\BibitemShut {NoStop}%
\bibitem [{\citenamefont {Moore}\ and\ \citenamefont {Read}(1991)}]{Moore1991}%
  \BibitemOpen
  \bibfield  {author} {\bibinfo {author} {\bibfnamefont {G.}~\bibnamefont
  {Moore}}\ and\ \bibinfo {author} {\bibfnamefont {N.}~\bibnamefont {Read}},\
  }\bibfield  {title} {\bibinfo {title} {Nonabelions in the fractional quantum
  hall effect},\ }\href {https://doi.org/10.1016/0550-3213(91)90407-O}
  {\bibfield  {journal} {\bibinfo  {journal} {Nuc. Phys. B}\ }\textbf {\bibinfo
  {volume} {360}},\ \bibinfo {pages} {362} (\bibinfo {year}
  {1991})}\BibitemShut {NoStop}%
\bibitem [{\citenamefont {Read}\ and\ \citenamefont {Rezayi}(1999)}]{Read1999}%
  \BibitemOpen
  \bibfield  {author} {\bibinfo {author} {\bibfnamefont {N.}~\bibnamefont
  {Read}}\ and\ \bibinfo {author} {\bibfnamefont {E.}~\bibnamefont {Rezayi}},\
  }\bibfield  {title} {\bibinfo {title} {Beyond paired quantum {Hall} states:
  {Parafermions} and incompressible states in the first excited {Landau}
  level},\ }\href {https://doi.org/10.1103/PhysRevB.59.8084} {\bibfield
  {journal} {\bibinfo  {journal} {Phys. Rev. B}\ }\textbf {\bibinfo {volume}
  {59}},\ \bibinfo {pages} {8084} (\bibinfo {year} {1999})}\BibitemShut
  {NoStop}%
\bibitem [{\citenamefont {Read}(2009)}]{Read2009}%
  \BibitemOpen
  \bibfield  {author} {\bibinfo {author} {\bibfnamefont {N.}~\bibnamefont
  {Read}},\ }\bibfield  {title} {\bibinfo {title} {Non-{Abelian} adiabatic
  statistics and {Hall} viscosity in quantum {Hall} states and px + ipy paired
  superfluids},\ }\href {https://doi.org/10.1103/PhysRevB.79.045308} {\bibfield
   {journal} {\bibinfo  {journal} {Phys. Rev. B}\ }\textbf {\bibinfo {volume}
  {79}},\ \bibinfo {pages} {045308} (\bibinfo {year} {2009})}\BibitemShut
  {NoStop}%
\bibitem [{Note1()}]{Note1}%
  \BibitemOpen
  \bibinfo {note} {This is up to an area dependant phase factor.}\BibitemShut
  {Stop}%
\bibitem [{\citenamefont {Bonderson}\ \emph {et~al.}(2011)\citenamefont
  {Bonderson}, \citenamefont {Gurarie},\ and\ \citenamefont
  {Nayak}}]{bonderson_plasma_2011}%
  \BibitemOpen
  \bibfield  {author} {\bibinfo {author} {\bibfnamefont {P.}~\bibnamefont
  {Bonderson}}, \bibinfo {author} {\bibfnamefont {V.}~\bibnamefont {Gurarie}},\
  and\ \bibinfo {author} {\bibfnamefont {C.}~\bibnamefont {Nayak}},\ }\bibfield
   {title} {\bibinfo {title} {Plasma analogy and non-{Abelian} statistics for
  {Ising}-type quantum {Hall} states},\ }\href
  {https://link.aps.org/doi/10.1103/PhysRevB.83.075303} {\bibfield  {journal}
  {\bibinfo  {journal} {Phys. Rev. B}\ }\textbf {\bibinfo {volume} {83}},\
  \bibinfo {pages} {075303} (\bibinfo {year} {2011})}\BibitemShut {NoStop}%
\bibitem [{\citenamefont {Wen}\ \emph {et~al.}(1994)\citenamefont {Wen},
  \citenamefont {Wu},\ and\ \citenamefont {Hatsugai}}]{Xiao-GangWen1994}%
  \BibitemOpen
  \bibfield  {author} {\bibinfo {author} {\bibfnamefont {X.~G.}\ \bibnamefont
  {Wen}}, \bibinfo {author} {\bibfnamefont {Y.~S.}\ \bibnamefont {Wu}},\ and\
  \bibinfo {author} {\bibfnamefont {Y.}~\bibnamefont {Hatsugai}},\ }\bibfield
  {title} {\bibinfo {title} {Chiral operator product algebra and edge
  excitations of a fractional quantum {Hall} droplet},\ }\href
  {https://doi.org/10.1016/0550-3213(94)90442-1} {\bibfield  {journal}
  {\bibinfo  {journal} {Nuc. Phys. B}\ }\textbf {\bibinfo {volume} {422}},\
  \bibinfo {pages} {476} (\bibinfo {year} {1994})}\BibitemShut {NoStop}%
\bibitem [{\citenamefont {Dubail}\ \emph
  {et~al.}(2012{\natexlab{b}})\citenamefont {Dubail}, \citenamefont {Read},\
  and\ \citenamefont {Rezayi}}]{Dubail2012}%
  \BibitemOpen
  \bibfield  {author} {\bibinfo {author} {\bibfnamefont {J.}~\bibnamefont
  {Dubail}}, \bibinfo {author} {\bibfnamefont {N.}~\bibnamefont {Read}},\ and\
  \bibinfo {author} {\bibfnamefont {E.~H.}\ \bibnamefont {Rezayi}},\ }\bibfield
   {title} {\bibinfo {title} {Edge-state inner products and real-space
  entanglement spectrum of trial quantum {Hall} states},\ }\href
  {https://doi.org/10.1103/PhysRevB.86.245310} {\bibfield  {journal} {\bibinfo
  {journal} {Phys. Rev. B}\ }\textbf {\bibinfo {volume} {86}},\ \bibinfo
  {pages} {1} (\bibinfo {year} {2012}{\natexlab{b}})}\BibitemShut {NoStop}%
\bibitem [{\citenamefont {Bernevig}\ \emph {et~al.}(2012)\citenamefont
  {Bernevig}, \citenamefont {Bonderson},\ and\ \citenamefont
  {Regnault}}]{bernevig_screening_2012}%
  \BibitemOpen
  \bibfield  {author} {\bibinfo {author} {\bibfnamefont {B.~A.}\ \bibnamefont
  {Bernevig}}, \bibinfo {author} {\bibfnamefont {P.}~\bibnamefont
  {Bonderson}},\ and\ \bibinfo {author} {\bibfnamefont {N.}~\bibnamefont
  {Regnault}},\ }\bibfield  {title} {\bibinfo {title} {Screening {Behavior} and
  {Scaling} {Exponents} from {Quantum} {Hall} {Wavefunctions}},\ }\href
  {http://arxiv.org/abs/1207.3305} {\bibfield  {journal} {\bibinfo  {journal}
  {arXiv preprint arXiv:1207.3305}\ } (\bibinfo {year} {2012})}\BibitemShut
  {NoStop}%
\bibitem [{\citenamefont {Fern}\ \emph
  {et~al.}(2018{\natexlab{a}})\citenamefont {Fern}, \citenamefont {Bondesan},\
  and\ \citenamefont {Simon}}]{Fern2018}%
  \BibitemOpen
  \bibfield  {author} {\bibinfo {author} {\bibfnamefont {R.}~\bibnamefont
  {Fern}}, \bibinfo {author} {\bibfnamefont {R.}~\bibnamefont {Bondesan}},\
  and\ \bibinfo {author} {\bibfnamefont {S.~H.}\ \bibnamefont {Simon}},\
  }\bibfield  {title} {\bibinfo {title} {Structure of edge-state inner products
  in the fractional quantum {Hall} effect},\ }\href
  {https://doi.org/10.1103/PhysRevB.97.155108} {\bibfield  {journal} {\bibinfo
  {journal} {Phys. Rev. B}\ }\textbf {\bibinfo {volume} {97}},\ \bibinfo
  {pages} {155108} (\bibinfo {year} {2018}{\natexlab{a}})}\BibitemShut
  {NoStop}%
\bibitem [{\citenamefont {Wu}\ \emph {et~al.}(2014)\citenamefont {Wu},
  \citenamefont {Estienne}, \citenamefont {Regnault},\ and\ \citenamefont
  {Bernevig}}]{wu_braiding_2014}%
  \BibitemOpen
  \bibfield  {author} {\bibinfo {author} {\bibfnamefont {Y.~L.}\ \bibnamefont
  {Wu}}, \bibinfo {author} {\bibfnamefont {B.}~\bibnamefont {Estienne}},
  \bibinfo {author} {\bibfnamefont {N.}~\bibnamefont {Regnault}},\ and\
  \bibinfo {author} {\bibfnamefont {B.~A.}\ \bibnamefont {Bernevig}},\
  }\bibfield  {title} {\bibinfo {title} {Braiding {Non}-{Abelian} {Quasiholes}
  in {Fractional} {Quantum} {Hall} {States}},\ }\href
  {https://link.aps.org/doi/10.1103/PhysRevLett.113.116801} {\bibfield
  {journal} {\bibinfo  {journal} {Phys. Rev. Lett.}\ }\textbf {\bibinfo
  {volume} {113}},\ \bibinfo {pages} {116801} (\bibinfo {year}
  {2014})}\BibitemShut {NoStop}%
\bibitem [{\citenamefont {Tserkovnyak}\ and\ \citenamefont
  {Simon}(2003)}]{tserkovnyak_monte_2003}%
  \BibitemOpen
  \bibfield  {author} {\bibinfo {author} {\bibfnamefont {Y.}~\bibnamefont
  {Tserkovnyak}}\ and\ \bibinfo {author} {\bibfnamefont {S.~H.}\ \bibnamefont
  {Simon}},\ }\bibfield  {title} {\bibinfo {title} {Monte {Carlo} {Evaluation}
  of {Non}-{Abelian} {Statistics}},\ }\href
  {https://link.aps.org/doi/10.1103/PhysRevLett.90.016802} {\bibfield
  {journal} {\bibinfo  {journal} {Phys. Rev. Lett.}\ }\textbf {\bibinfo
  {volume} {90}},\ \bibinfo {pages} {016802} (\bibinfo {year}
  {2003})}\BibitemShut {NoStop}%
\bibitem [{\citenamefont {Baraban}\ \emph {et~al.}(2009)\citenamefont
  {Baraban}, \citenamefont {Zikos}, \citenamefont {Bonesteel},\ and\
  \citenamefont {Simon}}]{baraban_numerical_2009}%
  \BibitemOpen
  \bibfield  {author} {\bibinfo {author} {\bibfnamefont {M.}~\bibnamefont
  {Baraban}}, \bibinfo {author} {\bibfnamefont {G.}~\bibnamefont {Zikos}},
  \bibinfo {author} {\bibfnamefont {N.}~\bibnamefont {Bonesteel}},\ and\
  \bibinfo {author} {\bibfnamefont {S.~H.}\ \bibnamefont {Simon}},\ }\bibfield
  {title} {\bibinfo {title} {Numerical {Analysis} of {Quasiholes} of the
  {Moore}-{Read} {Wave} {Function}},\ }\href
  {https://doi.org/10.1103/PhysRevLett.103.076801} {\bibfield  {journal}
  {\bibinfo  {journal} {Phys. Rev. Lett.}\ }\textbf {\bibinfo {volume} {103}},\
  \bibinfo {pages} {076801} (\bibinfo {year} {2009})}\BibitemShut {NoStop}%
\bibitem [{\citenamefont {Zaletel}\ and\ \citenamefont
  {Mong}(2012)}]{Zaletel2012}%
  \BibitemOpen
  \bibfield  {author} {\bibinfo {author} {\bibfnamefont {M.~P.}\ \bibnamefont
  {Zaletel}}\ and\ \bibinfo {author} {\bibfnamefont {R.~S.}\ \bibnamefont
  {Mong}},\ }\bibfield  {title} {\bibinfo {title} {Exact matrix product states
  for quantum {Hall} wave functions},\ }\href
  {https://doi.org/10.1103/PhysRevB.86.245305} {\bibfield  {journal} {\bibinfo
  {journal} {Phys. Rev. B}\ }\textbf {\bibinfo {volume} {86}},\ \bibinfo
  {pages} {245305} (\bibinfo {year} {2012})}\BibitemShut {NoStop}%
\bibitem [{\citenamefont {Estienne}\ \emph
  {et~al.}(2013{\natexlab{a}})\citenamefont {Estienne}, \citenamefont
  {Papi{\'c}}, \citenamefont {Regnault},\ and\ \citenamefont
  {Bernevig}}]{estienne_matrix_2013}%
  \BibitemOpen
  \bibfield  {author} {\bibinfo {author} {\bibfnamefont {B.}~\bibnamefont
  {Estienne}}, \bibinfo {author} {\bibfnamefont {Z.}~\bibnamefont {Papi{\'c}}},
  \bibinfo {author} {\bibfnamefont {N.}~\bibnamefont {Regnault}},\ and\
  \bibinfo {author} {\bibfnamefont {B.~A.}\ \bibnamefont {Bernevig}},\
  }\bibfield  {title} {\bibinfo {title} {Matrix product states for trial
  quantum {Hall} states},\ }\href
  {https://link.aps.org/doi/10.1103/PhysRevB.87.161112} {\bibfield  {journal}
  {\bibinfo  {journal} {Phys. Rev. B}\ }\textbf {\bibinfo {volume} {87}},\
  \bibinfo {pages} {161112} (\bibinfo {year} {2013}{\natexlab{a}})}\BibitemShut
  {NoStop}%
\bibitem [{\citenamefont {Estienne}\ \emph
  {et~al.}(2013{\natexlab{b}})\citenamefont {Estienne}, \citenamefont
  {Regnault},\ and\ \citenamefont {Bernevig}}]{estienne_fractional_2013}%
  \BibitemOpen
  \bibfield  {author} {\bibinfo {author} {\bibfnamefont {B.}~\bibnamefont
  {Estienne}}, \bibinfo {author} {\bibfnamefont {N.}~\bibnamefont {Regnault}},\
  and\ \bibinfo {author} {\bibfnamefont {B.~A.}\ \bibnamefont {Bernevig}},\
  }\bibfield  {title} {\bibinfo {title} {Fractional {Quantum} {Hall} {Matrix}
  {Product} {States} {For} {Interacting} {Conformal} {Field} {Theories}},\
  }\href {http://arxiv.org/abs/1311.2936} {\bibfield  {journal} {\bibinfo
  {journal} {arXiv preprint arXiv:1311.2936}\ } (\bibinfo {year}
  {2013}{\natexlab{b}})}\BibitemShut {NoStop}%
\bibitem [{\citenamefont {Jain}(2007)}]{Jain2007}%
  \BibitemOpen
  \bibfield  {author} {\bibinfo {author} {\bibfnamefont {J.}~\bibnamefont
  {Jain}},\ }\href {https://doi.org/10.1017/CBO9780511607561} {\emph {\bibinfo
  {title} {Composite fermions}}},\ Vol.\ \bibinfo {volume} {9780521862}\
  (\bibinfo  {publisher} {Cambridge University Press},\ \bibinfo {year}
  {2007})\ \bibinfo {note} {publication Title: Composite fermions}\BibitemShut
  {NoStop}%
\bibitem [{\citenamefont {Jain}(1989{\natexlab{a}})}]{Jain1989}%
  \BibitemOpen
  \bibfield  {author} {\bibinfo {author} {\bibfnamefont {J.~K.}\ \bibnamefont
  {Jain}},\ }\bibfield  {title} {\bibinfo {title} {Composite-{Fermion}
  {Approach} for the {Fractional} {Quantum} {Hall} {Effect}},\ }\href
  {https://journals.aps.org/prl/abstract/10.1103/PhysRevLett.63.199} {\bibfield
   {journal} {\bibinfo  {journal} {Phys. Rev. Lett.}\ }\textbf {\bibinfo
  {volume} {63}},\ \bibinfo {pages} {199} (\bibinfo {year}
  {1989}{\natexlab{a}})}\BibitemShut {NoStop}%
\bibitem [{\citenamefont {Hansson}\ \emph {et~al.}(2007)\citenamefont
  {Hansson}, \citenamefont {Chang}, \citenamefont {Jain},\ and\ \citenamefont
  {Viefers}}]{Hansson}%
  \BibitemOpen
  \bibfield  {author} {\bibinfo {author} {\bibfnamefont {T.~H.}\ \bibnamefont
  {Hansson}}, \bibinfo {author} {\bibfnamefont {C.~C.}\ \bibnamefont {Chang}},
  \bibinfo {author} {\bibfnamefont {J.~K.}\ \bibnamefont {Jain}},\ and\
  \bibinfo {author} {\bibfnamefont {S.}~\bibnamefont {Viefers}},\ }\bibfield
  {title} {\bibinfo {title} {Composite-fermion wave functions as correlators in
  conformal field theory},\ }\href {https://doi.org/10.1103/PhysRevB.76.075347}
  {\bibfield  {journal} {\bibinfo  {journal} {Phys. Rev. B}\ }\textbf {\bibinfo
  {volume} {76}},\ \bibinfo {pages} {075347} (\bibinfo {year} {2007})},\
  \bibinfo {note} {iSBN: 0753472007}\BibitemShut {NoStop}%
\bibitem [{\citenamefont {Suorsa}\ \emph
  {et~al.}(2011{\natexlab{a}})\citenamefont {Suorsa}, \citenamefont {Viefers},\
  and\ \citenamefont {Hansson}}]{Suorsa2011a}%
  \BibitemOpen
  \bibfield  {author} {\bibinfo {author} {\bibfnamefont {J.}~\bibnamefont
  {Suorsa}}, \bibinfo {author} {\bibfnamefont {S.}~\bibnamefont {Viefers}},\
  and\ \bibinfo {author} {\bibfnamefont {T.~H.}\ \bibnamefont {Hansson}},\
  }\bibfield  {title} {\bibinfo {title} {A general approach to quantum {Hall}
  hierarchies},\ }\href {https://doi.org/10.1088/1367-2630/13/7/075006}
  {\bibfield  {journal} {\bibinfo  {journal} {New J. Phys.}\ }\textbf {\bibinfo
  {volume} {13}},\ \bibinfo {pages} {75006} (\bibinfo {year}
  {2011}{\natexlab{a}})}\BibitemShut {NoStop}%
\bibitem [{\citenamefont {Kvorning}(2013)}]{Kvorning2013}%
  \BibitemOpen
  \bibfield  {author} {\bibinfo {author} {\bibfnamefont {T.}~\bibnamefont
  {Kvorning}},\ }\bibfield  {title} {\bibinfo {title} {Quantum {Hall} hierarchy
  in a spherical geometry},\ }\href
  {https://doi.org/10.1103/PhysRevB.87.195131} {\bibfield  {journal} {\bibinfo
  {journal} {Phy. Rev. B}\ }\textbf {\bibinfo {volume} {87}},\ \bibinfo {pages}
  {195131} (\bibinfo {year} {2013})}\BibitemShut {NoStop}%
\bibitem [{\citenamefont {Hansson}\ \emph {et~al.}(2017)\citenamefont
  {Hansson}, \citenamefont {Hermanns}, \citenamefont {Simon},\ and\
  \citenamefont {Viefers}}]{Hansson2017}%
  \BibitemOpen
  \bibfield  {author} {\bibinfo {author} {\bibfnamefont {T.~H.}\ \bibnamefont
  {Hansson}}, \bibinfo {author} {\bibfnamefont {M.}~\bibnamefont {Hermanns}},
  \bibinfo {author} {\bibfnamefont {S.~H.}\ \bibnamefont {Simon}},\ and\
  \bibinfo {author} {\bibfnamefont {S.~F.}\ \bibnamefont {Viefers}},\
  }\bibfield  {title} {\bibinfo {title} {Quantum {Hall} {Physics}-hierarchies
  and {CFT} techniques},\ }\href {https://arxiv.org/abs/1601.01697} {\bibfield
  {journal} {\bibinfo  {journal} {Rev. Mod. Phys.}\ }\textbf {\bibinfo {volume}
  {89}},\ \bibinfo {pages} {025005} (\bibinfo {year} {2017})}\BibitemShut
  {NoStop}%
\bibitem [{\citenamefont
  {Jain}(1989{\natexlab{b}})}]{jain_incompressible_1989}%
  \BibitemOpen
  \bibfield  {author} {\bibinfo {author} {\bibfnamefont {J.}~\bibnamefont
  {Jain}},\ }\bibfield  {title} {\bibinfo {title} {Incompressible quantum
  {Hall} states},\ }\href {https://link.aps.org/doi/10.1103/PhysRevB.40.8079}
  {\bibfield  {journal} {\bibinfo  {journal} {Phys. Rev. B}\ }\textbf {\bibinfo
  {volume} {40}},\ \bibinfo {pages} {8079} (\bibinfo {year}
  {1989}{\natexlab{b}})}\BibitemShut {NoStop}%
\bibitem [{\citenamefont {Balram}\ \emph
  {et~al.}(2018{\natexlab{a}})\citenamefont {Balram}, \citenamefont
  {Barkeshli},\ and\ \citenamefont {Rudner}}]{balram_parton_2018}%
  \BibitemOpen
  \bibfield  {author} {\bibinfo {author} {\bibfnamefont {A.~C.}\ \bibnamefont
  {Balram}}, \bibinfo {author} {\bibfnamefont {M.}~\bibnamefont {Barkeshli}},\
  and\ \bibinfo {author} {\bibfnamefont {M.~S.}\ \bibnamefont {Rudner}},\
  }\bibfield  {title} {\bibinfo {title} {Parton construction of a wave function
  in the anti-{Pfaffian} phase},\ }\href
  {https://link.aps.org/doi/10.1103/PhysRevB.98.035127} {\bibfield  {journal}
  {\bibinfo  {journal} {Phys. Rev. B}\ }\textbf {\bibinfo {volume} {98}},\
  \bibinfo {pages} {035127} (\bibinfo {year} {2018}{\natexlab{a}})}\BibitemShut
  {NoStop}%
\bibitem [{\citenamefont {Balram}\ \emph
  {et~al.}(2018{\natexlab{b}})\citenamefont {Balram}, \citenamefont
  {Mukherjee}, \citenamefont {Park}, \citenamefont {Barkeshli}, \citenamefont
  {Rudner},\ and\ \citenamefont {Jain}}]{balram_fractional_2018}%
  \BibitemOpen
  \bibfield  {author} {\bibinfo {author} {\bibfnamefont {A.~C.}\ \bibnamefont
  {Balram}}, \bibinfo {author} {\bibfnamefont {S.}~\bibnamefont {Mukherjee}},
  \bibinfo {author} {\bibfnamefont {K.}~\bibnamefont {Park}}, \bibinfo {author}
  {\bibfnamefont {M.}~\bibnamefont {Barkeshli}}, \bibinfo {author}
  {\bibfnamefont {M.~S.}\ \bibnamefont {Rudner}},\ and\ \bibinfo {author}
  {\bibfnamefont {J.~K.}\ \bibnamefont {Jain}},\ }\bibfield  {title} {\bibinfo
  {title} {Fractional {Quantum} {Hall} {Effect} at $\nu$ = 2 + 6 / 13 : {The}
  {Parton} {Paradigm} for the {Second} {Landau} {Level}},\ }\href
  {https://link.aps.org/doi/10.1103/PhysRevLett.121.186601} {\bibfield
  {journal} {\bibinfo  {journal} {Phys. Rev. Lett.}\ }\textbf {\bibinfo
  {volume} {121}},\ \bibinfo {pages} {186601} (\bibinfo {year}
  {2018}{\natexlab{b}})}\BibitemShut {NoStop}%
\bibitem [{\citenamefont {Balram}\ \emph {et~al.}(2019)\citenamefont {Balram},
  \citenamefont {Barkeshli},\ and\ \citenamefont
  {Rudner}}]{balram_parton_2019}%
  \BibitemOpen
  \bibfield  {author} {\bibinfo {author} {\bibfnamefont {A.~C.}\ \bibnamefont
  {Balram}}, \bibinfo {author} {\bibfnamefont {M.}~\bibnamefont {Barkeshli}},\
  and\ \bibinfo {author} {\bibfnamefont {M.~S.}\ \bibnamefont {Rudner}},\
  }\bibfield  {title} {\bibinfo {title} {Parton construction of
  particle-hole-conjugate {Read}-{Rezayi} parafermion fractional quantum {Hall}
  states and beyond},\ }\href
  {https://link.aps.org/doi/10.1103/PhysRevB.99.241108} {\bibfield  {journal}
  {\bibinfo  {journal} {Phys. Rev. B}\ }\textbf {\bibinfo {volume} {99}},\
  \bibinfo {pages} {241108} (\bibinfo {year} {2019})}\BibitemShut {NoStop}%
\bibitem [{\citenamefont {Balram}\ and\ \citenamefont
  {W{\'o}js}(2020)}]{balram_fractional_2020}%
  \BibitemOpen
  \bibfield  {author} {\bibinfo {author} {\bibfnamefont {A.~C.}\ \bibnamefont
  {Balram}}\ and\ \bibinfo {author} {\bibfnamefont {A.}~\bibnamefont
  {W{\'o}js}},\ }\bibfield  {title} {\bibinfo {title} {Fractional quantum
  {Hall} effect at $\nu$ = 2 + 4 / 9},\ }\href
  {https://doi.org/10.1103/PhysRevResearch.2.032035} {\bibfield  {journal}
  {\bibinfo  {journal} {Phys. Rev. Res.}\ }\textbf {\bibinfo {volume} {2}},\
  \bibinfo {pages} {032035} (\bibinfo {year} {2020})}\BibitemShut {NoStop}%
\bibitem [{\citenamefont
  {Balram}(2021{\natexlab{a}})}]{coimbatore_balram_non-abelian_2021}%
  \BibitemOpen
  \bibfield  {author} {\bibinfo {author} {\bibfnamefont {A.~C.}\ \bibnamefont
  {Balram}},\ }\bibfield  {title} {\bibinfo {title} {A non-{Abelian} parton
  state for the $\nu=2+3/8$ fractional quantum {Hall} effect},\ }\href
  {https://scipost.org/10.21468/SciPostPhys.10.4.083} {\bibfield  {journal}
  {\bibinfo  {journal} {SciPost Phys.}\ }\textbf {\bibinfo {volume} {10}},\
  \bibinfo {pages} {083} (\bibinfo {year} {2021}{\natexlab{a}})}\BibitemShut
  {NoStop}%
\bibitem [{\citenamefont {Faugno}\ \emph {et~al.}(2020)\citenamefont {Faugno},
  \citenamefont {Jain},\ and\ \citenamefont
  {Balram}}]{faugno_non-abelian_2020}%
  \BibitemOpen
  \bibfield  {author} {\bibinfo {author} {\bibfnamefont {W.~N.}\ \bibnamefont
  {Faugno}}, \bibinfo {author} {\bibfnamefont {J.~K.}\ \bibnamefont {Jain}},\
  and\ \bibinfo {author} {\bibfnamefont {A.~C.}\ \bibnamefont {Balram}},\
  }\bibfield  {title} {\bibinfo {title} {Non-{Abelian} fractional quantum
  {Hall} state at 3 / 7 -filled {Landau} level},\ }\href
  {https://doi.org/10.1103/PhysRevResearch.2.033223} {\bibfield  {journal}
  {\bibinfo  {journal} {Phys. Rev. Res.}\ }\textbf {\bibinfo {volume} {2}},\
  \bibinfo {pages} {033223} (\bibinfo {year} {2020})}\BibitemShut {NoStop}%
\bibitem [{\citenamefont {Faugno}\ \emph {et~al.}(2021)\citenamefont {Faugno},
  \citenamefont {Zhao}, \citenamefont {Balram}, \citenamefont {Jolicoeur},\
  and\ \citenamefont {Jain}}]{faugno_unconventional_2021}%
  \BibitemOpen
  \bibfield  {author} {\bibinfo {author} {\bibfnamefont {W.~N.}\ \bibnamefont
  {Faugno}}, \bibinfo {author} {\bibfnamefont {T.}~\bibnamefont {Zhao}},
  \bibinfo {author} {\bibfnamefont {A.~C.}\ \bibnamefont {Balram}}, \bibinfo
  {author} {\bibfnamefont {T.}~\bibnamefont {Jolicoeur}},\ and\ \bibinfo
  {author} {\bibfnamefont {J.~K.}\ \bibnamefont {Jain}},\ }\bibfield  {title}
  {\bibinfo {title} {Unconventional {Z} n parton states at $\nu$ = 7 / 3 :
  {Role} of finite width},\ }\href
  {https://doi.org/10.1103/PhysRevB.103.085303} {\bibfield  {journal} {\bibinfo
   {journal} {Phys. Rev. B}\ }\textbf {\bibinfo {volume} {103}},\ \bibinfo
  {pages} {085303} (\bibinfo {year} {2021})}\BibitemShut {NoStop}%
\bibitem [{\citenamefont {Wu}\ \emph {et~al.}(2017)\citenamefont {Wu},
  \citenamefont {Shi},\ and\ \citenamefont {Jain}}]{wu_non-abelian_2017}%
  \BibitemOpen
  \bibfield  {author} {\bibinfo {author} {\bibfnamefont {Y.~H.}\ \bibnamefont
  {Wu}}, \bibinfo {author} {\bibfnamefont {T.}~\bibnamefont {Shi}},\ and\
  \bibinfo {author} {\bibfnamefont {J.~K.}\ \bibnamefont {Jain}},\ }\bibfield
  {title} {\bibinfo {title} {Non-{Abelian} {Parton} {Fractional} {Quantum}
  {Hall} {Effect} in {Multilayer} {Graphene}},\ }\href
  {https://pubs.acs.org/doi/10.1021/acs.nanolett.7b01080} {\bibfield  {journal}
  {\bibinfo  {journal} {Nano Lett.}\ }\textbf {\bibinfo {volume} {17}},\
  \bibinfo {pages} {4643} (\bibinfo {year} {2017})}\BibitemShut {NoStop}%
\bibitem [{\citenamefont {Kim}\ \emph {et~al.}(2019)\citenamefont {Kim},
  \citenamefont {Balram}, \citenamefont {Taniguchi}, \citenamefont {Watanabe},
  \citenamefont {Jain},\ and\ \citenamefont {Smet}}]{kim_even_2019}%
  \BibitemOpen
  \bibfield  {author} {\bibinfo {author} {\bibfnamefont {Y.}~\bibnamefont
  {Kim}}, \bibinfo {author} {\bibfnamefont {A.~C.}\ \bibnamefont {Balram}},
  \bibinfo {author} {\bibfnamefont {T.}~\bibnamefont {Taniguchi}}, \bibinfo
  {author} {\bibfnamefont {K.}~\bibnamefont {Watanabe}}, \bibinfo {author}
  {\bibfnamefont {J.~K.}\ \bibnamefont {Jain}},\ and\ \bibinfo {author}
  {\bibfnamefont {J.~H.}\ \bibnamefont {Smet}},\ }\bibfield  {title} {\bibinfo
  {title} {Even denominator fractional quantum {Hall} states in higher {Landau}
  levels of graphene},\ }\href {https://doi.org/10.1038/s41567-018-0355-x}
  {\bibfield  {journal} {\bibinfo  {journal} {Nat. Phys.}\ }\textbf {\bibinfo
  {volume} {15}},\ \bibinfo {pages} {154} (\bibinfo {year} {2019})}\BibitemShut
  {NoStop}%
\bibitem [{\citenamefont {Timmel}\ and\ \citenamefont
  {Wen}(2023)}]{timmel_non-abelian_2023}%
  \BibitemOpen
  \bibfield  {author} {\bibinfo {author} {\bibfnamefont {A.}~\bibnamefont
  {Timmel}}\ and\ \bibinfo {author} {\bibfnamefont {X.-G.}\ \bibnamefont
  {Wen}},\ }\href {http://arxiv.org/abs/2308.09702} {\bibinfo {title}
  {Non-{Abelian} quantum {Hall} states in multi-layer rhombohedral stacked
  graphene}} (\bibinfo {year} {2023})\BibitemShut {NoStop}%
\bibitem [{\citenamefont {Balram}(2021{\natexlab{b}})}]{balram_abelian_2021}%
  \BibitemOpen
  \bibfield  {author} {\bibinfo {author} {\bibfnamefont {A.~C.}\ \bibnamefont
  {Balram}},\ }\bibfield  {title} {\bibinfo {title} {Abelian parton state for
  the $\nu$ = 4 / 11 fractional quantum {Hall} effect},\ }\href
  {https://doi.org/10.1103/PhysRevB.103.155103} {\bibfield  {journal} {\bibinfo
   {journal} {Phys. Rev. B}\ }\textbf {\bibinfo {volume} {103}},\ \bibinfo
  {pages} {155103} (\bibinfo {year} {2021}{\natexlab{b}})}\BibitemShut
  {NoStop}%
\bibitem [{\citenamefont {Balram}\ and\ \citenamefont
  {W{\'o}js}(2021)}]{balram_parton_2021}%
  \BibitemOpen
  \bibfield  {author} {\bibinfo {author} {\bibfnamefont {A.~C.}\ \bibnamefont
  {Balram}}\ and\ \bibinfo {author} {\bibfnamefont {A.}~\bibnamefont
  {W{\'o}js}},\ }\bibfield  {title} {\bibinfo {title} {Parton wave function for
  the fractional quantum {Hall} effect at $\nu$ = 6 / 17},\ }\href
  {https://doi.org/10.1103/PhysRevResearch.3.033087} {\bibfield  {journal}
  {\bibinfo  {journal} {Phys. Rev. Res.}\ }\textbf {\bibinfo {volume} {3}},\
  \bibinfo {pages} {033087} (\bibinfo {year} {2021})}\BibitemShut {NoStop}%
\bibitem [{\citenamefont {Balram}\ \emph {et~al.}(2022)\citenamefont {Balram},
  \citenamefont {Liu}, \citenamefont {Gromov},\ and\ \citenamefont
  {Papi{\'c}}}]{balram_very-high-energy_2022}%
  \BibitemOpen
  \bibfield  {author} {\bibinfo {author} {\bibfnamefont {A.~C.}\ \bibnamefont
  {Balram}}, \bibinfo {author} {\bibfnamefont {Z.}~\bibnamefont {Liu}},
  \bibinfo {author} {\bibfnamefont {A.}~\bibnamefont {Gromov}},\ and\ \bibinfo
  {author} {\bibfnamefont {Z.}~\bibnamefont {Papi{\'c}}},\ }\bibfield  {title}
  {\bibinfo {title} {Very-{High}-{Energy} {Collective} {States} of {Partons} in
  {Fractional} {Quantum} {Hall} {Liquids}},\ }\href
  {https://doi.org/10.1103/PhysRevX.12.021008} {\bibfield  {journal} {\bibinfo
  {journal} {Phys. Rev. X}\ }\textbf {\bibinfo {volume} {12}},\ \bibinfo
  {pages} {021008} (\bibinfo {year} {2022})}\BibitemShut {NoStop}%
\bibitem [{\citenamefont {Dora}\ and\ \citenamefont
  {Balram}(2022)}]{dora_nature_2022}%
  \BibitemOpen
  \bibfield  {author} {\bibinfo {author} {\bibfnamefont {R.~K.}\ \bibnamefont
  {Dora}}\ and\ \bibinfo {author} {\bibfnamefont {A.~C.}\ \bibnamefont
  {Balram}},\ }\bibfield  {title} {\bibinfo {title} {Nature of the anomalous 4
  / 13 fractional quantum {Hall} effect in graphene},\ }\href
  {https://doi.org/10.1103/PhysRevB.105.L241403} {\bibfield  {journal}
  {\bibinfo  {journal} {Phys. Rev. B}\ }\textbf {\bibinfo {volume} {105}},\
  \bibinfo {pages} {L241403} (\bibinfo {year} {2022})}\BibitemShut {NoStop}%
\bibitem [{\citenamefont {Wen}(1991)}]{Wen1991a}%
  \BibitemOpen
  \bibfield  {author} {\bibinfo {author} {\bibfnamefont {X.~G.}\ \bibnamefont
  {Wen}},\ }\bibfield  {title} {\bibinfo {title} {Non-{Abelian} statistics in
  the fractional quantum {Hall} states},\ }\href
  {https://doi.org/10.1103/PhysRevLett.66.802} {\bibfield  {journal} {\bibinfo
  {journal} {Phys. Rev. Lett.}\ }\textbf {\bibinfo {volume} {66}},\ \bibinfo
  {pages} {802} (\bibinfo {year} {1991})}\BibitemShut {NoStop}%
\bibitem [{\citenamefont {Bandyopadhyay}\ \emph {et~al.}(2018)\citenamefont
  {Bandyopadhyay}, \citenamefont {Chen}, \citenamefont {Ahari}, \citenamefont
  {Ortiz}, \citenamefont {Nussinov},\ and\ \citenamefont
  {Seidel}}]{bandyopadhyay_entangled_2018}%
  \BibitemOpen
  \bibfield  {author} {\bibinfo {author} {\bibfnamefont {S.}~\bibnamefont
  {Bandyopadhyay}}, \bibinfo {author} {\bibfnamefont {L.}~\bibnamefont {Chen}},
  \bibinfo {author} {\bibfnamefont {M.~T.}\ \bibnamefont {Ahari}}, \bibinfo
  {author} {\bibfnamefont {G.}~\bibnamefont {Ortiz}}, \bibinfo {author}
  {\bibfnamefont {Z.}~\bibnamefont {Nussinov}},\ and\ \bibinfo {author}
  {\bibfnamefont {A.}~\bibnamefont {Seidel}},\ }\bibfield  {title} {\bibinfo
  {title} {Entangled {Pauli} principles: {The} {DNA} of quantum {Hall}
  fluids},\ }\href {https://link.aps.org/doi/10.1103/PhysRevB.98.161118}
  {\bibfield  {journal} {\bibinfo  {journal} {Phys. Rev. B}\ }\textbf {\bibinfo
  {volume} {98}},\ \bibinfo {pages} {161118} (\bibinfo {year}
  {2018})}\BibitemShut {NoStop}%
\bibitem [{\citenamefont {Tanhayi~Ahari}\ \emph {et~al.}(2022)\citenamefont
  {Tanhayi~Ahari}, \citenamefont {Bandyopadhyay}, \citenamefont {Nussinov},
  \citenamefont {Seidel},\ and\ \citenamefont {Ortiz}}]{ahari_partons_2022}%
  \BibitemOpen
  \bibfield  {author} {\bibinfo {author} {\bibfnamefont {M.}~\bibnamefont
  {Tanhayi~Ahari}}, \bibinfo {author} {\bibfnamefont {S.}~\bibnamefont
  {Bandyopadhyay}}, \bibinfo {author} {\bibfnamefont {Z.}~\bibnamefont
  {Nussinov}}, \bibinfo {author} {\bibfnamefont {A.}~\bibnamefont {Seidel}},\
  and\ \bibinfo {author} {\bibfnamefont {G.}~\bibnamefont {Ortiz}},\ }\bibfield
   {title} {\bibinfo {title} {Partons as unique ground states of quantum hall
  parent hamiltonians: The case of fibonacci anyons},\ }\href
  {https://arxiv.org/abs/2204.09684} {\bibfield  {journal} {\bibinfo  {journal}
  {arXiv preprint arXiv:2204.09684}\ } (\bibinfo {year} {2022})}\BibitemShut
  {NoStop}%
\bibitem [{\citenamefont {Anand}\ \emph {et~al.}(2022)\citenamefont {Anand},
  \citenamefont {Patil}, \citenamefont {Balram},\ and\ \citenamefont
  {Sreejith}}]{anand_real-space_2022}%
  \BibitemOpen
  \bibfield  {author} {\bibinfo {author} {\bibfnamefont {A.}~\bibnamefont
  {Anand}}, \bibinfo {author} {\bibfnamefont {R.~A.}\ \bibnamefont {Patil}},
  \bibinfo {author} {\bibfnamefont {A.~C.}\ \bibnamefont {Balram}},\ and\
  \bibinfo {author} {\bibfnamefont {G.~J.}\ \bibnamefont {Sreejith}},\
  }\bibfield  {title} {\bibinfo {title} {Real-space entanglement spectra of
  parton states in fractional quantum {Hall} systems},\ }\href
  {https://link.aps.org/doi/10.1103/PhysRevB.106.085136} {\bibfield  {journal}
  {\bibinfo  {journal} {Phys. Rev. B}\ }\textbf {\bibinfo {volume} {106}},\
  \bibinfo {pages} {085136} (\bibinfo {year} {2022})}\BibitemShut {NoStop}%
\bibitem [{\citenamefont {Fuchs}\ and\ \citenamefont
  {Gepner}(1994)}]{fuchs_braiding_1994}%
  \BibitemOpen
  \bibfield  {author} {\bibinfo {author} {\bibfnamefont {J.}~\bibnamefont
  {Fuchs}}\ and\ \bibinfo {author} {\bibfnamefont {D.}~\bibnamefont {Gepner}},\
  }\bibfield  {title} {\bibinfo {title} {Braiding in conformal field theory and
  solvable lattice models},\ }\href
  {https://doi.org/10.1016/0550-3213(94)90004-3} {\bibfield  {journal}
  {\bibinfo  {journal} {Nuc. Phys. B}\ }\textbf {\bibinfo {volume} {413}},\
  \bibinfo {pages} {614} (\bibinfo {year} {1994})}\BibitemShut {NoStop}%
\bibitem [{\citenamefont {Tsuchiya}\ and\ \citenamefont
  {Kanie}(1987)}]{tsuchiya_vertex_1987}%
  \BibitemOpen
  \bibfield  {author} {\bibinfo {author} {\bibfnamefont {A.}~\bibnamefont
  {Tsuchiya}}\ and\ \bibinfo {author} {\bibfnamefont {Y.}~\bibnamefont
  {Kanie}},\ }\bibfield  {title} {\bibinfo {title} {Vertex operators in the
  conformal field theory on {P1} and monodromy representations of the braid
  group},\ }\href {https://doi.org/10.1007/BF00401159} {\bibfield  {journal}
  {\bibinfo  {journal} {Lett. Math. Phys.}\ }\textbf {\bibinfo {volume} {13}},\
  \bibinfo {pages} {303} (\bibinfo {year} {1987})}\BibitemShut {NoStop}%
\bibitem [{\citenamefont {Witten}(1989)}]{Witten1989}%
  \BibitemOpen
  \bibfield  {author} {\bibinfo {author} {\bibfnamefont {E.}~\bibnamefont
  {Witten}},\ }\bibfield  {title} {\bibinfo {title} {Quantum field theory and
  the {Jones} polynomial},\ }\href {https://doi.org/10.1007/BF01217730}
  {\bibfield  {journal} {\bibinfo  {journal} {Commun. Math. Phys.}\ }\textbf
  {\bibinfo {volume} {121}},\ \bibinfo {pages} {351} (\bibinfo {year}
  {1989})}\BibitemShut {NoStop}%
\bibitem [{\citenamefont {Schoutens}\ and\ \citenamefont
  {Wen}(2016)}]{schoutens_simple-current_2016}%
  \BibitemOpen
  \bibfield  {author} {\bibinfo {author} {\bibfnamefont {K.}~\bibnamefont
  {Schoutens}}\ and\ \bibinfo {author} {\bibfnamefont {X.-G.}\ \bibnamefont
  {Wen}},\ }\bibfield  {title} {\bibinfo {title} {Simple-current algebra
  constructions of 2+1-dimensional topological orders},\ }\href
  {https://doi.org/10.1103/PhysRevB.93.045109} {\bibfield  {journal} {\bibinfo
  {journal} {Phys. Rev. B}\ }\textbf {\bibinfo {volume} {93}},\ \bibinfo
  {pages} {045109} (\bibinfo {year} {2016})}\BibitemShut {NoStop}%
\bibitem [{\citenamefont {Di~Francesco}\ \emph {et~al.}(1997)\citenamefont
  {Di~Francesco}, \citenamefont {Mathieu},\ and\ \citenamefont
  {S{\'e}n{\'e}chal}}]{DiFrancesco1997}%
  \BibitemOpen
  \bibfield  {author} {\bibinfo {author} {\bibfnamefont {P.}~\bibnamefont
  {Di~Francesco}}, \bibinfo {author} {\bibfnamefont {P.}~\bibnamefont
  {Mathieu}},\ and\ \bibinfo {author} {\bibfnamefont {D.}~\bibnamefont
  {S{\'e}n{\'e}chal}},\ }\href {https://doi.org/10.1007/978-1-4612-2256-9}
  {\emph {\bibinfo {title} {Conformal {Field} {Theory}}}}\ (\bibinfo
  {publisher} {Springer New York},\ \bibinfo {address} {New York, NY},\
  \bibinfo {year} {1997})\ \bibinfo {note} {series Title: Graduate Texts in
  Contemporary Physics}\BibitemShut {NoStop}%
\bibitem [{\citenamefont {Ginsparg}(1988)}]{Ginsparg1988}%
  \BibitemOpen
  \bibfield  {author} {\bibinfo {author} {\bibfnamefont {P.}~\bibnamefont
  {Ginsparg}},\ }\bibfield  {title} {\bibinfo {title} {Applied {Conformal}
  {Field} {Theory}},\ }\href {http://arxiv.org/abs/hep-th/9108028} {\bibfield
  {journal} {\bibinfo  {journal} {arXiv preprint hep-th/9108028}\ } (\bibinfo
  {year} {1988})},\ \bibinfo {note} {arXiv: hep-th/9108028}\BibitemShut
  {NoStop}%
\bibitem [{\citenamefont {Cardy}(2010)}]{Cardy2008}%
  \BibitemOpen
  \bibfield  {author} {\bibinfo {author} {\bibfnamefont {J.}~\bibnamefont
  {Cardy}},\ }\bibfield  {title} {\bibinfo {title} {Conformal {Field} {Theory}
  and {Statistical} {Mechanics}},\ }in\ \href {http://arxiv.org/abs/0807.3472}
  {\emph {\bibinfo {booktitle} {Exact {Methods} in {Low}-{Dimensional}
  {Statistical} {Physics} and {Quantum} {Computing}}}}\ (\bibinfo  {publisher}
  {Oxford University Press},\ \bibinfo {year} {2010})\ pp.\ \bibinfo {pages}
  {65--98}\BibitemShut {NoStop}%
\bibitem [{\citenamefont {Moore}\ and\ \citenamefont
  {Seiberg}(1988)}]{Moore1988}%
  \BibitemOpen
  \bibfield  {author} {\bibinfo {author} {\bibfnamefont {G.}~\bibnamefont
  {Moore}}\ and\ \bibinfo {author} {\bibfnamefont {N.}~\bibnamefont
  {Seiberg}},\ }\bibfield  {title} {\bibinfo {title} {Polynomial equations for
  rational conformal field theories},\ }\href
  {https://doi.org/10.1016/0370-2693(88)91796-0} {\bibfield  {journal}
  {\bibinfo  {journal} {Phys. Lett. B}\ }\textbf {\bibinfo {volume} {212}},\
  \bibinfo {pages} {451} (\bibinfo {year} {1988})}\BibitemShut {NoStop}%
\bibitem [{\citenamefont {Moore}\ and\ \citenamefont
  {Seiberg}(1989)}]{Moore1989}%
  \BibitemOpen
  \bibfield  {author} {\bibinfo {author} {\bibfnamefont {G.}~\bibnamefont
  {Moore}}\ and\ \bibinfo {author} {\bibfnamefont {N.}~\bibnamefont
  {Seiberg}},\ }\bibfield  {title} {\bibinfo {title} {Classical and quantum
  conformal field theory},\ }\href {https://doi.org/10.1007/BF01238857}
  {\bibfield  {journal} {\bibinfo  {journal} {Commun. Math. Phys.}\ }\textbf
  {\bibinfo {volume} {123}},\ \bibinfo {pages} {177} (\bibinfo {year}
  {1989})}\BibitemShut {NoStop}%
\bibitem [{\citenamefont {Huang}(2005)}]{huang2005vertex}%
  \BibitemOpen
  \bibfield  {author} {\bibinfo {author} {\bibfnamefont {Y.-Z.}\ \bibnamefont
  {Huang}},\ }\bibfield  {title} {\bibinfo {title} {Vertex operator algebras,
  the verlinde conjecture, and modular tensor categories},\ }\href@noop {}
  {\bibfield  {journal} {\bibinfo  {journal} {Proceedings of the National
  Academy of Sciences}\ }\textbf {\bibinfo {volume} {102}},\ \bibinfo {pages}
  {5352} (\bibinfo {year} {2005})}\BibitemShut {NoStop}%
\bibitem [{\citenamefont {Frenkel}\ \emph {et~al.}(1989)\citenamefont
  {Frenkel}, \citenamefont {Lepowsky},\ and\ \citenamefont
  {Meurman}}]{frenkel1989vertex}%
  \BibitemOpen
  \bibfield  {author} {\bibinfo {author} {\bibfnamefont {I.}~\bibnamefont
  {Frenkel}}, \bibinfo {author} {\bibfnamefont {J.}~\bibnamefont {Lepowsky}},\
  and\ \bibinfo {author} {\bibfnamefont {A.}~\bibnamefont {Meurman}},\
  }\href@noop {} {\emph {\bibinfo {title} {Vertex operator algebras and the
  Monster}}}\ (\bibinfo  {publisher} {Academic press},\ \bibinfo {year}
  {1989})\BibitemShut {NoStop}%
\bibitem [{\citenamefont {Kac}\ and\ \citenamefont
  {Wang}(1994)}]{kac1994vertex}%
  \BibitemOpen
  \bibfield  {author} {\bibinfo {author} {\bibfnamefont {V.}~\bibnamefont
  {Kac}}\ and\ \bibinfo {author} {\bibfnamefont {W.}~\bibnamefont {Wang}},\
  }\bibfield  {title} {\bibinfo {title} {Vertex operator superalgebras and
  their},\ }\href@noop {} {\bibfield  {journal} {\bibinfo  {journal}
  {Mathematical aspects of conformal and topological field theories and quantum
  groups}\ }\textbf {\bibinfo {volume} {175}},\ \bibinfo {pages} {161}
  (\bibinfo {year} {1994})}\BibitemShut {NoStop}%
\bibitem [{\citenamefont {Fern}\ \emph
  {et~al.}(2018{\natexlab{b}})\citenamefont {Fern}, \citenamefont {Bondesan},\
  and\ \citenamefont {Simon}}]{Fern2018a}%
  \BibitemOpen
  \bibfield  {author} {\bibinfo {author} {\bibfnamefont {R.}~\bibnamefont
  {Fern}}, \bibinfo {author} {\bibfnamefont {R.}~\bibnamefont {Bondesan}},\
  and\ \bibinfo {author} {\bibfnamefont {S.~H.}\ \bibnamefont {Simon}},\
  }\bibfield  {title} {\bibinfo {title} {Effective edge state dynamics in the
  fractional quantum {Hall} effect},\ }\href
  {https://doi.org/10.1103/PhysRevB.98.155321} {\bibfield  {journal} {\bibinfo
  {journal} {Phys. Rev. B}\ }\textbf {\bibinfo {volume} {98}},\ \bibinfo
  {pages} {1} (\bibinfo {year} {2018}{\natexlab{b}})}\BibitemShut {NoStop}%
\bibitem [{\citenamefont {Ino}(1997)}]{ino1997chiral}%
  \BibitemOpen
  \bibfield  {author} {\bibinfo {author} {\bibfnamefont {K.}~\bibnamefont
  {Ino}},\ }\bibfield  {title} {\bibinfo {title} {Chiral vertices, fusion rules
  and vacua of fractional quantum hall systems},\ }\href
  {https://arxiv.org/abs/cond-mat/9702039} {\bibfield  {journal} {\bibinfo
  {journal} {arXiv preprint cond-mat/9702039}\ } (\bibinfo {year}
  {1997})}\BibitemShut {NoStop}%
\bibitem [{\citenamefont {Berry}(1997)}]{berry_quantal_1997}%
  \BibitemOpen
  \bibfield  {author} {\bibinfo {author} {\bibfnamefont {M.~V.}\ \bibnamefont
  {Berry}},\ }\bibfield  {title} {\bibinfo {title} {Quantal phase factors
  accompanying adiabatic changes},\ }\href
  {https://doi.org/10.1098/rspa.1984.0023} {\bibfield  {journal} {\bibinfo
  {journal} {Proc. Math. Phys. Eng. Sci.}\ }\textbf {\bibinfo {volume} {392}},\
  \bibinfo {pages} {45} (\bibinfo {year} {1997})}\BibitemShut {NoStop}%
\bibitem [{\citenamefont {Wilczek}\ and\ \citenamefont
  {Zee}(1984)}]{wilczek_appearance_1984}%
  \BibitemOpen
  \bibfield  {author} {\bibinfo {author} {\bibfnamefont {F.}~\bibnamefont
  {Wilczek}}\ and\ \bibinfo {author} {\bibfnamefont {A.}~\bibnamefont {Zee}},\
  }\bibfield  {title} {\bibinfo {title} {Appearance of {Gauge} {Structure} in
  {Simple} {Dynamical} {Systems}},\ }\href
  {https://doi.org/10.1103/PhysRevLett.52.2111} {\bibfield  {journal} {\bibinfo
   {journal} {Phys. Rev. Lett.}\ }\textbf {\bibinfo {volume} {52}},\ \bibinfo
  {pages} {2111} (\bibinfo {year} {1984})}\BibitemShut {NoStop}%
\bibitem [{\citenamefont {Jain}(1990)}]{jain_theory_1990}%
  \BibitemOpen
  \bibfield  {author} {\bibinfo {author} {\bibfnamefont {J.~K.}\ \bibnamefont
  {Jain}},\ }\bibfield  {title} {\bibinfo {title} {Theory of the fractional
  quantum {Hall} effect},\ }\href
  {https://link.aps.org/doi/10.1103/PhysRevB.41.7653} {\bibfield  {journal}
  {\bibinfo  {journal} {Phys. Rev. B}\ }\textbf {\bibinfo {volume} {41}},\
  \bibinfo {pages} {7653} (\bibinfo {year} {1990})}\BibitemShut {NoStop}%
\bibitem [{Note2()}]{Note2}%
  \BibitemOpen
  \bibinfo {note} {To see this we first not that, $i\partial \varphi (w) =
  \DOTSI \ointop \ilimits@ _{w} \protect \frac {dz}{2\pi i} \protect \frac
  {\protect \mathcal {R} [ V(z) V^\dagger (w)]}{z-w} = \DOTSI \ointop \ilimits@
  _{|z|>|w|} \protect \frac {dz}{2\pi i} \protect \frac { V(z) V^\dagger
  (w)}{z-w} + \DOTSI \ointop \ilimits@ _{|z|<|w|} \protect \frac {dz}{2\pi i}
  \protect \frac { V^\dagger (w) V(z)}{z-w} $ One then expands the $1/(z-w)$
  factors to express this as a sum of $V_k V^\dagger (w)$ products. Finally,
  one can integrate over $w$ to obtain the $a_n$ modes.}\BibitemShut {Stop}%
\bibitem [{\citenamefont {von Delft}\ and\ \citenamefont
  {Schoeller}(1998)}]{VonDelft1998}%
  \BibitemOpen
  \bibfield  {author} {\bibinfo {author} {\bibfnamefont {J.}~\bibnamefont {von
  Delft}}\ and\ \bibinfo {author} {\bibfnamefont {H.}~\bibnamefont
  {Schoeller}},\ }\bibfield  {title} {\bibinfo {title} {Bosonization for
  beginners - refermionization for experts},\ }\href
  {https://doi.org/10.1002/(sici)1521-3889(199811)7:4<225::aid-andp225>3.0.co;2-l}
  {\bibfield  {journal} {\bibinfo  {journal} {Ann. Phys.}\ }\textbf {\bibinfo
  {volume} {510}},\ \bibinfo {pages} {225} (\bibinfo {year}
  {1998})}\BibitemShut {NoStop}%
\bibitem [{\citenamefont {Kass}\ \emph {et~al.}(1990)\citenamefont {Kass},
  \citenamefont {Moody}, \citenamefont {Patera},\ and\ \citenamefont
  {Slansky}}]{kass_affine_1990}%
  \BibitemOpen
  \bibfield  {author} {\bibinfo {author} {\bibfnamefont {S.}~\bibnamefont
  {Kass}}, \bibinfo {author} {\bibfnamefont {R.~V.}\ \bibnamefont {Moody}},
  \bibinfo {author} {\bibfnamefont {J.}~\bibnamefont {Patera}},\ and\ \bibinfo
  {author} {\bibfnamefont {R.}~\bibnamefont {Slansky}},\ }\href@noop {} {\emph
  {\bibinfo {title} {Affine {Lie} {Algebras}, {Weight} {Multiplicities}, and
  {Branching} {Rules}}}}\ (\bibinfo  {publisher} {University of California
  Press},\ \bibinfo {year} {1990})\BibitemShut {NoStop}%
\bibitem [{\citenamefont {Witten}(1984)}]{witten_non-abelian_1984}%
  \BibitemOpen
  \bibfield  {author} {\bibinfo {author} {\bibfnamefont {E.}~\bibnamefont
  {Witten}},\ }\bibfield  {title} {\bibinfo {title} {Non-abelian bosonization
  in two dimensions},\ }\href {https://doi.org/10.1007/BF01215276} {\bibfield
  {journal} {\bibinfo  {journal} {Commun. Math. Phys.}\ }\textbf {\bibinfo
  {volume} {92}},\ \bibinfo {pages} {455} (\bibinfo {year} {1984})}\BibitemShut
  {NoStop}%
\bibitem [{\citenamefont {Affleck}(1986)}]{affleck_exact_1986}%
  \BibitemOpen
  \bibfield  {author} {\bibinfo {author} {\bibfnamefont {I.}~\bibnamefont
  {Affleck}},\ }\bibfield  {title} {\bibinfo {title} {Exact critical exponents
  for quantum spin chains, non-linear $\sigma$-models at $\theta$=$\pi$ and the
  quantum hall effect},\ }\href
  {https://www.sciencedirect.com/science/article/pii/0550321386901677}
  {\bibfield  {journal} {\bibinfo  {journal} {Nuc. Phys. B}\ }\textbf {\bibinfo
  {volume} {265}},\ \bibinfo {pages} {409} (\bibinfo {year}
  {1986})}\BibitemShut {NoStop}%
\bibitem [{\citenamefont {Nakanishi}\ and\ \citenamefont
  {Tsuchiya}(1992)}]{nakanishi_level-rank_1992}%
  \BibitemOpen
  \bibfield  {author} {\bibinfo {author} {\bibfnamefont {T.}~\bibnamefont
  {Nakanishi}}\ and\ \bibinfo {author} {\bibfnamefont {A.}~\bibnamefont
  {Tsuchiya}},\ }\bibfield  {title} {\bibinfo {title} {Level-rank duality of
  {WZW} models in conformal field theory},\ }\href
  {http://link.springer.com/10.1007/BF02101097} {\bibfield  {journal} {\bibinfo
   {journal} {Commun. Math. Phys.}\ }\textbf {\bibinfo {volume} {144}},\
  \bibinfo {pages} {351} (\bibinfo {year} {1992})}\BibitemShut {NoStop}%
\bibitem [{\citenamefont {Naculich}\ and\ \citenamefont
  {Schnitzer}(1990{\natexlab{a}})}]{naculich_duality_1990-1}%
  \BibitemOpen
  \bibfield  {author} {\bibinfo {author} {\bibfnamefont {S.~G.}\ \bibnamefont
  {Naculich}}\ and\ \bibinfo {author} {\bibfnamefont {H.~J.}\ \bibnamefont
  {Schnitzer}},\ }\bibfield  {title} {\bibinfo {title} {Duality relations
  between {SU} ({N})k and {SU} (k){N} {WZW} models and their braid matrices},\
  }\href {https://www.sciencedirect.com/science/article/pii/037026939090061A}
  {\bibfield  {journal} {\bibinfo  {journal} {Phys. Lett. B}\ }\textbf
  {\bibinfo {volume} {244}},\ \bibinfo {pages} {235} (\bibinfo {year}
  {1990}{\natexlab{a}})}\BibitemShut {NoStop}%
\bibitem [{\citenamefont {Naculich}\ and\ \citenamefont
  {Schnitzer}(1990{\natexlab{b}})}]{naculich_duality_1990}%
  \BibitemOpen
  \bibfield  {author} {\bibinfo {author} {\bibfnamefont {S.~G.}\ \bibnamefont
  {Naculich}}\ and\ \bibinfo {author} {\bibfnamefont {H.~J.}\ \bibnamefont
  {Schnitzer}},\ }\bibfield  {title} {\bibinfo {title} {Duality between
  {SU}({N})k and {SU}(k){N} {WZW} models},\ }\href
  {https://www.sciencedirect.com/science/article/pii/055032139090380V}
  {\bibfield  {journal} {\bibinfo  {journal} {Nuc. Phys. B}\ }\textbf {\bibinfo
  {volume} {347}},\ \bibinfo {pages} {687} (\bibinfo {year}
  {1990}{\natexlab{b}})}\BibitemShut {NoStop}%
\bibitem [{\citenamefont {Carpi}\ \emph {et~al.}(2018)\citenamefont {Carpi},
  \citenamefont {Kawahigashi}, \citenamefont {Longo},\ and\ \citenamefont
  {Weiner}}]{carpi_vertex_2018}%
  \BibitemOpen
  \bibfield  {author} {\bibinfo {author} {\bibfnamefont {S.}~\bibnamefont
  {Carpi}}, \bibinfo {author} {\bibfnamefont {Y.}~\bibnamefont {Kawahigashi}},
  \bibinfo {author} {\bibfnamefont {R.}~\bibnamefont {Longo}},\ and\ \bibinfo
  {author} {\bibfnamefont {M.}~\bibnamefont {Weiner}},\ }\href
  {https://www.ams.org/memo/1213/} {\emph {\bibinfo {title} {From {Vertex}
  {Operator} {Algebras} to {Conformal} {Nets} and {Back}}}},\ \bibinfo {series}
  {Memoirs of the {American} {Mathematical} {Society}}, Vol.\ \bibinfo {volume}
  {254}\ (\bibinfo  {publisher} {American Mathematical Society},\ \bibinfo
  {year} {2018})\BibitemShut {NoStop}%
\bibitem [{\citenamefont {Zhu}(1996)}]{zhu_modular_1996}%
  \BibitemOpen
  \bibfield  {author} {\bibinfo {author} {\bibfnamefont {Y.}~\bibnamefont
  {Zhu}},\ }\bibfield  {title} {\bibinfo {title} {Modular invariance of
  characters of vertex operator algebras},\ }\href
  {https://doi.org/10.1090/S0894-0347-96-00182-8} {\bibfield  {journal}
  {\bibinfo  {journal} {Journal of the American Mathematical Society}\ }\textbf
  {\bibinfo {volume} {9}},\ \bibinfo {pages} {237} (\bibinfo {year}
  {1996})}\BibitemShut {NoStop}%
\bibitem [{\citenamefont {Fukusumi}\ and\ \citenamefont
  {Yang}(2023)}]{fukusumi2023fermionic}%
  \BibitemOpen
  \bibfield  {author} {\bibinfo {author} {\bibfnamefont {Y.}~\bibnamefont
  {Fukusumi}}\ and\ \bibinfo {author} {\bibfnamefont {B.}~\bibnamefont
  {Yang}},\ }\bibfield  {title} {\bibinfo {title} {Fermionic fractional quantum
  hall states: A modern approach to systems with bulk-edge correspondence},\
  }\href {https://journals.aps.org/prb/abstract/10.1103/PhysRevB.108.085123}
  {\bibfield  {journal} {\bibinfo  {journal} {Phys. Rev. B}\ }\textbf {\bibinfo
  {volume} {108}},\ \bibinfo {pages} {085123} (\bibinfo {year}
  {2023})}\BibitemShut {NoStop}%
\bibitem [{\citenamefont {Cappelli}\ \emph {et~al.}(1993)\citenamefont
  {Cappelli}, \citenamefont {Trugenberger},\ and\ \citenamefont
  {Zemba}}]{cappelli_large_1993}%
  \BibitemOpen
  \bibfield  {author} {\bibinfo {author} {\bibfnamefont {A.}~\bibnamefont
  {Cappelli}}, \bibinfo {author} {\bibfnamefont {C.~A.}\ \bibnamefont
  {Trugenberger}},\ and\ \bibinfo {author} {\bibfnamefont {G.~R.}\ \bibnamefont
  {Zemba}},\ }\bibfield  {title} {\bibinfo {title} {Large {N} limit in the
  quantum {Hall} effect},\ }\href
  {https://doi.org/10.1016/0370-2693(93)91144-C} {\bibfield  {journal}
  {\bibinfo  {journal} {Phys. Lett. B}\ }\textbf {\bibinfo {volume} {306}},\
  \bibinfo {pages} {100} (\bibinfo {year} {1993})}\BibitemShut {NoStop}%
\bibitem [{\citenamefont {Diehl}(1997)}]{diehl_theory_1997}%
  \BibitemOpen
  \bibfield  {author} {\bibinfo {author} {\bibfnamefont {H.~W.}\ \bibnamefont
  {Diehl}},\ }\bibfield  {title} {\bibinfo {title} {The {Theory} of {Boundary}
  {Critical} {Phenomena}},\ }\href {https://doi.org/10.1142/S0217979297001751}
  {\bibfield  {journal} {\bibinfo  {journal} {Int. J. Mod. Phys. B}\ }\textbf
  {\bibinfo {volume} {11}},\ \bibinfo {pages} {3503} (\bibinfo {year}
  {1997})}\BibitemShut {NoStop}%
\bibitem [{\citenamefont {Cardy}(2015)}]{Cardy2015}%
  \BibitemOpen
  \bibfield  {author} {\bibinfo {author} {\bibfnamefont {J.}~\bibnamefont
  {Cardy}},\ }\href {https://doi.org/10.1017/CBO9781316036440} {\emph {\bibinfo
  {title} {Scaling and renormalization in statistical physics}}}\ (\bibinfo
  {publisher} {Cambridge University Press},\ \bibinfo {year}
  {2015})\BibitemShut {NoStop}%
\bibitem [{\citenamefont {Cardy}(2006)}]{Cardy2004}%
  \BibitemOpen
  \bibfield  {author} {\bibinfo {author} {\bibfnamefont {J.}~\bibnamefont
  {Cardy}},\ }\bibinfo {title} {Boundary conformal field theory},\ in\ \href
  {https://doi.org/10.1016/b0-12-512666-2/00398-9} {\emph {\bibinfo {booktitle}
  {Encyclopedia of Mathematical Physics}}}\ (\bibinfo  {publisher} {Elsevier},\
  \bibinfo {year} {2006})\ p.\ \bibinfo {pages} {333–340}\BibitemShut
  {NoStop}%
\bibitem [{\citenamefont {Andrei}\ \emph {et~al.}(2020)\citenamefont {Andrei},
  \citenamefont {Bissi}, \citenamefont {Buican}, \citenamefont {Cardy},
  \citenamefont {Dorey}, \citenamefont {Drukker}, \citenamefont {Erdmenger},
  \citenamefont {Friedan}, \citenamefont {Fursaev}, \citenamefont {Konechny},
  \citenamefont {Kristjansen}, \citenamefont {Makabe}, \citenamefont
  {Nakayama}, \citenamefont {O’Bannon}, \citenamefont {Parini}, \citenamefont
  {Robinson}, \citenamefont {Ryu}, \citenamefont {Schmidt-Colinet},
  \citenamefont {Schomerus}, \citenamefont {Schweigert},\ and\ \citenamefont
  {Watts}}]{Andrei2020}%
  \BibitemOpen
  \bibfield  {author} {\bibinfo {author} {\bibfnamefont {N.}~\bibnamefont
  {Andrei}}, \bibinfo {author} {\bibfnamefont {A.}~\bibnamefont {Bissi}},
  \bibinfo {author} {\bibfnamefont {M.}~\bibnamefont {Buican}}, \bibinfo
  {author} {\bibfnamefont {J.}~\bibnamefont {Cardy}}, \bibinfo {author}
  {\bibfnamefont {P.}~\bibnamefont {Dorey}}, \bibinfo {author} {\bibfnamefont
  {N.}~\bibnamefont {Drukker}}, \bibinfo {author} {\bibfnamefont
  {J.}~\bibnamefont {Erdmenger}}, \bibinfo {author} {\bibfnamefont
  {D.}~\bibnamefont {Friedan}}, \bibinfo {author} {\bibfnamefont
  {D.}~\bibnamefont {Fursaev}}, \bibinfo {author} {\bibfnamefont
  {A.}~\bibnamefont {Konechny}}, \bibinfo {author} {\bibfnamefont
  {C.}~\bibnamefont {Kristjansen}}, \bibinfo {author} {\bibfnamefont
  {I.}~\bibnamefont {Makabe}}, \bibinfo {author} {\bibfnamefont
  {Y.}~\bibnamefont {Nakayama}}, \bibinfo {author} {\bibfnamefont
  {A.}~\bibnamefont {O’Bannon}}, \bibinfo {author} {\bibfnamefont
  {R.}~\bibnamefont {Parini}}, \bibinfo {author} {\bibfnamefont
  {B.}~\bibnamefont {Robinson}}, \bibinfo {author} {\bibfnamefont
  {S.}~\bibnamefont {Ryu}}, \bibinfo {author} {\bibfnamefont {C.}~\bibnamefont
  {Schmidt-Colinet}}, \bibinfo {author} {\bibfnamefont {V.}~\bibnamefont
  {Schomerus}}, \bibinfo {author} {\bibfnamefont {C.}~\bibnamefont
  {Schweigert}},\ and\ \bibinfo {author} {\bibfnamefont {G.~M.~T.}\
  \bibnamefont {Watts}},\ }\bibfield  {title} {\bibinfo {title} {Boundary and
  defect cft: open problems and applications},\ }\href
  {https://doi.org/10.1088/1751-8121/abb0fe} {\bibfield  {journal} {\bibinfo
  {journal} {Journal of Physics A: Mathematical and Theoretical}\ }\textbf
  {\bibinfo {volume} {53}},\ \bibinfo {pages} {453002} (\bibinfo {year}
  {2020})}\BibitemShut {NoStop}%
\bibitem [{\citenamefont {Ishibashi}(1988)}]{Ishibashi1988}%
  \BibitemOpen
  \bibfield  {author} {\bibinfo {author} {\bibfnamefont {N.}~\bibnamefont
  {Ishibashi}},\ }\bibfield  {title} {\bibinfo {title} {The {Boundary} and
  {Crosscap} {States} in {Conformal} {Field} {Theories}},\ }\href
  {https://doi.org/https://doi.org/10.1142/S0217732389000320} {\bibfield
  {journal} {\bibinfo  {journal} {Mod. Phys. Lett.}\ }\textbf {\bibinfo
  {volume} {04}},\ \bibinfo {pages} {251} (\bibinfo {year} {1988})}\BibitemShut
  {NoStop}%
\bibitem [{\citenamefont {Cardy}(1989)}]{Cardy1989}%
  \BibitemOpen
  \bibfield  {author} {\bibinfo {author} {\bibfnamefont {J.~L.}\ \bibnamefont
  {Cardy}},\ }\bibfield  {title} {\bibinfo {title} {Boundary conditions, fusion
  rules and the {Verlinde} formula},\ }\href
  {https://doi.org/10.1016/0550-3213(89)90521-X} {\bibfield  {journal}
  {\bibinfo  {journal} {Nuc. Phys. B}\ }\textbf {\bibinfo {volume} {324}},\
  \bibinfo {pages} {581} (\bibinfo {year} {1989})},\ \bibinfo {note}
  {publisher: North-Holland}\BibitemShut {NoStop}%
\bibitem [{\citenamefont {Behrend}\ \emph {et~al.}(2000)\citenamefont
  {Behrend}, \citenamefont {Pearce}, \citenamefont {Petkova},\ and\
  \citenamefont {Zuber}}]{behrend_boundary_2000}%
  \BibitemOpen
  \bibfield  {author} {\bibinfo {author} {\bibfnamefont {R.~E.}\ \bibnamefont
  {Behrend}}, \bibinfo {author} {\bibfnamefont {P.~A.}\ \bibnamefont {Pearce}},
  \bibinfo {author} {\bibfnamefont {V.~B.}\ \bibnamefont {Petkova}},\ and\
  \bibinfo {author} {\bibfnamefont {J.-B.}\ \bibnamefont {Zuber}},\ }\bibfield
  {title} {\bibinfo {title} {Boundary conditions in rational conformal field
  theories},\ }\href
  {https://www.sciencedirect.com/science/article/pii/S055032130000225X}
  {\bibfield  {journal} {\bibinfo  {journal} {Nuc. Phys. B}\ }\textbf {\bibinfo
  {volume} {579}},\ \bibinfo {pages} {707} (\bibinfo {year}
  {2000})}\BibitemShut {NoStop}%
\bibitem [{\citenamefont {Li}\ \emph {et~al.}(2022)\citenamefont {Li},
  \citenamefont {Hsieh}, \citenamefont {Yao},\ and\ \citenamefont
  {Oshikawa}}]{li_boundary_2022}%
  \BibitemOpen
  \bibfield  {author} {\bibinfo {author} {\bibfnamefont {L.}~\bibnamefont
  {Li}}, \bibinfo {author} {\bibfnamefont {C.~T.}\ \bibnamefont {Hsieh}},
  \bibinfo {author} {\bibfnamefont {Y.}~\bibnamefont {Yao}},\ and\ \bibinfo
  {author} {\bibfnamefont {M.}~\bibnamefont {Oshikawa}},\ }\bibfield  {title}
  {\bibinfo {title} {Boundary conditions and anomalies of conformal field
  theories in 1+1 dimensions},\ }\href {http://arxiv.org/abs/2205.11190}
  {\bibfield  {journal} {\bibinfo  {journal} {arXiv preprint arXiv:2205.11190}\
  } (\bibinfo {year} {2022})}\BibitemShut {NoStop}%
\bibitem [{Note3()}]{Note3}%
  \BibitemOpen
  \bibinfo {note} {As pointed out in Ref. \cite {Dubail2012} this can be seen
  directly by expressing the states as modes of the $\phi _l$ fields applied on
  the vacuum. For example, consider the case of just a chiral boson where the
  modes of the field have the boundary condition $a_n \mathinner {|{B}\rangle }
  = \protect \bar {a}_{-n}\mathinner {|{B}\rangle }$. Now compute $\mathinner
  {\langle {0}|}a_1a_2\protect \bar {a}_1\protect \bar {a}_2 \mathinner
  {|{B}\rangle } = \mathinner {\langle {0}|}a_1a_2\protect \bar {a}_1a_{-2}
  \mathinner {|{B}\rangle } = 2 \mathinner {\langle {0}|}a_1\protect \bar
  {a}_{1}\mathinner {|{B}\rangle } = 2 \mathinner {\langle
  {0}|}a_1a_{-1}\mathinner {|{B}\rangle } = 2\mathinner {\langle {0|B}\rangle }
  = 2 = \mathinner {\langle {0}|}a_1a_2a_{-1}a_{-2}\mathinner {|{0}\rangle }$.
  One can generalize this approach to see that in the case where the $\phi _l$
  have half integral conformal dimension, which implies that the modes of the
  $\phi _l$ and $\protect \bar {\phi }_l$ must anti-commute, one can generally
  obtain an overall minus sign that depends on the number of $\phi _l$ modes in
  either of the states $\mathinner {|{v}\rangle }$ or $\mathinner {|{w}\rangle
  }$.}\BibitemShut {Stop}%
\bibitem [{Note4()}]{Note4}%
  \BibitemOpen
  \bibinfo {note} {Strictly speaking, we can have boundary operators which
  belong to other representations of $\protect \mathcal {A}$, however, in the
  present case we are only interested in correlation functions in this bCFT
  that only involve fields of $\protect \mathcal {A}$ and $\protect \overline
  {\protect \mathcal {A}}$.}\BibitemShut {Stop}%
\bibitem [{\citenamefont {Qi}\ \emph {et~al.}(2012)\citenamefont {Qi},
  \citenamefont {Katsura},\ and\ \citenamefont {Ludwig}}]{Qi2012}%
  \BibitemOpen
  \bibfield  {author} {\bibinfo {author} {\bibfnamefont {X.~L.}\ \bibnamefont
  {Qi}}, \bibinfo {author} {\bibfnamefont {H.}~\bibnamefont {Katsura}},\ and\
  \bibinfo {author} {\bibfnamefont {A.~W.}\ \bibnamefont {Ludwig}},\ }\bibfield
   {title} {\bibinfo {title} {General relationship between the entanglement
  spectrum and the edge state spectrum of topological quantum states},\ }\href
  {https://doi.org/10.1103/PhysRevLett.108.196402} {\bibfield  {journal}
  {\bibinfo  {journal} {Phys. Rev. Lett.}\ }\textbf {\bibinfo {volume} {108}},\
  \bibinfo {pages} {1} (\bibinfo {year} {2012})}\BibitemShut {NoStop}%
\bibitem [{\citenamefont {Rodr{\'i}guez}\ \emph {et~al.}(2013)\citenamefont
  {Rodr{\'i}guez}, \citenamefont {Davenport}, \citenamefont {Simon},\ and\
  \citenamefont {Slingerland}}]{Rodriguez2013}%
  \BibitemOpen
  \bibfield  {author} {\bibinfo {author} {\bibfnamefont {I.~D.}\ \bibnamefont
  {Rodr{\'i}guez}}, \bibinfo {author} {\bibfnamefont {S.~C.}\ \bibnamefont
  {Davenport}}, \bibinfo {author} {\bibfnamefont {S.~H.}\ \bibnamefont
  {Simon}},\ and\ \bibinfo {author} {\bibfnamefont {J.~K.}\ \bibnamefont
  {Slingerland}},\ }\bibfield  {title} {\bibinfo {title} {Entanglement spectrum
  of composite fermion states in real space},\ }\href
  {https://doi.org/10.1103/PhysRevB.88.155307} {\bibfield  {journal} {\bibinfo
  {journal} {Phys. Rev. B}\ }\textbf {\bibinfo {volume} {88}},\ \bibinfo
  {pages} {3} (\bibinfo {year} {2013})}\BibitemShut {NoStop}%
\bibitem [{\citenamefont {Davenport}\ \emph {et~al.}(2015)\citenamefont
  {Davenport}, \citenamefont {Rodr{\'i}guez}, \citenamefont {Slingerland},\
  and\ \citenamefont {Simon}}]{Davenport2015}%
  \BibitemOpen
  \bibfield  {author} {\bibinfo {author} {\bibfnamefont {S.~C.}\ \bibnamefont
  {Davenport}}, \bibinfo {author} {\bibfnamefont {I.~D.}\ \bibnamefont
  {Rodr{\'i}guez}}, \bibinfo {author} {\bibfnamefont {J.~K.}\ \bibnamefont
  {Slingerland}},\ and\ \bibinfo {author} {\bibfnamefont {S.~H.}\ \bibnamefont
  {Simon}},\ }\bibfield  {title} {\bibinfo {title} {Composite fermion model for
  entanglement spectrum of fractional quantum {Hall} states},\ }\href
  {https://doi.org/10.1103/PhysRevB.92.115155} {\bibfield  {journal} {\bibinfo
  {journal} {Phys. Rev. B}\ }\textbf {\bibinfo {volume} {92}},\ \bibinfo
  {pages} {1} (\bibinfo {year} {2015})}\BibitemShut {NoStop}%
\bibitem [{\citenamefont {Amico}\ \emph {et~al.}(2008)\citenamefont {Amico},
  \citenamefont {Fazio}, \citenamefont {Osterloh},\ and\ \citenamefont
  {Vedral}}]{Amico}%
  \BibitemOpen
  \bibfield  {author} {\bibinfo {author} {\bibfnamefont {L.}~\bibnamefont
  {Amico}}, \bibinfo {author} {\bibfnamefont {R.}~\bibnamefont {Fazio}},
  \bibinfo {author} {\bibfnamefont {A.}~\bibnamefont {Osterloh}},\ and\
  \bibinfo {author} {\bibfnamefont {V.}~\bibnamefont {Vedral}},\ }\bibfield
  {title} {\bibinfo {title} {Entanglement in many-body systems},\ }\href
  {https://doi.org/10.1103/RevModPhys.80.517} {\bibfield  {journal} {\bibinfo
  {journal} {Rev. Mod. Phys.}\ }\textbf {\bibinfo {volume} {80}},\ \bibinfo
  {pages} {517} (\bibinfo {year} {2008})}\BibitemShut {NoStop}%
\bibitem [{\citenamefont {Henderson}\ \emph {et~al.}(2021)\citenamefont
  {Henderson}, \citenamefont {Sreejith},\ and\ \citenamefont
  {Simon}}]{Henderson2021}%
  \BibitemOpen
  \bibfield  {author} {\bibinfo {author} {\bibfnamefont {G.~J.}\ \bibnamefont
  {Henderson}}, \bibinfo {author} {\bibfnamefont {G.~J.}\ \bibnamefont
  {Sreejith}},\ and\ \bibinfo {author} {\bibfnamefont {S.~H.}\ \bibnamefont
  {Simon}},\ }\bibfield  {title} {\bibinfo {title} {Entanglement action for the
  real-space entanglement spectra of chiral {Abelian} quantum {Hall} wave
  functions},\ }\href {https://doi.org/10.1103/PhysRevB.104.195434} {\bibfield
  {journal} {\bibinfo  {journal} {Phys. Rev. B}\ }\textbf {\bibinfo {volume}
  {104}},\ \bibinfo {pages} {195434} (\bibinfo {year} {2021})}\BibitemShut
  {NoStop}%
\bibitem [{\citenamefont {Gepner}\ and\ \citenamefont
  {Qiu}(1987)}]{gepner_modular_1987}%
  \BibitemOpen
  \bibfield  {author} {\bibinfo {author} {\bibfnamefont {D.}~\bibnamefont
  {Gepner}}\ and\ \bibinfo {author} {\bibfnamefont {Z.}~\bibnamefont {Qiu}},\
  }\bibfield  {title} {\bibinfo {title} {Modular invariant partition functions
  for parafermionic field theories},\ }\href
  {https://www.sciencedirect.com/science/article/pii/0550321387903488}
  {\bibfield  {journal} {\bibinfo  {journal} {Nuc. Phys. B}\ }\textbf {\bibinfo
  {volume} {285}},\ \bibinfo {pages} {423} (\bibinfo {year}
  {1987})}\BibitemShut {NoStop}%
\bibitem [{Note5()}]{Note5}%
  \BibitemOpen
  \bibinfo {note} {This is assuming that the $\alpha (N)$ of Eq. \ref
  {Eq:generalInProdAct} have no $N$ dependence.}\BibitemShut {Stop}%
\bibitem [{\citenamefont {Schellekens}\ and\ \citenamefont
  {Yankielowicz}(1989)}]{schellekens_extended_1989}%
  \BibitemOpen
  \bibfield  {author} {\bibinfo {author} {\bibfnamefont {A.~N.}\ \bibnamefont
  {Schellekens}}\ and\ \bibinfo {author} {\bibfnamefont {S.}~\bibnamefont
  {Yankielowicz}},\ }\bibfield  {title} {\bibinfo {title} {Extended chiral
  algebras and modular invariant partition functions},\ }\href
  {https://www.sciencedirect.com/science/article/pii/0550321389903106}
  {\bibfield  {journal} {\bibinfo  {journal} {Nuc. Phys. B}\ }\textbf {\bibinfo
  {volume} {327}},\ \bibinfo {pages} {673} (\bibinfo {year}
  {1989})}\BibitemShut {NoStop}%
\bibitem [{\citenamefont {Slingerland}\ and\ \citenamefont
  {Bais}(2001)}]{slingerland_quantum_2001}%
  \BibitemOpen
  \bibfield  {author} {\bibinfo {author} {\bibfnamefont {J.~K.}\ \bibnamefont
  {Slingerland}}\ and\ \bibinfo {author} {\bibfnamefont {F.~A.}\ \bibnamefont
  {Bais}},\ }\bibfield  {title} {\bibinfo {title} {Quantum groups and
  non-{Abelian} braiding in quantum {Hall} systems},\ }\href
  {https://doi.org/10.1016/S0550-3213(01)00308-X} {\bibfield  {journal}
  {\bibinfo  {journal} {Nuc. Phys. B}\ }\textbf {\bibinfo {volume} {612}},\
  \bibinfo {pages} {229} (\bibinfo {year} {2001})}\BibitemShut {NoStop}%
\bibitem [{\citenamefont {Gepner}\ and\ \citenamefont
  {Witten}(1986)}]{gepner_string_1986}%
  \BibitemOpen
  \bibfield  {author} {\bibinfo {author} {\bibfnamefont {D.}~\bibnamefont
  {Gepner}}\ and\ \bibinfo {author} {\bibfnamefont {E.}~\bibnamefont
  {Witten}},\ }\bibfield  {title} {\bibinfo {title} {String theory on group
  manifolds},\ }\href {https://doi.org/10.1016/0550-3213(86)90051-9} {\bibfield
   {journal} {\bibinfo  {journal} {Nuc. Phys. B}\ }\textbf {\bibinfo {volume}
  {278}},\ \bibinfo {pages} {493} (\bibinfo {year} {1986})}\BibitemShut
  {NoStop}%
\bibitem [{\citenamefont {Zamolodchikov}\ and\ \citenamefont
  {Fateev}(1986)}]{zamolodchikov_operator_1986}%
  \BibitemOpen
  \bibfield  {author} {\bibinfo {author} {\bibfnamefont {A.~B.}\ \bibnamefont
  {Zamolodchikov}}\ and\ \bibinfo {author} {\bibfnamefont {V.~A.}\ \bibnamefont
  {Fateev}},\ }\bibfield  {title} {\bibinfo {title} {Operator algebra and
  correlation functions in the two-dimensional {SU}(2) x {SU}(2) chiral
  {Wess}-{Zumino} model},\ }\href {https://www.osti.gov/biblio/5107985}
  {\bibfield  {journal} {\bibinfo  {journal} {Sov. J. Nucl. Phys. (Engl.
  Transl.); (United States)}\ }\textbf {\bibinfo {volume} {43:4}} (\bibinfo
  {year} {1986})}\BibitemShut {NoStop}%
\bibitem [{\citenamefont {Bonderson}(2007)}]{bonderson_non-abelian_2007}%
  \BibitemOpen
  \bibfield  {author} {\bibinfo {author} {\bibfnamefont {P.~H.}\ \bibnamefont
  {Bonderson}},\ }\emph {\bibinfo {title} {Non-{Abelian} {Anyons} and
  {Interferometry}}},\ \href {https://thesis.library.caltech.edu/2447/} {Ph.D.
  thesis},\ \bibinfo  {school} {California Institute of Technology} (\bibinfo
  {year} {2007})\BibitemShut {NoStop}%
\bibitem [{\citenamefont {Zamolodchikov}\ and\ \citenamefont
  {Fateev}(1985)}]{zamolodchikov_nonlocal_1985}%
  \BibitemOpen
  \bibfield  {author} {\bibinfo {author} {\bibfnamefont {A.~B.}\ \bibnamefont
  {Zamolodchikov}}\ and\ \bibinfo {author} {\bibfnamefont {V.~A.}\ \bibnamefont
  {Fateev}},\ }\bibfield  {title} {\bibinfo {title} {Nonlocal (parafermion)
  currents in two-dimensional conformal quantum field theory and self-dual
  critical points in {Z}/sub {N}/-symmetric statistical systems},\ }\href
  {https://www.osti.gov/biblio/5929972} {\bibfield  {journal} {\bibinfo
  {journal} {Sov. Phys. - JETP (Engl. Transl.); (United States)}\ }\textbf
  {\bibinfo {volume} {62:2}} (\bibinfo {year} {1985})}\BibitemShut {NoStop}%
\bibitem [{\citenamefont {Barkeshli}\ and\ \citenamefont
  {Wen}(2010)}]{barkeshli2010effective}%
  \BibitemOpen
  \bibfield  {author} {\bibinfo {author} {\bibfnamefont {M.}~\bibnamefont
  {Barkeshli}}\ and\ \bibinfo {author} {\bibfnamefont {X.-G.}\ \bibnamefont
  {Wen}},\ }\bibfield  {title} {\bibinfo {title} {Effective field theory and
  projective construction for z k parafermion fractional quantum hall states},\
  }\href {https://journals.aps.org/prb/abstract/10.1103/PhysRevB.81.155302}
  {\bibfield  {journal} {\bibinfo  {journal} {Phys. Rev. B}\ }\textbf {\bibinfo
  {volume} {81}},\ \bibinfo {pages} {155302} (\bibinfo {year}
  {2010})}\BibitemShut {NoStop}%
\bibitem [{\citenamefont {Barkeshli}\ and\ \citenamefont
  {Wen}(2011)}]{barkeshli2011bilayer}%
  \BibitemOpen
  \bibfield  {author} {\bibinfo {author} {\bibfnamefont {M.}~\bibnamefont
  {Barkeshli}}\ and\ \bibinfo {author} {\bibfnamefont {X.-G.}\ \bibnamefont
  {Wen}},\ }\bibfield  {title} {\bibinfo {title} {Bilayer quantum hall phase
  transitions and the orbifold non-abelian fractional quantum hall states},\
  }\href {https://journals.aps.org/prb/abstract/10.1103/PhysRevB.84.115121}
  {\bibfield  {journal} {\bibinfo  {journal} {Physical Review B}\ }\textbf
  {\bibinfo {volume} {84}},\ \bibinfo {pages} {115121} (\bibinfo {year}
  {2011})}\BibitemShut {NoStop}%
\bibitem [{\citenamefont {Fuji}\ and\ \citenamefont
  {Lecheminant}(2017)}]{fuji2017non}%
  \BibitemOpen
  \bibfield  {author} {\bibinfo {author} {\bibfnamefont {Y.}~\bibnamefont
  {Fuji}}\ and\ \bibinfo {author} {\bibfnamefont {P.}~\bibnamefont
  {Lecheminant}},\ }\bibfield  {title} {\bibinfo {title} {Non-abelian s u (n-
  1)-singlet fractional quantum hall states from coupled wires},\ }\href
  {https://journals.aps.org/prb/abstract/10.1103/PhysRevB.95.125130} {\bibfield
   {journal} {\bibinfo  {journal} {Phys. Rev. B}\ }\textbf {\bibinfo {volume}
  {95}},\ \bibinfo {pages} {125130} (\bibinfo {year} {2017})}\BibitemShut
  {NoStop}%
\bibitem [{\citenamefont {Gepner}(1987)}]{gepner1987new}%
  \BibitemOpen
  \bibfield  {author} {\bibinfo {author} {\bibfnamefont {D.}~\bibnamefont
  {Gepner}},\ }\bibfield  {title} {\bibinfo {title} {New conformal field
  theories associated with lie algebras and their partition functions},\ }\href
  {https://www.sciencedirect.com/science/article/pii/0550321387901763}
  {\bibfield  {journal} {\bibinfo  {journal} {Nuc. Phys. B}\ }\textbf {\bibinfo
  {volume} {290}},\ \bibinfo {pages} {10} (\bibinfo {year} {1987})}\BibitemShut
  {NoStop}%
\bibitem [{\citenamefont {Fern}\ \emph {et~al.}(2017)\citenamefont {Fern},
  \citenamefont {Kombe},\ and\ \citenamefont {Simon}}]{fern_how_2017}%
  \BibitemOpen
  \bibfield  {author} {\bibinfo {author} {\bibfnamefont {R.}~\bibnamefont
  {Fern}}, \bibinfo {author} {\bibfnamefont {J.}~\bibnamefont {Kombe}},\ and\
  \bibinfo {author} {\bibfnamefont {S.}~\bibnamefont {Simon}},\ }\bibfield
  {title} {\bibinfo {title} {How {SU}(2)\$\_4\$ {Anyons} are {Z}\$\_3\$
  {Parafermions}},\ }\href {https://doi.org/10.21468/SciPostPhys.3.6.037}
  {\bibfield  {journal} {\bibinfo  {journal} {SciPost Phys.}\ }\textbf
  {\bibinfo {volume} {3}},\ \bibinfo {pages} {037} (\bibinfo {year}
  {2017})}\BibitemShut {NoStop}%
\bibitem [{\citenamefont {Fendley}(2014)}]{fendley_free_2014}%
  \BibitemOpen
  \bibfield  {author} {\bibinfo {author} {\bibfnamefont {P.}~\bibnamefont
  {Fendley}},\ }\bibfield  {title} {\bibinfo {title} {Free parafermions},\
  }\href {https://doi.org/10.1088/1751-8113/47/7/075001} {\bibfield  {journal}
  {\bibinfo  {journal} {J. Phys. A Math. Theor.}\ }\textbf {\bibinfo {volume}
  {47}},\ \bibinfo {pages} {075001} (\bibinfo {year} {2014})}\BibitemShut
  {NoStop}%
\bibitem [{\citenamefont {Suorsa}\ \emph
  {et~al.}(2011{\natexlab{b}})\citenamefont {Suorsa}, \citenamefont {Viefers},\
  and\ \citenamefont {Hansson}}]{Suorsa2011}%
  \BibitemOpen
  \bibfield  {author} {\bibinfo {author} {\bibfnamefont {J.}~\bibnamefont
  {Suorsa}}, \bibinfo {author} {\bibfnamefont {S.}~\bibnamefont {Viefers}},\
  and\ \bibinfo {author} {\bibfnamefont {T.~H.}\ \bibnamefont {Hansson}},\
  }\bibfield  {title} {\bibinfo {title} {Quasihole condensates in quantum
  {Hall} liquids},\ }\href {https://doi.org/10.1103/PhysRevB.83.235130}
  {\bibfield  {journal} {\bibinfo  {journal} {Phys. Rev. B}\ }\textbf {\bibinfo
  {volume} {83}},\ \bibinfo {pages} {235130} (\bibinfo {year}
  {2011}{\natexlab{b}})}\BibitemShut {NoStop}%
\bibitem [{\citenamefont {Wen}(1999)}]{wen_projective_1999}%
  \BibitemOpen
  \bibfield  {author} {\bibinfo {author} {\bibfnamefont {X.~G.}\ \bibnamefont
  {Wen}},\ }\bibfield  {title} {\bibinfo {title} {Projective construction of
  non-{Abelian} quantum {Hall} liquids},\ }\href
  {https://link.aps.org/doi/10.1103/PhysRevB.60.8827} {\bibfield  {journal}
  {\bibinfo  {journal} {Phys. Rev. B}\ }\textbf {\bibinfo {volume} {60}},\
  \bibinfo {pages} {8827} (\bibinfo {year} {1999})}\BibitemShut {NoStop}%
\bibitem [{\citenamefont {Bandyopadhyay}\ \emph {et~al.}(2020)\citenamefont
  {Bandyopadhyay}, \citenamefont {Ortiz}, \citenamefont {Nussinov},\ and\
  \citenamefont {Seidel}}]{Bandyopadhyay2020}%
  \BibitemOpen
  \bibfield  {author} {\bibinfo {author} {\bibfnamefont {S.}~\bibnamefont
  {Bandyopadhyay}}, \bibinfo {author} {\bibfnamefont {G.}~\bibnamefont
  {Ortiz}}, \bibinfo {author} {\bibfnamefont {Z.}~\bibnamefont {Nussinov}},\
  and\ \bibinfo {author} {\bibfnamefont {A.}~\bibnamefont {Seidel}},\
  }\bibfield  {title} {\bibinfo {title} {Local {Two}-{Body} {Parent}
  {Hamiltonians} for the {Entire} {Jain} {Sequence}},\ }\href
  {https://doi.org/10.1103/PhysRevLett.124.196803} {\bibfield  {journal}
  {\bibinfo  {journal} {Phys. Rev. Lett.}\ }\textbf {\bibinfo {volume} {124}},\
  \bibinfo {pages} {196803} (\bibinfo {year} {2020})}\BibitemShut {NoStop}%
\bibitem [{Note6()}]{Note6}%
  \BibitemOpen
  \bibinfo {note} {We have $V^\dagger _1(z) = V^\dagger _{22}(z)V^\dagger
  _{12}(z)$, $V^\dagger _0(z) = [V^\dagger _{21}(z)V^\dagger _{12}(z) +
  V^\dagger _{22}(z)V^\dagger _{11}(z)]/\protect \sqrt {2}$ and $V^\dagger
  _{-1}(z) = V^\dagger _{21}(z)V^\dagger _{11}(z)$.}\BibitemShut {Stop}%
\bibitem [{\citenamefont {Jain}(2020)}]{jain_thirty_2020}%
  \BibitemOpen
  \bibfield  {author} {\bibinfo {author} {\bibfnamefont {J.~K.}\ \bibnamefont
  {Jain}},\ }\href {https://doi.org/10.1142/11751} {\emph {\bibinfo {title}
  {Thirty {Years} of {Composite} {Fermions} and {Beyond}}}}\ (\bibinfo {year}
  {2020})\ \bibinfo {note} {arXiv:2011.13488 [cond-mat]}\BibitemShut {NoStop}%
\end{thebibliography}%
